Федеральное государственное автономное
образовательное учреждение высшего образования
«Московский физико-технический институт
(национальный исследовательский университет)»
(МФТИ, Физтех)

На правах рукописи

Мещеринов Вячеслав Вячеславович

**ИССЛЕДОВАНИЕ ХИМИЧЕСКОГО И ИЗОТОПНОГО СОСТАВА ГАЗОВЫХ СМЕСЕЙ
НА ОСНОВЕ ЛАЗЕРНОЙ СПЕКТРОСКОПИИ ВЫСОКОГО РАЗРЕШЕНИЯ**

1.3.19 Лазерная физика

Диссертация на соискание ученой степени
кандидата физико-математических наук

Научный руководитель:
доктор физико-математических наук
Родин Александр Вячеславович

Москва, 2024 г.

# Оглавление

# ВВЕДЕНИЕ

Спектроскопия высокого разрешения в колебательно-вращательной области спектра длительное время остается одним из наиболее чувствительных и точных методов детектирования и диагностики состояния вещества в газообразной фазе. Спектральные методы анализа являются основным методом современной наблюдательной астрономии, широко применяются в исследовании живых систем, в геофизике, диагностике климата, экологическом мониторинге, а также имеют важное прикладное значение в различных технологических процессах, сфере безопасности и иных видах деятельности.

Одним из фундаментальных ограничений спектроскопического метода в оптической области спектра является спектральное разрешение применяемой аппаратуры (фурье- и эшелле-спектрометров), которое, как правило, не позволяет разрешать контуры индивидуальных спектральных линий молекулярного поглощения. Применение лазеров с распределенной обратной связью позволяет преодолеть этот недостаток за счет прецизионного управления частотой зондирующего излучения, которое характеризуется высокой степенью монохроматичности. Однако применение методов лазерной спектроскопии для исследования и диагностики природных сред и антропогенных объектов в рамках *in situ* или дистанционных методик сталкивается с необходимостью развития новых подходов.

## Актуальность темы диссертации

Спектр поглощения вещества представляет собой его уникальную характеристику, обусловленную энергетической структурой. Используя соответствующие методики, можно идентифицировать газовые компоненты в смеси по их спектрам поглощения. Методы лазерной спектроскопии позволяют анализировать газовые смеси путем стимуляции переходов между колебательно-вращательными уровнями энергетического спектра выбранного газа. Это позволяет точно идентифицировать отдельные газовые компоненты и определить их концентрации или изотопные отношения. Особенно эффективно такие методы работают с молекулярными газами неорганической природы и простыми органическими соединениями в газовой фазе.

Лазерная спектроскопия обладает высоким быстродействием и позволяет определять концентрацию веществ за доли секунды, что является значительным преимуществом по сравнению с хроматографией. Аналитические устройства, разработанные на основе лазерной спектроскопии, могут найти широкое применение благодаря малым массогабаритным характеристикам, низкому энергопотреблению и высокой устойчивости к внешним воздействиям. Это выгодно отличает их от масс-спектрометров или фурье-спектрометров, которые более подходят для лабораторного анализа. Диапазон применения спектроскопических методов включает экологический мониторинг, исследование атмосфер и подвергнутого пиролизу грунта планет и иных тел Солнечной системы, анализ солнечной активности, медицину, геофизику, контроль качества воздуха в замкнутых пространствах и многое другое. Таким образом, развитие методов прецизионного измерения химического и изотопного состава газовых смесей на основе лазерной спектроскопии представляет на сегодняшний день значительный интерес как в фундаментальных так и прикладных аспектах.

Одним из предложенных методов прецизионного измерения химического и изотопного состава газовых смесей является описанное развитие метода диодно-лазерной спектроскопии, адаптированного к разработке компактного спектрометра для анализа лунного реголита *in situ*. Эта задача представляется фундаментально важной, исследование Луны как геологического объекта, включая изучение ее строения и химического состава, поможет реконструировать раннюю историю Земли, так как первые 500 млн лет эволюции нашей планеты невозможно исследовать напрямую, но считается, что свидетельства этого периода сохранились на Луне, чья ранняя история тесно связана с историей Земли. Одной из ключевых задач является изучение летучих веществ в лунных поверхностных породах и неглубоких слоях, которые хранят историю развития лунных недр, ударных событий и взаимодействия с солнечным ветром ввиду отсутствия у Луны плотной атмосферы.

Также предложено развитие метода модуляционной спектроскопии в комбинации с квадратурным приемом сигнала, реализованное в газоанализаторе для дистанционного мониторинга метана в атмосферном воздухе. Поскольку с 1983 по 2023 год содержание метана в атмосфере увеличилось примерно на 20%, при том, что метан значительно эффективнее $CO_2$ в качестве парникового газа – его кумулятивный потенциал глобального потепления превышает аналогичный показатель для $CO_2$ в 28 раз за 100 лет и в 84 раза за 20 лет, необходимость в проведении мониторинга вблизи естественных и антропогенных источников метана представляется очевидной.

Источниками метана в атмосферном воздухе, помимо природного газа, могут быть биогенные и эндогенные (геологические) процессы. Например, болотистая местность является важным естественным источником метана. Еще одним фактором неопределенности является таяние многолетней мерзлоты в Арктическом регионе. Вблизи поверхности арктической многолетней мерзлоты содержится большое количество органических соединений углерода и остатков растительной и животной органики, которые не разлагаются из-за низких температур.

Глубинные слои многолетней мерзлоты содержат минеральные грунты. При таянии многолетней мерзлоты происходит высвобождение метана и углекислого газа в атмосферу, что усугубляет проблему изменения климата.

Однако в настоящее время 50-65% глобальных выбросов метана связаны с антропогенной деятельностью. За последние два десятилетия основной причиной роста атмосферных выбросов метана стало увеличение антропогенных выбросов, как от сельского хозяйства и отходов в Южной и Юго-Восточной Азии, Южной Америке и Африке, так и от ископаемого топлива в Китае, Российской Федерации и США. Таким образом, существенное сокращение выбросов метана может быть достигнуто за счет устранения утечек в трубопроводах и промышленных установках в районах добычи нефти и газа, для чего необходимо внедрение соответствующего оборудования, примером которого может служить описанный в данной работе дистанционный газоанализатор метана.

## Цели и задачи диссертационной работы

Целью данной работы является исследование состава газовых смесей и определение изотопных отношений отдельных газовых составляющих на основе предложенных методов оперативного прецизионного анализа, не требующих использования дорогостоящего лабораторного оборудования. Предлагается для получения данных о составе анализируемой газовой смеси применение спектроскопических методов, позволяющих добиться высокой чувствительности к выбранным газовым составляющим. Рассматривается дистанционное зондирование атмосферы и изучение газовой пробы *in situ*.

Для изучения газовых проб в аналитическом объеме менее 2 мл разработан и испытан многоканальный диодно-лазерный спектрометр ДЛС-Л, входящий в состав газового хроматографа ГХ-Л полярной посадочной станции «Луна-27» отечественной миссии «Луна-Ресурс». Прибор позволяет определять содержание $CO_2$ и $H_2O$ и их изотопные отношения с субпроцентной точностью.

Для дистанционного мониторинга концентрации выбранного газа в составе атмосферного воздуха создан компактный дистанционный газоанализатор ГИМЛИ на основе модуляционной абсорбционной лазерной спектроскопии, подходящий для мониторинга содержания метана в атмосферном воздухе на промышленных территориях и вблизи других возможных антропогенных источников метана – вблизи мусорных полигонов, животноводческих ферм или сетей газоснабжения, – а также в зонах естественных источников метана – болотистой местности, вблизи водоемов, в областях таяния многолетней мерзлоты.

**Для достижения поставленных целей были решены следующие задачи:**

1. Предварительное определение рабочих спектральных диапазонов для анализа содержания и изотопных отношений водяного пара и углекислого газа в режиме *in situ*, а также для дистанционного мониторинга содержания метана;

2. Исследование выходных пучков лазерного излучения с применением матричного фотодетектора и их прецизионная юстировка с целью подавления интерференции в принимаемом сигнале многоканального диодно-лазерного спектрометра ДЛС-Л. Разработка вариативной методики обработки аналитических данных спектрометра ДЛС-Л для определения содержания выбранного газа в смеси и его изотопных отношений, позволяющей значительно подавлять вклад паразитных составляющих сигнала. Проведение испытаний компактного сенсора ДЛС-Л, показавших возможность определения изотопных отношений с субпроцентной точностью;

3. Разработка и апробация методики модуляционной лазерной спектроскопии поглощения с квадратурным детектированием принимаемого сигнала на подвижной платформе в лабораторных и полевых условиях с последующим исследованием свойств данной методики – отсутствия зависимости при дистанционных измерениях с регистрацией рассеянного сигнала от типа рассеивающей поверхности, высокой точности определения содержания метана на уровне долей естественного фонового содержания в атмосфере благодаря алгоритму стабилизации частоты зондирующего излучения, возможности независимого вычисления дистанции до отражающей поверхности – и проведением калибровки разработанного на основе указанной методики дистанционного газоанализатора ГИМЛИ с демонстрацией зависимости чувствительности устройства от дистанции до рассеивающей поверхности.

## Научная новизна

Научная новизна работы состоит в следующем:

1. Впервые разработан многоканальный лазерный спектрометр для исследования содержания и изотопного состава продуктов пиролиза лунного реголита *in situ* в приполярной области Луны на основе абсорбционной лазерной спектроскопии, который позволит получить первые данные прямого исследования полярного лунного грунта без необходимости в доставке образцов на Землю для лабораторного анализа, что позволит избежать связанных с этим возможного загрязнения проб примесями земного происхождения и проанализировать наличие загрязнения земной изотопией образцов лунного грунта, исследованного ранее в лабораторных условиях, по измеренным обилию и изотопии летучих веществ в лунном реголите;

2. Разработана и апробирована методика обработки данных сверхкомпактного лазерного спектрометра с аналитическим объемом менее 2 мл, позволяющая значительно подавлять вклад паразитных составляющих сигнала и получать изотопные отношения исследуемых летучих соединений с субпроцентной точностью, что подтверждено результатами кросс-калибровки;

3. Разработан и апробирован на подвижной платформе метод модуляционной лазерной спектроскопии с квадратурным приемом сигнала и режимом динамической стабилизации частоты сканирующего лазерного излучения, изучены характерные особенности данной методики;

4. На основе модуляционной лазерной спектроскопии с квадратурным приемом сигнала и режимом динамической стабилизации частоты сканирующего лазерного излучения впервые разработан компактный лидарный газоанализатор для дистанционного зондирования атмосферных примесей, пригодный к установке на малые беспилотные летательные аппараты, превосходящий известные аналоги по совокупности характеристик.

## Практическая значимость

Практическая ценность работы заключается в следующем:

1. Разработан и апробирован метод обработки данных сверхкомпактного лазерного спектрометра, позволяющая значительно снижать вклад паразитных составляющих сигнала, неизбежных для приборов с подобным форм-фактором, и определять в аналитическом объеме менее 2 мл содержание заданных газовых составляющих и их изотопные отношения с субпроцентной точностью;

2. Разработанный многоканальный диодно-лазерный спектрометр на основе абсорбционной лазерной спектроскопии ДЛС-Л позволит получить первые данные прямого исследования содержания углекислого газа и воды и их изотопного состава в продуктах пиролиза лунного реголита. Интерпретация результатов позволит судить об основных возможных источниках летучих веществ на Луне и их запасенном содержании в реголите;

3. Продемонстрирована эффективность метода модуляционной лазерной спектроскопии в сочетании с квадратурным приемом сигнала и возможность реализации данного метода на базе компактного легкого прибора с малым электропотреблением, пригодного к установке на беспилотный летательный аппарат или иную подвижную платформу, движущуюся со скоростью 10-20 м/с, благодаря высокой частоте цикла обработки данных и цифровому усилению принимаемого сигнала;

4. Показана возможность стабилизации длины волны излучения, генерируемого лазерным источником прибора, с точностью до $10^{-4}$ см$^{-1}$, что значительно увеличивает чувствительность прибора в сравнении с существующими аналогами;

5. Показана возможность независимого определения оптического пути при помощи спектроскопического лидара с непрерывной модуляцией длины волны.

## Методология и методы исследования

Для получения экспериментальных результатов было разработано экспериментальное оборудование, реализующее различные методы лазерной спектроскопии. Для определения свойств зондируемой среды, в частности, содержания метана, углекислого газа и водяного пара, экспериментальные данные анализировались с помощью детальных численных моделей молекулярного поглощения.

Основой лазерной ИК-спектроскопии является принцип прецизионного управления частотой перестраиваемого лазера в диапазоне, обеспечивающем покрытие характерных спектральных особенностей исследуемого вещества. При этом спектральное разрешение измерений определяется шириной полосы излучения лазера, а также точностью стабилизации и перестройки его частоты.

В случае классической лазерной спектроскопии поглощения лазерное излучение пропускается через однопроходную аналитическую кювету с пробой газа и измеряется зависимость коэффициента пропускания анализируемого объема от частоты, что дает возможность исследовать спектральные свойства пробы. Высокая точность оптической юстировки и правильно подобранный режим работы лазерного спектрометра при стабилизации лазерного излучения по спектральным линиям поглощения исследуемых газов в реперном объеме позволяет получить высокую точность определения содержания выбранных газовых составляющих и их изотопных отношений. Такой подход был реализован при создании многоканального диодно-лазерного спектрометра ДЛС-Л, входящего в состав газового хроматографа ГХ-Л полярной посадочной станции «Луна-27» отечественной миссии «Луна-Ресурс».

В приложениях, связанных с экологическим мониторингом и промышленной безопасностью, физический доступ к анализируемому объёму газовой смеси зачастую бывает невозможен в силу удалённости объекта, необходимости обследования значительной территории либо агрессивного характера анализируемой среды. Это делает особенно актуальными методы дистанционного газоанализа, как пассивные, основанные на спектральном анализе естественного излучения объекта, так и активные, предполагающие подсветку объекта в заданном диапазоне и анализ рассеянного излучения. Дистанционное измерение концентрации атмосферных газовых составляющих возможно посредством непрерывного измерения глубины линии поглощения этого газа. В случае установки газоанализатора на движущуюся платформу (например, беспилотный летательный аппарат) с типичной скоростью 15-20 м/с и требуемого пространственного разрешения зондирования не хуже единиц метров необходима методика, допускающая частоту обработки принимаемого сигнала на уровне 10-100 кГц. При этом чувствительность такой методики должна обеспечивать определение превышения естественного фона выбранной газовой составляющей атмосферного воздуха на десятки процентов.

Подобным требованиям соответствует модуляционная лазерная абсорбционная спектроскопия в сочетании с квадратурным детектированием рассеянного от подстилающей поверхности излучения. Использование отношения первой и второй гармонической составляющих принимаемого сигнала для определения концентрации газовых составляющих известны в литературе, однако применение прецизионной динамической стабилизации длины волны лазерного излучения по третьей гармонической составляющей модуляции сигнала, что эквивалентно обратной связи по положению реперной спектральной линии, в автономных газоанализаторах до сих пор не применялось. Такой подход позволяет значительно повысить чувствительность прибора, сохраняя малые габариты, массу и энергопотребление.

## Положения, выносимые на защиту

1. Диодно-лазерная спектроскопия в сочетании с предложенным алгоритмом подавления паразитных составляющих аналитического сигнала, основанным на методике ортогональных полиномов, позволяет определять содержание газовой составляющей в смеси, а также изотопные отношения для водяного пара с отклонением изотопных сигнатур от поверочных значений на 0.92‰ для $\delta_{VSMOW}^{17}O$, 6.90‰ для $\delta_{VSMOW}^{18}O$ и 12.34‰ для $\delta_{VSMOW}D$ в спектральном окне 2639 нм и для углекислого газа с отклонением 0.91‰ для $\delta_{VPDB}^{13}C$ в спектральном окне 2786 нм в случае применения сверхкомпактного *in situ* спектрометра с аналитическим объемом 1.34 мл без динамической стабилизации длины волны лазерного излучения, что позволит исследовать изотопные отношения летучих соединений лунного реголита;

2. Предложенный алгоритм на основе последовательного применения методики ортогональных полиномов к измеренному аналитическому спектру и к невязке измеренного и синтетического спектров позволяет уменьшить не только низкочастотный вклад паразитных составляющих аналитического сигнала с периодом $\sim 1.7$ см$^{-1}$, соответствующий биениям на оптическом пути $\sim 0.3$ см между выходным окном диодного лазера и плоской поверхностью коллимирующей излучение линзы, на $\sim 5 \times 10^{-3}$ см$^{-1}$, но и среднечастотный с периодом $\sim 0.2$ см$^{-1}$, соответствующий биениям на оптическом пути $\sim 2.5$ см, на $\sim 1.4 \times 10^{-5}$ см$^{-1}$ при уровне поглощения выбранной спектральной линии HDO $\sim 1.6 \times 10^{-4}$ см$^{-1}$;

3. Модуляционная лазерная спектроскопия с квадратурным приемом сигнала, выраженном в цифровом усилении принимаемого сигнала, позволяют исключить проблему базовой линии, присущую классической диодно-лазерной спектроскопии, а при применении алгоритма динамической стабилизации длины волны генерируемого лазерным источником излучения с точностью до $10^{-4}$ см$^{-1}$ достигают чувствительности дистанционного измерения интегрального содержания метана в столбе атмосферы высотой 50 м 15 ppm·м или 13.6 % естественного фонового содержания и приводят к линейной зависимости чувствительности от дистанции до рассеивающей излучение поверхности.

**Степень достоверности и апробация результатов**

Достоверность полученных результатов подтверждается их внутренней непротиворечивостью и непротиворечивостью публикациям других авторов, а также сопоставлением экспериментальных данных с теоретическими расчетами. Основные положения и результаты работы докладывались и обсуждались на следующих международных и всероссийских конференциях:

1. XVI Конференция молодых ученых «Фундаментальные и прикладные космические исследования» (г. Москва, 2019);

2. Двадцать пятая Всероссийская научная конференция студентов-физиков и молодых ученых (г. Ростов-на-Дону, 2019);

3. XIX Symposium on High Resolution Molecular Spectroscopy HighRus-2019, (г. Нижний Новгород, 2019) – 2 доклада;

4. European Geosciences Union General Assembly (online, 2020);

5. The Eleventh Moscow Solar System Symposium 11M-S3 (г. Москва, 2020);

6. Земля и космос: Всероссийская научная конференция с международным участием к столетию академика РАН К. Я. Кондратьева (г. Санкт-Петербург, 2020);

7. 19th International Conference on Laser Optics (г. Санкт-Петербург, 2020);

8. Научно-практическая конференция с международным участием и элементами школы молодых ученых "Перспективы развития металлургии и машиностроения с использованием завершенных фундаментальных исследований и НИОКР" (г. Екатеринбург, 2020);

9. XXVII Международный Симпозиум «Оптика атмосферы и океана. Физика атмосферы» (г. Москва, 2021) – 2 доклада;

10. The Twelfth Moscow Solar System Symposium 12M-S3 (г. Москва, 2021);

11. Конференция молодых ученых «ТЕХНОГЕН-2021» (г. Екатеринбург, 2021);

12. 20th International Conference on Laser Optics (г. Санкт-Петербург, 2022) – 2 доклада;

13. The Thirteenth Moscow Solar System Symposium 13M-S3 (г. Москва, 2022);

14. XX Symposium on High Resolution Molecular Spectroscopy HighRus-2023 (г. Иркутск, 2023) – 2 доклада;

15. The Fourteenth Moscow Solar System Symposium 14M-S3 (г. Москва, 2023).

## Публикации

Основное содержание диссертации отражено в 1 патенте РФ, в 3 рецензируемых научных публикациях (индексируются SCOPUS и WoS), в 2 сборниках трудов конференций (индексируются SCOPUS и WoS) и 19 тезисах конференций, 7 из которых включены в РИНЦ. Список публикаций приведен в конце диссертации перед списком литературы.

### Внедрение результатов исследования

На основе метода диодно-лазерной абсорбционной спектроскопии был разработан образец многоканального диодно-лазерного спектрометра ДЛС-Л, который должен быть отправлен на полярной посадочной станции «Луна-27» к южному полюсу Луны в 2028 году для анализа содержания летучих веществ в лунном реголите и их изотопного состава в рамках отечественной космической миссии «Луна-Ресурс». Данные датчика ДЛС-Л помогут в дальнейшем понимании физики и химии лунного тела, поскольку это первые данные прямого исследования полярного лунного грунта.

Разработанный на основе сочетания модуляционной лазерной спектроскопии с квадратурным приемом сигнала дистанционный газоанализатор метана лидарного типа ГИМЛИ применяется для мониторинга содержания метана в атмосферном воздухе в рамках деятельности Научно-технического центра мониторинга окружающей среды и экологии МФТИ.

## Личный вклад автора

Основные результаты диссертации получены лично автором, либо при его непосредственном участии.

## Объем и структура диссертации

Диссертация состоит из введения, четырех глав и заключения. Материал изложен на 257 страницах, содержит 155 рисунков и 11 таблиц. Список литературы содержит 300 источников.

Первая глава посвящена анализу известных из литературы спектроскопических методов определения составляющих газовых смесей, а также оптических систем, применяемых в лазерной спектроскопии, и их проектированию.

Вторая глава посвящена разработке спектрометра ДЛС-Л. Обоснована необходимость получения результатов прямого изучения обилия и изотопного состава летучих соединений лунного реголита, минуя стадии доставки и хранения для его исследования в лабораторных условиях. Подробно описаны детали юстировки и настройки разработанного устройства. Обоснован выбор спектрального диапазона. Представлена методика обработки аналитических данных сверхкомпактного лазерного спектрометра, позволяющая значительно снижать вклад паразитных составляющих полезного сигнала, неизбежных для приборов подобного форм-фактора, и определять содержание углекислого газа и водяного пара и их изотопные отношения. Представлен анализ результатов, полученных в результате лабораторных функциональных испытаний прибора.

Третья глава посвящена разработке и апробации методики модуляционной лазерной спектроскопии с квадратурным приемом сигнала и режимом динамической стабилизации частоты сканирующего лазерного излучения на лабораторном макете. Обоснована значимость задачи экологического мониторинга содержания метана вблизи источников естественной эмиссии и антропогенных выбросов. Продемонстрирована работоспособность предложенной методики и потенциальная возможность ее применения для разработки полевого устройства по дистанционному мониторингу содержания метана в атмосферном воздухе.

Четвертая глава посвящена разработке инфракрасного дистанционного газоанализатора лидарного типа для мониторинга содержания метана в атмосферном воздухе. Представлены существующие аналоги разработанного устройства. Описаны оптическая схема, электроника и корпус прибора, управляющее программное обеспечение. Продемонстрированы результаты проведения полевых испытаний прототипа прибора, его калибровки, а также обнаруженные особенности используемых фотонных компонетов, предложены методы борьбы с солнечными засветками, представляющими заметную проблему для эксплуатации прототипа газоанализатора. Описана разработка новой версии устройства с использованием большей апертуры принимающей рассеянное излучение оптики и большей мощностью генерируемого зондирующего лазерного излучения, ожидаемая чувствительность которого вырастет более чем на порядок величины.

В приложении представлено описание патента РФ на изобретение способа и устройства для автономного дистанционного определения концентрации атмосферных газовых составляющих.

## Благодарности

за разработку аналоговой электроники, без которой половина описанных далее приборов не состоялась бы, за билет в лазерную спектроскопию, возможность прочувствовать многие эффекты «на пальцах», направляющие подталкивания и сотни мудрых историй. Также хочется поблагодарить моего научного руководителя Александра Родина за полученные в цикле авторских лекций знания об устройстве атмосфер, за возможность кропотливо разрабатывать научное оборудование и представлять получаемые результаты миру, совместные мозговые штурмы и навыки самостоятельной работы.

Отдельно хочется поблагодарить свою семью за безусловные любовь и поддержку, без которых не было бы сил на всю проделанную работу и ощущения ее осмысленности.

# ГЛАВА 1. ЛАЗЕРНАЯ СПЕКТРОСКОПИЯ ВЫСОКОГО РАЗРЕШЕНИЯ

Спектр поглощения является уникальной характеристикой вещества, связанной с его энергетической структурой, по которой при использовании соответствующей методики можно идентифицировать газовую составляющую в составе смеси. Методы лазерной спектроскопии, ограниченные доступным обилием источников излучения – лазеров, работающих преимущественно в ближнем и среднем ИК-диапазонах, позволяют анализировать газовые смеси, стимулируя переходы между колебательно-вращательными уровнями энергетического спектра выбранного газа, тем самым идентифицируя отдельные газовые составляющие и с высокой точностью определяя их концентрации или изотопные отношения. Такому анализу хорошо поддаются молекулярные газы неорганической природы и простые органические соединения в газовой фазе.

Лазерная спектроскопия обладает высоким быстродействием и позволяет вычислять концентрацию веществ за доли секунды в отличие от хроматографии. Разработанные на основе лазерной спектроскопии аналитические устройства могут иметь широкое применение в силу малых массогабаритных характеристик и невысокого энергопотребления, а также высокой устойчивости к внешним воздействиям в отличии от масс-спектрометров или фурье-спектрометров, подходящих скорее для лабораторного анализа.

Диапазон применения спектроскопических методов включает в себя экологический мониторинг, изучение атмосфер и подвергнутого пиролизу грунта планет Солнечной системы, анализ солнечной активности, медицину, геофизику, контроль качества воздуха в замкнутых пространствах и многое другое.

В данной главе будет приведено краткое рассмотрение некоторых методик лазерной спектроскопии с преимущественной опорой на опыт их применения автором – оптико-акустической спектроскопии, классической диодно-лазерной спектроскопии, модуляционной лазерной спектроскопии, спектроскопии полного внутрирезонаторного выхода, спектроскопии с затуханием излучения в резонаторе, гетеродинной спектроскопии; а также ряда оптических систем, наиболее часто применяющихся для работы в рамках спектроскопических методов – однопроходных систем, многоходовых систем Уайта и Чернина, системы Эрриотта, а также зеркально-кольцевой системы.

## 1.1. Спектроскопические методики анализа составляющих газовых смесей

Методы лазерной спектроскопии можно разделить на два типа по положению объекта изучения – методы дистанционного газоанализа и методы анализа газовых проб *in situ*. К методам дистанционного газоанализа можно в меньшей мере отнести классическую методику диодно-лазерной спектроскопии, в большей мере – методику модуляционной спектроскопии и гетеродинную спектроскопию.

К методам анализа газовых проб *in situ* можно отнести более широкий спектр применений классической диодно-лазерной спектроскопии, в некоторых случаях методику модуляционной спектроскопии, а также методику спектроскопии полного внутрирезонаторного выхода и спектроскопии с затуханием излучения в резонаторе.

С методологической точки зрения, методы лазерной спектроскопии можно разделить на три группы. В первую группу входят методы, в которых сведения о свойствах образца получают за счет изменений параметров излучения, проходящего через образец, таких как интенсивность, время затухания, поляризация или фаза. К этой категории относятся классическая диодно-лазерная спектроскопия, модуляционная спектроскопия, внутрирезонаторная спектроскопия, спектроскопия с затуханием в резонаторе и гетеродинная спектроскопия.

Методы второй группы используют информацию о свойствах образца, содержащуюся в переизлученном свете, как это характерно для комбинационного рассеяния и флуоресценции. И наконец в третью группу входят калориметрические методы, основанные на прямом измерении поглощенной мощности в образце через изменения его собственных параметров.

Далее рассмотрим подробнее методику оптико-акустической спектроскопии, классическую методику диодно-лазерной спектроскопии, модуляционную спектроскопию, методику спектроскопии с затуханием излучения в резонаторе и спектроскопии полного внутрирезонаторного выхода, а также гетеродинную спектроскопию.

### 1.1.1. Оптико-акустическая спектроскопия

Особенность калориметрических методов заключается в том, что информация о свойствах среды извлекается посредством прямого измерения поглощенной мощности оптического излучения по изменениям физических и термодинамических характеристик среды. В зависимости от измеряемой физической характеристики среды – давления, нагрева или изменения электрических свойств – среди калориметрических методов выделяют оптико-акустический, оптико-термический и оптико-гальванический подходы.

Оптико-акустическая (ОА) спектроскопия представляет собой метод анализа веществ, основанный на регистрации акустических волн, возникающих в результате поглощения оптического излучения веществом. Этот метод базируется на эффекте, впервые описанном в 1881 году, и используется для анализа газов благодаря его высокой чувствительности [1].

Формирование полезного сигнала в ОА-спектроскопии проходит три основные стадии. Сперва происходит изменение заселенности колебательно-вращательных уровней молекул – при прохождении оптического излучения через газ, молекулы возбуждаются, то есть происходит переход на более высокие вращательные энергетические уровни на фоне установления больцмановского распределения по вращательным и колебательным уровням из-за столкновений между молекулами.

Затем происходит преобразование энергии возбужденных молекул в тепловую через возвращение молекул в исходное состояние по окончании импульса. Этот процесс происходит путем перераспределения поглощенной энергии в пределах данного колебательного состояния через вращательную релаксацию, а затем через более медленную колебательную релаксацию. В результате часть энергии переходит в тепло, что вызывает изменение давления в среде и порождает акустические колебания.

После чего колебания давления, возникающие в результате поглощения газовой средой энергии оптического излучения, регистрируются высокочувствительным микрофоном. Микрофон преобразует акустический сигнал в электрический, интенсивность которого пропорциональна количеству поглощенной энергии, что позволяет определить концентрацию исследуемого вещества [2].

На рисунке 1.1 представлена форма импульса излучения $I(t)$ и последующая форма импульса давления $p(t)$ в газовой среде в случае, когда импульс излучения значительно короче процессов релаксации в газе, а интервал между импульсами существенно превышает их длительность.

Увеличение давления, которое происходит в отсутствии акустического резонанса, можно выразить как

$$p(t) = \frac{2\beta E K(\nu) N l \left(1 - exp\left[-t(\tau_{col}^{-1} + \tau_{rad}^{-1})\right]\right)}{3V(1 + \tau_{col}/\tau_{rad})}, \qquad (1.1)$$

где $N$ – плотность газа; $V$, $l$ – объем и длина ячейки; $K(\nu)$ – коэффициент поглощения; $E$ – энергия оптического импульса, $\tau_{col}$ – постоянная времени столкновительной релаксации, $\tau_{rad}$ – постоянная времени радиационной релаксации. Общая энергия газа в оптико-акустическом детекторе возрастает из-за столкновительной релаксации, часть которой преобразуется в поступательную энергию молекул – *β*. При наличии в исследуемом газе нескольких типов молекул, поглощающих излучение, общее давление в детекторе можно рассматривать как суперпозицию их индивидуальных вкладов. В случае наличия акустических резонансов в ОА-ячейке, сигнал давления *p*(*t*) должен быть свернут с аппаратной функцией устройства.

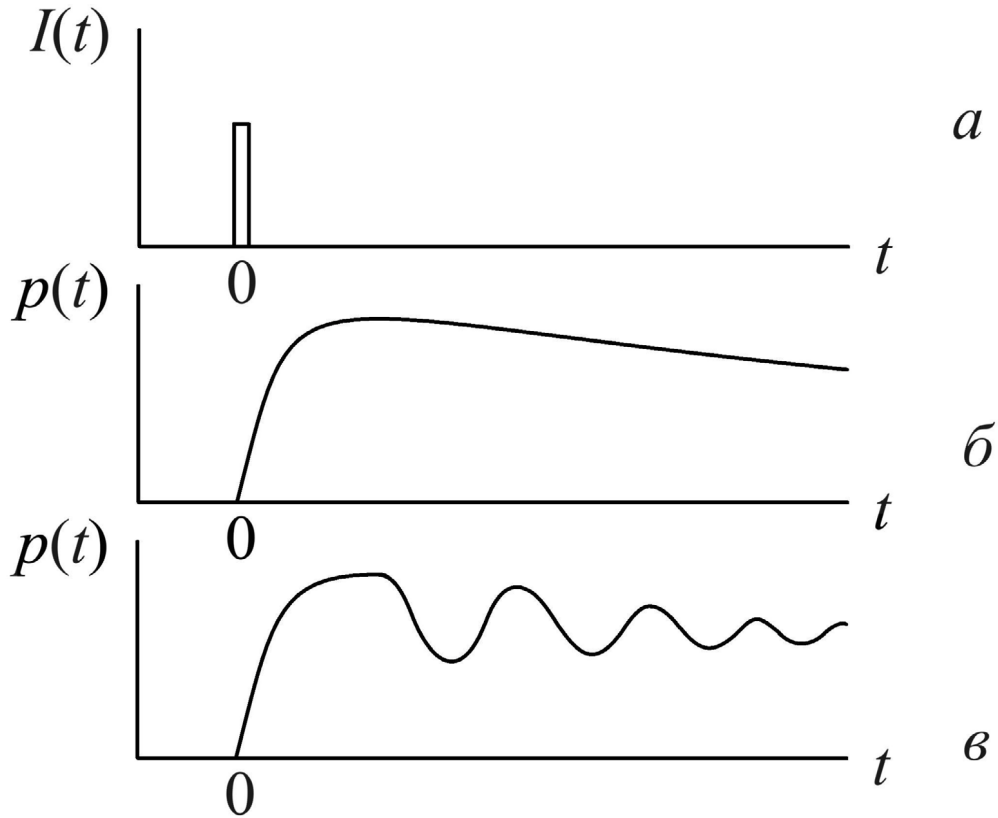

Рисунок 1.1 – Импульсы излучения **(а)** и давления в ОА-детекторе в отсутствии акустических резонансов **(б)** и при их наличии **(в)**.

Как следует из выражения (1.1), ОА-сигнал пропорционален мощности или энергии оптического излучения и коэффициенту поглощения $K(\nu)$, что позволяет напрямую измерять концентрацию анализируемого газа. Это отличает ОА-метод от спектрофотометрического, для которого сигнал пропорционален $\exp[-K(\nu)]$, как будет показано в следующей главе.

Чувствительность ОА-спектрометров такова, что при использовании импульсного излучения с энергией 1 Дж или непрерывного излучения мощностью 1 Вт предельно малые значения регистрируемого коэффициента поглощения могут достигать значений порядка $10^{-9}$ см$^{-1}$.

Типичная схема ОА-детектора, представленная на рисунке 1.2, включает следующие компоненты [2-5]:

- источник модулированного оптического излучения – оптическое излучение модулируется для создания периодических изменений интенсивности света, что необходимо для генерации акустических сигналов;

- ОА-ячейка – анализируемый газ вводится в ячейку длиной полтора-два десятка сантиметров через специальные вводы;

- датчики давления – высокочувствительные датчики, такие как конденсаторные или электретные микрофоны, используются для регистрации импульсов давления в ОА-ячейке, преобразуя изменения давления в электрические сигналы, которые затем обрабатываются и анализируются.

Для повышения чувствительности измерений оптическое излучение должно модулироваться так, чтобы поглощение света прерывалось на время, достаточное для полного затухания тепловых эффектов в газе. С этой целью применяется два подхода – модуляции интенсивности и частоты излучения.

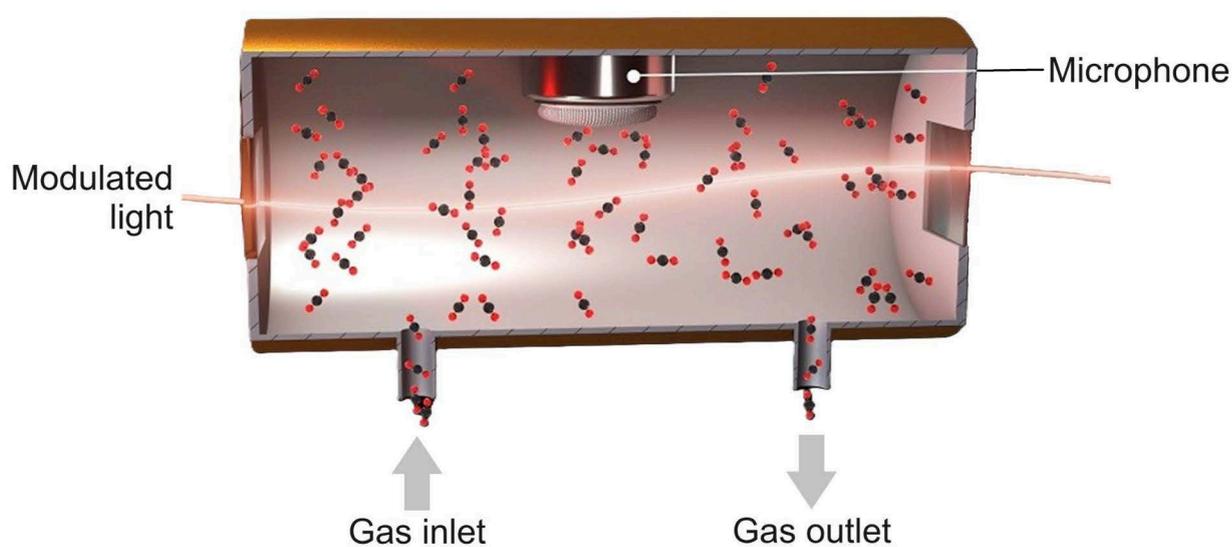

Рисунок 1.2 – Схема оптико-акустического детектора.

Таким образом, оптико-акустическая спектроскопия является мощным и высокочувствительным инструментом для анализа газов, обладающим широким спектром применений благодаря своей способности детектировать низкие концентрации веществ и обеспечивать точные количественные измерения.

### 1.1.2. Диодно-лазерная спектроскопия поглощения

Спектроскопия поглощения с применением перестраиваемых лазеров в качестве источников излучения является распространенным методом оптического газоанализа, основанным на зависимости частоты генерации лазера от тока инжекции и температуры лазерного кристалла. [6]. Возможность непрерывного изменения частоты излучения лазера позволяет различать контуры отдельных вращательных линий в колебательно-вращательном спектре поглощения молекулярных газов. Это позволяет помимо определения состава и концентрации отдельных составляющих в исследуемой газовой смеси вычислять температуру, давление, скорость потока и другие характеристики газообразной среды в отсутствии необходимости в калибровке [7,8].

Характерные спектральные линии, соответствующие колебательно-вращательным переходам ряда молекулярных газов, представляющих интерес для изучения в качестве маркеров различных естественных фотохимических или антропогенных процессов, находятся в инфракрасном диапазоне спектра. На рисунке 1.3 представлены интенсивности спектральных линий перехода между двумя колебательно-вращательными состояниями для углекислого газа и ряда малых газовых составляющих атмосферного воздуха в ближнем и среднем инфракрасных диапазонах спектра при температуре 296 К и давлении 1 атм.

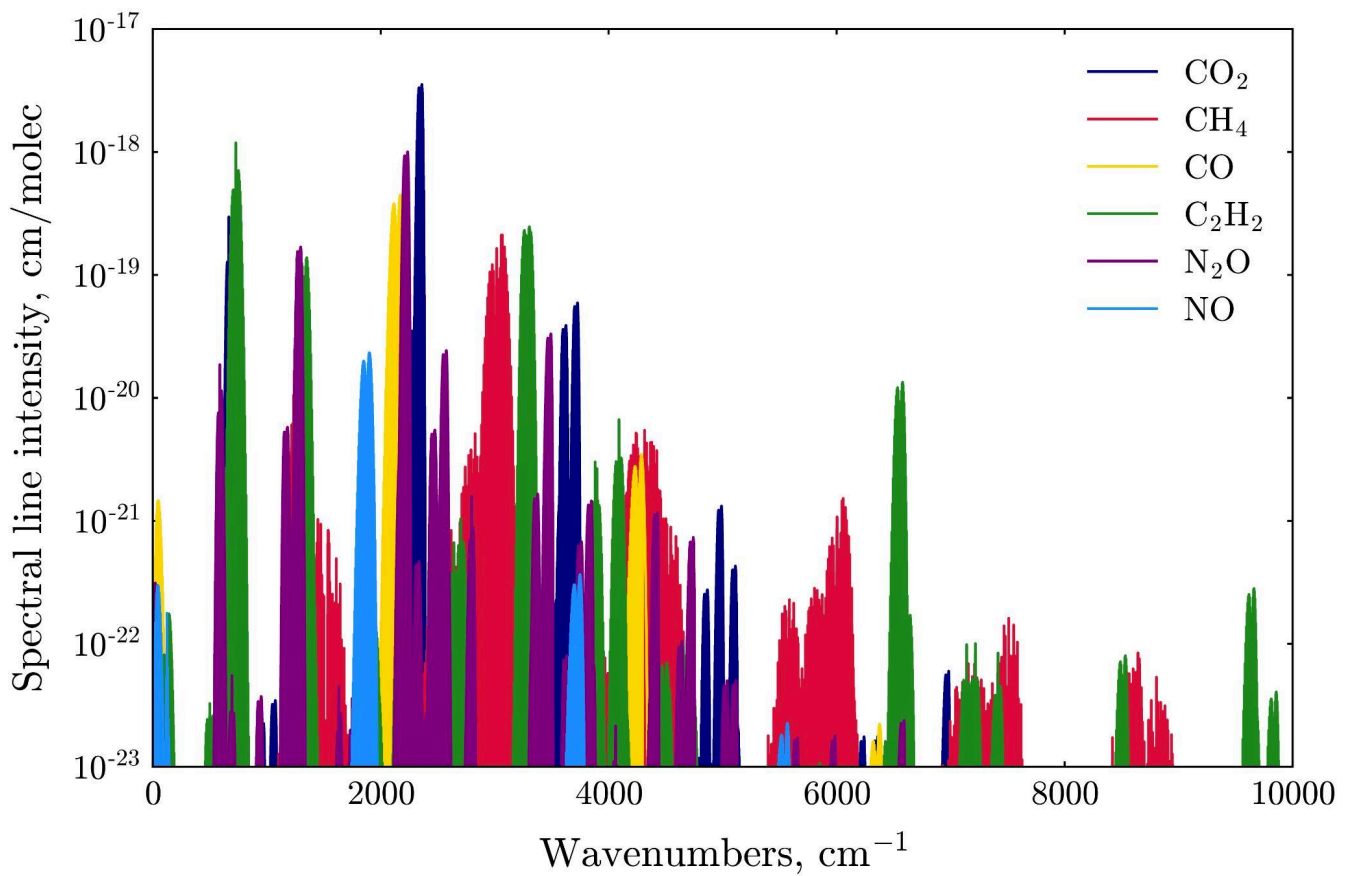

Рисунок 1.3 – Спектр поглощения ряда молекулярных газов в ИК-диапазоне, соответствующего колебательно-вращательным переходам.

Спектральное разрешение абсорбционной диодно-лазерной спектроскопии (ДЛС) во многом определяется точностью стабилизации и перестройки частоты лазерного источника излучения. Поскольку лазеры с резонатором Фабри-Перо имеют тенденцию к скачкообразным переходам между модами [9], для спектроскопических задач преимущественно используются полупроводниковые лазеры с распределенной обратной связью (РОС-лазеры), которые обеспечивают узкую линию генерации < 1 МГц. Перестройка длины волны излучения в стандартных РОС-лазерах осуществляется путем изменения температуры активной среды лазера при помощи элемента Пельтье с постоянной времени до нескольких секунд или регулированием тока инжекции при характерной частоте модуляции, достигающей единиц МГц [10]. Регулирование температуры кристалла позволяет добиться перестройки частоты излучения лазера вследствие изменения геометрии резонатора активной среды в диапазоне десятков см$^{-1}$, необходимой для вывода частоты в необходимый спектральный диапазон. Токовая перестройка частоты осуществляется в диапазоне единиц см$^{-1}$ благодаря изменению показателя преломления активной среды в ходе роста концентрации носителей заряда в активной области [11].

Как правило, диодный лазер функционирует в квазинепрерывном режиме, при котором ток накачки задается непрерывными пилообразными импульсами, что позволяет в дальнейшем получить частотную развертку. Динамическая стабилизация циклов токовой перестройки частоты излучения лазера осуществляется с помощью пропорционально-интегрального регулятора по обратной связи между термоэлементом Пельтье и термистором, встроенными в корпус диодного лазера.

Типичный диодно-лазерный спектрометр состоит из генератора когерентного излучения, объема анализируемой газовой смеси и приемника излучения, прошедшего через аналитический объем. Стабилизация частоты лазерного излучения является ключевым элементом метода ДЛС, поскольку она во многом определяет чувствительность и точность спектрометра. Качество стабилизации накладывает ограничения на длительность накопления сигнала и, как следствие, на достижимый уровень отношения сигнал/шум. При наличии температурных колебаний или дрейфа электроники стабилизация излучения, основанная лишь на температурных данных, позволяет копить сигнал лишь в течение единиц секунд.

Для устранения этой проблемы был предложен метод стабилизации частоты диодного лазера с использованием спектральной информации [12,13]. В этом подходе, при исследовании газов с низким уровнем поглощения в выбранном спектральном диапазоне, в спектрометр вводится реперный канал.

Этот канал представляет собой оптическую кювету, наполненную изучаемым газом при низком давлении, порядка нескольких десятков миллибар. Параллельно с основными измерениями проводятся дополнительные измерения в реперном канале. На основе положения узкой линии поглощения в этом канале в цикл стабилизации частоты лазера вводится дополнительная обратная связь. Это позволяет достичь стабилизации частоты с точностью до нескольких МГц, что уменьшает ограничения на время накопления сигнала. При работе с сильным поглощением аналогичный алгоритм может быть реализован на основе данных аналитического канала.

Подробное описание методологии классической ДЛС, а также современные принципы анализа данных в рамках этой методики представлены в разделе 2.4.

### 1.1.3. Модуляционная лазерная спектроскопия

Диодные лазеры сочетают хорошую стабильность интенсивности с быстрой перестройкой частоты, однако в ходе развития методики абсорбционной спектроскопии стало ясно, что сочетание методов лазерной абсорбции с модуляционными техниками дает метод, подходящий для работы с более слабым полезным сигналом, нежели в случае классической ДЛС. В течение многих лет изучалось множество модуляционных техник. Кроме того, был сделан уклон в сторону модуляции длины волны относительно модуляции мощности, а плавная и синусоидальная модуляция оказалась более эффективной, чем модуляция типа "вкл-выкл", использовавшаяся до этого в приложениях с импульсными лазерами.

Существует два основных модуляционных метода, подходящих для абсорбционной спектроскопии – модуляция длины волны (Wavelength Modulation Spectroscopy, WMS) [14] и модуляция частоты (Frequency Modulation Spectroscopy, FMS) [15]. Они различаются по соотношению между амплитудой и частотой модуляции. Если в в первом случае амплитуда модуляции больше ее частоты, то во втором – все наоборот [16,17]. Однако, поскольку эти методы имеют разные режимы работы, а техника модуляции длины волны проще, более надежна и более распространена, в данной работе рассматривается только этот метод.

Как подробно показано в разделе 3.3, применение метода модуляционной спектроскопии в сочетании с регистрацией сигнала синхронным или квадратурным приемником позволяет работать со спектральными линиями поглощения молекулярных газов малой интенсивности, не доступных для анализа методом классической ДЛС. Это связано с возможностью увеличения отношения сигнала к шуму регистрируемого сигнала в синхронном приемнике более чем на порядок и уходом от низкочастотного шума $1/f$. К тому же этот метод позволяет устранить проблемы, связанные с базовой линией, поскольку принимаемый фотоприемником сигнал имеет нулевую базовую линию [18].

Недостатком данного метода является сложность решения обратной задачи для аналитического определения содержания исследуемого газа, что приводит к необходимости калибровки приборов, основанных на этой методике. Для преодоления этой проблемы несколькими исследовательскими группами был предложен метод модуляционной спектроскопии, который избавляет от необходимости калибровки [19-21].

Не вдаваясь в подробности описания метода спектроскопии с модуляцией длины волны, приведенные в главе 3, отметим, что в ряде задач может быть оправданным использование комбинации перестраиваемой диодно-лазерной спектроскопии и модуляционной спектроскопии. В таком подходе ток лазера обычно одновременно модулируется низкочастотным пилообразным сигналом и более высокочастотным синусоидальным сигналом, как показано на рисунке 1.4.

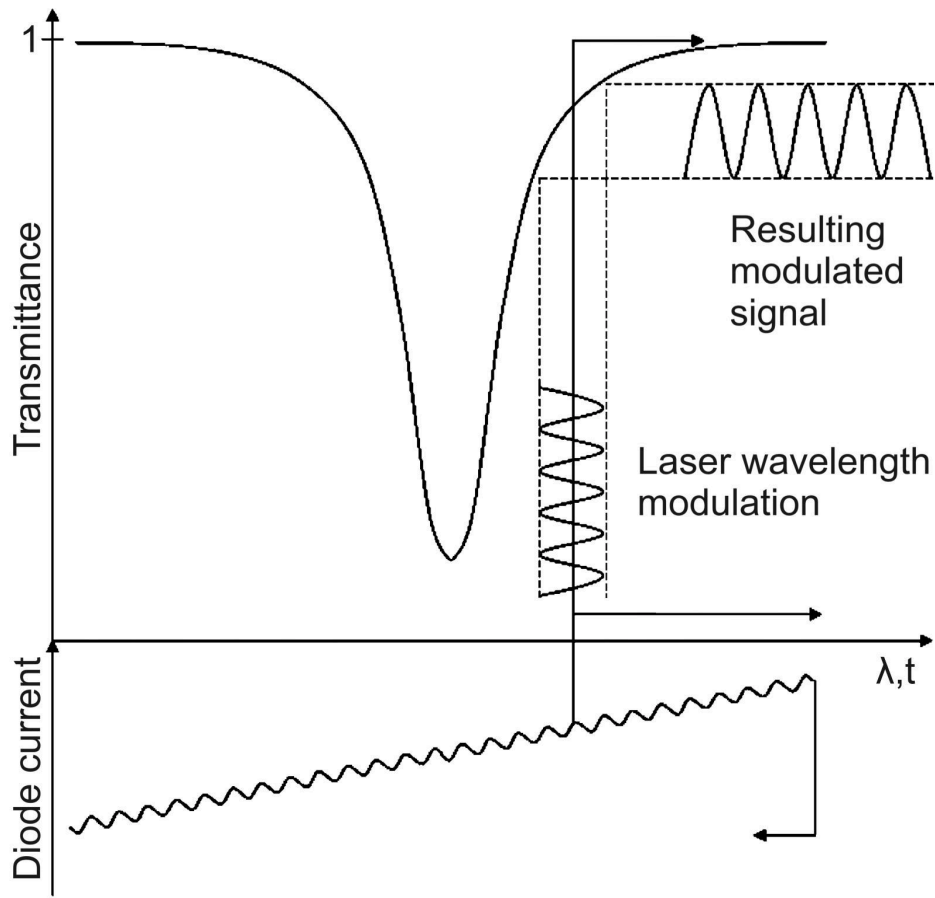

Рисунок 1.4 – Принцип работы комбинации перестраиваемой диодно-лазерной спектроскопии и модуляционной спектроскопии.

Пилообразный сигнал позволяет медленно сканировать спектральный диапазон вблизи линии либо же нескольких линий поглощения для анализа их формы с типичными частотами от единиц Гц до десятков Гц. Синусоидальная модуляция тока, обычно на частоте порядка 10 кГц, приводит к модуляцие частоты с интенсивностью, связанной с амплитудой модуляции тока накачки лазера. Наряду со сканированием и частотной модуляцией форма изменения тока накачки лазерного диода вызывает аналогичное изменение мощности излучения лазера. Амплитуда синусоидальной модуляции интенсивности определяется наклоном зависимости мощности лазера от тока, который слабо изменяется в ходе сканирования [22,23].

Существует и упрощенный подход в рамках комбинирования перестраиваемой диодно-лазерной спектроскопии и модуляционной спектроскопии, подходящий для увеличения частоты модуляции или же для адаптации к менее высокочастотной электронике [24]. В рамках такого подхода используется четырехточечное приближение синусоиды, показанное на рисунке 1.5, которого достаточно для вычисления второй квазипроизводной принимаемого сигнала. Период тока инжекции состоит из трех областей: область отсутствия накачки (а), область стационарного лазерного излучения (b) и область сканирования и модуляции частоты лазера (с).

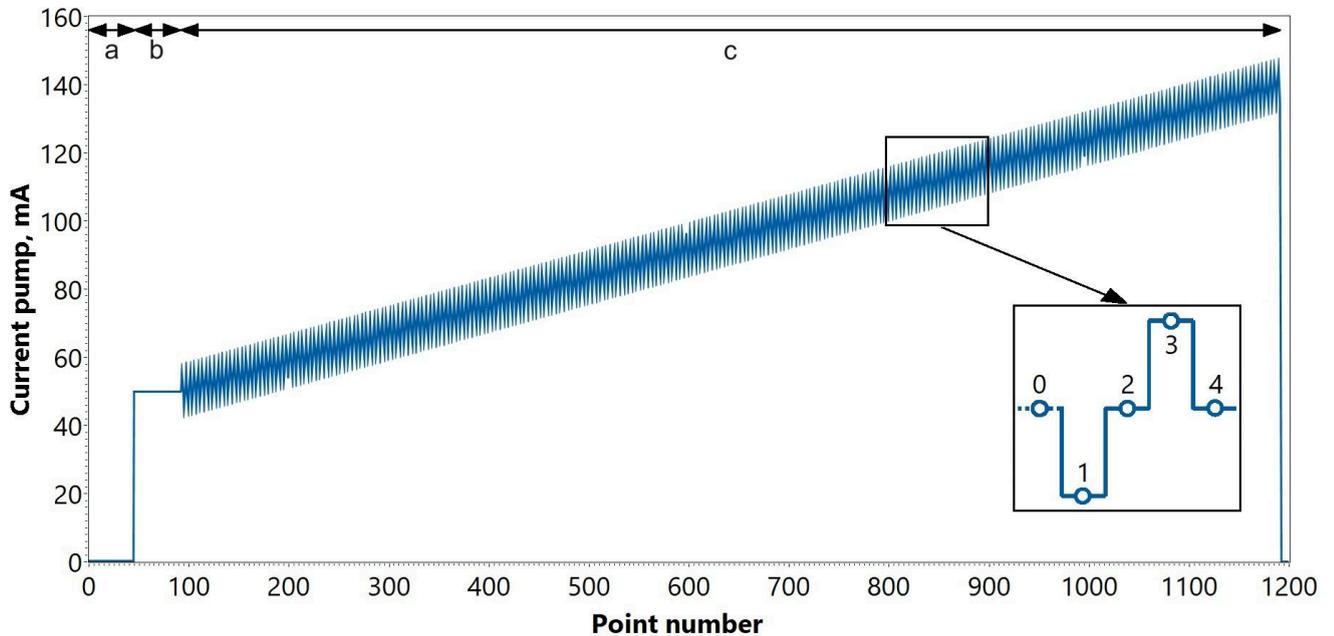

Рисунок 1.5 – Форма одного периода тока инжекции. На вставке показана форма одного периода модуляции тока инжекции.

Модуляция достигается путем периодического уменьшения тока инжекции в первой точке и его увеличения в третьей точке на величину амплитуды модуляции. Вторая и четвертая точки имеют одинаковые значения тока, что приводит к схожим интенсивностям лазерного излучения в этих точках при отсутствии поглощающей среды. Однако частоты излучения лазера в этих точках различаются.

Это связано с тем, что из-за уменьшения тока инжекции в первой точке снижается температура активной области лазера, а во второй точке температура ниже, чем в нулевой, поскольку она еще не успела стабилизироваться. В свою очередь, температура в третьей точке повышается, а в четвертой точке она оказывается выше, чем в нулевой. Небольшая разница в интенсивности излучения лазера между второй и четвертой точками, обусловленная температурной зависимостью порогового тока лазера при записи спектров, представляет собой эффект высшего порядка малости [25].

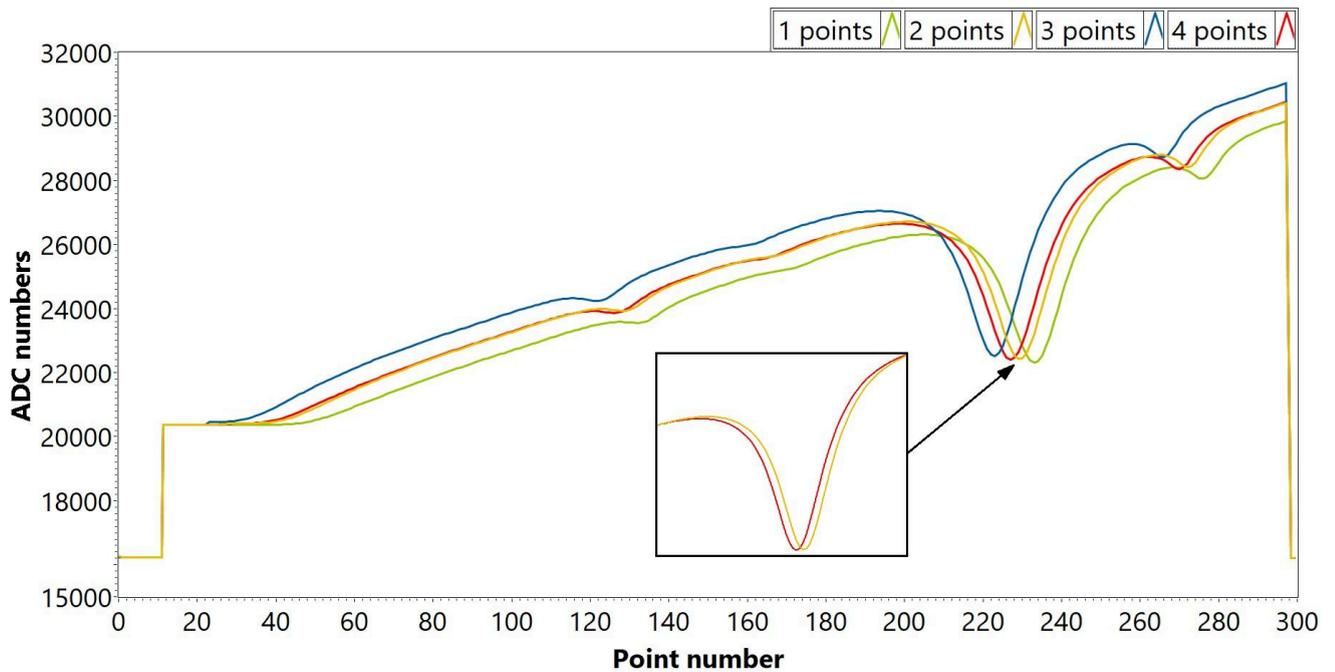

Рисунок 1.6 – Регистрируемый сигнал, разбитый на массивы соответствующих точек.

Регистрируемый сигнал разбивается на четыре массива данных. На рисунке 1.6 представлены эти четыре массива. Интенсивности излучения в массивах, соответствующих второму и четвертому пунктам модуляции, практически идентичны. Исследование разницы частот между массивами первого и третьего пунктов соответствует традиционному методу модуляции, в то время как массивы второго и четвертого пунктов относятся к методу модуляции с нестационарным нагревом и охлаждением активной области лазера [24,25]. Глубина модуляции тока инжекции в этом методе должна быть значительно больше, чем в традиционном методе модуляции, для возможности достижения необходимой глубины модуляции частоты, сопоставимой с шириной линии поглощения.

Такой подход был использован при разработке в МФТИ *in situ* газоанализатора с открытой кюветой для регистрации потоков парниковых газов методом турбулентных пульсаций. На момент написания диссертации цикл разработки этого прибора еще не завершен и опубликованные результаты апробации этого подхода для измерений в полевых условиях отсутствуют.

### 1.1.4. Спектроскопия затухания излучения в резонаторе

Метод спектроскопии затухания излучения в резонаторе (ЗИР) является техникой лазерной спектроскопии поглощения, которая может быть реализована с использованием импульсных или непрерывных источников излучения и обладает значительно более высокой чувствительностью по сравнению с традиционной абсорбционной спектроскопией.

ЗИР-метод основан на измерении скорости поглощения, а не величины поглощения светового импульса, заключенного в замкнутый объем с высоким коэффициентом добротности. Преимущества по сравнению с классической спектроскопией поглощения заключаются, во-первых, в присущей этому методу нечувствительности к флуктуациям интенсивности источника света и, во-вторых, в чрезвычайно длинных эффективных путях (многие километры), которые можно реализовать в стабильных оптических ячейках. По сравнению с другими чувствительными методами спектроскопии поглощения, особенно теми, которые используют модуляцию, ЗИР-спектроскопия имеет дополнительное преимущество, заключающееся в измерении поглощения в абсолютной шкале. Было показано, что метод ЗИР-спектроскопии эффективен в газоанализе для измерения как сильного поглощения малых газовых составляющих, так и слабого поглощения основных компонентов газовых смесей [26].

В 1980 году было предложено использовать оптическую ячейку для измерения отражательной способности зеркальных покрытий, поскольку потери на отражение аналогичны поглощению газовой средой в оптической кювете [27]. Модулируя непрерывное излучение и измеряя фазовый сдвиг, вносимый оптической ячейкой, получилось точно определить высокую отражательную способность зеркал. В 1984 году было продемонстрировано, что отражательная способность может быть измерена еще точнее, если в определенный момент отключать источник непрерывного излучения и регистрировать последующее затухание интенсивности света в оптической ячейке [28]. В обоих методах свет заводился в ячейку при случайных совпадениях частоты лазера с частотой одной из мод резонатора.

Методика ЗИР-спектроскопии (Cavity Ring-Down Spectroscopy, CRDS) была впервые продемонстрирован Энтони О'Кифом и Дэвидом А.Г. Диаконом в 1988 году [29]. В ЗИР-методе лазерный импульс вводится в оптический резонатор с высокой добротностью, который содержит исследуемое вещество, и наблюдается его ослабление при выходе из резонатора. ЗИР-спектроскопия представляет собой метод лазерной спектроскопии поглощения, обладающий высоким потенциалом для точного количественного определения молекулярного поглощения.

Принцип работы ЗИР-спектроскопии заключается в измерении времени затухания излучения $t_p$, которое происходит в результате поглощения вещества в оптическом резонаторе, а не в измерении ослабления интенсивности излучения при его прохождении через кювету с поглощающим веществом. Импульс света направляется в резонатор, ограниченный двумя высокоотражающими зеркалами, из которого при каждом проходе импульса через резонатор выходит миллионная доля излучения. Скорость уменьшения интенсивности выходящего излучения определяется полными потерями резонатора и поглощением исследуемого образца. Измерение поглощения основывается на времени затухания сигнала выходящего излучения и численной подгонке скорости затухания сигнала. В последние годы спектрометры, в которых свет вводится в моду оптического резонатора, отличающиеся исключительно высокой чувствительностью к поглощению, достигающей $10^{-7}–10^{-9}$ см$^{-1}$, приобрели широкое применение [30].

Типичная установка ЗИР-спектрометра (рисунок 1.7.1) состоит из импульсного лазерного источника, оптического резонатора с согласующей оптикой и фотодетектора. Стабильный резонатор формируется двумя конфокальными зеркалами с высоким коэффициентом отражения, которые также служат окнами для газовой кюветы. Лазерный импульс вводится в резонатор через одно из зеркал и ограничивается ирисовой диафрагмой для формирования моды TEM$_{00}$ в резонаторе. Затем импульс циркулирует в резонаторе с минимальными дифракционными потерями.

В дальнейшем была представлена методика ЗИР-спектроскопии с фазовым сдвигом (Phase-Shift Cavity Ring-Down Spectroscopy, PS-CRDS) [31]. В этой модификации спектр поглощения извлекается из измерения величины фазового сдвига модулированного по интенсивности непрерывного излучения при прохождении через оптический резонатор. Для измерения времени затухания света в резонаторе в этом методе требуется электрооптический модулятор (рисунок 1.7.2).

Следующим этапом развития ЗИР-спектроскопии стало использование перестраиваемых непрерывных диодных лазеров (Continuous-Wave Cavity Ring-Down Spectroscopy, CW-CRDS) [32,33]. В данной методике при совпадении частоты перестраиваемого лазерного излучения с модой резонатора излучение эффективно проходит в резонатор после чего осуществляется стандартное измерение затухания сигнала во времени. Для этого подхода необходимо использование оптического прерывателя, что проиллюстрировано на рисунке 1.7.3. Преимуществом спектроскопии с использованием лазерного источника непрерывного излучения является лучшее спектральное разрешение – чувствительность в этом случае может быть улучшена путем тщательного проектирования сочетания мод лазерного излучения с модами резонатора.

В методе резонаторного увеличения поглощения (РУП, Cavity-Enhanced Absorption Spectroscopy, CEAS) не используется ни прерыватель, ни модулятор (рисунок 1.7.4). В этом подходе лазерное излучение попадает в резонатор благодаря совпадению частоты излучения с собственными модами резонатора, а измеряется интегрированная по времени интенсивность выходящего из резонатора излучения. Из картины интенсивности излучения, вышедшего из резонатора, как функции длины волны определяется поглощение вещества [34].

В стандартном ЗИР-эксперименте с непрерывным лазером информация о поглощении извлекается из временной зависимости интенсивности излучения, выходящего из внешнего резонатора, при настройке частоты излучения лазера на моду резонатора. Здесь используется триггерная система активного контроля, когда необходимо начать регистрацию данных с помощью высокоскоростной электроники. Для регистрации всего спектра лазер медленно сканируется непрерывно или шагами с одной частоты на другую [30].

РУП-спектроскопия может быть реализована двумя эквивалентными способами: либо медленное сканирование лазерного излучения при стабильной модовой структуре, либо сканирование модовой структуры резонатора при неперестраиваемом лазере. Перестройка мод при этом может быть реализована за счет быстрого движения одного из зеркал на пьезоэлектрической керамике. Необходимым условием экспериментов является то, что лазерное излучение должно быть существенно долго в резонансе с модой резонатора, чтобы излучение внутри резонаторной моды достигло своего предельного значения.

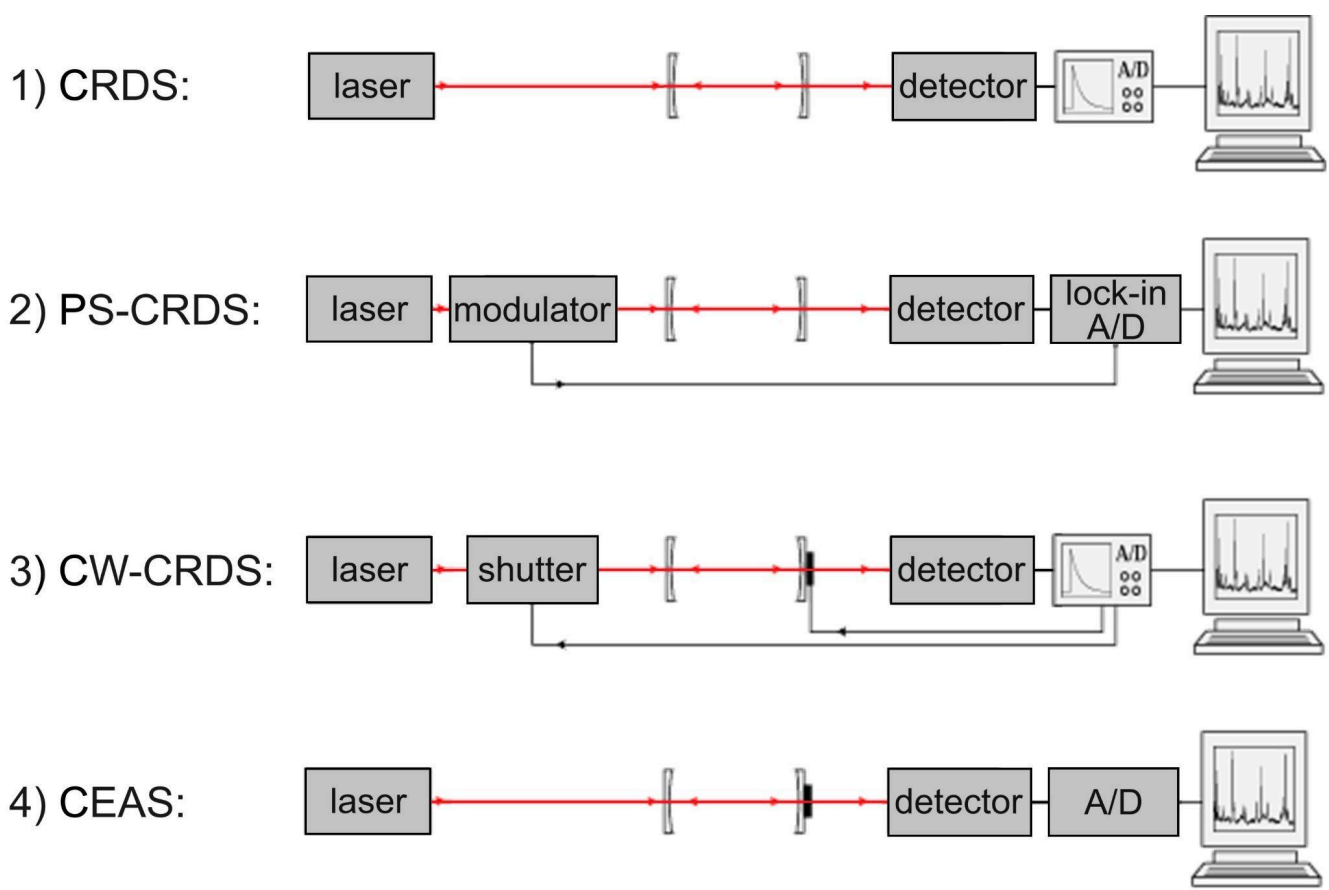

Рисунок 1.7 – Блок-схема экспериментальных установок для 1) ЗИР-спектроскопии (CRDS), 2) ЗИР-спектроскопии с фазовым сдвигом (PS-CRDS), 3) ЗИР-спектроскопии с непрерывными источниками лазерного излучения (CW-CRDS) и 4) РУП-спектроскопии (CEAS).

Энергия лазерного импульса уменьшается при каждом его проходе через резонатор из-за потерь на зеркалах и поглощения излучения исследуемым газом. Уменьшение энергии регистрируется по интенсивности света, прошедшего через выходное зеркало резонатора, как функция времени. Сигнал $S(t)$ представляет собой произведение интенсивности света $I(t)$ на выходном зеркале на пропускание зеркала $T$.

Для короткого лазерного импульса профиль интенсивности выходящего из резонатора излучения напоминает последовательность импульсов с убывающей интенсивностью. Если длительность лазерного импульса превышает время, необходимое для полного обхода излучения через резонатор, импульсы накладываются, образуя непрерывно затухающий сигнал. При возбуждении нескольких поперечных мод биения между ними могут вызвать модуляцию сигнала, зависящую от структуры продольных мод в резонаторе [30].

Рассмотрим затухание сигнала в отсутствие биений. За полный проход резонатора интенсивность света $I(t)$ уменьшается пропорционально квадрату коэффициента отражения зеркала R

$$I(t + t_r) = R^2 I(t), \tag{1.2}$$

где $t_r$ – время одного обхода резонатора лазерным импульсом. Для излучения, совершившего $n$ полных проходов резонатора с момента времени $t$, сигнал $S$ примет вид

$$S(t + nt_r) = R^{2n} S(t) = S(t) exp[2n \cdot ln(R)] = S(t) exp[-2n(1 - R)], \tag{1.3}$$

если величина $R$ близка к 1. Тогда, обозначив коэффициент потерь за один обход резонатора как $L_0 = 2(1-R)$, получим

$$S(t + nt_r) = S(t) exp[-nL_0]. \tag{1.4}$$

Таким образом, метод ЗИР-спектроскопии использует выражение модифицированного закона Бугера-Ламберта-Бера, согласно которому можно определить коэффициенты потерь резонатора и отражения зеркал, решая обратную задачу по времени затухания сигнала (рисунок 1.8).

При заполнении резонатора поглощающей средой, возникает дополнительное ослабление сигнала на каждом проходе излучения по резонатору. Согласно закону Бугера-Ламберта-Бера для поглощения в веществе, это ослабление $L_a$ определяется поглощением образца $L_a = 2K(\nu) \cdot l_a$ где $K(\nu)$ – частотно-зависимый коэффициент поглощения среды с длиной $l_a$ внутри резонатора. Следовательно, коэффициент потерь полного прохода излучения в резонаторе с образцом примет форму

$$L = L_0 + L_a = 2[(1 - R) + K(\nu)l_a], \tag{1.5}$$

тогда для сигнала после n обходов резонатора получим

$$S(t + nt_r) = S(t) exp[-nL] = S(t) exp\left[-\frac{t}{t_r}L\right], \tag{1.6}$$

откуда для постоянной времени затухания получим

$$\tau = \frac{t_r}{L} = \frac{t_r}{2[(1-R) + K(\nu)l_a]}. \tag{1.7}$$

В общем случае полученное выражение выполняется, если ширина линии поглощения внутри резонатора больше ширины полосы излучения внутри резонатора. Это условие является общим для всех методов поглощения, для которых выполняется закон Бугера. На ЗИР-спектроскопию накладывается еще одно условие – потери на поглощении исследуемого вещества должны быть много меньше общих резонаторных потерь. При выполнении этого условия эффективный коэффициент поглощения может быть измерен даже в случае, если ширина полосы лазерного излучения больше, чем ширина линии поглощения образца [30].

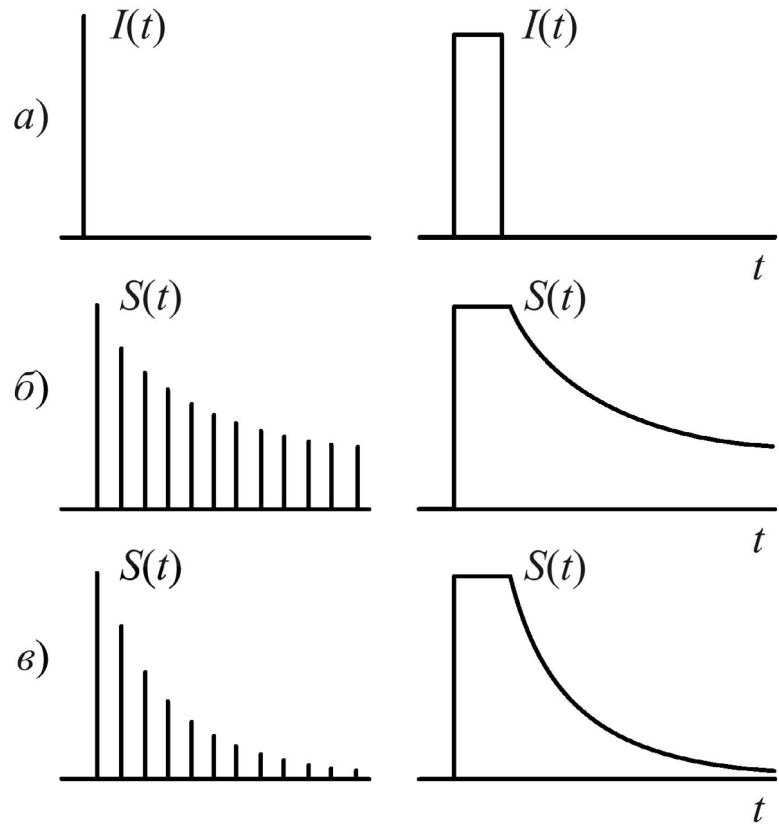

Рисунок 1.8 – Временные развертки излучения, входящего в резонатор (*а*) и выходящего из него при отсутствии (*б*) и наличии (*в*) в резонаторе поглощающего вещества при $t_p < t_r$ (слева) $t_p > t_r$ (справа).

ЗИР-спектр поглощения определяется как разница коэффициентов резонаторных потерь

$$K(\nu)l_r = \frac{L - L_0}{2}. \tag{1.8}$$

Выразив коэффициенты потерь через времена затухания излучения $\tau = t_r/L$ и $\tau_0 = t_r/L_0$, $(\tau - \tau_0) = \Delta\tau$, получим

$$K(\nu)l_a = \frac{L_0}{2}\frac{\Delta\tau}{\tau} = \frac{\Delta\tau}{\tau}(1-R). \tag{1.9}$$

Исходя из полученного выражения, можно оценить минимальное поглощение, измеряемое методикой ЗИР-спектроскопии. Так при коэффициенте отражения зеркал R = 0.9999 и при точности определения времени затухания $(\Delta\tau/\tau) \sim 10^{-3}$ минимальное регистрируемое поглощение может достигать $\sim 10^{-7}$ [30,35]. Более подробное рассмотрение этой методики представлено в обзоре [26].

### 1.1.5. Спектроскопия полного внутрирезонаторного выхода

Описанная выше методика спектроскопии с затуханием излучения в резонаторе является широко используемым методом обнаружения газов. Но несмотря на высокую чувствительность данный метод имеет ряд ограничений, таких как низкая устойчивость к помехам и высокие требования к высокочастотной электронике [36-38].

Методика внеосевой спектроскопии полного внутрирезонаторного выхода (Off-Axis Integrated Cavity Output Spectroscopy, OA-ICOS) представляет собой одну из разновидностей РУП-спектроскопии, отличающуюся высокой чувствительностью и простой структурой, предложенную в качестве альтернативы уже хорошо известному методу ЗИР-спектроскопии [39]. В отличие от ЗИР-спектроскопии, в рамках данной методики измеряется не время затухания, а непосредственно интенсивность пропускания излучения исследуемой газовой пробой в высокодобротном оптическом резонаторе, позволяющем достигать больших значений эффективного оптического пути и следовательно высокой чувствительности. Это упрощает обработку сигнала и обеспечивает линейный отклик в широком динамическом диапазоне. Технология OA-ICOS также менее требовательна к точности оптической юстировки и менее чувствительна к изменениям температуры и давления, что позволяет использовать более простые и недорогие компоненты [40-42].

Время формирования равновесного состояния резонансных мод высокодобротного резонатора можно записать как

$$\tau = \frac{L}{c(1-R)},$$  (1.10)

где $L$ – длина резонатора, $R$ – коэффициент отражения зеркал, $c$ – скорость света. В случае наличия поглощающей среды внутри резонатора это выражение примет вид

$$\tau = \frac{L}{c[1 - R + K(\nu)L]},$$  (1.12)

где $K(\nu)$ – коэффициент поглощения исследуемого газа. В таком случае для эффективного оптического пути получим выражение

$$L_{eff} = \frac{L}{1 - R + K(\nu)L},\tag{1.13}$$

таким образом, и величина эффективного оптического пути и постоянной времени резонатора определяются длиной резонатора и коэффициентом отражения зеркал [13].

Классическая конфигурация метода с расположением источника лазерного излучения на одной оптической оси с осью резонатора и регистрацией возбуждающихся продольных мод с областью свободной дисперсии (free spectral range, FSR), связанной с длиной резонатора как

$$FSR = \frac{c}{2L},\tag{1.14}$$

обладает рядом недостатков, среди которых высокий уровень оптических шумов. По этой причине была предложена внеосевая методика, в рамках которой излучение заводится в оптический резонатор под углом к основной оптической оси и на расстоянии от центра зеркала. Излучение многократно отражается от зеркал, формируя на зеркалах паттерн из точек отражений лазерного луча. Этот паттерн может быть как эллиптическим, так и иметь форму фигур Лиссажу [43]. При такой конфигурации метода область свободной дисперсии определяется как

$$FSR_{eff} = \frac{c}{2nL},\tag{1.15}$$

где $n$ – количество проходов лазерного луча, соответствующих совмещению паттерна самого с собой. Для идеального резонатора $n$ может быть определено из выражения (1.13), однако на практике количество проходов определяется качеством юстировки [13,44]. Иллюстрация оптического пути лазерного луча в такой схеме представлена на рисунке 1.9.

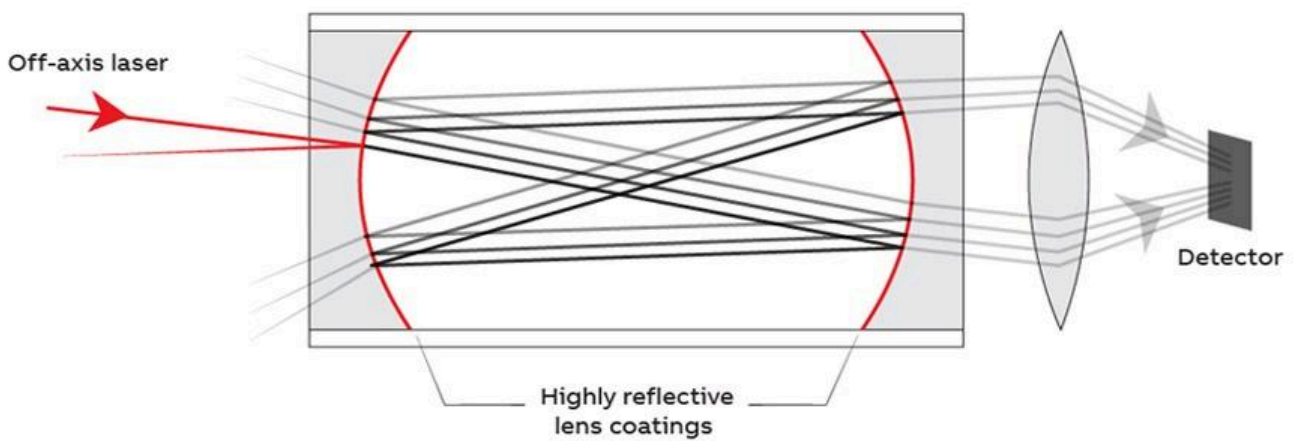

Рисунок 1.9 – Оптический путь лазерного луча в схеме внеосевой спектроскопии полного внутрирезонаторного выхода.

Высокая чувствительность, обусловленная близостью значения коэффициента отражения зеркал $R$ к единице, достигается ценой малой оптической мощности на выходе резонатора. В случае отсутствия поглощающей газовой среды в резонаторе при пренебрежении поглощением подложкой зеркал

$$T = 1 - R, \tag{1.16}$$

и для величины интенсивности излучения, выходящей из резонатора, получим

$$I \approx I_0 \frac{1 - R}{2}, \tag{1.17}$$

где $I_o$ – интенсивность излучения, попадающего в резонатор, а коэффициент 2 связан с тем, что потери происходят на обоих зеркалах резонатора.

При проектировании высокодобротных резонаторов для работы в рамках описываемой методики большое внимание необходимо уделять форме паттерна. Геометрия резонатора должна быть такова, чтобы пятна паттерна не пересекались друг с другом во избежание возникновения оптических шумов в регистрируемом сигнале и не приближались к границам зеркал для предотвращения сокращения чувствительности. Диаметр пятна паттерна $s$ можно оценить из геометрических параметров резонатора и длины волны $\lambda$ лазерного излучения [13,45]:

$$s = \sqrt{\frac{2\lambda L}{\pi ln 2} \sqrt{\frac{2R_m^2}{L(R_m - \frac{1}{2})}}}, \tag{1.18}$$

где $R_m$ – радиус кривизны зеркала. Угловое смещение точек паттерна на зеркале после однократного прохождения резонатора определяется как

$$\theta = arccos\left(1 - \frac{L}{R_m}\right). \tag{1.19}$$

Полученное выражение задает характер паттерна на сферическом зеркале резонатора, в частности, с его помощью определяются количество проходов за один оборот паттерна вокруг оптической оси $m$ и количество проходов лазерного пучка, при которых паттерн совмещается сам с собой $n$. Условие для значений $m$ и $n$ выглядит следующим образом:

$$\frac{\pi}{\theta} - \frac{m}{n} < \frac{s}{\pi D}, \tag{1.20}$$

где $D$ – диаметр самого паттерна на сферическом зеркале.

Таким образом, плотность паттерна задается последними тремя выражениями и радиусом отдаления точки ввода излучения от оптической оси резонатора. Форма паттерна регулируется юстировкой оптических элементов – идеальной юстировке соответствует круглый паттерн [13].

При выборе скорости сканирования частоты лазерного излучения необходимо учитывать постоянную времени резонатора $\tau$. Как было показано эмпирически [46], длительность перестройки излучения на величину полуширины сканируемой линии поглощения исследуемого газа на ее полувысоте должна быть не меньше утроенной величины $\tau$. При несоблюдении этого условия регистрируемые линии поглощения будут иметь несимметричный контур, что приведет к ухудшению спектрального разрешения.

Стоит также отметить, что в случае слабого поглощения исследуемой газовой среды ($K(\nu) \cdot L \ll 1\text{-}R$), регистрируемые данные могут быть описаны с помощью закона Бугера-Ламберта-Бера. Для случая $K(\nu) \cdot L \approx 1\text{-}R$ можно вывести общее выражение для затухания излучения в резонаторе, учитывающее поглощение газовой среды и потери на зеркалах и описывающее как слабые, так и сильные линии поглощения. Для этого необходимо рассмотреть интенсивность излучения на выходе резонатора после разного количества проходов лазерного излучения внутри резонатора – $I_1, I_2 \ldots I_n$:

$$I_1 = I_0 C_p (1-R)^2 exp[-K(\nu)L], \tag{1.21}$$

$$I_2 = I_0 C_p (1-R)^2 R^2 (1-a)^2 exp[-3K(\nu)L], \tag{1.22}$$

$$I_n = I_0 C_p (1-R)^2 \sum \left( R^{2n} (1-a)^{2n} exp[-(2n+1)K(\nu)L] \right) =$$
$$= I_0 C_p (1-R)^2 exp[-K(\nu)t] \sum \left( R^2 (1-a)^2 exp[-2K(\nu)L] \right)^n. \tag{1.23}$$

где $C_p$ – параметр эффективности совпадения мод резонатора с пространственным распределением вводимого пучка излучения, имеющий значение от 0 до 1, $a$ – параметр, учитывающий дифракционные потери и качество юстировки резонатора, также принимающий значения от 0 до 1. Используя выражение для суммы геометрической прогрессии, можно получить

$$I_K = \frac{I_0 C_p (1-R)^2 exp[-K(\nu)L]}{1 - R^2 (1-a)^2 exp[-2K(\nu)L]}. \tag{1.24}$$

Для функции пропускания в отсутствии поглощения находим:

$$I = \frac{I_0 C_p (1-R)^2}{1 - R^2 (1-a)^2}. \tag{1.25}$$

Тогда для функции пропускания получим

$$T = \frac{I_K}{I} = \frac{(1 - R^2 (1-a)^2) \, exp[-K(\nu)L]}{1 - R^2 (1-a)^2 exp[-2K(\nu)L]}, \tag{1.26}$$

откуда можно получить выражение для коэффициента поглощения исследуемого газа:

$$K(\nu) = -\frac{ln\left[\dfrac{R^2(1-a)^2-1+\sqrt{[1-R^2(1-a)^2]^2+4R^2(1-a)^2T^2}}{2R^2(1-a)^2T}\right]}{L}.$$ (1.27)

### 1.1.6. Гетеродинная лазерная спектроскопия

Термин «гетеродинирование» был введен в радиотехнике для обозначения процесса преобразования частоты исходного сигнала при его квадратичном детектировании с более мощным опорным сигналом другой частоты [47]. Впервые эта техника была применена канадским исследователем Реджинальдом Фессендом в 1901 году, когда он использовал два близких радиосигнала для получения акустического сигнала, слышимого человеком, с целью упрощения восприятия передаваемого по радиотелеграфным сетям кода Морзе [48]. После окончания Второй мировой войны метод гетеродинного приема использовался практически во всех радиоприемниках.

В 1947 году советский ученый-радиофизик Габриэль Симонович Горелик и американец Теодор Форрестер предсказали возможность применения гетеродинирования в спектроскопических исследованиях [49,50]. Возможность повсеместного использования гетеродинного приема излучения в спектроскопии появилась после создания лазерного источника излучения. Уже в 1961 году в работах с первым газовым лазером наблюдались биения между его модами [51]. В том же году Форрестер теоретически обосновал преимущества применения метода гетеродинирования в спектроскопических исследованиях с использованием лазерных источников излучения [52]. В последующей работе по наблюдению межмодовых биений лазерного излучения, проведенной Эрриоттом, выводы Форрестера были подтверждены экспериментально [53]. В 1969 году гетеродинный прием был использован в первом эксперименте по применению метода диодно-лазерной спектроскопии [54].

Дальнейшее развитие лазерного гетеродинирования было связано с успехами в создании высокостабильного лазерного источника с узкой полосой генерации. Исследованиями в этом направлении занимались Владилен Степанович Летохов и Вениамин Павлович Чеботаев в СССР, а также их американские коллеги. В 1972 году Джон МакЭлрой впервые предложил использование гетеродинной техники для наблюдения атмосфер [55], поскольку атмосфера является наиболее информативным источником данных о состоянии планеты и ее эволюции – исследование спектральных свойств атмосферы позволяет определить среднюю концентрацию основных и малых газовых составляющих, их вертикальное распределение, а также вертикальные профили давления и температуры [56-58].

Измерения с высоким спектральным разрешением позволяют напрямую определять скорость ветра, основываясь на регистрации доплеровского смещения отдельных спектральных линий, вызванного движением воздушных масс в атмосфере [59,60]. Это, в свою очередь, дает возможность восстановления энергетического баланса и картины глобальной циркуляции атмосферы планеты. Такие данные могут быть использованы для верификации трехмерных численных моделей атмосферы [57,61].

Для детального изучения атмосфер планет требуются длительные непрерывные наблюдения, которые редко могут быть реализованы в рамках космических миссий с ограниченной продолжительностью. Наличие определенной орбиты и ограниченность поля зрения прибора приводят к невозможности полного покрытия диска планеты в ходе наблюдений с межпланетной станции. К тому же, в этом случае существуют значительные ограничения на габариты, вес и энергопотребление научной аппаратуры на борту. Наземные наблюдения не имеют этих недостатков. Возможность наземных наблюдений атмосфер планет в окнах прозрачности атмосферы Земли существует благодаря эффекту Доплера, возникающему вследствие относительного движения планет. В результате этого эффекта спектральные линии молекул, наблюдаемые в атмосфере удаленной планеты, сдвинуты относительно соответствующих линий земной атмосферы. Всвязи с этим с начала 1970-х годов для решения обозначенных задач применяется метод гетеродинной спектроскопии. Это единственный известный метод, позволяющий осуществлять измерения с высоким спектральным разрешением $\lambda/\Delta\lambda \sim 10^8$ при сохранении приемлемого отношения сигнала к шуму.

Кроме того, в случае гетеродинного метода приема излучения апертура используемого телескопа не имеет определяющего значения, то есть нет необходимости в использовании телескопов с большим диаметром входной линзы. Таким образом, становится возможным проведение длительных наблюдений с использованием широко распространенных телескопов метрового класса, которые не столь активно используются научным сообществом, как немногочисленные современные телескопы с большой апертурой.

В этом методе принимаемый телескопом сигнал совмещается с излучением локального осциллятора – лазерного источника – и регистрируется квадратичным детектором. Нелинейность приема детектора приводит к переносу спектра оптического сигнала в область радиочастот, как показано на рисунке 1.10, что позволяет применять для анализа сигнала обычные радиочастотные (РЧ) спектроанализаторы. При этом мощность анализируемого РЧ-сигнала увеличивается пропорционально мощности локального осциллятора. Спектральное разрешение гетеродинного приема ограничено разрешением РЧ-спектроанализатора, регистрирующего сигнал на промежуточной частоте, а также частотной стабильностью локального осциллятора. Подробное описание основных принципов лазерного гетеродинирования дано в книге Владимира Всеволодовича Протопопова и Николая Дмитриевича Устинова [62].

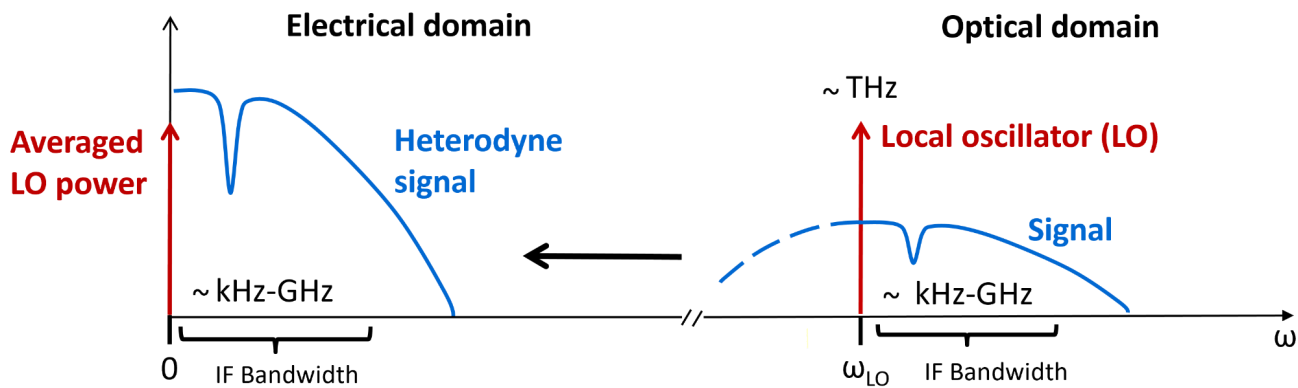

Рисунок 1.10 – Перенос исследуемого сигнала из терагерцового диапазона в радиочастотный.

Гетеродинные радиометры – приборы, построенные на принципах гетеродинного приема, – получили распространение благодаря своей способности регистрировать некогерентное во времени излучение с шириной спектра, обычно на несколько порядков превышающей ширину полосы пропускания фотодетектора. Таким образом, вклад принимаемого излучения в электрический сигнал на выходе фотодетектора статистически подобен собственному шуму регистрирующей системы [63]. Можно считать, что радиометр измеряет изменение уровня шумового сигнала в тракте, вызванное наличием принимаемого излучения на его входе.

Помимо высокого спектрального разрешения, преимуществом гетеродинного метода приема является его лучшая чувствительность по сравнению с методами прямого детектирования. В этих методах предельная чувствительность, связанная с дробовым шумом фототока, не достигается из-за малой мощности падающего излучения. В то время как в гетеродинных радиометрах регистрируемая мощность может быть увеличена за счет мощности опорного сигнала.

Совокупность преимуществ гетеродинного метода приема излучения играет определяющую роль в задачах наземного наблюдения планетных атмосфер. В настоящее время только эта методика позволяет решить указанный круг задач благодаря высокому спектральному разрешению и одновременно большой чувствительности. Однако гетеродинные радиометры имеют ряд недостатков, связанных главным образом со сложностью конструкции приборов и строгими требованиями к юстировке оптических элементов [64]. Другой проблемой до недавнего времени был крайне ограниченный выбор стабильного источника излучения, подходящего в качестве гетеродина. Тем не менее, интенсивное развитие лазерной техники и волоконной оптики в последние два десятилетия позволило по-новому взглянуть на традиционные подходы в гетеродинной спектроскопии оптического диапазона [61,65].

Так в 2014 году в МФТИ был разработан прототип гетеродинного спектрорадиометра ближнего ИК-диапазона, подходящего для зондирования земной атмосферы. С его помощью впервые было произведено измерение поглощения атмосферных $CO_2$ и $CH_4$ вблизи 1.61 мкм и 1.65 мкм со спектральным разрешением $\lambda/\delta\lambda$ ~$10^8$. В качестве локального осциллятора использовался диодный РОС-лазер, а смеситель был реализован на основе Y-разветвителя из кварцевого одномодового волокна. Полоса пропускания приемника составила ~3 МГц. При времени экспозиции 10 минут был записан спектр поглощения атмосферы над Москвой с отношением сигнал/шум ~120, ограниченным дробовым пределом. Примененный к полученному спектру алгоритм восстановления данных позволил получить вертикальный профиль метана с максимальной концентрацией 2148 ± 10 ppbv у поверхности и интегральной плотностью $(4.59 \pm 0.02) \times 10^{22}$ см$^{-2}$ [66,67].

В 2020 году развитие этого направления работы в МФТИ позволило реализовать новую методику дистанционных измерений ветра, основанную на доплеровском анализе линии поглощения $CO_2$ в диапазоне 1.605 мкм, измеренной в режиме прямых солнечных наблюдений – форма профиля линии позволила установить однозначную зависимость между доплеровским смещением линии и высотой над поверхностью, на которой формируется соответствующая часть профиля линии [68].

В работах Теодора Костюка – создателя первого гетеродинного спектрометра для астрономических наблюдений THIS, – было показано, что получение спектральных характеристик планет земной группы методом наземного гетеродинного приема возможно по наблюдению тепловой эмиссии поверхности планеты [56,69,70].

В 2014 году в Университете Тохоку был сконструирован единственный на сегодняшний день активно используемый для исследования планетных атмосфер гетеродинный спектрометр высокого разрешения MILAHI (Mid-Infrared LAser Heterodyne Instrument). В данном приборе в качестве локальных осцилляторов используются квантово-каскадные лазеры с длинами волн 7.7 мкм, 9.6 мкм и 10.3 мкм, которые были подобраны для наблюдения молекул $CO_2$, $CH_4$, $H_2O_2$, $H_2O$ и HDO. Детекторами выступают фотодиоды на основе HgCdTe с диаметром активной зоны 0.1 мм и полосой приема 3 ГГц. Для анализа сигнала применяется спектроанализатор с полосой 1 ГГц и спектральным разрешением 61 кГц. Достигнутый уровень шума составляет в температурном эквиваленте 3000 К. Разрешение гетеродинного спектрометра превышает $10^6$ благодаря узкой (менее 20 МГц) ширине линии генерации используемого квантово-каскадного лазера [71].

Среди научных задач, которые потенциально могут быть решены с помощью MILAHI, создатели прибора отмечают определение температурных профилей на высотах 65-90 км для Венеры и от поверхности планеты до 30 км для Марса, получение профилей ветра на высотах 75-90 км для Венеры и на высотах 5-25 км для Марса, прямое измерение мезосферных ветров и температур по доплеровскому сдвигу эмиссионных линий на высоте 110 км на Венере и на 75 км на Марсе, наблюдение малых газовых составляющих и определение изотопного соотношения для атмосфер этих планет. Этот гетеродинный спектрометр используется для наблюдений с вершины Халеакала на Гавайях. С его помощью успешно были получены спектр поглощения озона в земной атмосфере, спектр неравновесной эмиссии $CO_2$ в атмосфере Венеры [61], определены скорость ветра на высоте 85 км и профили температуры на высотах 70–100 км в атмосфере Венеры [72], а также проведены прямые измерения зональных ветров на высоте около 80 км во время глобального пылевого шторма 2018 года в атмосфере Марса [73].

Попытка создания такого устройства была предпринята при участии автора в МФТИ в 2015 году [74]. Был разработан прототип гетеродинного спектрометра для изучения атмосфер планет земной группы в Солнечной системе на телескопе метрового класса, внешний вид которого представлен на рисунке 1.11, и подготовлено ПО для управления всеми блоками прибора.

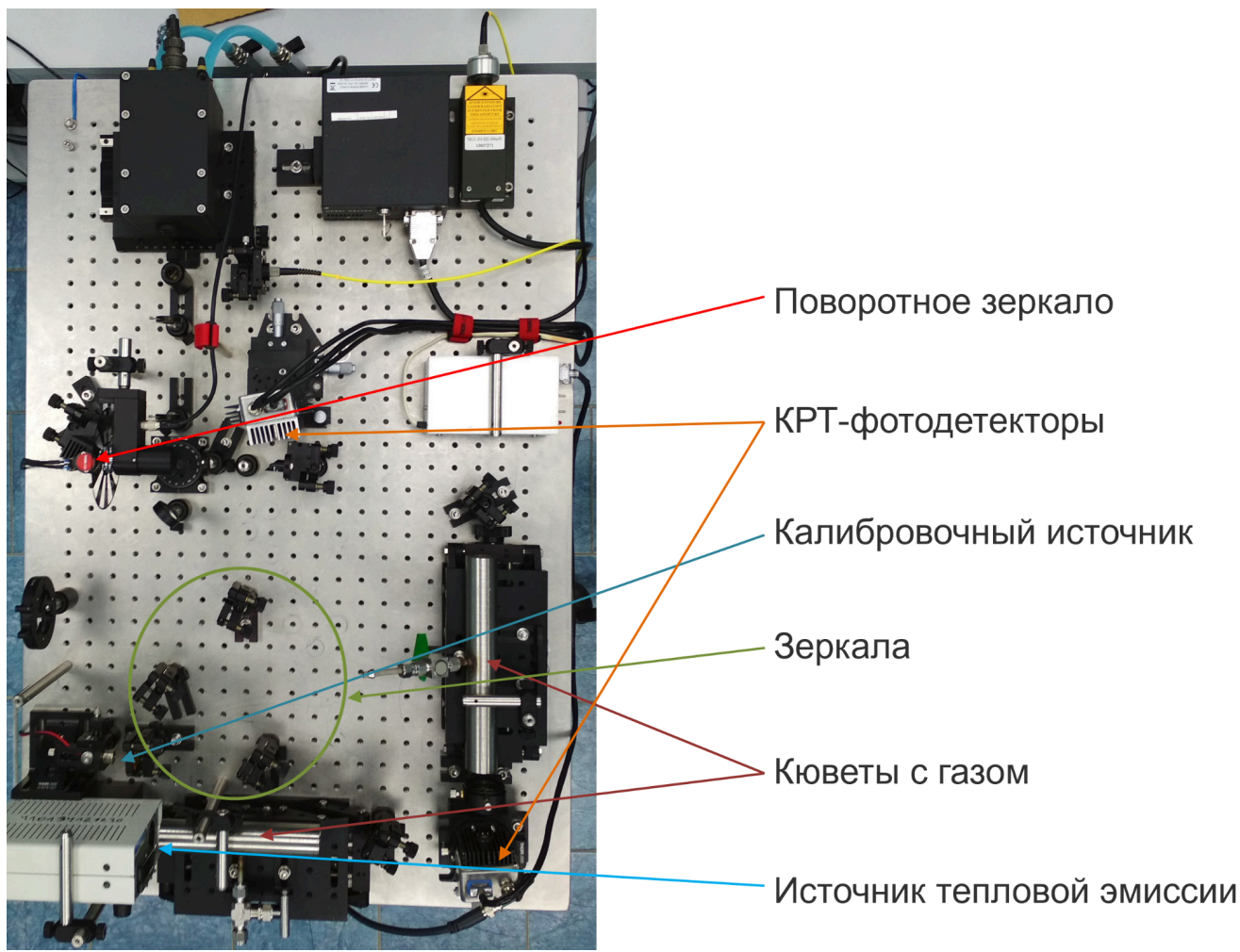

Рисунок 1.11 – Внешний вид разработанного в МФТИ гетеродинного спектрометра среднего ИК-диапазона.

Характеристики разработанного гетеродинного спектрометра среднего ИК-диапазона представлены в таблице 1.1. Данный прибор потенциально должен был не уступать своему действующему японскому аналогу.

Таблица 1.1 – Характеристики гетеродинного спектрометра среднего ИК-диапазона.

| Спектральный диапазон | 5.4 мкм, 7.8 мкм, 10.4 мкм |
|---|---|
| Полоса пропускания CdHgTe-детектора | 850 МГц |
| Анализатор спектра | Цифровой FFT-спектрометр (разрешение – 76 кГц, полоса приема до 5 ГГц) |
| Сигнал/шум | 10 дробовых пределов ККЛ |

В лабораторных условиях была проведена имитация наблюдения линии поглощения в континуальном эмиссионном спектре в диапазоне 7.8 мкм (рисунок 1.12), для чего излучение абсолютно черного тела пропускалось сквозь кювету с исследуемым газом – ацетиленом – при давлении около 10 мбар, после чего совмещалось с лазерным излучением излучением.

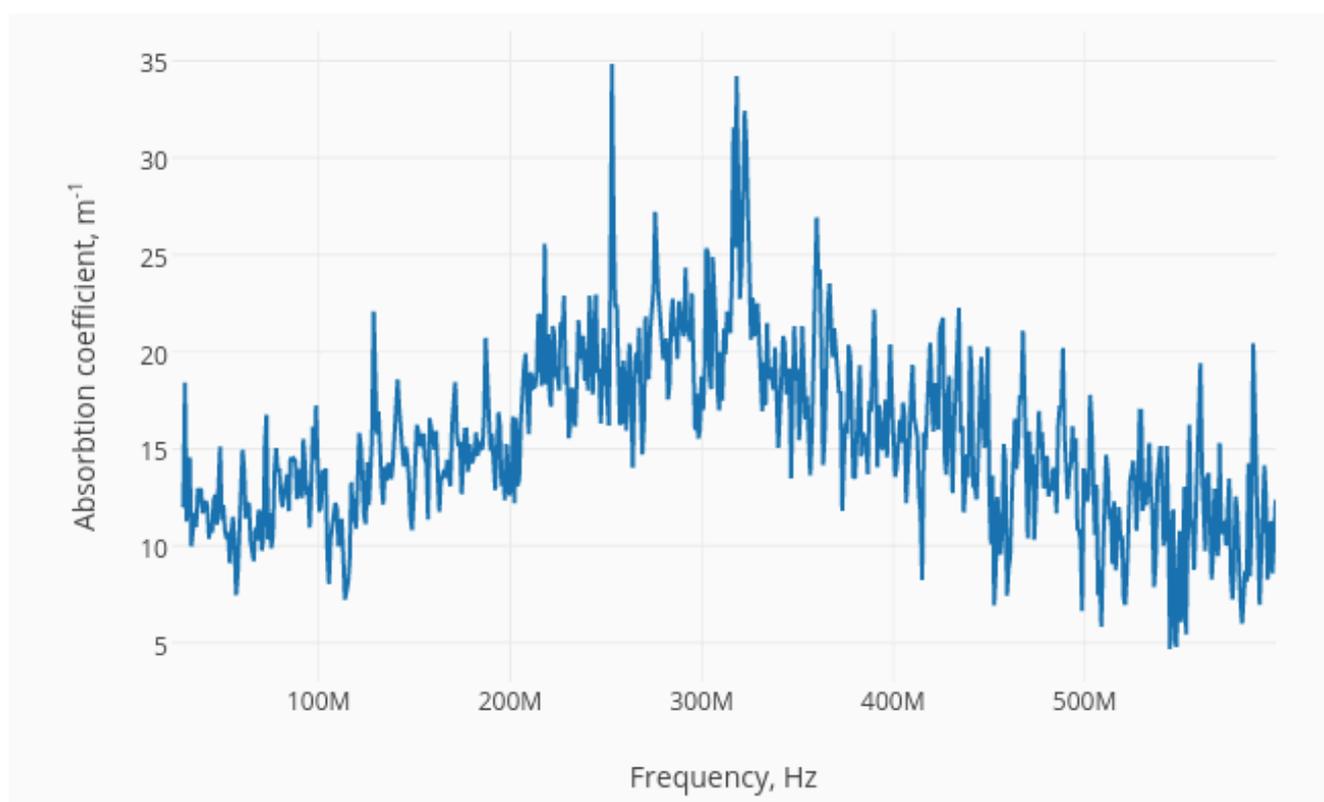

Рисунок 1.12 – Полученный спектр поглощения ацетилена.

Однако в силу постепенного выхода из строя дорогостоящих квантово-каскадных лазерных источников излучения и отсутствия на тот момент возможности переоснащения проект был приостановлен в 2019 году.

## 1.2. Оптические системы, применяемые в лазерной спектроскопии

Оптические системы, используемые для спектроскопического анализа газовых смесей делятся на однопроходные и многопроходные. За исключением случая сильного поглощения, редко используемого в аналитических задачах, однопроходные системы удобны для применения в реперных кюветах, необходимых для корректной стабилизации лазерного излучения по частоте энергетического перехода искомого газа. Оптические требования к таким кюветам обычно не очень высоки. Реже встречаются задачи по анализу слабо поглощающего в выбранном спектральном диапазоне, для которых для реперных кювет требуются многопроходные системы. В большинстве случаев многопроходные системы применяются в аналитических кюветах – таким образом, требования по минимизации оптических и спектральных искажений к таким системам весьма высоки.

В большинстве задач спектроскопии поглощения простых молекулярных газов используются вращательно-колебательные переходы анализируемых газов, таким образом рабочим спектральным диапазоном выступают ближний и средний инфракрасный диапазоны спектра. Помимо выбора подходящего источника излучения – полупроводникового лазера (диодного лазера [75-77], межполосного каскадного лазера [78] или квантово-каскадного лазера [79]), реже газового лазера или лазера на красителях, необходимо выбрать подходящие оптические окна в случае применения закрытых кювет и иные оптические элементы системы – светоделители, фильтры верхних/нижних частот или полосовые, серые фильтры и т.д.

График пропускания боросиликатного стекла (NBK-7), кварцевого стекла (fused silica), сапфира, ZnSe, BaF$_2$, CaF$_2$, MgF$_2$ и Ge, используемых для прозрачных окон оптических систем в ИК-диапазоне, приведен на рисунке 1.13. Окна располагаются под углом к входящему в кювету лазерному лучу для уменьшения паразитных отражений. Поскольку оба окна расположены под углом друг к другу, любое смещение луча от первого окна компенсируется вторым окном. Для устранения эффекта оптического эталона сами окна зачастую имеют клиновидную форму.

На оптические поверхности, например, на рабочие поверхности окон кювет, могут наноситься различные покрытия, необходимые для изменения их оптических свойств. Спектральные характеристики и другие свойства наносимых оптических тонких пленок определяются структурой и количеством слоев в покрытии, показателями преломления используемых материалов и оптическими свойствами подложки. Структура большинства оптических покрытий напоминает ряд дискретных чередующихся слоев материалов с высокими и низкими показателями преломления.

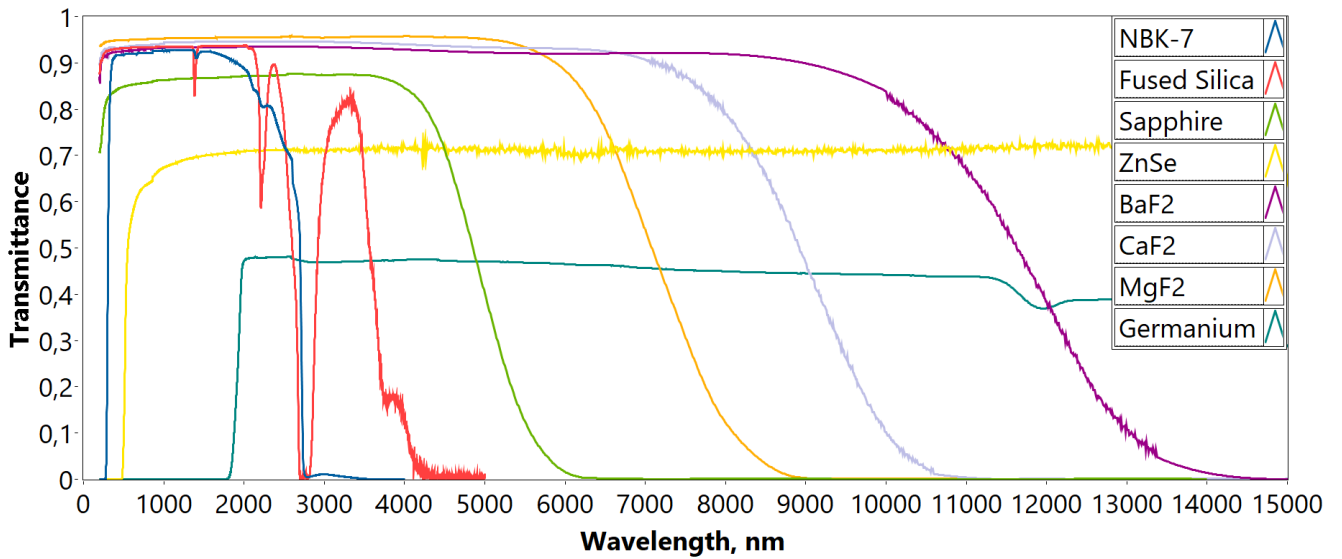

Рисунок 1.13 – Сравнение пропускания излучения боросиликатного стекла (NBK-7), кварцевого стекла (fused silica), сапфира, селенида цинка, фторида бария, фторида кальция, фторида магния и германия при толщине образца 10 мм.

Среди наиболее типичных примеров покрытий – антибликовые или просветляющие – твердые преломляющие покрытия, которые минимизируют поверхностные отражения в определенных диапазонах длин волн при нанесении на оптическую поверхность. Использование просветленной оптики может уменьшить потери на отражениях примерно на порядок. Широкополосные просветляющие покрытия состоят из нескольких слоев – чередующихся материалов с высоким показателем преломления и низким. Слои наносятся на подложку с помощью электронно-лучевого осаждения. Толщина слоев оптимизируется для создания деструктивной интерференции между отраженными волнами и конструктивной интерференции между пропущенными волнами. В результате получается оптика с улучшенными характеристиками в заданном диапазоне длин волн, а также с минимальными внутренними отражениями. Также применяются V-покрытия – многослойные, диэлектрические, тонкопленочные, антибликовые покрытия, которые разработаны для минимизации отражательной способности в коротком диапазоне длин волн. Отражательная способность поверхности быстро возрастает по обе стороны от минимума, что придает кривой отражения V-образную форму.

В случае обратной задачи применяются диэлектрические высокоотражающие покрытия, представляющие собой твердые, тугоплавкие, оксидные покрытия, которые максимизируют отражение поверхности в определенных диапазонах длин волн и под определенными углами падения. Эти покрытия так же состоят из чередующихся слоев материалов с высокими и низкими показателями преломления.

# 1.2.1. Однопроходные системы

Однопроходные кюветы, предполагающие однократное наполнение газовой смесью с последующей запайкой, чаще всего изготавливают из боросиликатного стекла или плавленого кварца с наклонными окнами из того же материала или из сапфира. Такие кюветы являются наиболее герметичными и позволяют работать с ними на протяжении многих лет в случае успешной процедуры наполнения газовой смесью. Пример кюветы этого типа приведен на рисунке 1.14.

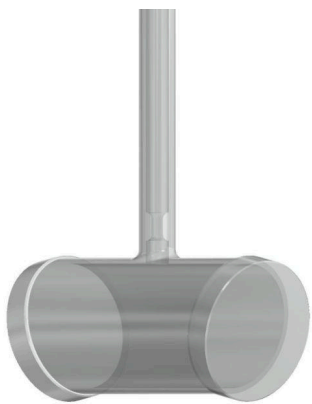

Рисунок 1.14 – Однопроходная кювета из кварцевого стекла с окнами, расположенными под углом 11°.

Металлические кюветы с окнами из сапфира, ZnSe, BaF$_2$, CaF$_2$, MgF$_2$ или Ge используются при необходимости обновления газовой пробы с какой-то периодичностью при помощи регулирующих и запорных клапанов. Подобные кюветы в большей степени подвержены протеканию обменных процессов компонентов газовой смеси со стенками кюветы и имеют обыкновенно худшую герметичность из-за несовершенной системы крепления окон.

Пороговая чувствительность лазерного спектрометра определяется длиной пути лазерного излучения в поглощающей среде, поэтому для работы со слабыми линиями поглощения активно применяются многоходовые газовые кюветы (МГК). Поскольку в многоходовых зеркальных оптических системах световой луч многократно проходит через ограниченный объем между зеркалами, это дает возможность значительно увеличить оптический путь и зарегистрировать слабое поглощение исследуемых газовых составляющих, находящихся между зеркалами.

Пусть при каждом отражении от зеркала интенсивность луча уменьшается в ($1-\varepsilon$) раз, тогда после $N$ отражений при дистанции между зеркалами $L$ световой луч пройдет оптический путь длиной $N{\cdot}L$, с задержкой по времени $\tau = NL/c$, где $c$ – скорость света. При этом коэффициент передачи системы $T$, определяемый отношением интенсивности луча, прошедшего через кювету с оптическим путем $NL$, к интенсивности входящего луча, выразится как

$$T = (1 - \varepsilon)^N. \tag{1.28}$$

Рассмотрим возможность использования оптической системы, состоящей из двух плоских зеркал. Введение плоского зеркала в оптическую систему не изменяет параметры пучка, поэтому размер пятна на выходе МГК с $N$ отражениями от плоского зеркала будет эквивалентен размеру пятна в открытом пространстве без отражений на той же длине $L$. Размер пятна луча с входным диаметром $D_0$ на зеркале после $N$ проходов определяется углом дифракционной расходимости

$$\Theta = \frac{\lambda}{D_0}. \tag{1.29}$$

При диаметре лазерного пучка на входе в систему $D_0 = 1$ мм и длине волны $\lambda = 1$ мкм угол дифракции составляет 0.001, а размер лазерного пятна на расстоянии $L = 100$ м будет равен 10 см. Следовательно при таких больших габаритах использование оптических систем с плоскими зеркалами в МГК неприемлемо [30].

Из этого следует, что для измерений, проводимых в рамках методики лазерной спектроскопии, направленных на детектирование слабых спектров поглощения различных газов, необходимы более сложные оптические системы МГК.

## 1.2.2. Многоходовая система Уайта

В 1942 году Джоном Уайтом было опубликовано описание многоходовой системы [80], предназначенной для исследования поглощения в парах соединений с высокими температурами кипения. Система Уайта получила широкое распространение и продолжает использоваться в спектроскопии, претерпев незначительные изменения.

Одним из ключевых достижений Уайта стало устранение виньетирования наклонных лучей, что долго оставалось проблемой в многоходовых оптических системах. Он добился этого, заменив плоское коллективное зеркало на вогнутое сферическое зеркало, что позволило системе Уайта обеспечить высокий коэффициент пропускания, в основном ограниченный только потерями на поглощение и рассеяние на зеркалах.

Основу системы Уайта составляют три вогнутых сферических зеркала с одинаковыми радиусами кривизны (рисунок 1.15). Два зеркальных объектива, обозначенные как $A$ и $B$, находятся на двойном фокусном расстоянии от формирующихся промежуточных изображений входной щели на полевом зеркале *C*. Центры кривизны объективов располагаются на поверхности полевого зеркала, а центр кривизны самого полевого зеркала находится ровно посередине между *A* и *B*. На рисунке 1.15 центры кривизны зеркал *A*, *B* и *C* указаны строчными буквами *a*, *b* и *c* соответственно. Подобная конфигурация зеркал образует систему сопряженных фокусов: световые лучи, исходящие из любой точки зеркала А, фокусируются полевым зеркалом в соответствующей точке зеркала *B*, а все лучи из этой точки вновь фокусируются в исходной точке на зеркале *A* [81].

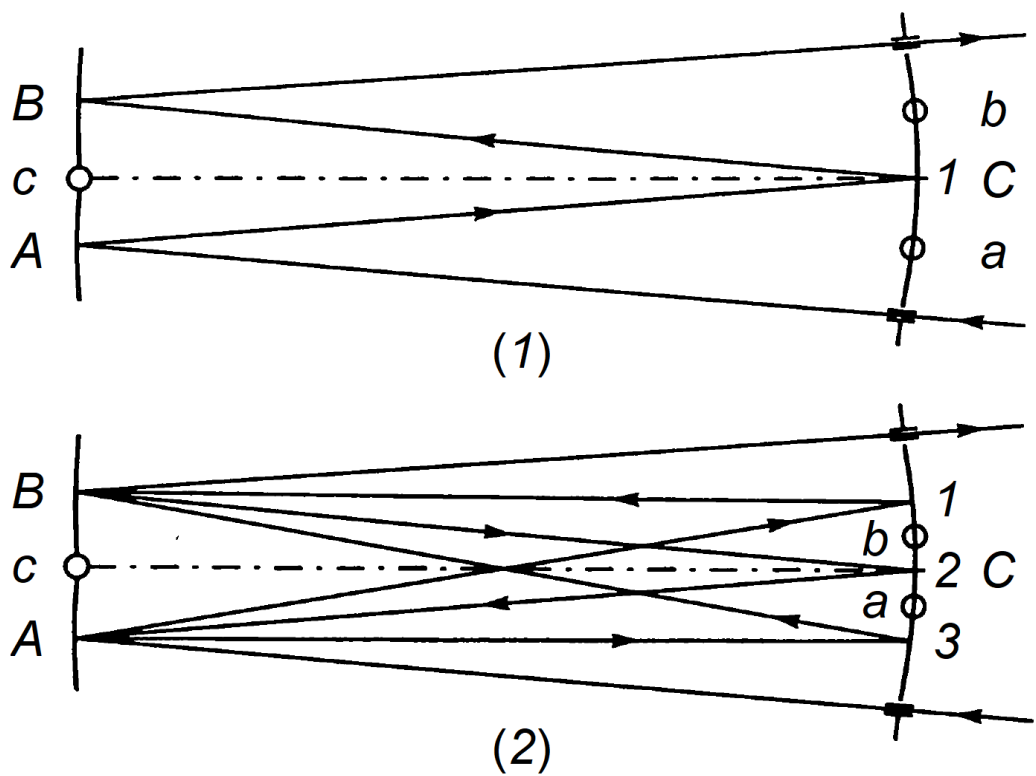

Рисунок 1.15 – Классическая многоходовая система Уайта на четыре (1) и восемь (2) проходов. *A* и *B* – объективные зеркала, *C* – полевое зеркало.

Свет попадает в систему через щель вблизи зеркала *C*, затем отражается последовательно от зеркала *A* на зеркало *C*, оттуда на *B*, обратно на *C* и снова на *A*. Этот цикл отражений продолжается до тех пор, пока следующее изображение входной щели не достигнет выходной щели, расположенной на противоположном краю зеркала *C*.

Положение каждого следующего изображения определяется правилом: вблизи центра кривизны сферического зеркала точка объекта и её изображение лежат на концах отрезка, середина которого совпадает с центром кривизны. Таким образом, первое изображение входной щели (обозначенное как *1* на рисунке 1.15), созданное зеркалом *A*, находится на таком же расстоянии от его центра кривизны, как и сама входная щель. Поскольку центр кривизны зеркала *C* находится посередине между зеркалами *A* и *B*, зеркало *C* формирует изображение зеркала *A* на зеркале *B*.

Сферические зеркала, расположенные на двойном фокусном расстоянии от полевого зеркала, формируют на нём промежуточные изображения источника с единичным увеличением, при этом каждое следующее изображение инвертируется относительно предыдущего. Световые пятна на сферических зеркалах также инвертированы друг относительно друга, поэтому любые дефекты на этих зеркалах не накапливаются с каждым проходом света, точно совмещаясь с их изображениями [81].

Число проходов регулируется поворотом зеркал $A$ и $B$ относительно друг друга в горизонтальной плоскости, что изменяет расстояние между их центрами кривизны и, соответственно, увеличивает или уменьшает число проходов. Если сферические зеркала симметричны относительно зеркала $C$ и его центра кривизны, то расстояние между последовательными изображениями на этом зеркале равно расстоянию между центрами кривизны зеркал $A$ и $B$. Длина зеркала, поделенная на это расстояние, определяет число проходов $N$, которое может быть рассчитано по формуле:

$$N = 2(n + 1), \tag{1.30}$$

где $n$ – число промежуточных изображений на полевом зеркале $C$ – 1, 3, 5, 7 и т.д.

Система Уайта малочувствительна к юстировке. При горизонтальном смещении зеркальных объективов изображения на полевом зеркале будут появляться парами, а при вертикальном смещении они будут чередоваться, поднимаясь и опускаясь. В обоих случаях потерь интенсивности не происходит.

Если юстировка полевого зеркала нарушена, первое изображение зеркала $A$ не будет полностью отображаться на зеркале $B$, что приведет к потере части света. Однако после первого отражения дальнейших потерь не будет, так как свет, достигший $B$, продолжит отражаться между одинаковыми точками на зеркалах $A$ и $B$ [81].

В системе Уайта сферические зеркала функционируют как линзы, разделенные удвоенным фокусным расстоянием, при этом луч многократно проходит расстояние, равное четырем фокусным расстояниям. Это делает ячейку Уайта особенно подходящей для создания длинного оптического пути в небольшом объеме образца. Такие ячейки обычно оснащаются зеркалами с защитным покрытием из серебра (protected Ag), обеспечивающими хорошее отражение в широком диапазоне длин волн от 450 нм до 8 мкм. Пример такой кюветы представлен на рисунке 1.16.

Однако система Уайта не лишена недостатков. Во-первых, из-за многократных отражений от одного и того же зеркала в выходной пучок попадает рассеянный свет, прошедший различное число проходов, что приводит к искажениям выходного сигнала, особенно заметным при использовании когерентных источников света.

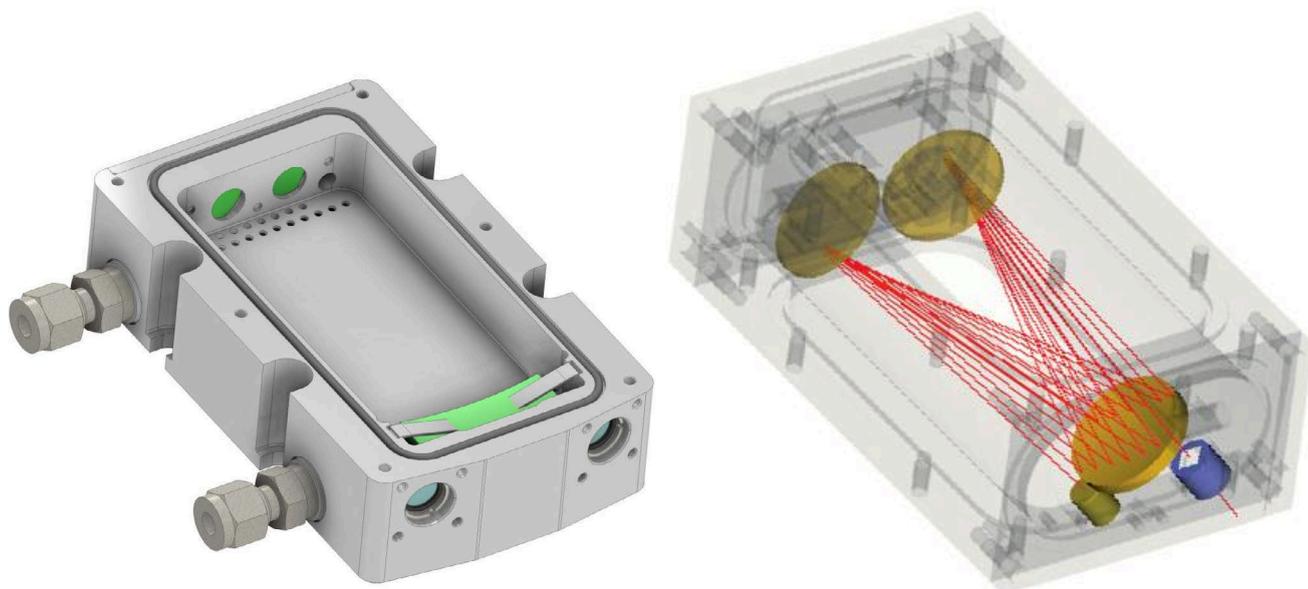

Рисунок 1.16 – Примеры МГК для задач газоанализа, построенных на системе Уайта.

Во-вторых, при многократном отражении луча ошибки в юстировке каждого сферического зеркала могут приводить к перекрытию изображений, что ограничивает максимальную длину пути.

Основным же недостатком системы Уайта является индивидуальное крепление зеркальных объективов – повреждение юстировочной оснастки кюветы приведет к накоплению ошибки в положениях изображений.

### 1.2.3. Многопроходные кюветы Эрриотта

Первое детальное исследование систем, состоящих из двух коаксиальных вогнутых зеркал, которые фактически представляют собой оптический резонатор с внеосевым ходом лучей, было выполнено в 1964 году Дональдом Эрриоттом [82].

Если внеосевой лазерный пучок вводится в резонатор со сферическими зеркалами, на зеркалах формируются эллипсы световых пятен. Правильно выбрав точку входа и углы наклона пучка, можно расположить пятна вдоль окружностей с заданным радиусом. В таком случае лучи, движущиеся внутри резонатора, будут располагаться на поверхности гиперболоида, как показано на рисунке 1.17.

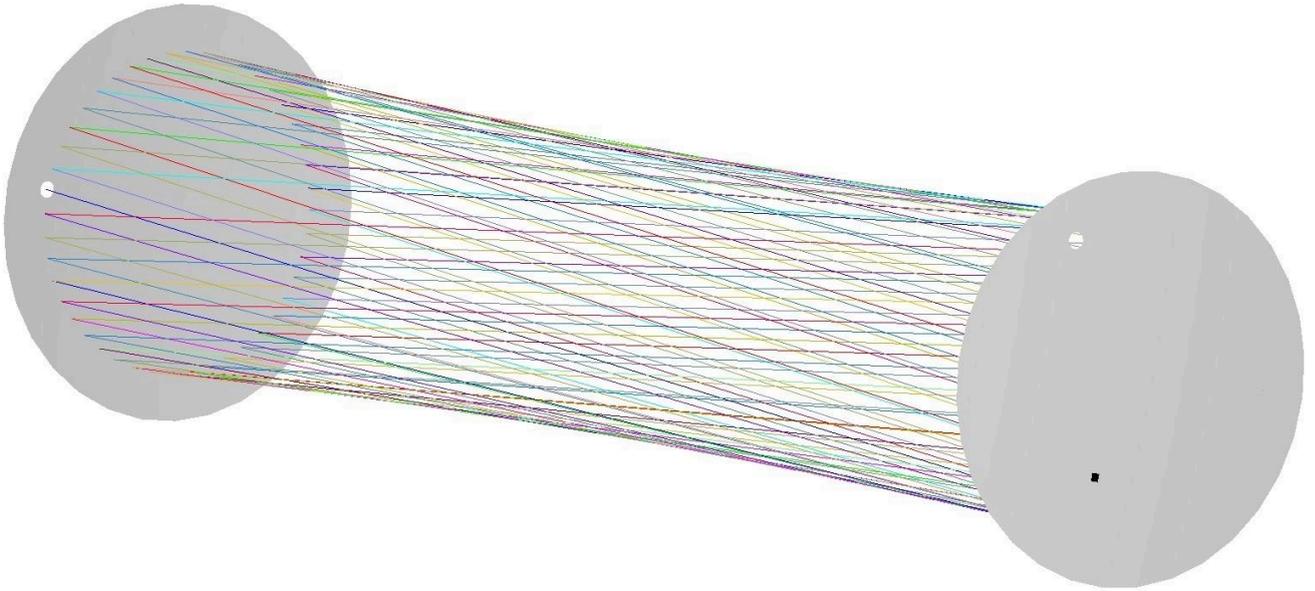

Рисунок 1.17 – Ход лучей внутри резонатора Эрриотта.

Если выполняется условие

$$2n\theta = 2\pi,$$ (1.31)

где *n* – целое число, *2θ* – угловое смещение светового пятна на одном из зеркал после каждого двойного прохода, лучи будут двигаться по замкнутым траекториям, возвращаясь в начальную точку после *n* двойных проходов. Это позволяет организовать ввод и вывод излучения в одной точке, разделяя пучки по направлению. Угловое смещение на каждом проходе определяется по формуле

$$cos\theta = 1 - \frac{d}{2f},$$ (1.32)

где *d* – расстояние между зеркалами, *f* – фокусное расстояние зеркал [82]. То есть замкнутость траектории зависит лишь от отношения *d/f*. Луч вернется в исходную точку независимо от угла ввода в резонатор, а траектории световых пятен в общем случае будут эллиптическими. Однако при эллиптических траекториях значение *θ* утрачивает простую геометрическую интерпретацию. После *n* циклов пятно на зеркале будет идентичным по размеру входному пятну, хотя промежуточные изображения могут изменять размер в зависимости от параметров входного пучка. Для минимизации размера всех пятен кривизна волнового фронта пучка совпадала с кривизной зеркала, а апертура ограничивалась диаметром основной моды на зеркале.

Если выполняется условие

$$2n\theta = 2m\pi,$$ (1.33)

где *n* и *m* – целые числа при условии *m* ≠ *n*, возможны более сложные замкнутые траектории. В этом случае точка пересечения пучка с зеркалом совершает *m* оборотов по эллипсу, прежде чем вернуться в исходную позицию. Такие конфигурации дают большое количество проходов, но используются редко из-за близости промежуточных изображений к выходному и сложности настройки [81].

МГК Эрриотта со сравнительно большим оптическим путем используются в основном для спектроскопии поглощения молекулярных газов со слабым поглощением, например, для мониторинга окружающей среды, процессов горения, медицинской диагностики, фундаментальной атомной и молекулярной физики. По сравнению с классическими системами с большим оптическим путем, такими как кюветы Уайта, ячейка Эрриотта более устойчива к малым возмущениям. Основные преимущества ячейки Эрриотта заключаются в том, что она обеспечивает достаточно длинный оптический путь в относительно компактном корпусе, при этом она проще в применении чем высокоточные оптические резонаторы, которые обычно требуют пространственного согласования мод, точной оптической юстировки или резонансного возбуждения. Пример закрытой МГК Эрриотта представлен на рисунке 1.18.

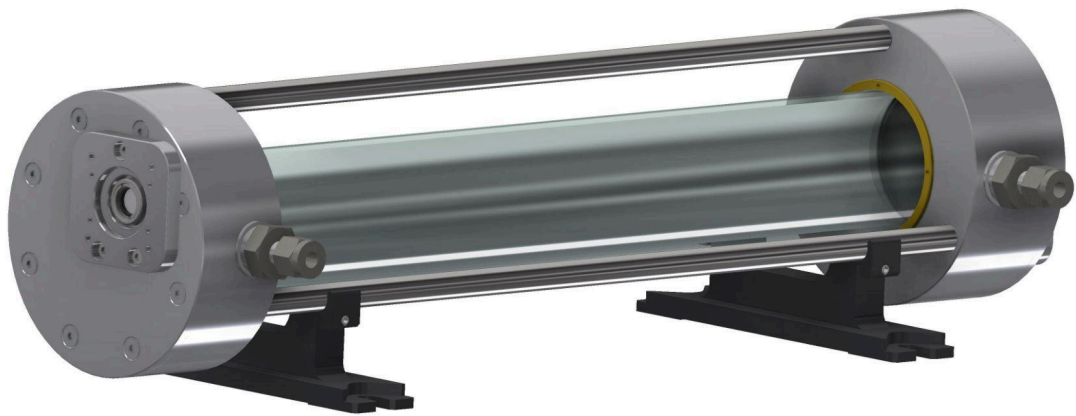

Рисунок 1.18 – Внешний вид закрытой МГК Эрриотта.

Однако кюветы Эрриотта имеют и недостатки: помимо требований к лазерным источникам, выполняющихся в большинстве случаев при использовании современных полупроводниковых лазеров, эти системы позволяют изменять число проходов не более чем в 1.2 раза при довольно трудоемком процессе перестройки. Они также неэффективны с точки зрения использования внутреннего объема из-за расположения световых пятен на окружностях или эллипсах.

Поскольку в кюветах Эрриотта происходит многократное отражение, они чувствительны к небольшим тепловым сдвигам, которые могут повлиять на юстировку выходного сигнала. Поэтому сферические зеркала, используемые в таких системах, обыкновенно изготавливаются на подложках из кварцевого стекла, что обеспечивает повышенную термостабильность и помогает сохранить оптическую центровку выходного сигнала.

Рассмотрим на примере проекта, не вошедшего в диссертационную работу, реальный расчет такой системы в среде проектирования и моделирования оптических систем Zemax. Резонатор Эрриотта создается с помощью двух сферических зеркал с одинаковым фокусным расстоянием, причем по крайней мере одно из зеркал содержит отверстие для ввода/вывода излучения (центрированное или внеосевое). Принципиальная схема резонатора Эрриотта приведена на рисунке 1.19.

В случае применения сферических зеркал фокусное расстояние будет соответствовать половине радиуса кривизны. Примем $f = 500$ мм, $d = 400.7$ мм, $n = 44$, $m = 13$. Положение пятен на зеркалах определяется углом падения и положением зеркал. Когда световой луч входит в систему под определенным углом, он формирует последовательность пятен на поверхности зеркал, которые расположены по окружности.

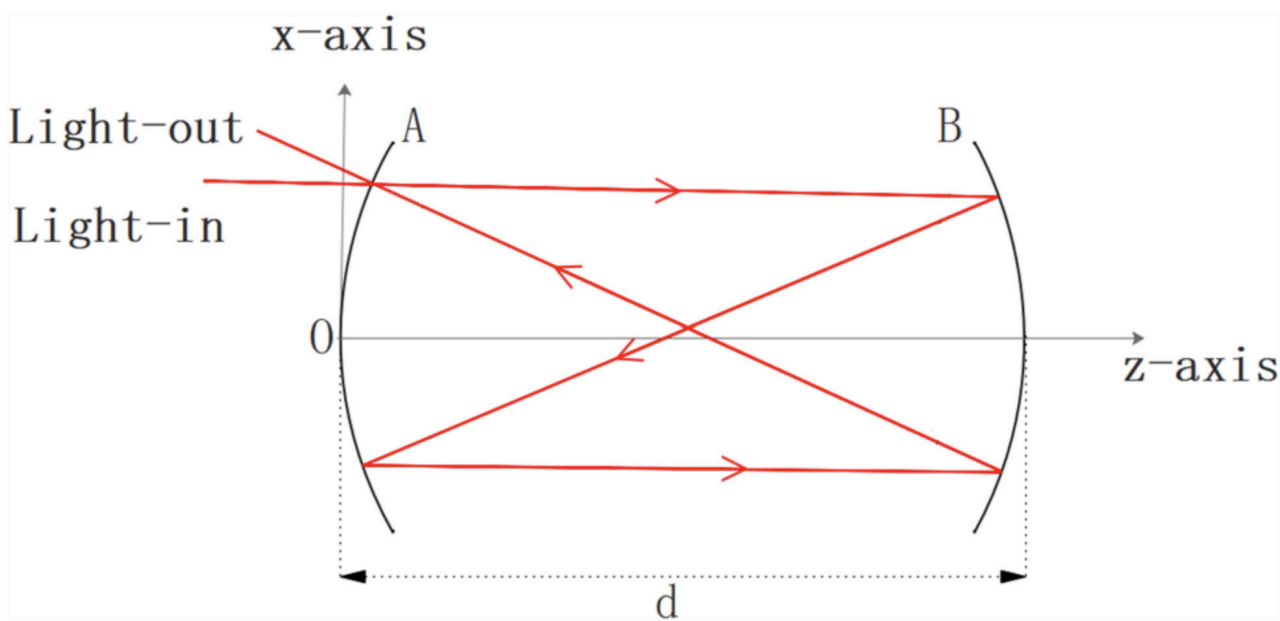

Рисунок 1.19 – Принципиальная схема многопроходной системы Эрриотта. Луч отражается двумя одинаковыми соосными сферическими зеркалами А и В в плоскости x-z декартовой системы координат [83].

Согласно расчету Zemax для паттерна с диаметром 76 мм выбранную длину базы 400.7 мм надо скорректировать до 402 мм, поскольку формула (1.32) работает в параксиальном приближении, а в данном случае угол входа луча достаточно большой ($2.2^0$ и $-4.4^0$ по двум осям), как показано на рисунке 1.20.

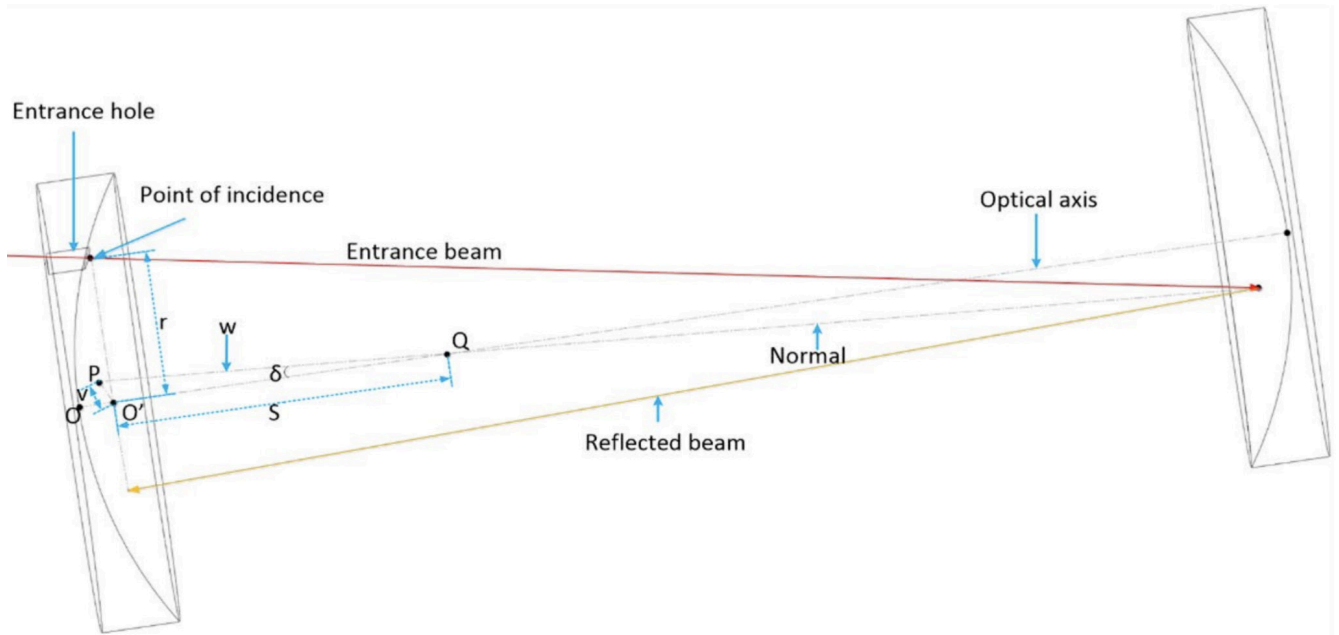

Рисунок 1.20 – Геометрическое представление входящего луча и первого отражения от сферического зеркала [83].

Пятна распределяются так, что каждый последующий отраженный луч попадает в новую точку на зеркале, формируя характерный рисунок, показанный на рисунке 1.21.

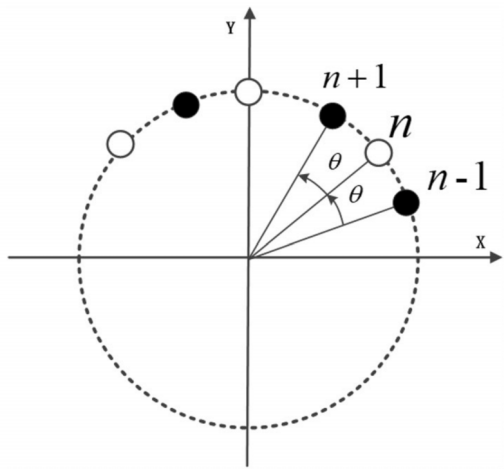

Рисунок 1.21 – Диаграмма распределения пятен на выходном зеркале. Сплошные пятна представляют собой точки отражений на выходном зеркале, а кружки – отражения на входном зеркале [83].

Для обеспечения равномерного распределения пятен необходимо точно контролировать углы падения и положения зеркал. Расстояние между зеркалами также играет важную роль в формировании пятен, так как оно влияет на количество отражений и, следовательно, на общую оптическую длину пути. Для оптимизации распределения пятен и обеспечения стабильной работы системы с переменной оптической длиной пути важно учитывать взаимодействие всех параметров, включая угол падения, радиус кривизны зеркал и расстояние между ними. Это позволяет достичь точного контроля над оптическими характеристиками системы и обеспечивает высокую точность измерений [83]. Расположение пятен лазерного луча на входном и выходном зеркалах показано на рисунке 1.22.

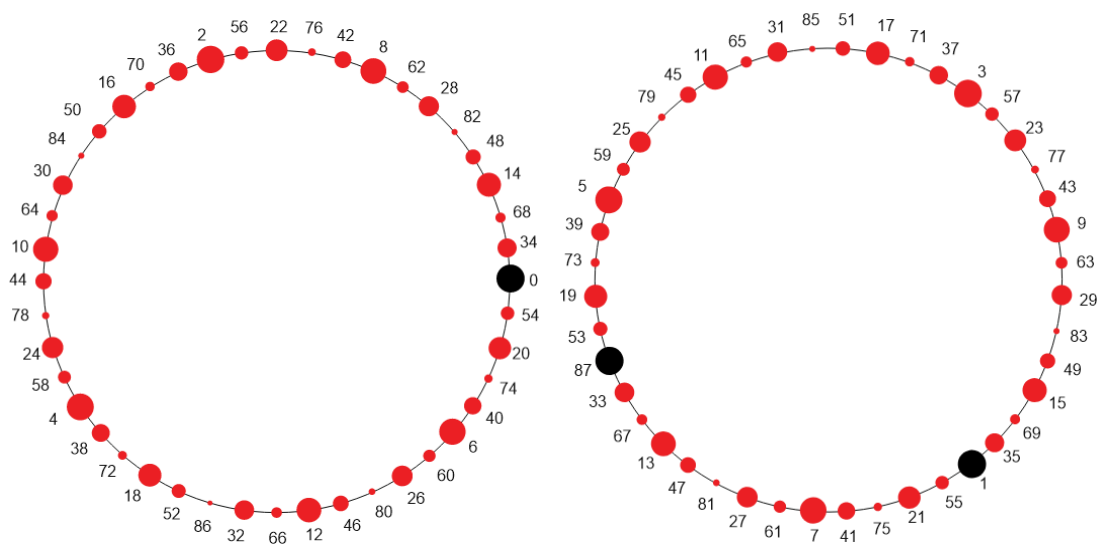

Рисунок 1.22 – Входное зеркало (слева) и выходное зеркало (справа). Выходное зеркало повернуто на 53° относительно входного. Диаметр пятен иллюстрирует интенсивность падающего излучения, черным выделены положение входящего лазерного пучка, первого пятна и выхода луча из системы.

При такой конфигурации пятен разность хода между соседними пятнами – 34 прохода, интерференция между пятнами паттерна при такой разности хода будет крайне мала, рядом с выходным пятном 87 находится не очень яркое пятно 33. Получаем схему с 87 проходами: оптический путь 34.97 м при диаметре паттерна 76 мм.

Система Эрриотта позволяет работать одновременно с несколькими источниками излучения при использовании коаксиальных паттернов. Поэтому неэффективность использования внутреннего объема кюветы можно отчасти компенсировать, добавив в рассчитанную систему внутренний паттерн.

Пусть расстояние между соседними пятнами будет большим, а диаметр паттерна составит 64 мм. Для такого диаметра 40 пятен на зеркале обеспечивает 5 мм между центрами соседних пятен. Замкнем паттерн на 33 пятне. Тогда будем иметь длину оптического пути 13.17 м при угле ввода $1.85^0$ и $-3.6^0$ по двум осям. Вид внутреннего паттерна на обоих зеркалах показан на рисунке 1.23.

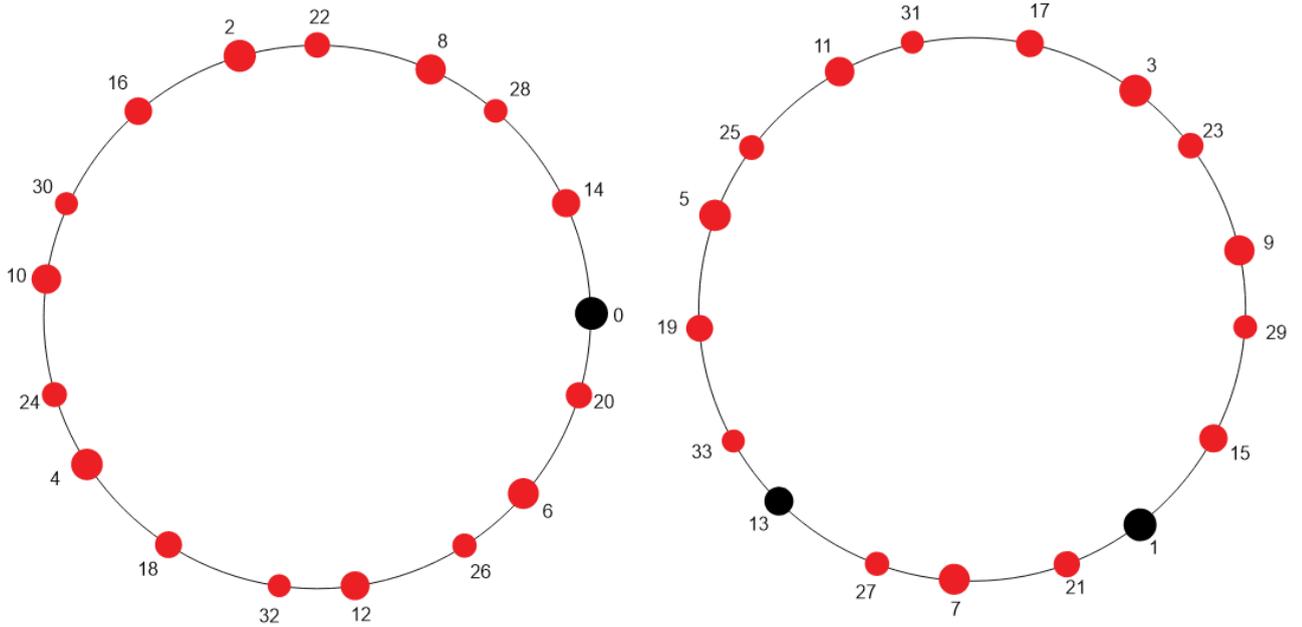

Рисунок 1.23 – Входное зеркало (слева) и выходное зеркало (справа). Выходное зеркало повёрнуто на 43° относительно входного.

Иллюстрация внешнего и внутреннего паттернов на одном зеркале представлена на рисунке 1.24. Таким образом, получен расчет входного и выходного зеркал диаметром 90 мм, радиусом кривизны 1000 мм, чистой апертурой >85.5 мм и с двумя отверстиями 3.5 мм для двух паттернов диаметрами 64 мм и 76 мм.

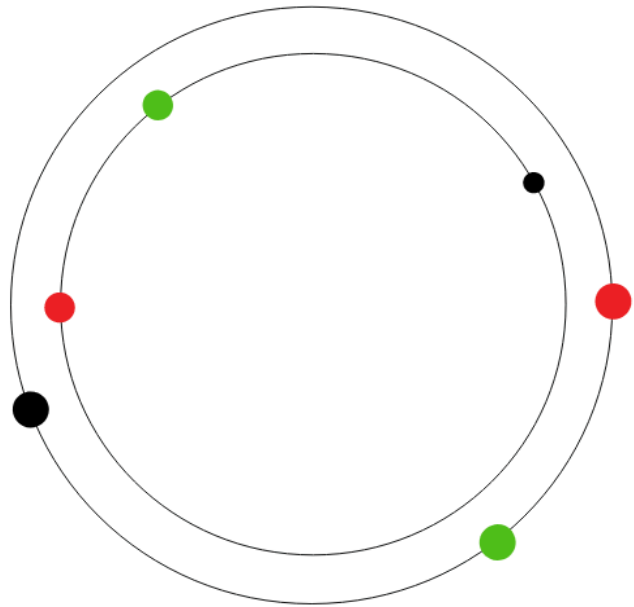

Рисунок 1.24 – Внешний и внутренний паттерны. Черное пятно – вход, красное – выход, зеленое – точка прицеливания. Внутренний паттерн повернут на 180 градусов.

Материалом подложки для работы в ИК диапазоне спектра выбрано кварцевое стекло, напыленного на подложку покрытия – золото с защитным слоем (protected Au). Также для удобства работы предусмотрено окошко для юстировки. Визуализация рассчитанного зеркала показана на рисунке 1.25.

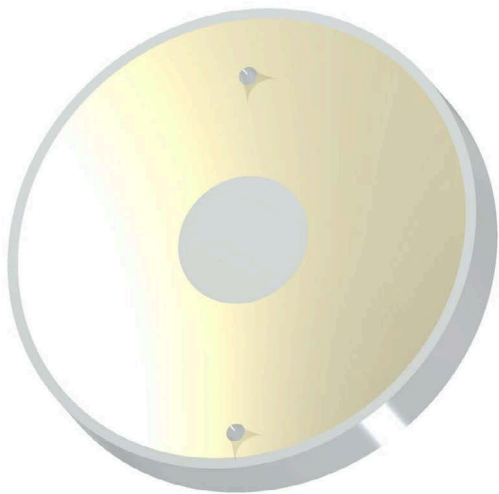

Рисунок 1.25 – Внешний вид рассчитанного входного зеркала для многопроходной системы Эрриотта.

Примером современного применения системы Эрриотта в широко используемых спектрометрах может служить представленный в 2011 году анализатор метана LI-7700 (LI-COR Biosciences) с открытой кюветой Эрриотта [84]. Это полевой прибор для измерения потоков метана методом турбулентных пульсаций [85], в котором используется перестраиваемый диодный лазер в сочетании с открытой конфигурацией кюветы Эрриотта (рисунок 1.26) с базой 0.47 м и длиной оптического пути 30 м. Содержание метана измеряется при помощи методики модуляционной спектроскопии.

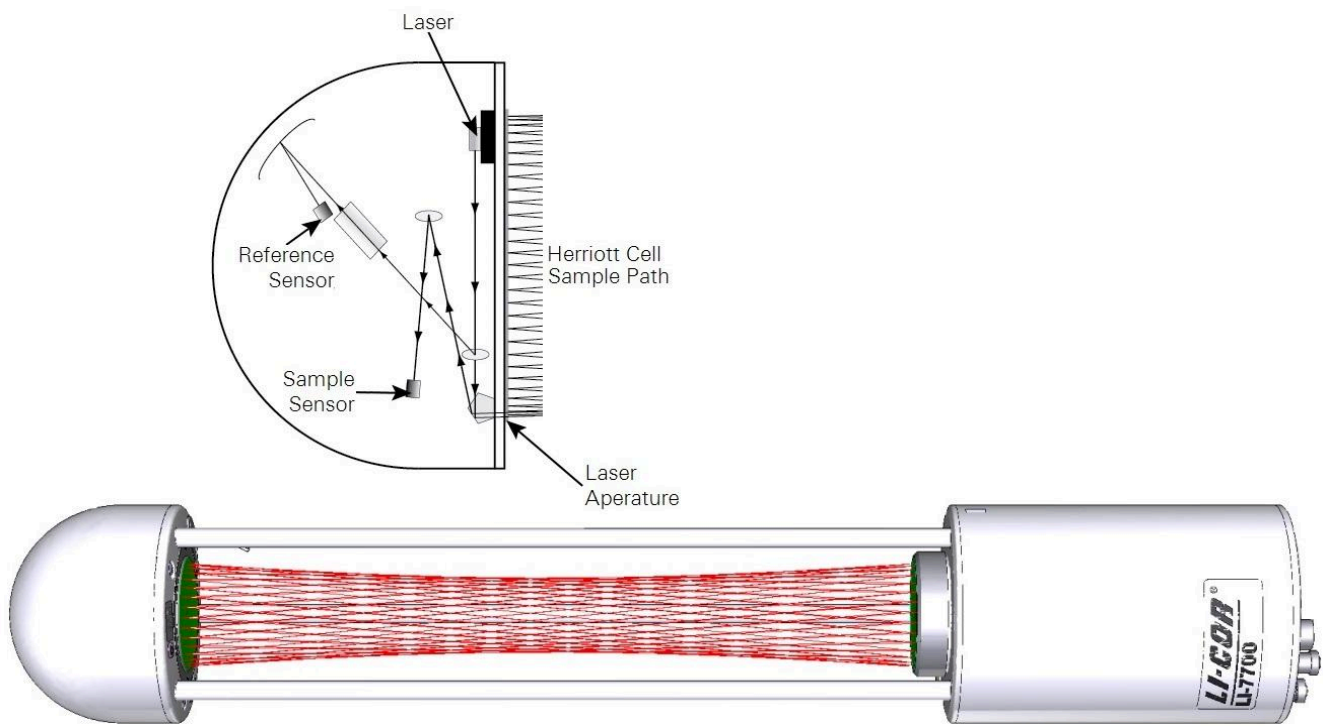

Рисунок 1.26 – Иллюстрация корпуса LI-7700 с открытой кюветой Эрриотта.

### 1.2.4. Многоходовая система Чернина

Многоходовые матричные системы (ММС), обладающие чрезвычайно большим числом отражений, отличающиеся высокой стабильностью и компактным расположением изображений в виде прямоугольной матрицы, были разработаны в СССР Семеном Моисеевичем Черниным совместно с Евгенией Григорьевной Барской в 1982 году [86]. Эти системы включают в себя конфигурации с тремя и четырьмя сферическими зеркалами, представляющие собой комбинации нескольких несогласованных, но взаимно связанных конфокальных сферических резонаторов, с заранее заданным положением выходного излучения. Основные принципы разработанных матричных систем [87] таковы:

1. Изображения на полевых зеркалах формируются в виде многорядной матрицы. Вертикальное смещение изображений осуществляется за счет оптической связи полевых и объективных зеркал.

2. Зеркальные объективы жестко закреплены на общей поворотной плате, что обеспечивает их устойчивость к вибрациям. Компактная многострочная матрица промежуточных изображений на полевых зеркалах позволяет существенно увеличить длину оптического пути при минимальном числе зеркал. Снижение количества зеркал в системе улучшает стабильность изображений по сравнению с многоходовыми многозеркальными системами Ханца [88], Накамуры [89] и Уайта [90]. Матричные системы унаследовали рациональные элементы системы Уайта, такие как зеркальные объективы и полевые зеркала с одинаковым радиусом кривизны, что позволяет работать с источниками большой угловой апертуры.

Вертикальное смещение изображений в ММС осуществляется исключительно с помощью отражающей оптики, что отличает их от системы Шульца-Дюбуа [91], в которой используется уголковый призменный отражатель с фокусирующей линзой для вертикального разнесения изображений [81].

Высокая виброустойчивость матричных систем объясняется жесткой механической связью зеркальных объективов, что предотвращает накопление ошибок, характерных для системы Уайта, где объективы механически разделены. Эта концепция жесткой связи берет начало от первых разработок спектральных приборов, таких как работы Уодсворта [92].

На рисунке 1.27 показана схема трехобъективной матричной системы. Изображения на поверхности полевых зеркал обозначены пунктирными кружками, а номера в этих кружках указывают последовательность формирования изображений при настройке системы на 30 проходов. Спаренные зеркальные объективы 3 и 4 жестко закреплены на поворотной плате составного подвижного держателя. Объективы соединены под фиксированным углом, таким образом, что расстояние между центрами их кривизны равно половине шага изображения по горизонтали.

Принцип работы системы следующий: излучение от источника попадает на зеркальный объектив 3 через входное отверстие 1, формируя промежуточное изображение на поверхности основного полевого зеркала 8. Поворот платы 5 задает горизонтальное смещение, а наклон держателя 6 — вертикальное смещение изображения. После попадания на полевое зеркало 8 излучение отражается на объектив 4, создавая второе изображение. Спаренные объективы 3 и 4 поочередно создают смещенные изображения до тех пор, пока не образуются две строки изображений. После отражения от вспомогательного полевого зеркала 9 излучение направляется на дополнительный зеркальный объектив 10, затем возвращается на полевое вспомогательное зеркало 9 с вертикальным смещением. Излучение снова попадает на зеркальный объектив 3, и процесс повторяется, пока последнее изображение не выйдет через выходное отверстие 2 [81].

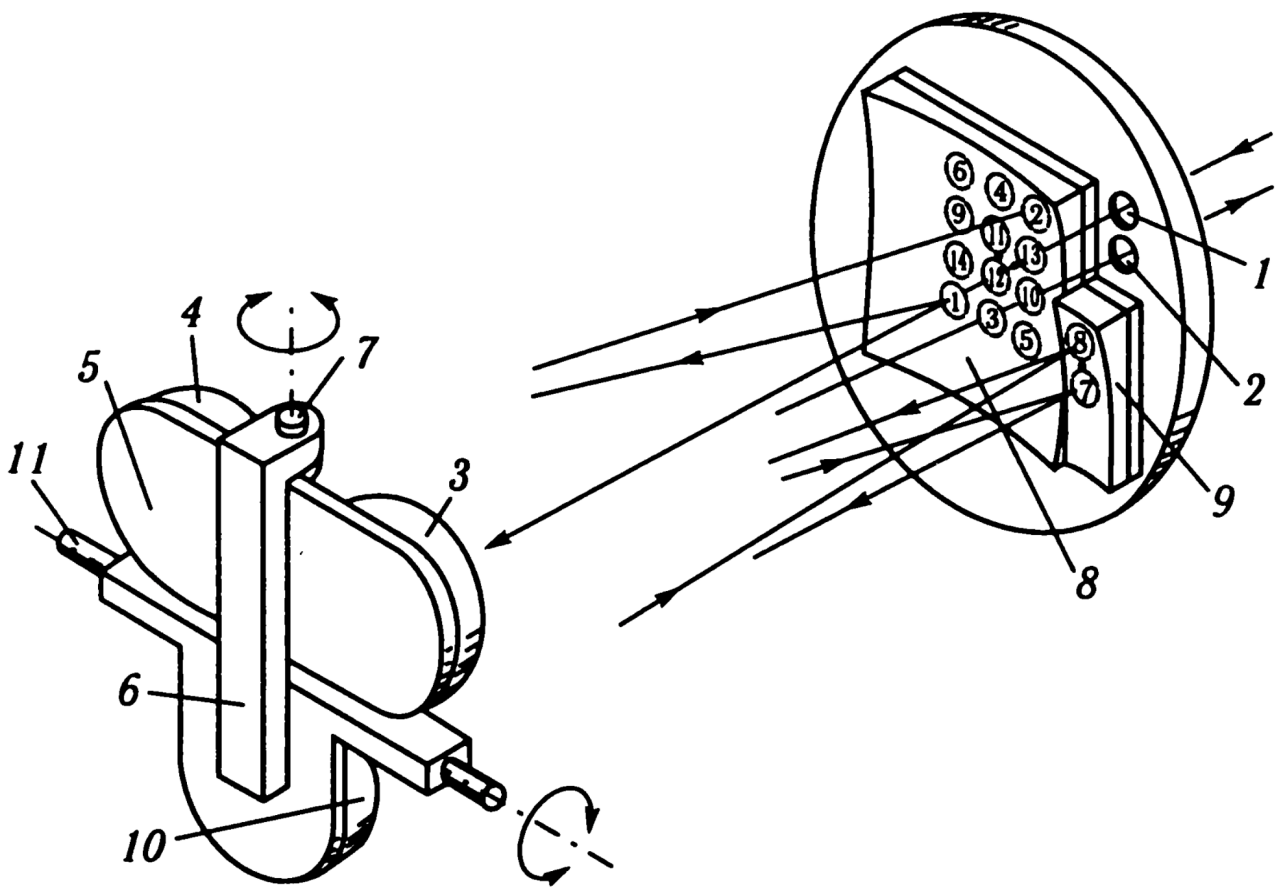

Рисунок 1.27 – Многоходовая матричная 3-объективная система: 1 – входное отверстие; 2 – выходное отверстие; 3, 4 – спаренные зеркальные объективы; 5 – поворотная плата; 6 – подвижный держатель; 7 – вертикальная ось вращения; 8 – основное полевое зеркало; 9 – вспомогательное полевое зеркало; 10 – дополнительное полевое зеркало; 11 – горизонтальная ось вращения [81].

Число строк в матрице изменяется наклоном держателя 6 с объективами, а число столбцов – поворотом платы 5 с жестко закрепленными объективами. При этом центры кривизны объективов скользят по поверхности полевого зеркала, оставаясь на постоянном расстоянии друг от друга. Важно отметить, что последнее изображение достигает выходного отверстия 2 только в фиксированных положениях платы 5 и держателя 6.

Число проходов можно рассчитать по формуле

$$N = 2(mn - 1),\qquad(1.34)$$

где $m$ – число строк, кратное двум, а $n$ – число столбцов $\geq 2$.

В случае четырехобъективной ММС, пример реализации которой представлен на рисунке 1.28, объективы установлены на держателе, который может поворачиваться вокруг вертикальной и горизонтальной осей.

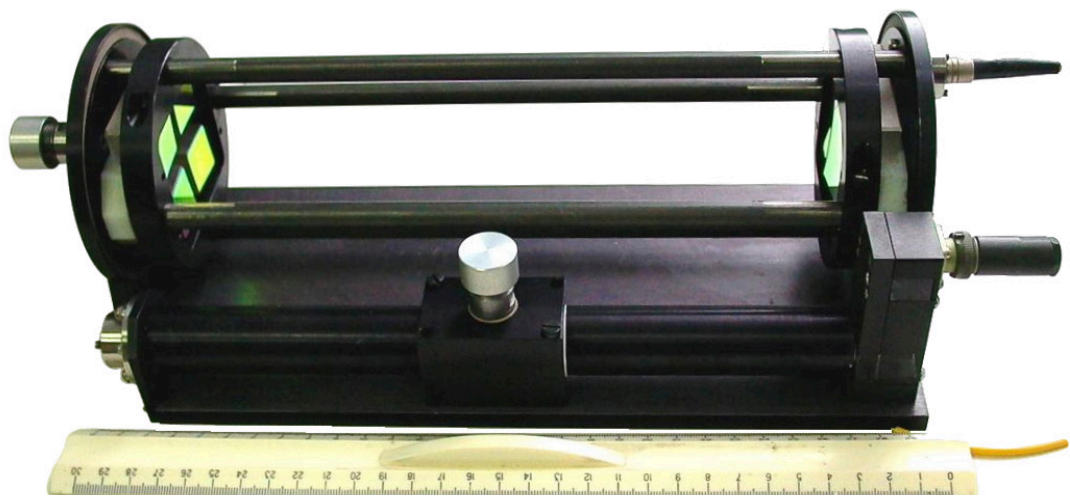

Рисунок 1.28 – Четырехобъективная многоходовая матричная кювета ИОФ РАН – «Полюс» с базовой длиной 25 см.

Работа системы аналогична трехобъективной, но включает две пары спаренных объективов друг под другом. Изменение числа строк и столбцов в матрице осуществляется наклоном и поворотом держателя соответственно. Последнее изображение достигает выходного отверстия так же только при фиксированных смещениях первого изображения.

Формула для расчета числа проходов в четырехобъективной системе будет иметь следующий вид

$$N = (m - 1)(4n - 2),\qquad(1.35)$$

где $m$ – число строк, кратное двум, а $n$ – число столбцов $\geq 2$. Пример расчетной и экспериментальной матриц изображений входного лазерного луча на полевых зеркалах показан на рисунке 1.29.

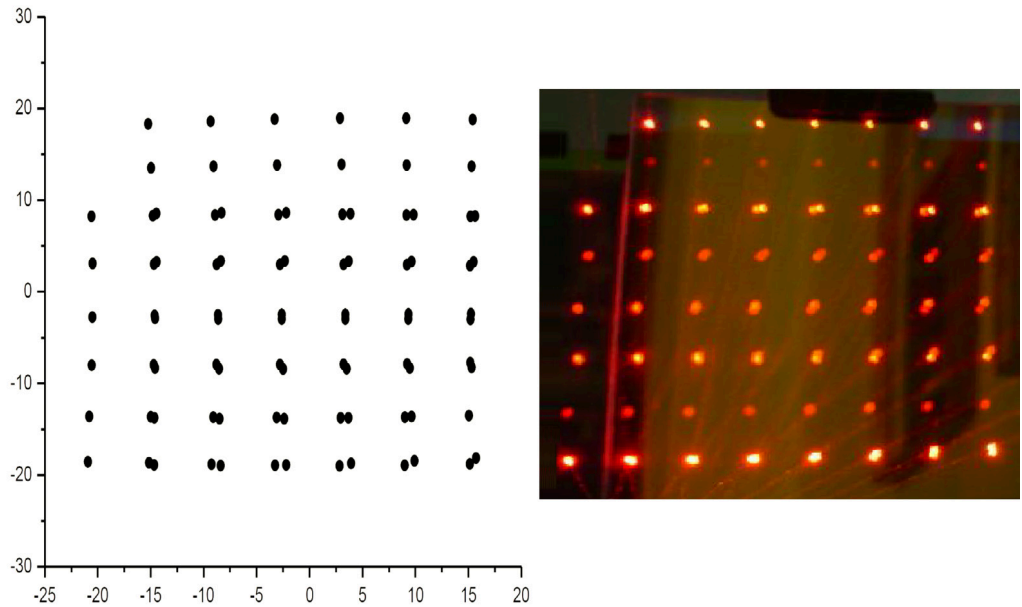

Рисунок 1.29 – Моделирование матрицы изображений на полевых зеркалах (слева) и экспериментальная картина (справа) [93].

Четырехобъективная система обладает рядом преимуществ, включая повышенную точность и надежность, полную самокомпенсацию ошибок и уменьшенные аберрации поля зрения. Система позволяет значительно увеличить длину оптического пути, достигая нескольких сотен проходов луча (рисунок 1.30), сохраняя стабильность изображений в условиях вибраций.

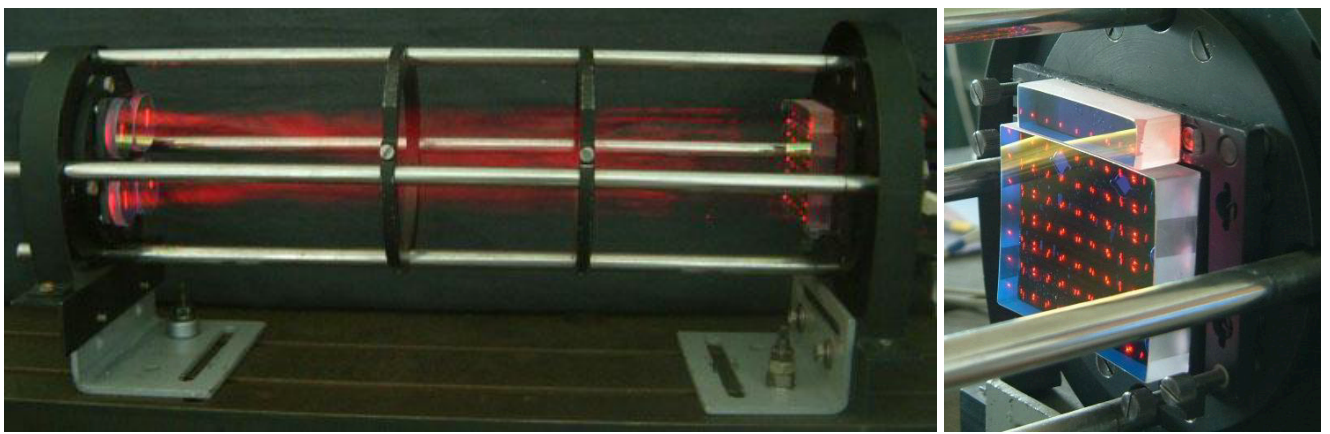

Рисунок 1.30 – МГК, построенная по четырехобъективной матричной схеме Чернина (слева). Матрица изображений на полевых зеркалах при длине оптического пути 150 м (справа).

## 1.2.5. Зеркально-кольцевая многопроходная система

Отказавшись от использования традиционных зеркальных систем, С.М. Чернин совместно с коллегами из Института химической физики АН СССР в 1987 году разработал кольцевой рефлектор с многократным отражением. Его конструкция претерпела несколько этапов изменений, начиная с пустотелого многогранника и вогнутого цилиндра [94], и в конечном итоге была преобразована в цилиндрическое кольцо с внутренней сферической поверхностью [95]. Ключевым элементом наиболее усовершенствованной версии кольцевой системы с внутренней сферической отражающей поверхностью, разработанной С.М. Черниным [96], является кольцевой рефлектор, представляющий собой симметричный вогнутый сферический пояс с внутренней зеркальной поверхностью. Излучение от источника, направленное в кольцевой рефлектор, многократно отражается от его сферической поверхности. На рисунке 1.31 слева от кольцевого рефлектора показана осветительная часть оптической системы, а справа — приемная часть [81].

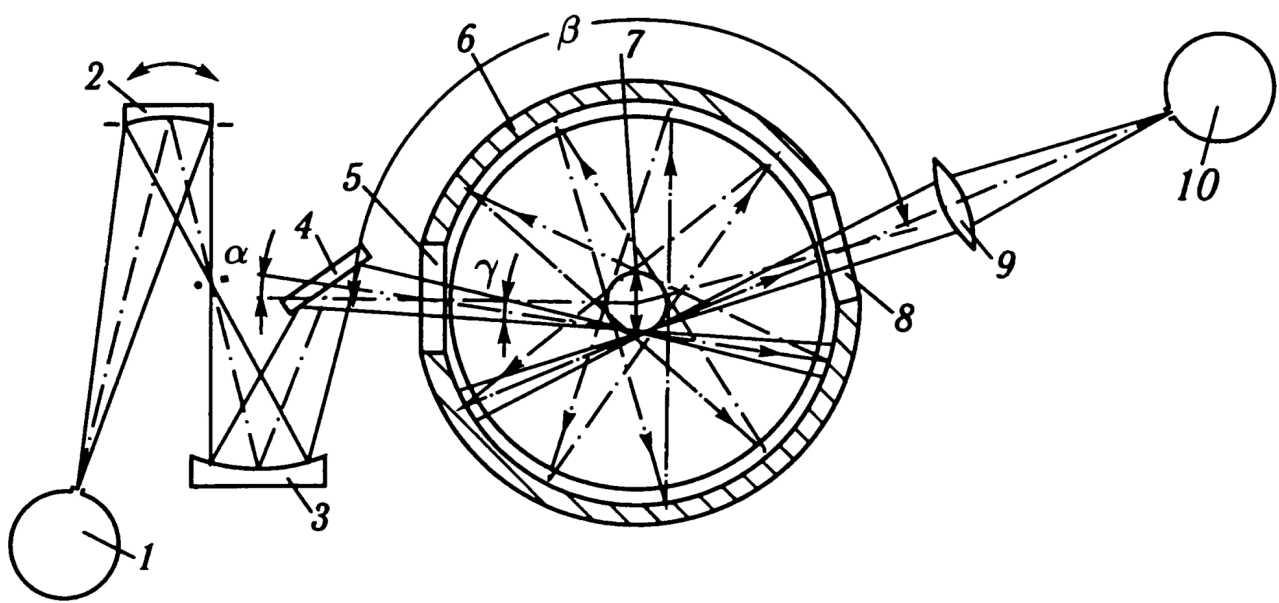

Рисунок 1.31 – Многоходовая зеркально-кольцевая система: 1 – источник излучения; 2 – подвижное сферическое зеркало; 3 – неподвижное сферическое зеркало; 4 – неподвижное плоское зеркало; 5 – входное окно; 6 – кольцевой рефлектор; 7 – окружность, на которой располагаются фокусы («кольцо фокусировки»); 8 – выходное окно; 9 – фокусирующая линза; 10 – приемник излучения [81].

Излучение, направленное от источника к рефлектору, многократно отражается от зеркальной поверхности, что увеличивает длину оптического пути. Расходящийся пучок от источника высокой яркости 1 попадает на подвижное сферическое зеркало 2, а отраженный сходящийся пучок направляется на неподвижное сферическое зеркало 3. Сферическое зеркало 2 формирует пространственное изображение источника между зеркалами 2 и 3. Сходящийся пучок от сферического зеркала 3 направляется на неподвижное плоское зеркало 4, откуда он отражается и поступает в входное окно 5 зеркального рефлектора 6. Сферическое зеркало 3 формирует изображение источника в центре зеркального рефлектора 6. Расходящийся пучок, попадающий на сферическую зеркальную поверхность рефлектора 6, многократно отражается. Все отраженные сходящиеся пучки пересекаются в пространстве, создавая кольцо фокусировки 7. После определенного числа проходов через кольцевое пространство пучок выходит из зеркального рефлектора через выходное окно 8. Затем излучение фокусируется на приемнике 10 с помощью линзы 9 [81].

Сферическое зеркало 2 установлено так, что оно может углово смещаться в меридиональной плоскости. Его крепление выполняет роль входного зрачка оптической системы. Поворот сферического зеркала 2 изменяет угол падения пучка α на входе в рефлектор, что, в свою очередь, влияет на количество проходов пучка через систему. Чем больше проходов, тем меньший диаметр кольца фокусировки 7. Постоянное положение пучка в плоскости входного окна 5 при изменении числа проходов обеспечивается передачей изображения входного зрачка на входное окно [81].

Число $N$ проходов пучка в кольцевом рефлекторе многопроходной системы может быть определено по следующему уравнению:

$$N = \frac{2\pi - \gamma}{4\alpha}, \tag{1.36}$$

$N$ принимает четные целые значения, и максимальное число $N_{max}$ проходов ограничено следующим образом:

$$N_{max} = \frac{\beta}{\gamma}, \tag{1.37}$$

где $\gamma$ – угловая апертура в направлении входного окна 5, $\alpha$ – угол между осью входного окна и оптической осью системы формирования пучка, а $\beta$ – угол между осями входного окна 5 и выходного окна 8 кольцевого рефлектора.

Размеры кольцевого рефлектора и угол падения входного пучка определяются следующими взаимосвязанными уравнениями:

$$\beta = \pi - arcsin\left(\frac{O_{in}}{2R}\right), \tag{1.38}$$

$$d = 2R \cdot sin\alpha, \tag{1.39}$$

где $O_{in}$ – диаметр пучка (перенесенное и всегда фиксированное изображение входного зрачка) на входном окне рефлектора 5, $R$ – радиус кривизны сферы рефлектора и $d$ – диаметр пространственного кольца фокусировки 7.

В 1994 году аналогичную систему представили в Германии [97]. Уже на новом витке технологического развития в 2018 году коллектив швейцарских разработчиков из Лаборатории загрязнения воздуха и экологических технологий научно-исследовательского института Empa представил собственную разработку кольцевой многопроходной системы для применения в лазерных спектрометрах [98]. Поскольку в компактных спектрометрах МГК представляет собой критический элемент с точки зрения достижимого размера и чувствительности, ими была представлена концепцию такой системы, объединившей в себе компактность, механическую жесткость и оптическую стабильность.

Представленная на рисунке 1.32 сегментированная круговая многоходовая система с длиной оптического пути до 10 м при общей массе менее 200 г позволяет эффективно и без помех складывать пучки. В отсутствии дополнительной оптики для предварительного формирования пучка, эта система демонстрирует нормированный уровень шума на уровне $10^{-4}$ ($2\sigma$) при частоте 1 Гц.

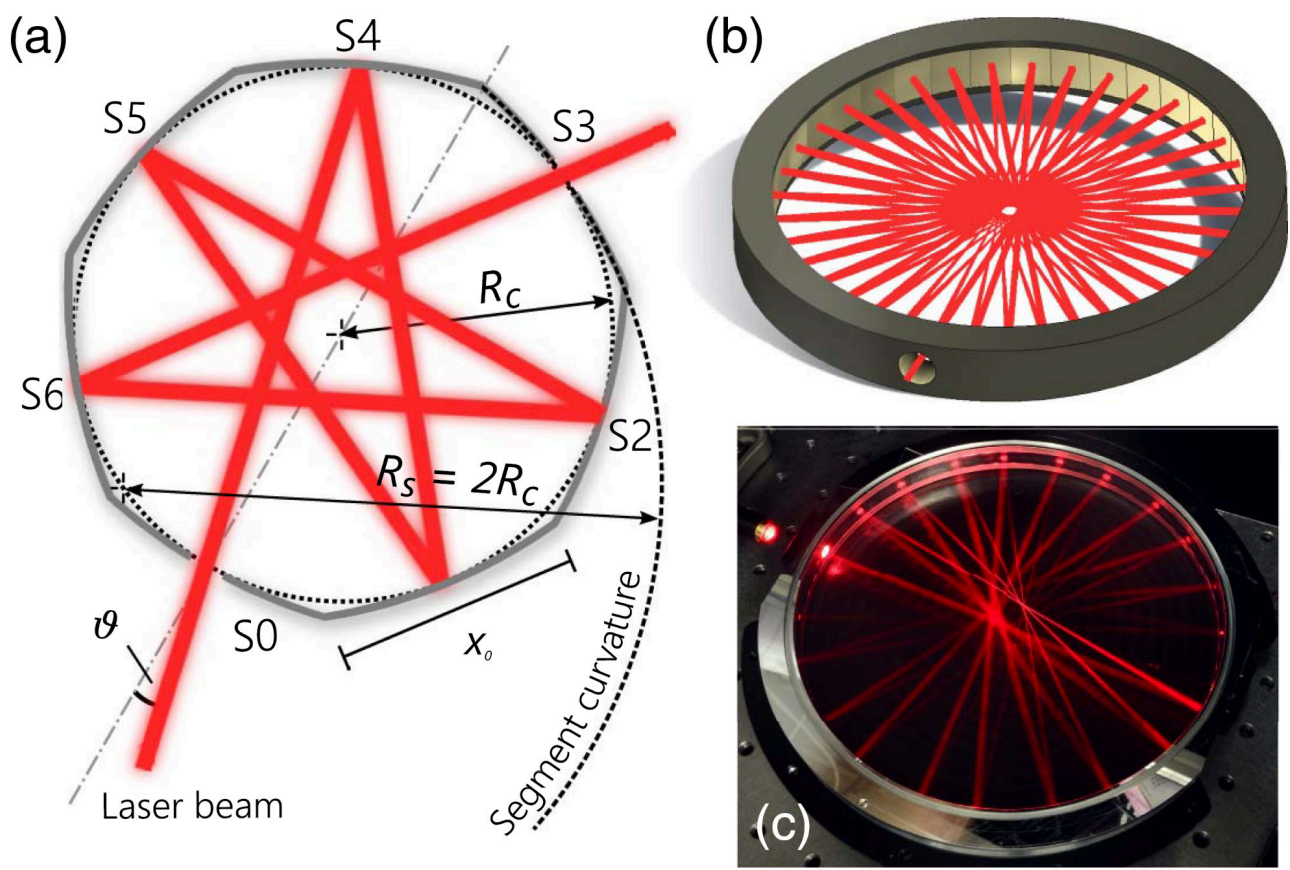

Рисунок 1.32 – **(a)** Схематическое изображение внутренней поверхности многопроходной кольцевой ячейки с семью конфокальными сегментами. Угол сопряжения $\vartheta$ пучка определяет картину отражения по правилам N-звездной полиграммы. **(b)** Концептуальная модель и моделирование трассировки лучей. **(c)** Фотография прототипа ячейки. Принцип складывания луча визуализируется с помощью красной трассировки лазера. Угол $\vartheta$ был выбран таким образом, чтобы реализовать 22 отражения [98].

Поскольку основным вызовом при создании МГК является объединение компактности конструкции с механической жесткостью, был предложен сегментированный круглый дизайн ячейки, который значительно уменьшает размеры устройства, сохраняя его структурную устойчивость.

Высокая оптическая стабильность, критически важная для точных измерений, была достигнута за счёт применения технологии алмазного точения, что обеспечило высокую точность изготовления оптических поверхностей и минимизацию дефектов, способных вызвать оптические искажения. Круговой дизайн обеспечивает эффективное и свободное от интерференции прохождение луча через кювету за счет оптимального размещения зеркал и точного контроля углов отражения.

По сравнению с ячейкой Эрриотта основными преимуществами предложенного дизайна являются:

- Компактность – кюветы Эрриотта, несмотря на их широкое применение, часто ограничены в возможностях миниатюризации из-за необходимости точного согласования оптических элементов и сложной настройки оптического пути луча, предложенная круговая сегментированная МГК значительно компактнее благодаря более рациональному использованию пространства и эффективному складыванию луча.

- Механическая устойчивость – кюветы Эрриотта, как правило, менее устойчивы к механическим воздействиям, что может привести к искажению оптического пути и снижению точности измерений, предложенный дизайн обеспечивает высокую механическую жёсткость, что минимизирует влияние вибраций и других механических факторов.

- Оптическая эффективность – ячейки Эрриотта обладают хорошей оптической эффективностью, но могут быть подвержены интерференционным эффектам, представленный дизайн демонстрирует более высокую оптическую стабильность и низкий уровень шума, что делает ее более подходящей для высокоточных измерений малых газовых составляющих.

- Бо́льшая длина оптического пути – предложенная ячейка обеспечивает длину оптического пути до 10 м при общей массе менее 200 г, кюветы Эрриотта с аналогичной длиной оптического пути обычно имеют большую массу и размеры.

Рассчитать длину оптического пути $L$ луча в предложенной круговой МГК можно согласно выражению

$$L = p \cdot d \sim \sqrt{\frac{d^3}{\lambda}}, \tag{1.40}$$

где $p$ – число сегментов, $d$ – диаметр круговой кюветы.

В 2020 году был представлен прибор (рисунок 1.33), построенный на базе сегментированной круговой МГК с применением энергоэффективного подхода к управлению лазерным источником – кусочно-непрерывного режима [99]. Этот компактный спектрометр на основе применения квантово-каскадного лазера [79] в рамках методики спектроскопии поглощения в среднем инфракрасном диапазоне, был разработан в качестве мобильной платформы для высокоточных измерений атмосферного метана на борту небольших беспилотных летательных аппаратов (БПЛА).

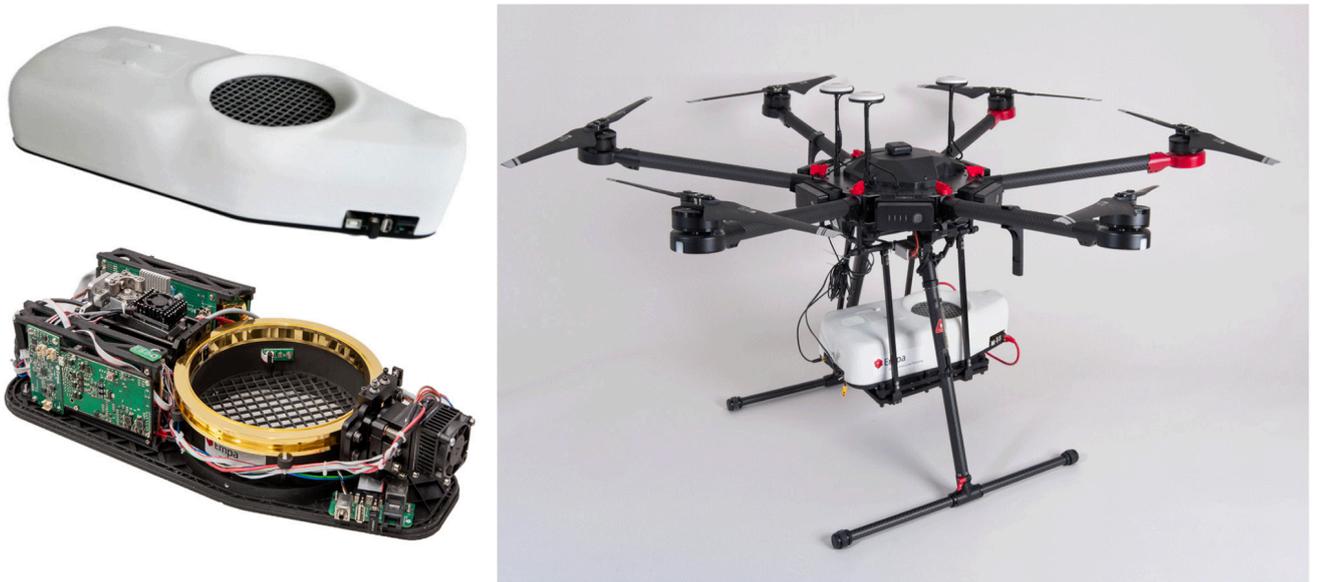

Рисунок 1.33 – Лазерный спектрометр, основанный на применении сегментированной круговой МГК (слева). Прибор, установленный на гексакоптер DJI Matrice 600 (справа) [99].

# ГЛАВА 2. ПРИМЕНЕНИЕ МЕТОДОВ ИЗМЕРЕНИЯ СОСТАВА ГАЗОВЫХ ПРОБ IN SITU В ИССЛЕДОВАНИЯХ КОСМИЧЕСКИХ ОБЪЕКТОВ

Метод абсорбционной лазерной спектроскопии находит свое применение не только в случае лабораторного изучения газовых смесей, анализа состава земного атмосферного воздуха и определения интегральной картины распределения параметров состояния молекулярных газов по высоте над поверхностью. С равным успехом различные вариации этой методики используются для анализа атмосфер других планет или подвергнутых пиролизу образцов грунта космических объектов. Возможно создание приборов, устанавливаемых на межпланетные космические аппараты: орбитальные системы наблюдения с возможностью покрытия широких областей поверхности планет или их естественных спутников, а также спускаемые поверхностные аппараты, обладающие возможностью проведения более чувствительных измерений *in situ* в выбранной точке поверхности.

Так приборы, основанные на методе абсорбционной лазерной спектроскопии, планируются к применению и уже использовались в ряде миссий к внеземным объектам. Для изучения состава атмосферы и обнаружения газовых примесей в атмосфере Марса используется лазерный спектрометр TLS-SAM, установленный на марсоходе Curiosity [100], который был отправлен на Марс в рамках миссии Марсианской научной лаборатории NASA MSL (Марсианская научная лаборатория). В случае проведения поверхностных и геологических исследований ядер комет такие приборы могут дать информацию о составе кометного вещества во время пролетов или при контактных измерениях. Также возможны наблюдения на наземных телескопах метрового класса атмосфер планет Солнечной системы и в перспективе экзопланет при использовании метода гетеродинной абсорбционной лазерной спектроскопии [61,65,71-74].

На базе Института космических исследований РАН осуществлена разработка трех приборов, основанных на принципах диодно-лазерной спектроскопии: спектрометра ДЛС в составе газового хроматографа ХМС-1Ф, разработанного в рамках неудачной космической экспедиции «Фобос-Грунт» [101], многоканального диодно-лазерного спектрометра М-ДЛС для *in situ* измерений состава атмосферы Марса, созданного в рамках второго этапа совместной миссии Европейского космического агентства и госкорпорации Роскосмос «ExoMars-2022» [102], и многоканального диодно-лазерного спектрометра ДЛС-Л, входящего в состав газового хроматографа ГХ-Л полярной посадочной станции «Луна-27» отечественной миссии «Луна-Ресурс» [103,104]. Автор принимал непосредственное участие в работе над двумя последними проектами.

Марсианский многоканальный диодно-лазерный спектрометр М-ДЛС входит в состав научной полезной нагрузки посадочной платформы «Казачок» миссии «ЭкзоМарс». Лазерный спектрометр, работающий в среднем ИК-диапазоне, предназначался для долгосрочного мониторинга в приповерхностной атмосфере изотопных соотношений марсианских летучих соединений – углекислого газа и водяного пара.

Концепция прибора М-ДЛС, внешний вид которого представлен на рисунке 2.1, заключалась в применении метода спектроскопии полного внутрирезонаторного выхода (ICOS) для увеличения эффективной длины оптического пути до ~50-100 м. При этом основанный на данном методе прибор будет сочетать в себе высокую чувствительность и точность измерений с относительно простой и надежной конструкцией.

Для достижения прецизионного уровня измерения давления газовой смеси в кювете в различных диапазонах давления использовались два типа датчиков давления: мембранные датчики давления Heimann Sensor HVS Vac04 (Германия), которые представляют собой миниатюрные датчики типа Пирани, и миниатюрные абсолютные пьезорезистивные датчики давления Kulite XT-140 (США) на основе миниатюрной кремниевой диафрагмы. XT-140 должен использоваться для измерения давления выше 30 мбар, а HVS Vac 04 – для измерения меньших давлений. Эти датчики были откалиброваны при различных температурах с помощью высокоточных лабораторных датчиков давления: Atovac ACM200 (Корея) и Inficon DI200 (Швейцария) [13].

Для измерения температуры также используются два типа датчиков: высокоточный полупроводниковый датчик температуры с отрицательным температурным коэффициентом сопротивления Amphenol Corporation NTC MC65F103A (США) и платиновый датчик температуры Honeywell International 701-102BAB-B00 (США) [13].

Датчики давления двух типов установлены внутри кюветы ICOS, два датчика температуры MC65F103A и один 701-102BAB-B00 установлены на поверхности кюветы ICOS для точного измерения значений давления и температуры. Кроме того, внутри корпуса спектрометра установлены датчики давления и температуры, позволяющие отслеживать состояние прибора во время работы.

Таким образом, система подготовки анализируемой газовой смеси позволяет сформировать образец марсианского воздуха с контролируемыми значениями давления и температуры, что важно для дальнейших спектроскопических измерений. Положение датчиков давления и температуры схематически изображено на рисунке 2.1.

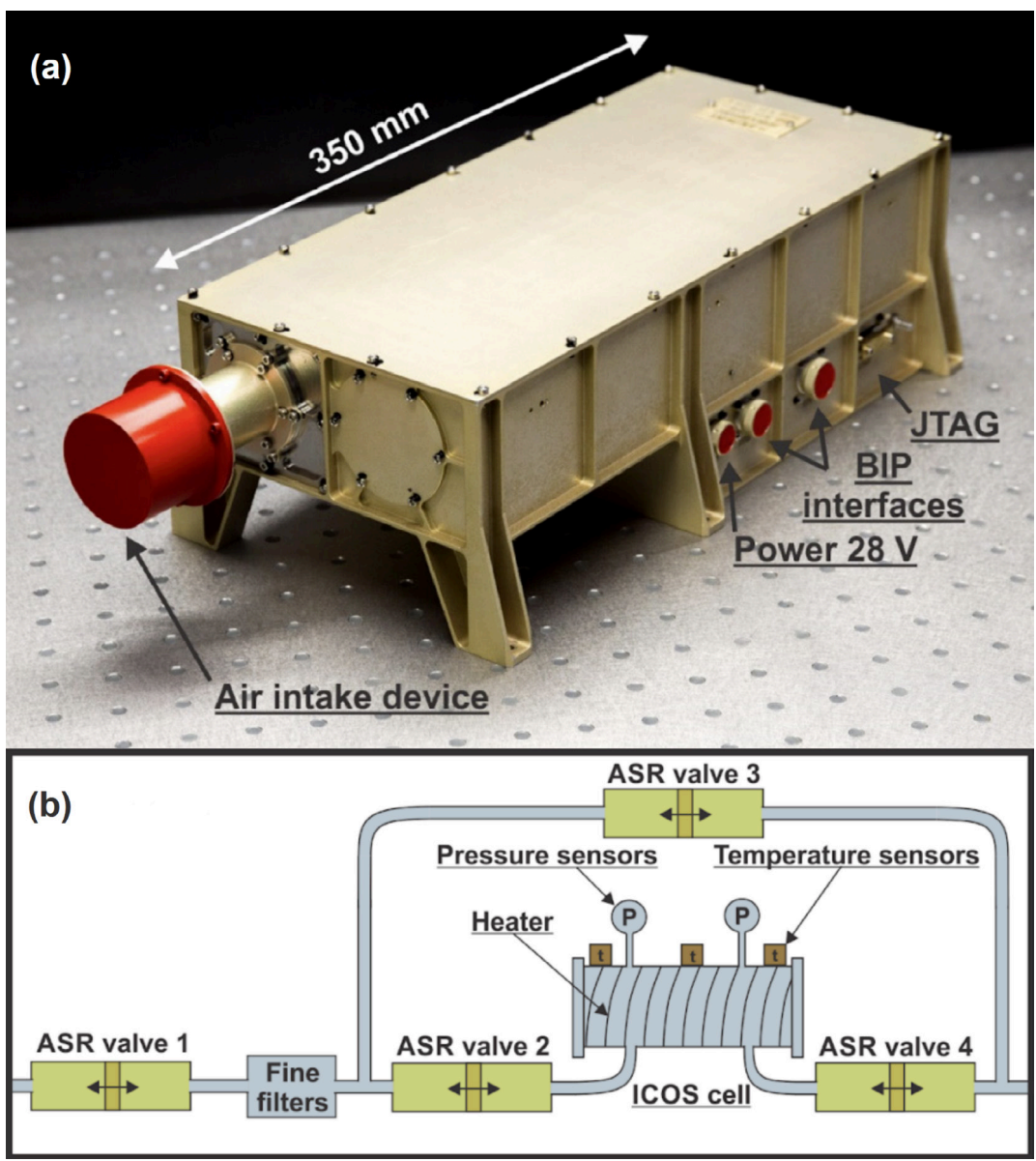

Рисунок 2.1 – **(a)** Внешний вид спектрометра М-ДЛС с внешними интерфейсами и справочными размерами; **(b)** принципиальная схема газовой системы прибора М-ДЛС с указанием положений датчиков давления и температуры кюветы ICOS, обмотанной нихромовой нитью (нагревателем) для поддержания температуры. [102].

Основной вклад автора в работе со спектрометром М-ДЛС заключался в проведении калибровки перечисленных датчиков давления и температуры, для чего был собран калибровочный стенд и разработано ПО, сборке конструкторско-доводочного (КДО) и штатного образцов (ШО) прибора, проведении термовакуумных, климатических и электромагнитных испытаний и участии в подготовке статьи по результатам проведенных функциональных испытаний прибора [102].

Далее глава будет посвящена подробному рассмотрению разработки прибора ДЛС-Л, описанию его структуры и методики измерений. Экспериментальные результаты, полученные в ходе предполетных функциональных испытаний, приведены в конце главы.

Личный вклад автора заключается в юстировке и сборке оптического блока прибора, проведении предполетных функциональных испытаний, анализе режимов работы прибора, разработке методики обработки экспериментальных данных, проведении анализа спектроскопических данных прибора ДЛС-Л при помощи разработанного автором комплекса программного обеспечения, получении итоговых результатов по оценке точности определения изотопных отношений при помощи прибора ДЛС-Л. В настоящем разделе используются материалы статей [103,104], подготовленных автором.

## 2.1. Изучение изотопного состава летучих в полярных районах Луны

По словам академика РАН Э.М. Галимова, Луна – «новый континент, ждущий своего освоения» [105]. Освоение Луны может в будущем помочь в решении экологических и энергетических проблем человечества. Она может превратиться в экономичный космодром – аванпост исследований Солнечной системы и дальнего космоса, контроля астероидной опасности. Ее можно использовать и как платформу для размещения глобальных информационных систем [105].

Крайне важным аспектом освоения Луны является ее исследование как геологического тела, в первую очередь ее строения и химического состава. Такие исследования являются в том числе и способом реконструкции ранней истории нашей планеты, поскольку не представляется возможным изучить первые 500 млн лет эволюции Земли, когда формировались океан и атмосфера, а также сложные органические соединения. Однако считается, что свидетельства этого периода сохранились на Луне, ранняя история эволюции которой тесно связана с историей Земли [105].

Одной из важных вех в этом направлении является изучение летучих веществ в лунных поверхностных породах и неглубокого, доступного для непосредственного анализа, залегания, хранящих историю развития лунных недр, ударных событий и взаимодействия с солнечным ветром ввиду отсутствия у Луны плотной атмосферы.

Сохранение на лунной поверхности попавших на нее летучих соединений сильно осложняется неблагоприятными для этого условиями: в зависимости от уровня освещенности перепад температур лежит в диапазоне от -160℃ до +120℃, при этом вода в конденсированном состоянии может находиться на поверхности Луны при температуре не выше -166℃, а для твердой фазы $CO_2$ она должна быть еще ниже [106]. Однако еще в 1960-х высказывались предположения о возможном сохранении летучих в конденсированном состоянии в неосвещенных участках на дне полярных кратеров, в так называемых холодных ловушках [107]. Такое соображение представлялось хорошо обоснованным, поскольку из-за малого наклона оси вращения Луны к плоскости эклиптики дно многих кратеров в полярных областях находится в постоянном затенении, благодаря чему температура там может не подниматься выше -230℃ [108].

Это предположение окончательно подтвердилось только спустя полвека в ходе эксперимента LCROSS (Lunar CRater Observation and Sensing Satellite – Спутник для наблюдения и зондирования лунных кратеров) на орбитальной станции NASA LRO (Lunar Reconnaissance Orbiter) [109]. В ходе эксперимента разгонный блок Centaur врезался в поверхность Луны с высокой скоростью, за ним следовал блок сопровождения с измерительной аппаратурой, которая отслеживала последствия удара. Облако выбросов от удара, достигшее 10 км в поперечнике и более 1 км в высоту, было видно на снимках камер блока сопровождения. За этим столкновением наблюдали и приборы аппарата LRO.

Обнаружение заметной концентрации водорода в полярных областях Луны в ходе этого эксперимента при помощи российского нейтронного детектора LEND (Lunar Exploration Neutron Detector) стало одним из наиболее заметных достижений в ее исследованиях за последние годы. Самым вероятным веществом, связывающим водород в таких количествах, является вода. Место падения аппарата LCROSS внутри постоянно затененного дна кратера Кабеус недалеко от южного полюса Луны демонстрирует самую высокую концентрацию водорода в южной полярной области Луны, что соответствует расчетному массовому содержанию водяного льда от 0.5 до 4.0%, в зависимости от толщины вышележащего слоя сухого реголита [109].

Благодаря дистанционному обследованию поверхности Луны стало возможным оценить содержание воды в верхнем слое лунного реголита и построить первые глобальные количественные карты распределения воды на лунной поверхности (рисунок 2.2) на основе данных о спектрах отражения в ближнем инфракрасном диапазоне прибора NASA Moon Mineralogy Mapper, установленного на орбитальном аппарате «Чандраян-1» Индийской организации космических исследований [110]. Было обнаружено, что содержание воды увеличивается в зависимости от широты, приближаясь в полярных областях к значениям от ~500 до ~750 ppm (частей на миллион) вблизи полюса. В нескольких местах было обнаружено аномально высокое содержание воды, указывающее на возможные внутренние магматические источники, но глобальной корреляции между составом поверхности и содержанием воды найдено не было. Также было показано, что содержание поверхностной воды может меняться на ~200 ppm в течение лунных суток, а верхний метровый слой реголита может содержать в общей сложности ~$1.2 \times 10^{14}$ г воды.

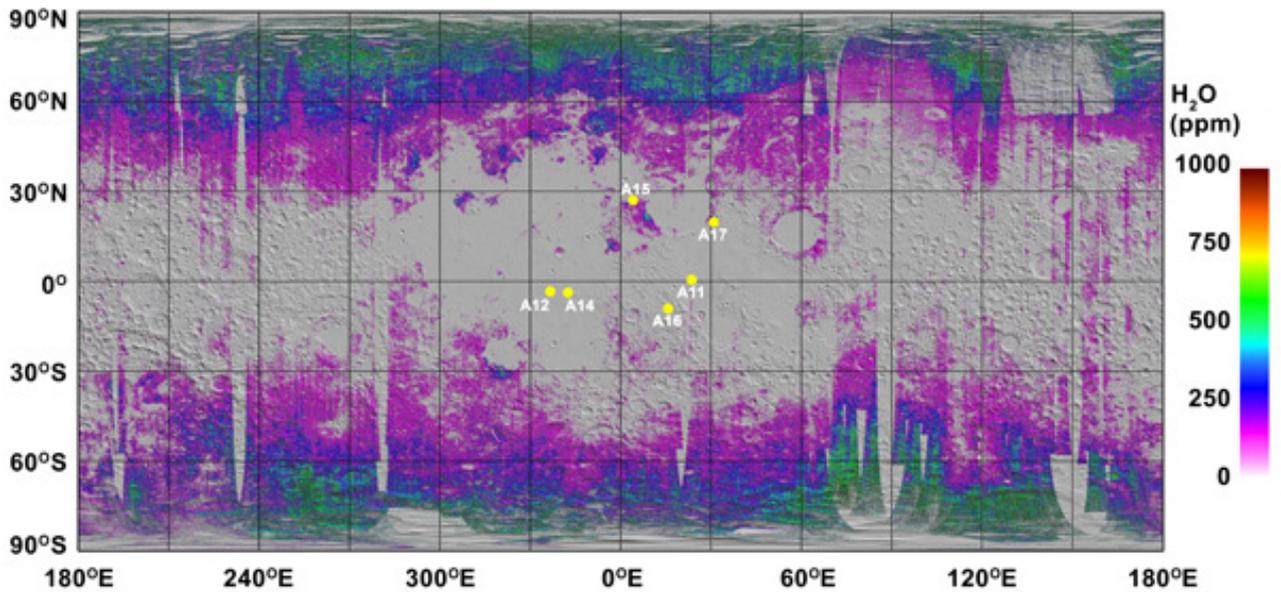

Рисунок 2.2 – Глобальная карта предполагаемого содержания воды на поверхности Луны. Места посадки «Аполлонов» отмечены желтыми точками [110].

Детальное исследование образцов лунного грунта лабораторными методами стало возможным благодаря их доставке на Землю в рамках советских миссий «Луна-16», «Луна-20» и «Луна-24» (рисунок 2.3), а также серии миссий NASA «Аполлон». Изучение образцов лунного грунта позволило получить более полную информацию о формировании Луны и ранних стадиях развития Солнечной системы. Широкий диапазон доставленных геологических пород свидетельствует о богатой вулканической истории, частых ударных событиях и взаимодействии с солнечным ветром и космическими лучами.

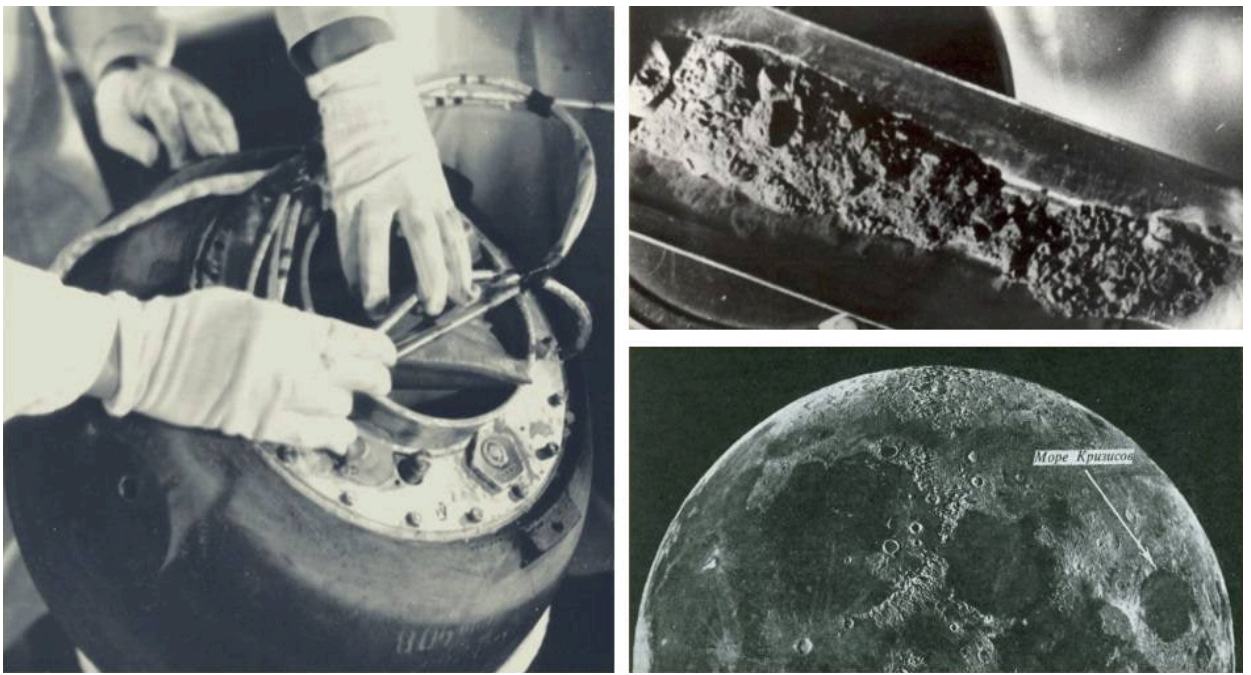

Рисунок 2.3 – Лунный реголит, доставленный в ходе миссии «Луна-24» на Землю (из архива ГК «Роскосмос»).

Говоря об источниках летучих соединений на Луне, рассматривают три основных возможных сценария: дегазация лунной мантии, взаимодействие протонов солнечного ветра с поверхностными породами и ударная дегазация падающих метеоритов и комет [111]. Состав образующихся летучих и их изотопологов в каждом случае будет уникальным.

В первом случае источник летучих веществ обеспечивается дегазацией горячей магмы внутри Луны с выходом газов по трещинам коры. Для таких газовых смесей характерно доминирование $H_2O$ и $CO_2$ по сравнению с другими компонентами, содержащими H и C. Изотопное соотношение летучих элементов должно быть таким же, как и в среднем для Луны [112-114].

В случае «бомбардировки» реголита энергичными протонами солнечного ветра при отсутствии атмосферного щита возможна их реакция с кислородом минералов пород поверхности и синтез молекул воды как конечного продукта. Такой механизм обеспечивает источник воды с солнечным составом изотопов водорода и лунной изотопией кислорода [115,116].

При высокоскоростном ударе метеоритов и комет по поверхности Луны их летучие вещества выделяются в формирующийся ударный плюм, который при газодинамическом расширении распространяется на значительные расстояния и конденсированные продукты которого затем рассеиваются по большой площади лунной поверхности. Состав формирующихся летучих веществ определяется термохимическим равновесием с разогретым силикатным веществом, только при значительно более высоких температурах [117,118].

Вопрос об устойчивости летучих веществ в полярном реголите в условиях космического вакуума и других факторов остается открытым. Летучие вещества могут присутствовать в адсорбированном состоянии на поверхности частиц холодного реголита либо в виде свободного льда, либо в химически связанном состоянии в минералах (гидратах, карбонатах, нитратах и т.д.). В первых двух случаях летучие вещества, особенно низкокипящие, могут теряться даже при небольшом нагреве. В последнем случае летучие соединения стабильны даже при значительном нагревании, поскольку для их извлечения требуется большая энергия.

Для молекулярной воды и ОН-групп в поверхностном слое лунного грунта рассматриваются два основных источника привнесения [106]:

1. межпланетная пыль, представляющая собой частицы размером менее 100 мкм, образовавшиеся в основном при разрушении ядер комет и астероидов, в состав вещества которых могут входить летучие;

2. взаимодействие силикатов поверхности с протонами солнечного ветра и космическим излучением, в результате чего образуются молекулы гидросиликатов.

Одним из наиболее информативных параметров, позволяющих сделать предположения о природе происхождения воды в различных космических объектах, является соотношение изотопов водорода D/H, которое сравнивается со стандартным значением D/H для Венской стандартной средней океанической воды (Vienna Standard Mean Ocean Water) D/H$_{VSMOW}$ = (155.60 ± 0.12) × 10$^{-6}$ [119], а также с известными значениями D/H в молекулах воды, OH-группах или в молекулах водорода, свойственных для иных объектов Солнечной системы. Анализ характеризующих летучие эндогенного происхождения, то есть связанные с процессами в недрах, лунных образцов показал повышенное значение δD относительно VSMOW [120,121], где символ δ означает изотопную сигнатуру, определяемою как

$$\delta = \left( \frac{R_{sample}}{R_{standard}} - 1 \right) \cdot 1000‰, \tag{2.1}$$

и в данном случае $R$ = D/H. Повышенное значение δD в лунной воде может объясняться дегазацией излившегося магматического расплава с потерей ~ 90% изначально содержавшейся в лаве воды [122], D-H дифференциацией еще в горячем протолунном диске [123], или привнесением воды с высоким δD уже после аккреции Луны в ходе ударных событий [124], однако ни одна из приведенных гипотез пока не может считаться полностью доказанной.

В пользу того, что источником привнесения воды является межпланетная пыль свидетельствуют данные по изотопному составу водорода в агглютинатах (агломератах из спекшегося в результате ударных воздействий частиц реголита): значения D/H охватывают диапазон от кометных значений 3-5 × 10$^{-4}$ до величин, характерных для вещества углистых хондритов и земных значений D/H$_{VSMOW}$ [125]. Однако в тех же агглютинатах были зафиксированы значения D/H в несколько раз ниже VSMOW, что говорит о существенной роли солнечного ветра и космических лучей [126], особенно с учетом того, что глубина проникновения последних может достигать глубин порядка единиц метров, где температуры постоянны и составляют -35°C, что обеспечивает стабильность сформировавшихся водосодержащих соединений.

Также проводились оценки изотопного состава ювенильной воды, вышедшей из недр лунного тела, при изучении пирокластических стекол (рисунок 2.4). Их исследование показало, что если в расплавных включениях в кристаллах оливина δD в молекуле H$_2$O составляет ~200-300‰ (рисунок 2.5), что мало отличается от земных значений и укладывается в соответствующий интервал значений δD для углистых хондритов всех типов, то в стекле δD составляет от 400 до 850‰ [114].

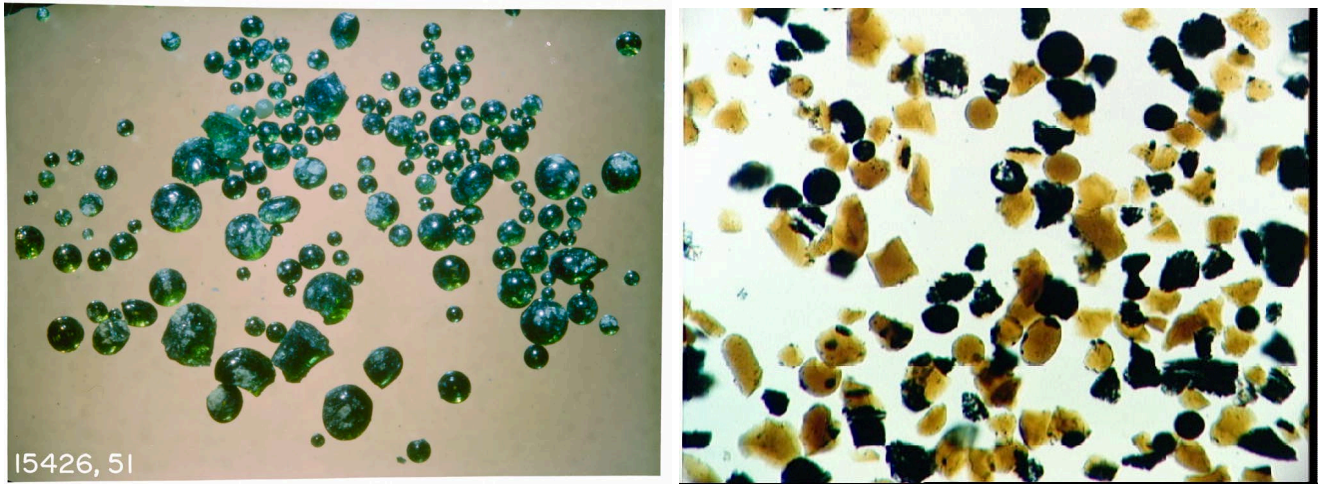

Рисунок 2.4 – Крупный план зеленых (слева) и оранжевых (справа) стеклянных сфер из образцов, доставленных в ходе миссий «Аполлон-15» и «Аполлон-17».

Изучение лунных образцов апатитов, кристаллизовавшихся из магм, показало разброс значений δD в очень широких пределах от -200 до 1000‰ [111]. При этом типичные значения δD для земных базальтов лежат в диапазоне от -110 до 10‰, а для их магм – от -90 до -30‰ [127].

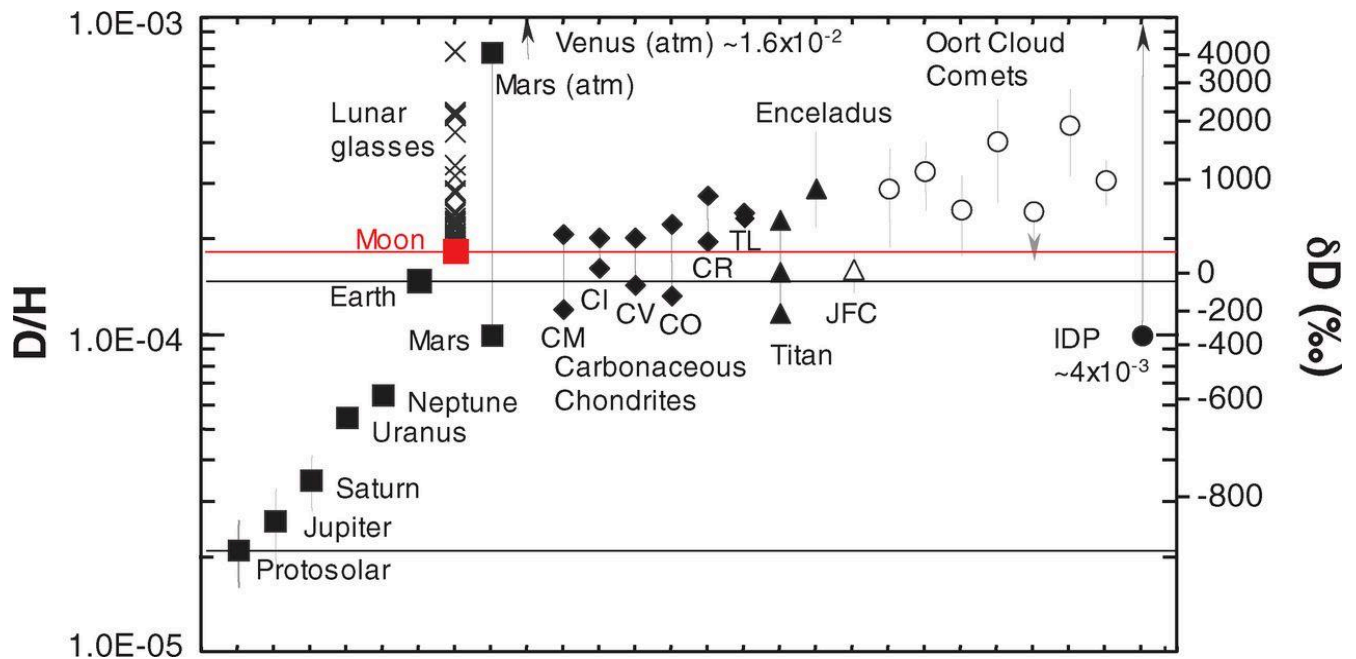

Рисунок 2.5 – Вариации изотопного состава водорода в объектах Солнечной системы [114]. Символы Х представляют собой значения δD отдельных стеклянных шариков и расплавных включений. Значение для лунной мантии (красный квадрат) представляет собой самое низкое δD и самое высокое содержание воды, измеренное в лунных расплавных включениях. IDP – частицы межпланетной пыли; JFC – кометы семейства Юпитера.

Помимо δD значимыми параметрами, позволяющими обосновывать гипотезы о природе происхождения воды в различных космических объектах, являются величины $\delta^{18}O$ и $\delta^{17}O$, определяемые согласно формуле (2.1), где $R = {}^{18}O/{}^{16}O$ или $R = {}^{17}O/{}^{16}O$ соответственно. Этот подход обусловлен тем, что на ранних стадиях формирования Солнечной системы в разных ее частях сложились несколько отличающиеся соотношения трех изотопов кислорода $^{16}O$, $^{17}O$ и $^{18}O$, которые были унаследованы космическими телами, образовавшимися в той или иной зоне. При дальнейшем изменении изотопного состава соединений, образующих тело сформировавшихся планет, их спутников или комет, пары $^{18}O/^{16}O$ и $^{17}O/^{16}O$ менялись пропорционально [128]. Известно, что для земного вещества вне зависимости от его природы существует зависимость $\delta^{17}O_{VSMOW}$ = 0.52 × $\delta^{18}O_{VSMOW}$ [129] при стандартных значениях $^{18}O/^{16}O_{VSMOW}$ = (2005.20 ± 0.45) × $10^{-6}$ [130] и $^{17}O/^{16}O_{VSMOW}$ = (379.9 ± 0.8) × $10^{-6}$ [131].

Изотопные составы кислорода разных веществ разных тел Солнечной системы в координатах $\delta^{18}O$-$\delta^{17}O$ выстраиваются в линию, которая называется линией масс-зависимого фракционирования. Имеющие общий космохимический генезис вещества ложатся на единую линию фракционирования на диаграмме $\delta^{18}O$-$\delta^{17}O$ (рисунок 2.6), так величины $\delta^{18}O$ и $\delta^{17}O$, определенные для разных минералов, воды и газов на Земле ложатся на линию земного фракционирования, на которую также укладываются величины $\delta^{18}O$-$\delta^{17}O$ образцов Луны, что свидетельствует об общем источнике вещества Луны и Земли [105].

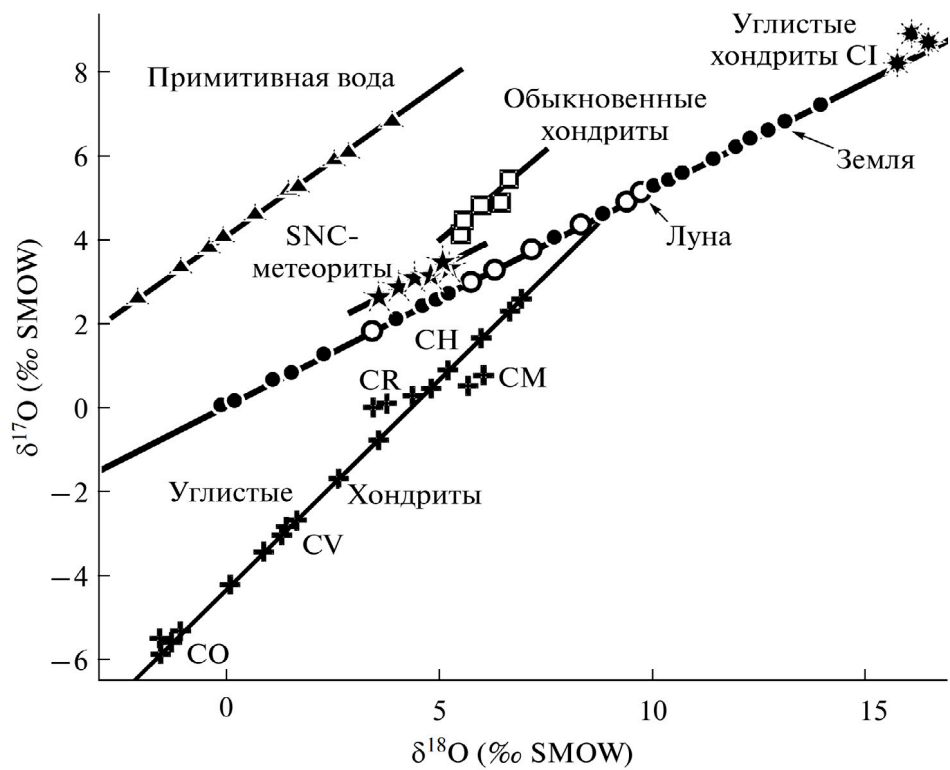

Рисунок 2.6 – Соотношение $\delta^{18}O$–$\delta^{17}O$ для разных космических тел [128].

Еще одним показателем, дающим понимание природы происхождения лунного вещества, является изотопная сигнатура $\delta^{13}C$, определяемая согласно (2.1), где R = $^{13}C/^{12}C$. Углерод в лунных образцах, доставленных на Землю, распределен неоднородно (~145 ppm в пыли и ~260 ppm в брекчии – грубообломочной осадочной горной породе) и имеет уникально высокий уровень $\delta^{13}C \approx 19‰$ для лунной пыли и $\delta^{13}C \approx 11‰$ для брекчии [132]. Эти значения не похожи на свойственные природному углероду Земли или метеоритов [133] (земной стандарт Vienna Pee Dee Belemnite $^{13}C/^{12}C_{VPDB}$ = (11224 ± 28) × $10^{-6}$ [133,134]). Распространение изотопа $^{13}C$ в Солнечной системе представлено на рисунке 2.7 [135].

Возможной причиной таких величин $\delta^{13}C$ в лунных образцах является вклад углерода солнечного ветра, поскольку существует корреляция между количеством углерода и водорода в лунных образцах. Однако если углерод, извлеченный из образцов, представляет собой смесь лунного углерода и углерода солнечного ветра, они оба присутствуют в практически одинаковой химической форме. В противном случае можно было бы ожидать, что углерод, экстрагированный при разных температурах, будет иметь разные значения $\delta^{13}C$.

Объяснить же такие значения $\delta^{13}C$ обусловленностью земным загрязнением представляется затруднительным, поскольку возможные земные источники углеродных загрязнителей, такие как смазка и ракетное топливо, имеют низкие значения $\delta^{13}C$ [132].

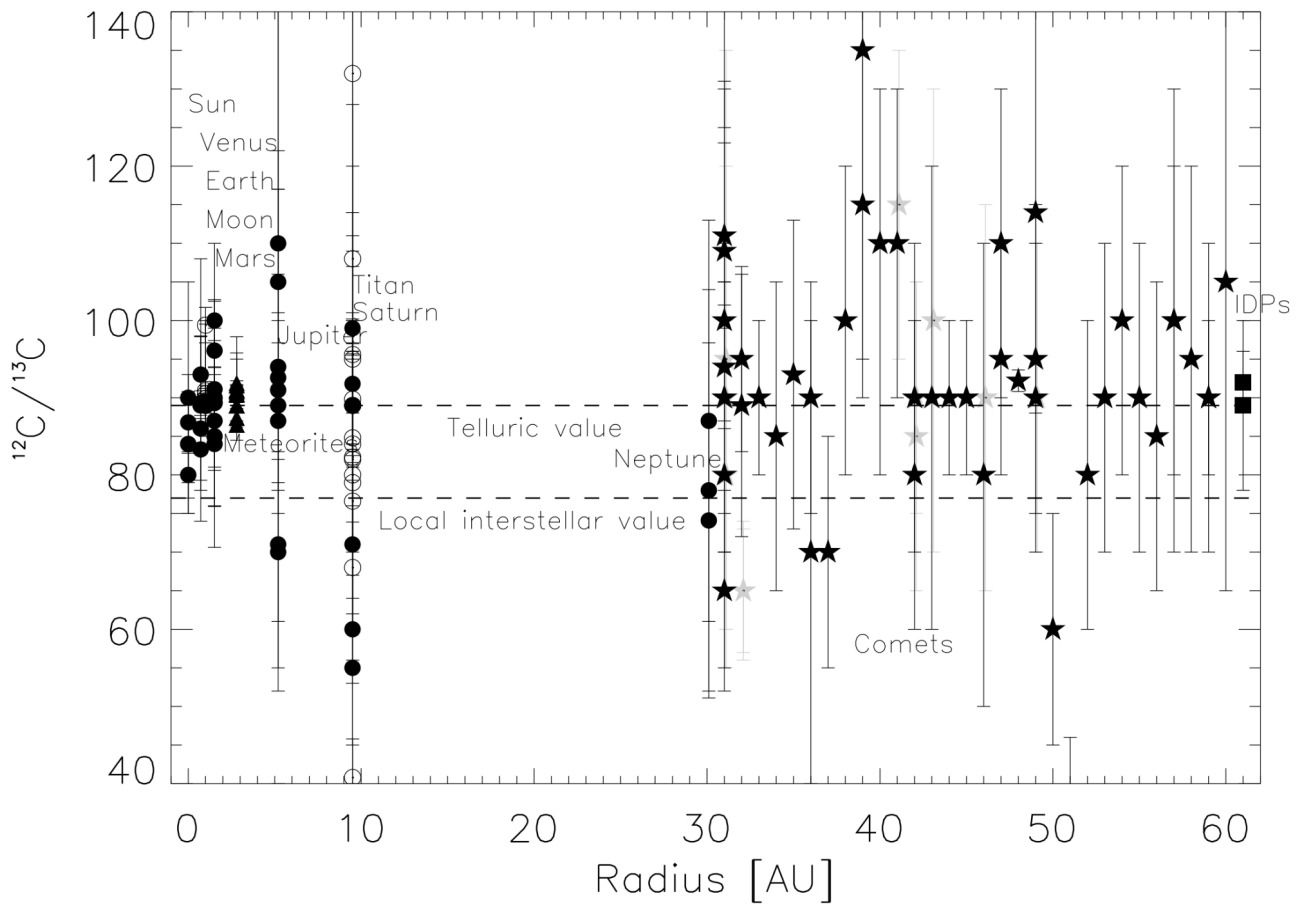

Рисунок 2.7 – Измерения соотношения $^{12}C/^{13}C$ в различных объектах Солнечной системы [135]. Черные кружки обозначают планеты и Солнце, пустые кружки – спутники планет. Треугольники обозначают измерения объемных изотопов соотношения $^{12}C/^{13}C$ в метеоритах и расположены по радиусу пояса астероидов. Кометы обозначены черными звездами, а IDP (частицы межпланетной пыли) – закрашенными квадратами.

Стоит упомянуть и об очевидных ограничениях анализа лунного реголита в условиях земных лабораторий. При значительно более высокой чувствительности возможных лабораторных методик исследования в сравнении с дистанционными наблюдениями с орбитальных аппаратов нельзя забывать про фактор высокой неопределенности в первом случае, связанный с возможностью загрязнения образцов после их забора на поверхности Луны. Еще в ранних работах по лабораторному изучению доставленных на Землю образцов авторы отмечали вероятность того, что обнаруженная в исследуемом веществе вода могла оказаться результатом загрязнения в ходе транспортировки и хранения или же непосредственно в процессе лабораторного анализа [136,137]. Этим фактором, а также ростом чувствительности используемой аналитической аппаратуры объяснялось обнаружение воды в образцах из колонки реголита с «Луны-24» при отрицательных результатах ее поиска в образцах более ранних миссий [137]. Исследования реголита, доставленного станциями «Луна-16» и «Луна-20», также показывали обоснованность опасений контаминации лунного грунта [138].

Сводная иллюстрация распространения в Солнечной системе четырех наиболее изученных изотопных отношений по данным [139] приведена на рисунке 2.8.

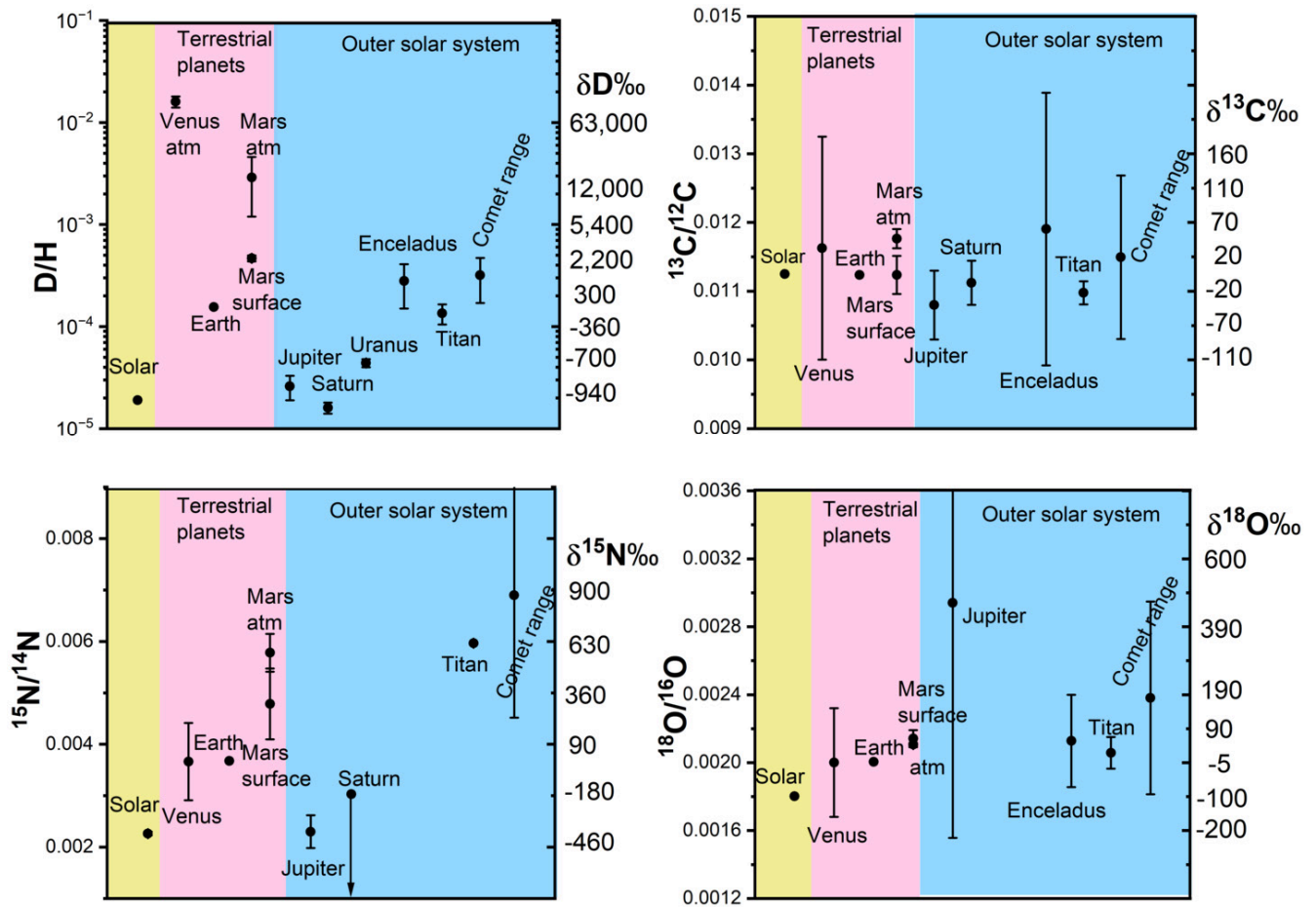

Рисунок 2.8 – Измерения соотношения стабильных изотопов в Солнечной системе [139].

Так или иначе, с уверенностью можно сказать, что вода в виде молекул $H_2O$ и OH-групп, а также другие летучие присутствуют на Луне в трех основных резервуарах: в магматических породах, в тонком миллиметровом слое большей части лунной поверхности, а также в реголите полярных областей Луны. Однако вопрос о полном понимании количества, разнообразия и глубины накопления воды и других летучих веществ, а также их источников остается открытым.

Таким образом, важной и по-прежнему актуальной задачей при изучении летучих соединений в наиболее богатых ими полярных районах Луны является комплексное изучение их химического и изотопного состава, а также их запасенного содержания в реголите. Возможность проведения *in situ* исследований лунного грунта, минуя этапы его доставки на Землю, многолетнего хранения и изучения в земных лабораториях, связанных с вероятностью загрязнения исследуемых образцов, позволила бы приблизиться к убедительному ответу на фундаментальные вопросы о процессах формирования Луны и источниках летучих веществ на ней, равно как помогла бы в прикладных вопросах доступности воды и других летучих для будущих межпланетных миссий [140,141].

## 2.2. Многоканальный диодно-лазерный спектрометр ДЛС-Л в составе газового хроматографа ГХ-Л

В рамках программы возобновления советского цикла непосредственных исследований Луны в 2005 году были анонсированы планы запусков космических аппаратов (КА) «Луна-25», «Луна-26» и «Луна-27». Автоматическая посадочная станция «Луна-25» должна была сесть в полярном районе Луны, преследуя основную научную задачу – определение наличия водяного льда в грунте. В качестве возможного места посадки станции был выбран район, расположенный к северу от южного полярного кратера Богуславский. Резервный район посадки расположен к юго-западу от кратера Манцини, также находящегося в южном полярном регионе.

11 августа 2023 года с космодрома Восточный был произведен успешный запуск ракеты-носителя «Союз-2.1б», несущей на борту автоматическую станцию «Луна-25». 16 августа станция была выведена на окололунную орбиту. В соответствии с программой полета автоматической станции «Луна-25», 19 августа была предусмотрена выдача импульса для ее выведения на предпосадочную эллиптическую орбиту. Около 14:57 по московскому времени связь с автоматической станцией прервалась.

По результатам предварительного анализа специалисты АО «НПО Лавочкина» пришли к выводу, что в связи с отклонением фактических параметров импульса от расчетных, автоматическая станция перешла на нерасчетную орбиту и прекратила свое существование в результате столкновения с поверхностью Луны. Один из предлагаемых вариантов продолжения отечественной программы изучения Луны предполагает возможность повторения миссии по посадке на Южный полюс Луны в 2025-2026 гг. Сроки запуска КА миссии «Луна-Ресурс-1» обозначены как 2027г для орбитального аппарата «Луна-26» и 2028г для посадочной станции «Луна-27» (рисунок 2.9).

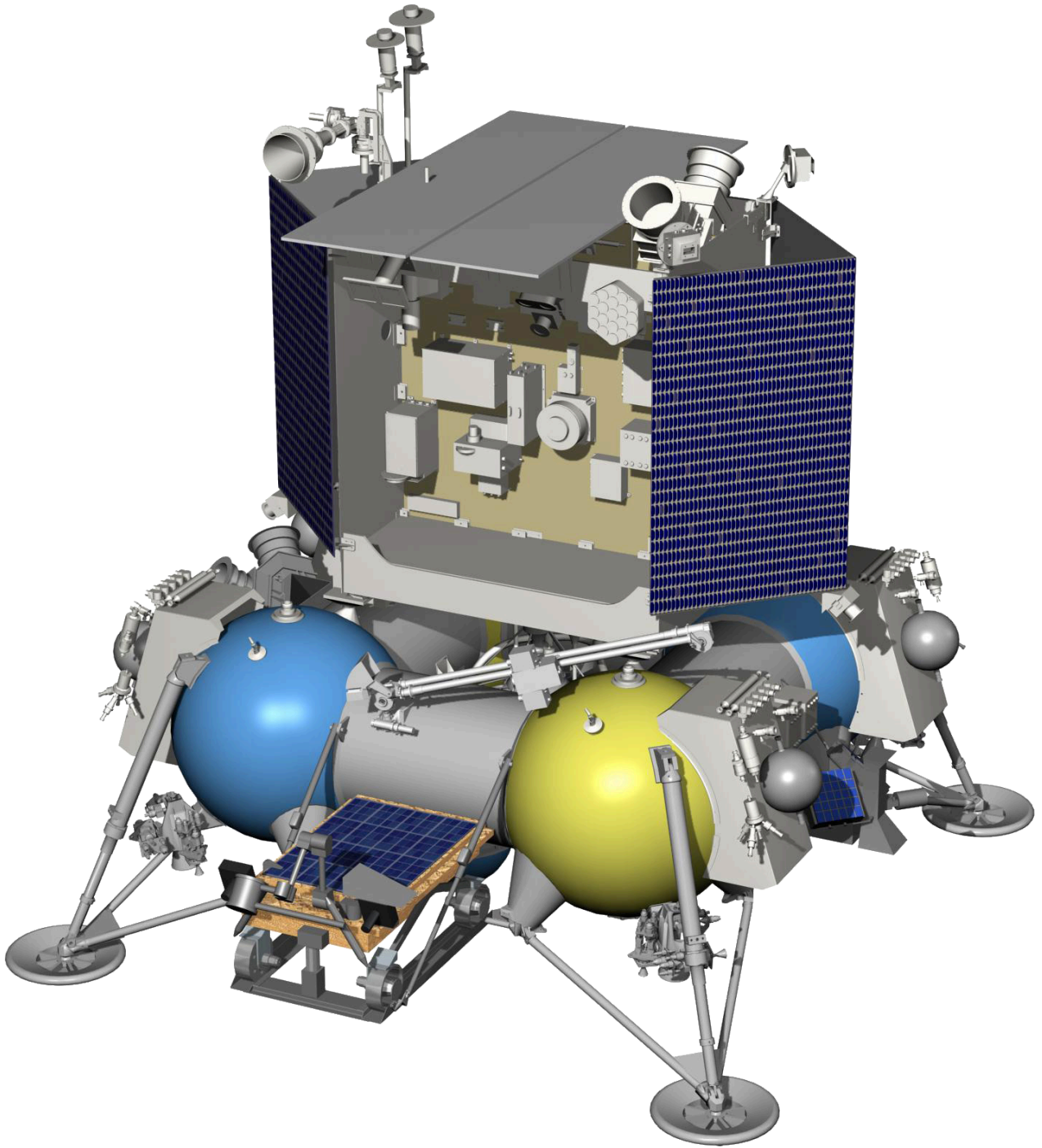

Рисунок 2.9 – Общий вид посадочного модуля миссии «Луна-27».

«Луна-27» – тяжелая посадочная станция, предназначенная для выполнения широкого круга научных задач: отработки технологии высокоточной и безопасной посадки, контактных исследований Луны в районе Южного полюса, включая бурение на глубину до 2 м с помощью криогенной глубинной бурильной установки. Место посадки будет выбрано с учетом результатов работы орбитальной станции «Луна-26» [142].

В качестве научных целей миссии рассматриваются следующие задачи:

- проведение контактных исследований грунта в районе южного полюса Луны;
- поиск водяного льда при помощи нейтронного детектора, внедренного в лунный грунт;
- изучение магнитных аномалий на поверхности Луны, а также воздействия факторов межпланетной среды на поверхность Луны;
- исследования минералогического, химического, элементного и изотопного состава лунного реголита в образцах, доставляемых с различных глубин от поверхности до десятков сантиметров из 2-3 разных мест около космического аппарата, и в образцах реголита поверхностного слоя;
- исследования физических (механических, тепловых и др.) свойств грунта лунной поверхности;
- исследования ионной, нейтральной и пылевой составляющих экзосферы Луны и эффектов взаимодействия поверхности Луны с межпланетной средой и солнечным ветром;
- изучение внутреннего строения Луны и ее глобального движения методами сейсмологии и небесной механики.

Помимо научных экспериментов, в рамках миссии будут отработаны средства и методики обеспечения высокоточной и безопасной посадки, которые в дальнейшем будут применяться для перспективных лунных экспедиций на посадочных платформах для доставки на поверхность целевых комплексов лунной базы [142].

В состав научной аппаратуры станции «Луна-27» входит Газовый аналитический комплекс (ГАК), разрабатываемый в ИКИ РАН для проведения исследований лунного грунта с борта полярной посадочной станции. Газоаналитический комплекс выполняет задачу по всестороннему прямому исследованию летучих компонентов в доступной толще лунного реголита в месте посадки КА «Луна–27». Непосредственными научными задачами, решаемыми комплексом, являются:

- исследование химического и изотопного состава и количественные измерения концентрации летучих соединений (вода, $CO_2$, органические соединения, благородные газы и др.) в слое реголита Луны;
- получение информации о составе намороженных летучих соединениях в реголите;

- получение информации о составе и формах связи летучих соединений с минеральным веществом реголита;
- исследование органического вещества в полярном реголите;
- измерение изотопного состава основных летучих элементов: C, H, O;
- исследование динамики намораживания летучих веществ в поверхностном слое реголита в суточном цикле Луны;
- исследование химического и изотопного состава лунной экзосферы [143].

При исследовании летучих веществ в реголите Луны *in situ* большое значение имеет отбор образцов грунта для анализа. Поскольку при миграционном накоплении летучие вещества концентрируются на поверхности частиц реголита, предпочтительным является забор мелкой раздробленной фракции реголита, а не крупных камней. Также возможно формирование льда при очень большой концентрации летучих веществ, тогда размер частиц не столь важен, но всегда существенным условием для сохранения летучих веществ в забираемой пробе является сохранение температуры пробы во время всей цепочки манипуляций при ее перемещении до анализа.

Эффективным методом исследования содержания летучих в силикатах является метод термического анализа твердого вещества, сопряженный с анализом выделяющихся при нагреве газов. Для всестороннего исследования химического состава выделяемых газов предполагается применять следующие методы: термоанализ, газовую хроматографию, масс-спектрометрию и лазерную ИК-спектроскопию. Информация о температуре выхода того или иного газа позволяет судить о форме связи соответствующего летучего компонента с силикатным веществом. Адсорбированные летучие вещества, как правило, отделяются при температурах до 300°C, а минерализованные компоненты разлагаются с выделением летучих в газовую фазу в широком диапазоне температур от ~400°C до ~1300°C. Количество газов, выделенных за всё время нагрева пробы, дает информацию об общем содержании летучих в исследуемом веществе. При этом температура нагрева должна быть достаточно высокой для обеспечения наиболее полной дегазации пробы [143].

ГАК представляет собой набор устройств, выполняющих приём отобранной для анализа порции лунного грунта, его пиролиз, сбор выделяемых при этом летучих газовых компонентов и их распределение по измерительным системам комплекса – газовому хроматографу и масс-спектрометру. ГАК состоит из термического анализатора (ТА-Л) и газового хроматографа (ГХ-Л). Прототипом комплекса является газоаналитический комплекс КА «Фобос-грунт», состоявший из термического анализатора (ТДА), газового хроматографа (ХМС-1Ф) и масс-спектрометра (МАЛ-1Ф) [144].

Все приборы комплекса компактно установлены на балконе КА «Луна-27» (рисунок 2.10) в непосредственной близости к лунному манипуляторному комплексу (ЛМК) так, что загрузочные отверстия прибора ТА-Л доступны для загрузки из грунтозаборного устройства манипуляторного комплекса. Между прибором ГХ-Л и прибором ТА-Л имеется пневматическая и электрическая связь. Прибор ГХ-Л играет главную роль в управлении комплексом [145]. Он задаёт циклограмму работы комплекса в целом и управляет потоком газа-носителя He, переносящим анализируемые газы между приборами комплекса [143].

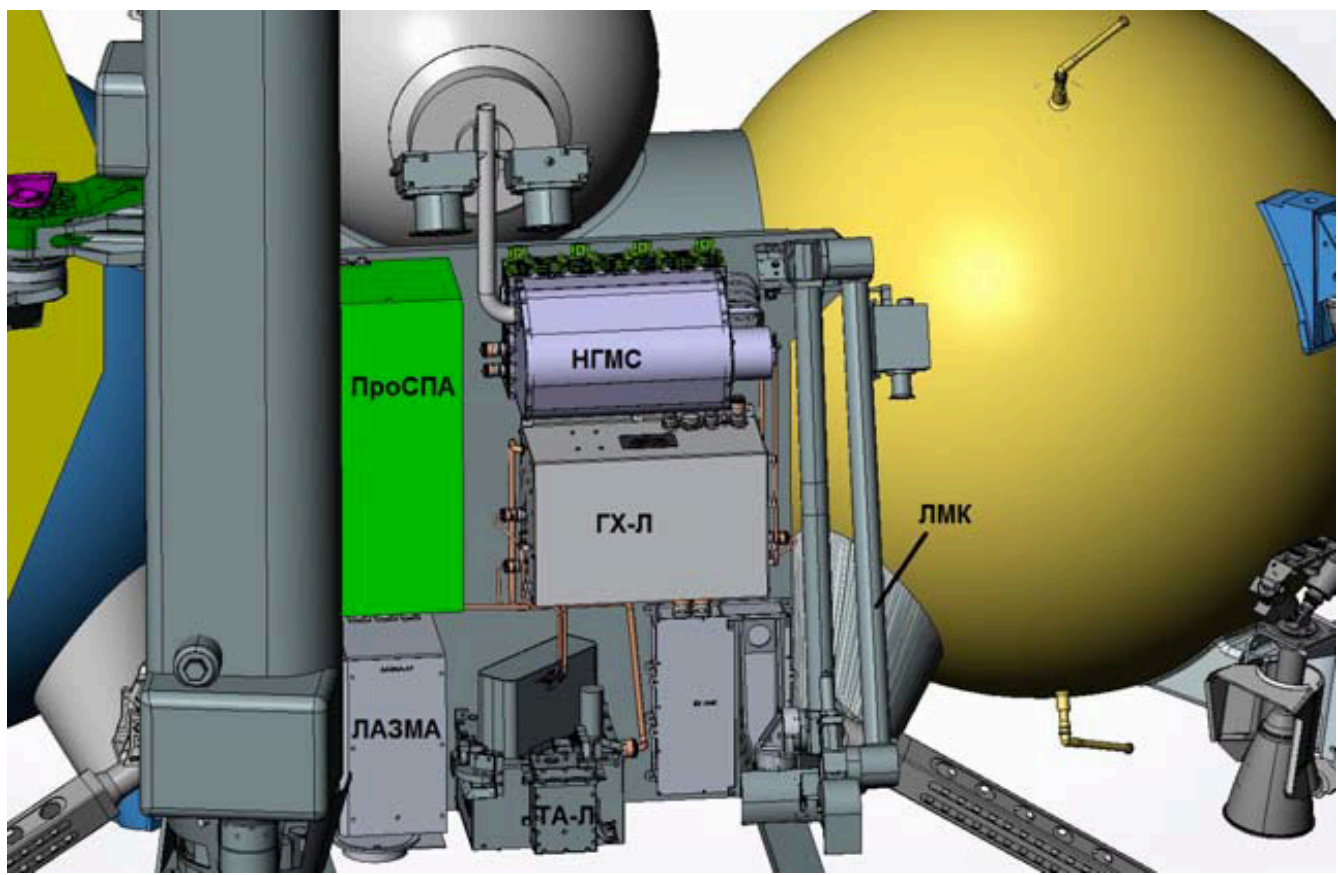

Рисунок 2.10 – Предварительное размещение ГАК на балконе КА «Луна-27».

Газоаналитический комплекс получает грунт от грунтозаборного устройства манипуляторного комплекса ЛМК, аналогичного соответствующему комплексу аппарата «Луна-25» [146]. Связь с КА «Луна-27» производится через блок управления научной информацией (БУНИ).

Прибор ТА-Л (рисунок 2.11) предназначен для прямого термического анализа отобранных образцов реголита, а также для мобилизации летучих компонентов грунта в газовую фазу для их последующего анализа в приборе ГХ-Л.

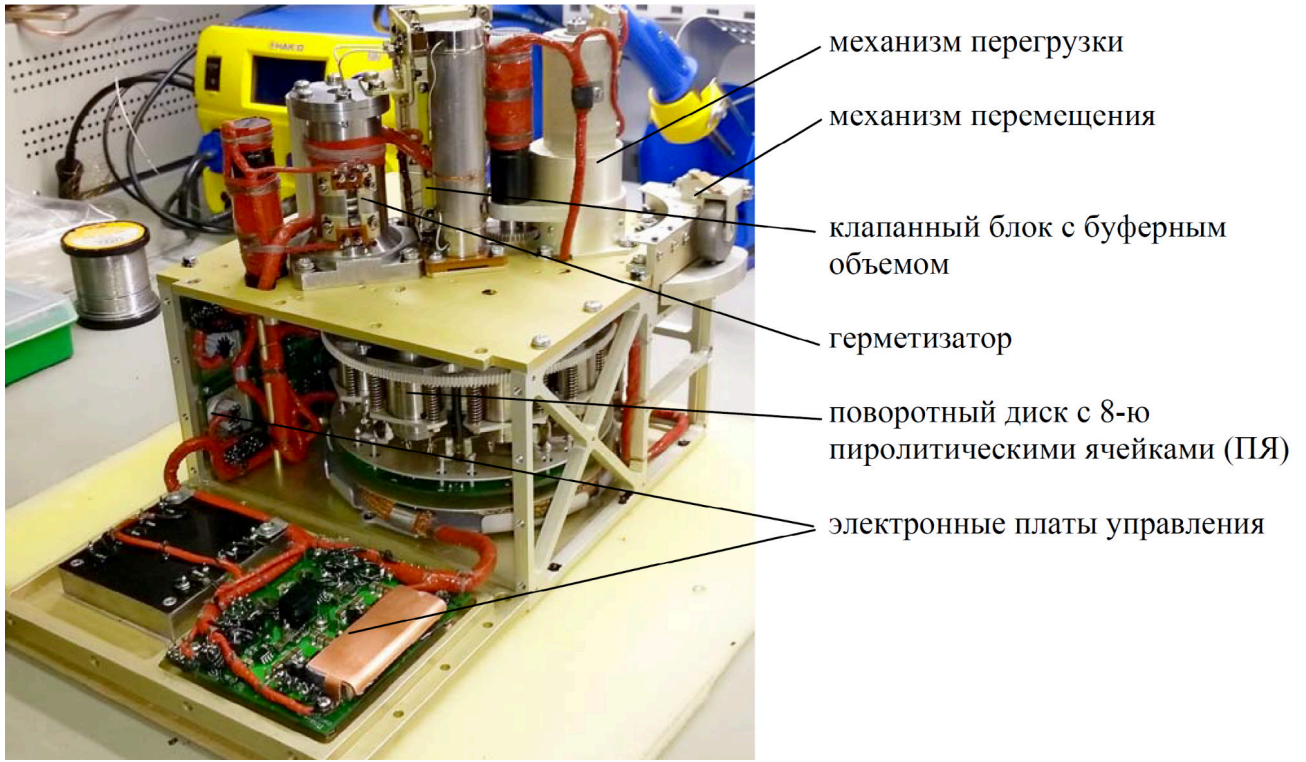

механизм перегрузки

механизм перемещения

клапанный блок с буферным объемом

герметизатор

поворотный диск с 8-ю пиролитическими ячейками (ПЯ)

электронные платы управления

Рисунок 2.11 – Внешний вид прибора ТА-Л со снятыми стенками.

Задачами прибора являются:

- прием образцов грунта от манипуляторного комплекса;
- подготовка полученных образцов к последующему анализу (дозировка, герметизация, уплотнение);
- термическая десорбция летучих компонентов ($H_2O$, $CO_2$ и другие), намерзших на поверхностном слое реголита, включая легколетучие органические соединения;
- термическое разложение (пиролиз до 1000ºС) образцов реголита с целью высвобождения из минеральных соединений связанных летучих веществ в газообразную форму для последующего анализа;
- пиролиз высокополимеризованных органических соединений [143].

Принцип действия ТА-Л основан на использовании стандартных методов термического анализа. Методика анализа состоит в плавном нагреве исследуемой пробы и контроле температуры и потребляемой мощности нагрева. При отсутствии фазовых превращений в исследуемой пробе рост мощности и температуры имеет гладкий характер. При возникновении фазовых переходов в пробе мощность нагревателя ячейки при ровном ходе температуры будет либо возрастать, если переход связан с поглощением энергии, либо уменьшаться, если переход связан с выделением энергии. Информация о температуре фазового перехода позволяет соотнести его с табличным значением температуры фазового перехода для соответствующего минерала, а величина теплового эффекта перехода, соотнесенная с его удельным значением, позволяет определить количество данного минерала в пробе грунта [143].

Так как главное назначение ГАК связано с анализом летучих веществ реголита, то основной интерес при термоанализе направлен на переходы, возникающие при отделении летучих компонентов. Информативность прибора ТА-Л при этом существенно расширяется, если его данные дополняются анализом выделяющихся газов в реальном времени. Информация о профиле выделяющихся газов в зависимости от температуры образца обеспечивает более надежную интерпретацию фазовых переходов, связанных с реакцией терморазложения минералов, выделяющих летучие компоненты. Высокая чувствительность анализа выделяющихся газов в приборе ГХ-Л может дать информацию о терморазложении малых количеств соответствующих минералов в условиях, когда чувствительность самого метода термоанализа недостаточна [143].

Прибор ГХ-Л (рисунок 2.12) в составе хроматографического комплекса предназначен для проведения хромато-масс-спектрометрического анализа газовых компонентов из образцов лунного реголита, подготовленных прибором ТА-Л, а также для изотопного анализа основных летучих элементов – H, C, O – при помощи встроенного диодного лазерного спектрометра ДЛС.

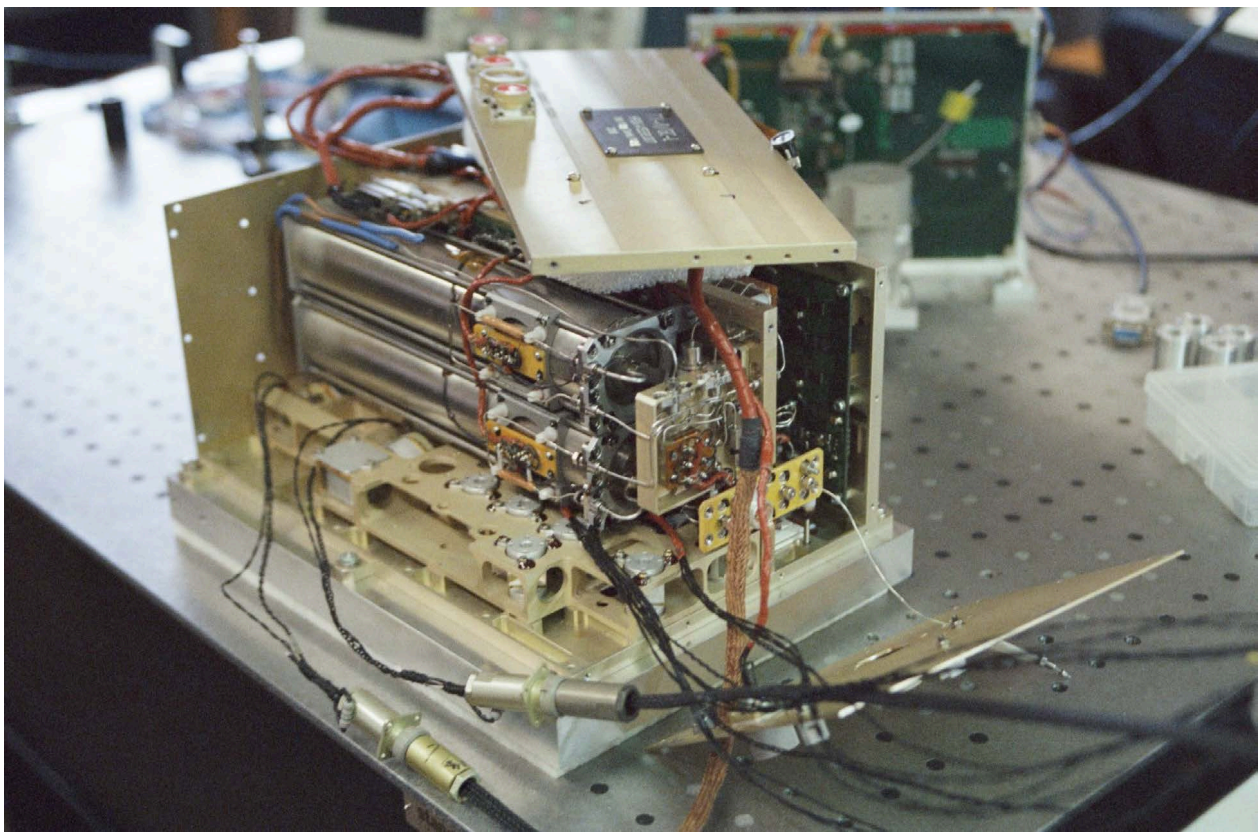

Рисунок 2.12 – Прибор ГХ-Л со снятыми стенками с установленным спектрометром ДЛС-Л.

Задачами прибора ГХ-Л являются [147]:

- сбор газов, выделяемых из образца грунта в пиролитической ячейке при нагреве;
- распределение газов разных типов (постоянные газы, органика и др.) для анализа на соответствующих хроматографических колонках;
- разделение газовой смеси на отдельные компоненты;

- измерение количеств каждой из газовых компонент;
- измерение изотопии элементов H, C, O в молекулах $H_2O$ и $CO_2$ [143].

Газовый хроматограф ГХ-Л состоит из следующих основных частей: баллоны с газом-носителем и системой газоподачи, модули хроматографических капиллярных колонок (КК) с детекторами, адсорбционные накопители (АН), диодно-лазерный спектрометр (ДЛС), электронные платы управления [143]. Пневматическая схема газоаналитического комплекса показана на рисунке 2.13.

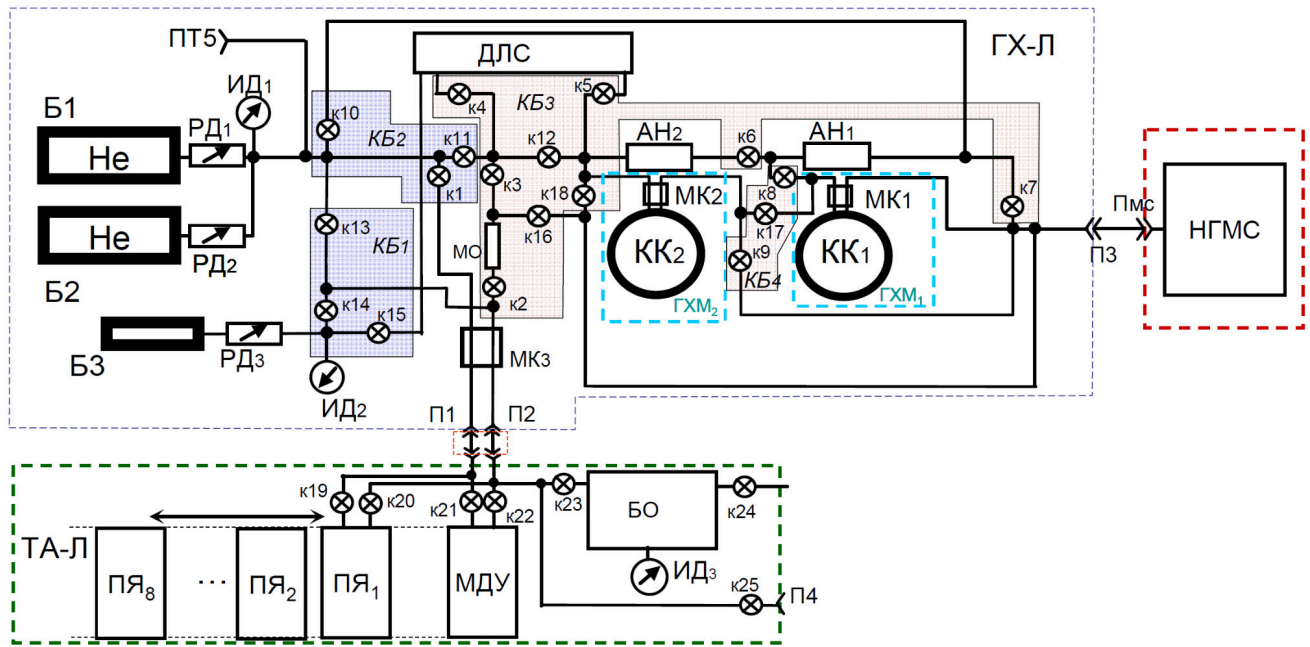

Рисунок 2.13 – Пневматическая схема газоаналитического комплекса. ДЛС – кювета диодно-лазерного спектрометра; $ПЯ_{1-8}$ – пиролитические ячейки; МДУ – многоразовое десорбционное устройство; $ГХМ_{1,2}$ – модули капиллярных колонок; Б1,2 - баллоны с газом-носителем (гелий); Б3- баллон с тестовой газовой смесью; $РД_{1-3}$ - регуляторы давления; $ИД_{1,3}$ - измерители давления; $МК_{1-3}$ – микрокатарометры; $АН_{1,2}$ - адсорбционные накопители; $КК_{1,2}$- капиллярные колонки; МО - мерный объём; к1-25 – микроклапаны; П1,2,3,4,мс – газовые порты; ПТ5 – технологический газовый порт; $КБ_{1-4}$ – клапанные блоки, БО – буферный объем.

В баллонах Б1 и Б2 хранится газ-носитель (гелий) при максимальном давлении ~80 бар, а в баллоне Б3 тестовая газовая смесь при давлении ~30 бар. Система газоподачи обеспечивает редуцирование давления газа-носителя до заданной рабочей величины и подачу его по магистралям прибора в зависимости от текущей моды работы. Редуцирование давления осуществляется соответствующим регулятором давления, который имеет обратную связь с измерителем давления. Давление газа-носителя и тестовой газовой смеси может поддерживаться в диапазоне от 50 до 2500 мбар с шагом 50 мбар. Точность поддержания давления ~0.3 мбар [143].

Прокачка газа-носителя и увлекаемых им измеряемых газов происходит по системе капиллярных трубок из нержавеющей стали (внешний диаметр 0.8 мм и внутренний диаметр 0.25 и 0.125 мм). Капиллярные трубки, по которым прокачивается исследуемая смесь газов, на всех участках – от пиролитической ячейки до входного штуцера масс-спектрометра – имеют намотку электроизолированной нихромовой проволоки, которая обеспечивает нагрев этих трубок до 70ºС [143]. Микроклапаны к1 – к25 электромагнитного типа расположены в четырех клапанных блоках, из которых КБ$_1$ и КБ$_2$ не имеют прогрева, так как управляют только потоком газа-носителя, а блоки КБ$_3$ и КБ$_4$ имеют прогрев до 70ºС для предотвращения конденсации ряда высококипящих газов. В клапанном блоке КБ$_3$ имеется мерный объем (МО) величиной 5 мкл, который используется для дозировки тестовой смеси из баллона Б3 для калибровочных целей и для анализа порции летучих веществ из буферного объема.

Особенность газового хроматографа ГХ-Л – использование многоканального диодно-лазерного спектрометра ДЛС-Л – обеспечивает селективную идентификацию нескольких компонентов внутренней атмосферы аналитического объема, выделяемых при пиролизе образцов грунта вблизи от места посадки лунного зонда. ДЛС-Л предназначен для независимого измерения динамики пиролитического выхода и интегрального содержания $H_2O$ и $CO_2$, а также для определения изотопных соотношений D/H, $^{18}O/^{17}O/^{16}O$, $^{13}C/^{12}C$ для изотопологов $H_2O$ и $CO_2$. Данные датчика ДЛС-Л помогут в дальнейшем понимании физики и химии лунного тела, поскольку это первые данные прямого исследования полярного лунного грунта.

## 2.2.1. Принципиальная схема прибора ДЛС-Л

Устройство ДЛС-Л основано на использовании перестраиваемых по длине волны излучения диодных лазеров с распределенной обратной связью, генерирующих монохроматическое излучение в узких спектральных интервалах регистрации молекулярного поглощения в ближнем ИК-диапазоне спектра, соответствующего колебательно-вращательным переходам молекулярных газов. Такой подход хорошо зарекомендовал себя в космическом приборостроении [101,102,148,149].

Внешний вид спектрометра ДЛС-Л, установленного в вакуумную камеру для проведения цикла физических испытаний по определению концентрации $CO_2$ и $H_2O$, а также отношений их изотопологов, представлен на рисунках 2.14 и 2.15.

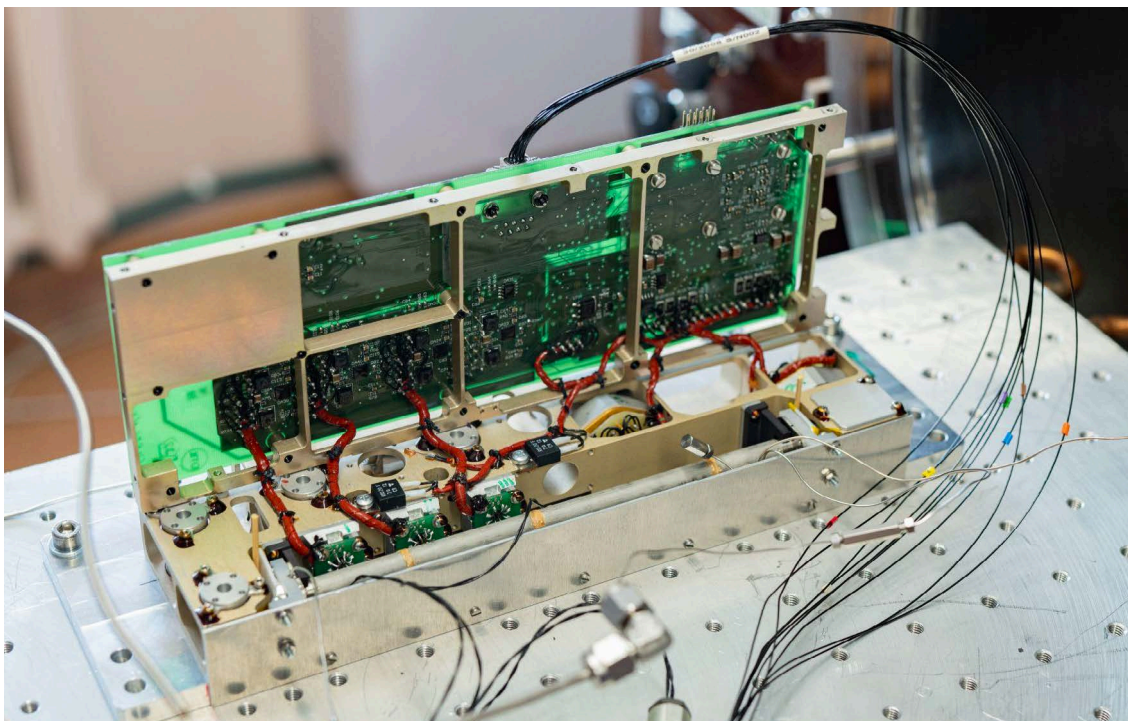

Рисунок 2.14 – Сборка лазерного спектрометра ДЛС-Л со стороны аналитической кюветы.

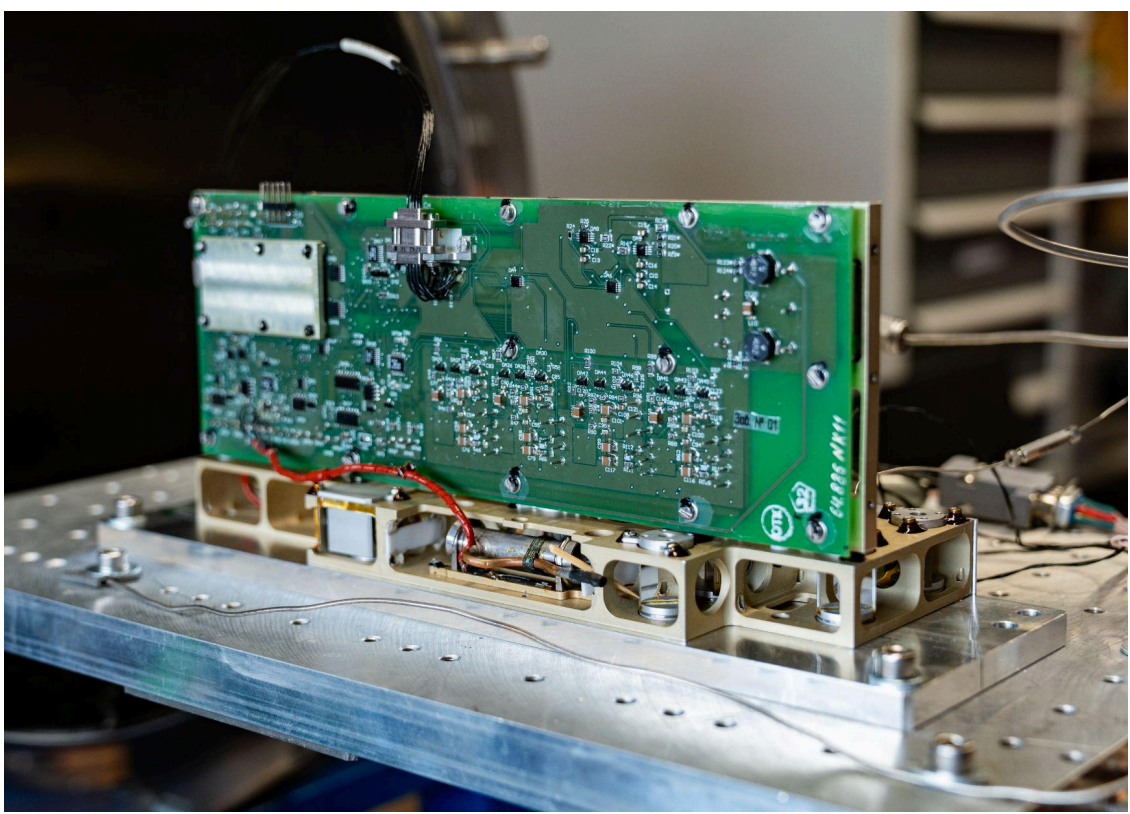

Рисунок 2.15 – Сборка лазерного спектрометра ДЛС-Л со стороны реперной кюветы.

В спектрометре ДЛС-Л предусмотрено объединение до четырёх диодных лазеров с полупроводниковыми структурами на основе многокомпонентных твердых растворов соединений GaAs, InP, обогащенных Sb. Встроенный в конструкцию лазерного модуля термоэлемент Пельтье позволяет выбирать и прецизионно стабилизировать частоту генерируемого лазерного излучения, задание параметров периодического пилообразного тока накачки лазера определяет спектральный интервал сканирования частоты излучения шириной 1-2 см$^{-1}$, охватывающий несколько линий молекулярного поглощения [143].

Определяющими условиями опытно-конструкторской работы по созданию комплекса научной аппаратуры для космического аппарата «Луна-Ресурс-1» были ограничения по электропотреблению, габаритам и массе. Результат проектирования спектрометра, удовлетворяющего указанным высоким требованиям, представлен на рисунках 2.16 и 2.17.

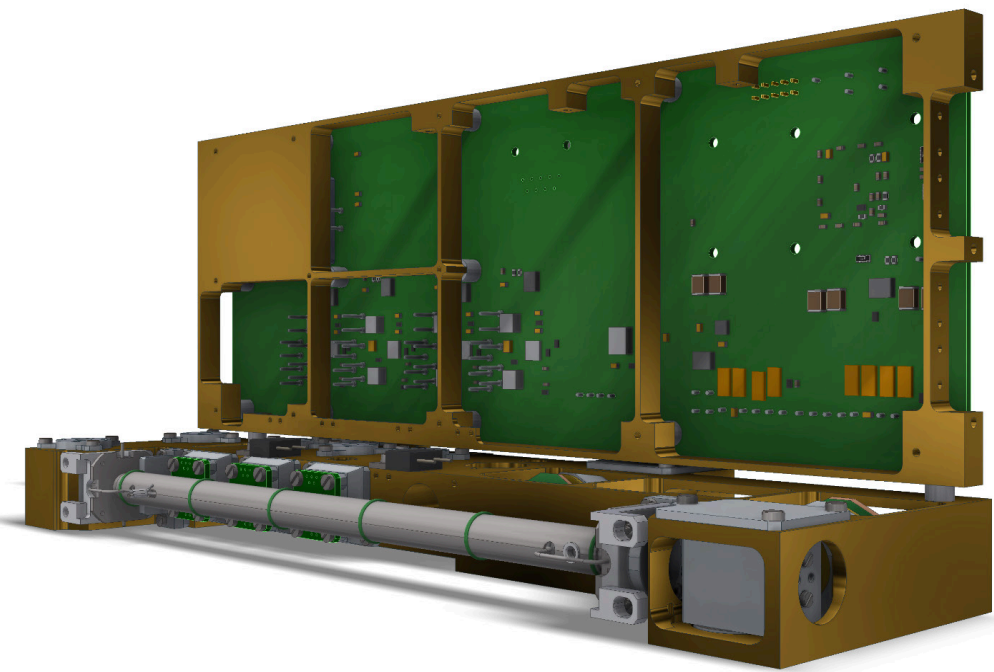

Рисунок 2.16 – 3д-модель лазерного спектрометра ДЛС-Л со стороны аналитической кюветы.

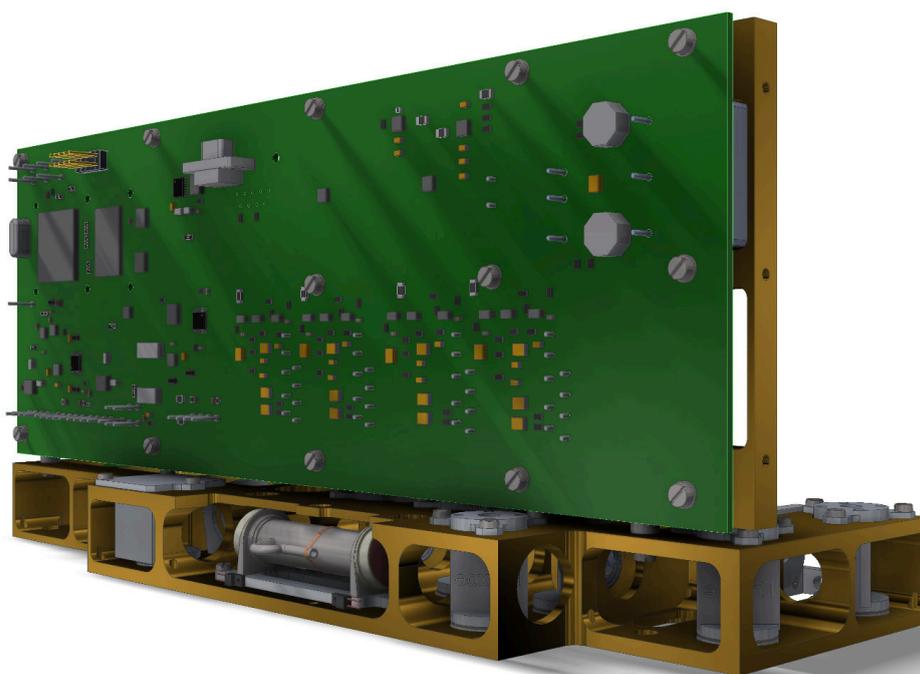

Рисунок 2.17 – 3д-модель лазерного спектрометра ДЛС-Л со стороны реперной кюветы.

Также в ходе разработки спектрометра ДЛС-Л к массогабаритным ограничениям было добавлено дополнительное ограничение внутреннего объема аналитической кюветы, связанное с пропускной способностью и объемом капиллярной системы напуска прибора ГХ-Л. Внешний вид кюветы аналитического канала прибора с длиной 19 см и диаметром внутреннего сечения 3 мм представлен на рисунке 2.18.

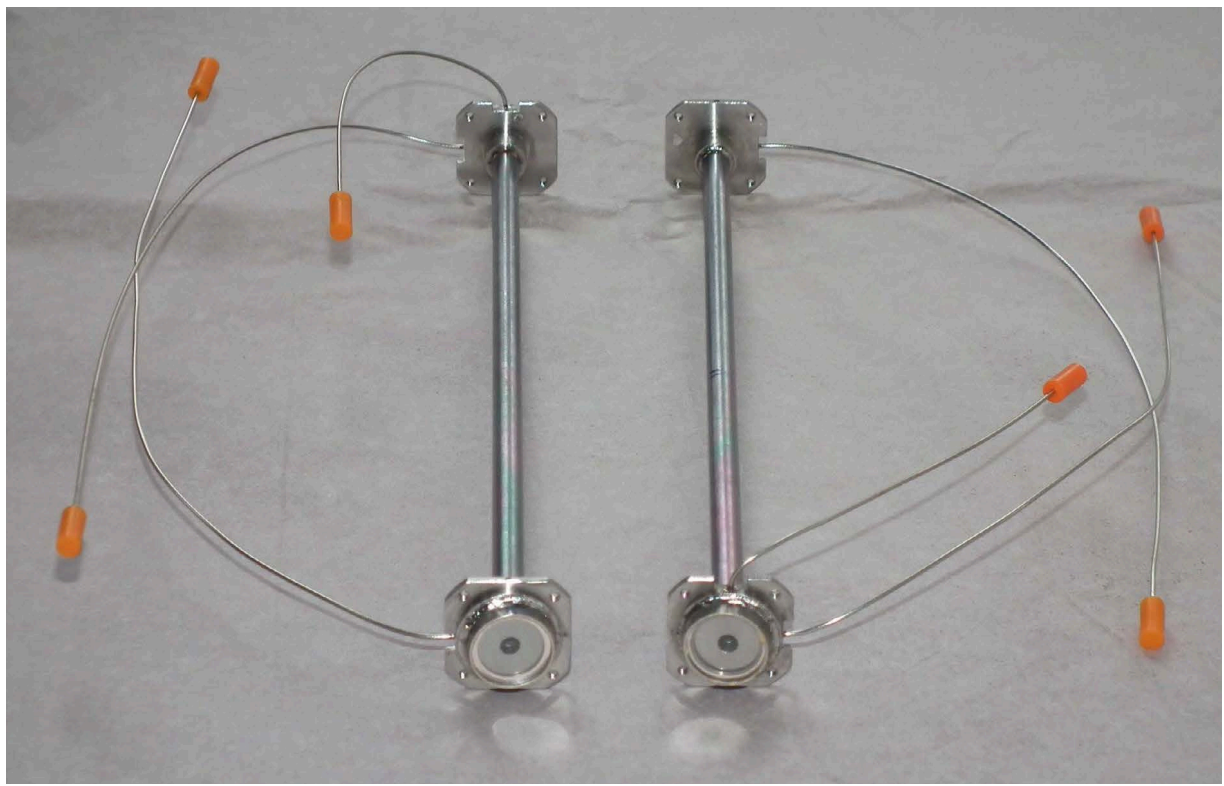

Рисунок 2.18 – Внешний вид кюветы аналитического канала спектрометра ДЛС-Л.

В результате поставленных ограничений спектрометр ДЛС-Л, результаты физических испытаний которого будут описаны далее, обладает характеристиками, представленными в таблице 2.1.

Таблица 2.1 – Характеристики спектрометра ДЛС-Л.

| Габариты с/без блока электроники | 258×88×115 / 258×88×22 |
|---|---|
| Масса | 650 грамм |
| Энергопотребление | 5-7 Вт |
| Объем аналитической кюветы | 1.34 мл |

Как можно видеть из представленных характеристик, среди семейства лазерных спектрометров поглощения прибор ДЛС-Л можно отнести к сверхкомпактным сенсорам. Несмотря на это, спектрометр должен позволять измерять как вариации концентрации выбранных газов в процессе пиролитического выхода, так и изотопные отношения в режиме долговременных измерений. Стоит отметить, что последнее является крайне нетривиальной задачей для оборудования нелабораторного класса, тогда как ДЛС-Л в составе ГХ-Л должен отвечать всем требованиям космического приборостроения: радиационной стойкости, высоким прочностным, электромагнитным характеристикам; прибор должен быть оборудован локальными электронагревателями, расположенными по его объему, сохранять работоспособность при воздействии на элементы его поверхности электростатических разрядов и др.

Функционально ДЛС-Л можно разделить на три основные системы:

1. Внешняя система пробоподготовки прибора ГХ-Л, описанная выше;

2. Оптоэлектронная система, в состав которой входят следующие элементы: однопроходная аналитическая кювета, однопроходная реперная кювета, диодные лазеры, фотодиод, зеркала, светоделители;

3. Электронная система управления, состоящая из электронной платы ДЛС управления лазерными диодами, опроса фотодиодов и датчиков температуры, передачи информации в прибор ГХ-Л.

Ниже каждая из внутренних систем прибора ДЛС-Л будет рассмотрена отдельно.

## 2.2.2. Устройство оптической схемы прибора и ее юстировка

Все используемые диодные лазеры с РОС-структурой в корпусах ТО5 производства компании Nanoplus (Германия) специально изготовлены под выбранные спектральные диапазоны с выходной мощностью 2-5 мВт. Встроенный в конструкцию лазерного модуля термоэлемент Пельтье позволяет выбирать и прецизионно стабилизировать частоту лазерного излучения, а задание параметров периодического пилообразного тока накачки лазера определяет спектральный интервал сканирования частоты излучения шириной около 1-2 см$^{-1}$, охватывающий несколько линий молекулярного поглощения. Подобные компактные диодно-лазерные модули характеризуются гибкостью настроек режимов измерений, отсутствием перескоков между модами генерации и шумов в диапазоне непрерывной перестройки порядка нескольких обратных сантиметров при мгновенной ширине линии лазерного излучения не более 10 МГц [143].

Пучки излучения каждого диодного лазера коллимируются асферическими широкоапертурными микролинзами C036TME-D производства Thorlabs Inc и при помощи дихроичных светоделителей и плоских алюминиевых зеркал с защитным покрытием направляются на параболическое зеркало Thorlabs MPD127127-90-M01, фокусирующее излучение на InAs-фотоприемник через аналитическую кювету ГХ-Л, представляющую собой трубку-капилляр длиной 190 мм и внутренним диаметром 3 мм с наклоненными под углом 8° торцевыми сапфировыми оптическими окнами, через которую газом-носителем He протягивается газовая смесь, выделяемая при пиролизе пробы исследуемого грунта. При этом крайне малый объем аналитической кюветы ~1.34 мл позволяет получать сравнительно высокие концентрации продуктов пиролиза. Оптическая схема прибора представлена на рисунке 2.19.

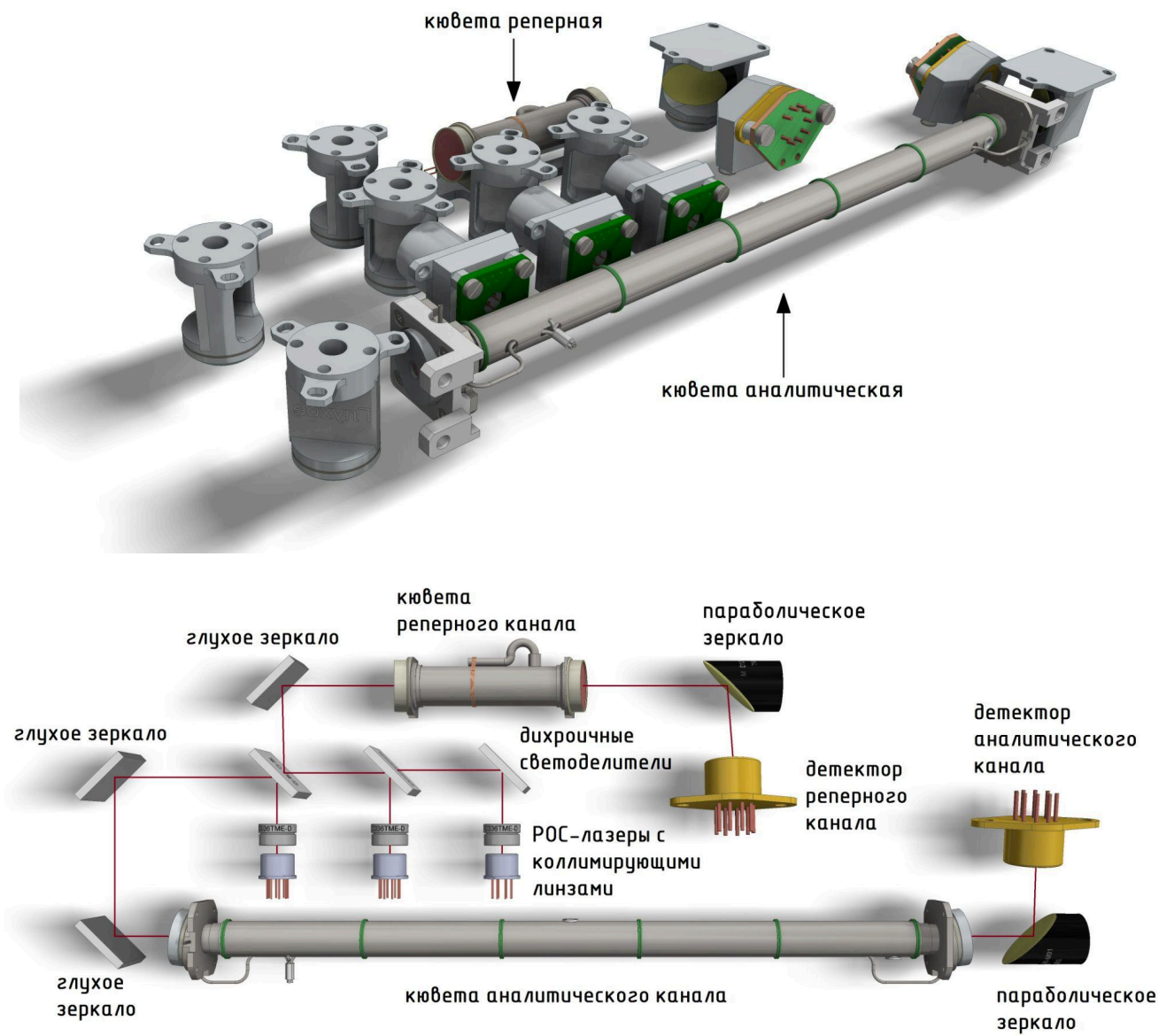

Рисунок 2.19 – Функциональная схема и взаимное расположение фотонных и оптико-механических компонентов ДЛС-Л.

Используемые фотодиоды J12TE3-66D-R01M производства Teledyne Judson Technologies (США) с диаметром чувствительной пластины 1 мм имеют максимальный отклик на падающее излучение с диапазоном длин волн 2.6-3.2 мкм на уровне $2 \times 10^{11}$ см$\frac{\sqrt{\text{Гц}}}{\text{Вт}}$. Зависимость обнаружительной способности семейства InAs фотодиодов J12 от длины волны представлена на рисунке 2.20.

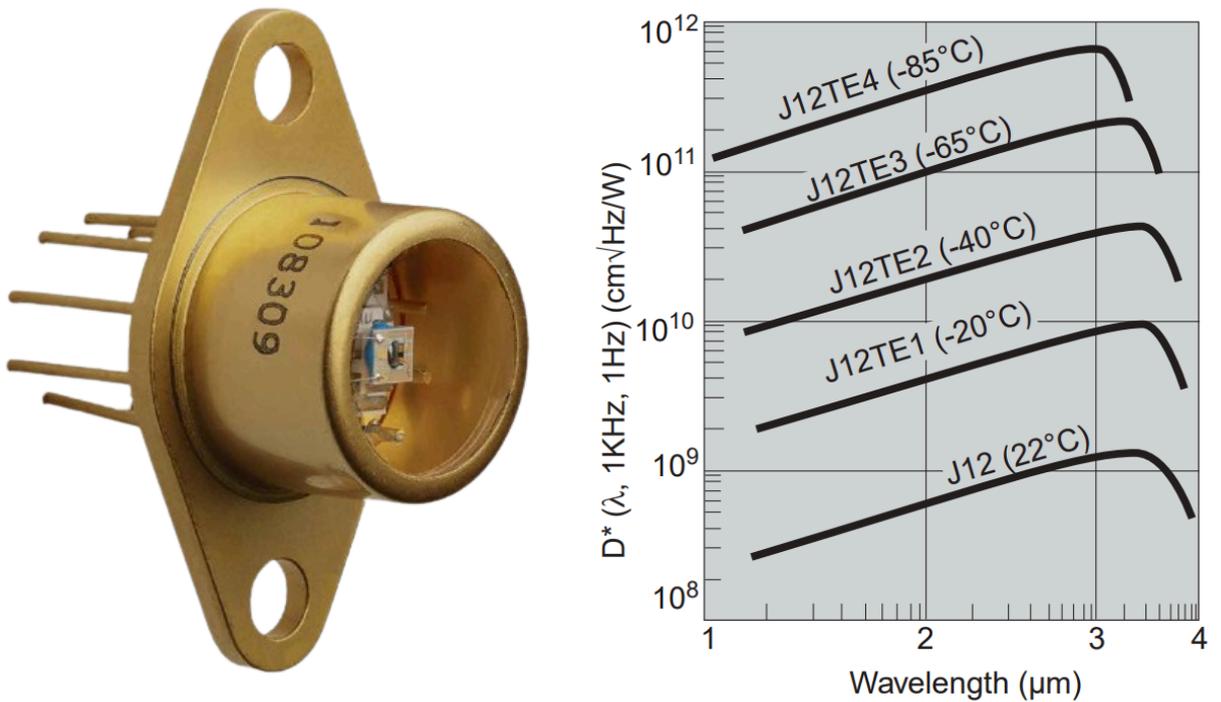

Рисунок 2.20 – Внешний вид (слева) и зависимость обнаружительной способности семейства InAs фотодиодов J12 от длины волны (справа).

Согласно спецификации диодных лазеров (ДЛ) положение излучающего кристалла в корпусе ТО5 может быть не вполне симметричным, в связи с чем лазерные диоды монтировались в конструкцию лазерных модулей для формирования плоскопараллельного пучка лазерного излучения. Лазерный модуль (рисунок 2.21) состоит из алюминиевого основания с монтажной электрической платой и алюминиевого крепления с плоско-выпуклой коллимирующей линзой C036TME-D с диаметром 9.24 мм и фокусным расстоянием 4 мм.

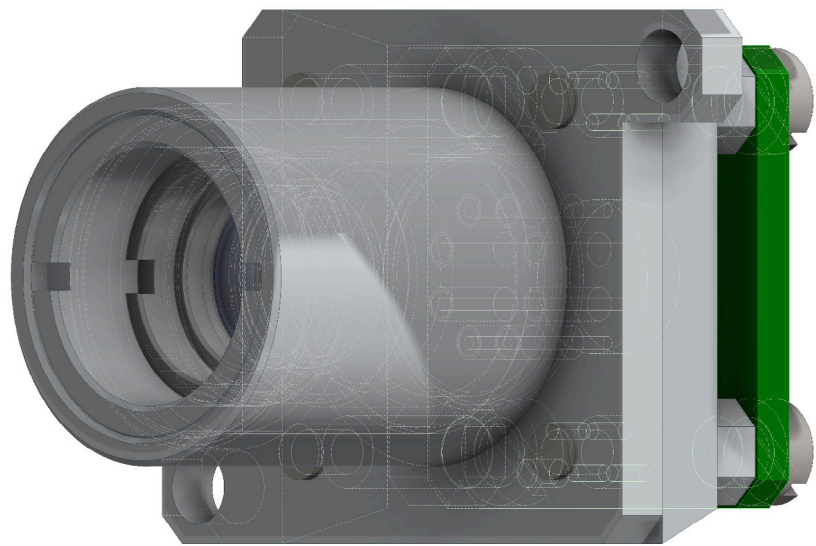

Рисунок 2.21 – Трехмерная модель лазерного модуля.

Расстояние между коллимирующей линзой и лазерным диодом может регулироваться с помощью резьбового соединением оправки линзы с корпусом лазерного модуля. Алюминиевое основание, на которое устанавливается ДЛ, позволяет выравнивать оптическую ось коллимирующей линзы с оптической осью ДЛ. Процесс юстировки лазерного модуля на изготовленной для этой процедуры оснастке представлен на рисунке 2.22.

Все конструктивные детали прибора ДЛС-Л были спроектированы штатными конструкторами Института космических исследований РАН и изготовлены на опытном производстве ИКИ РАН.

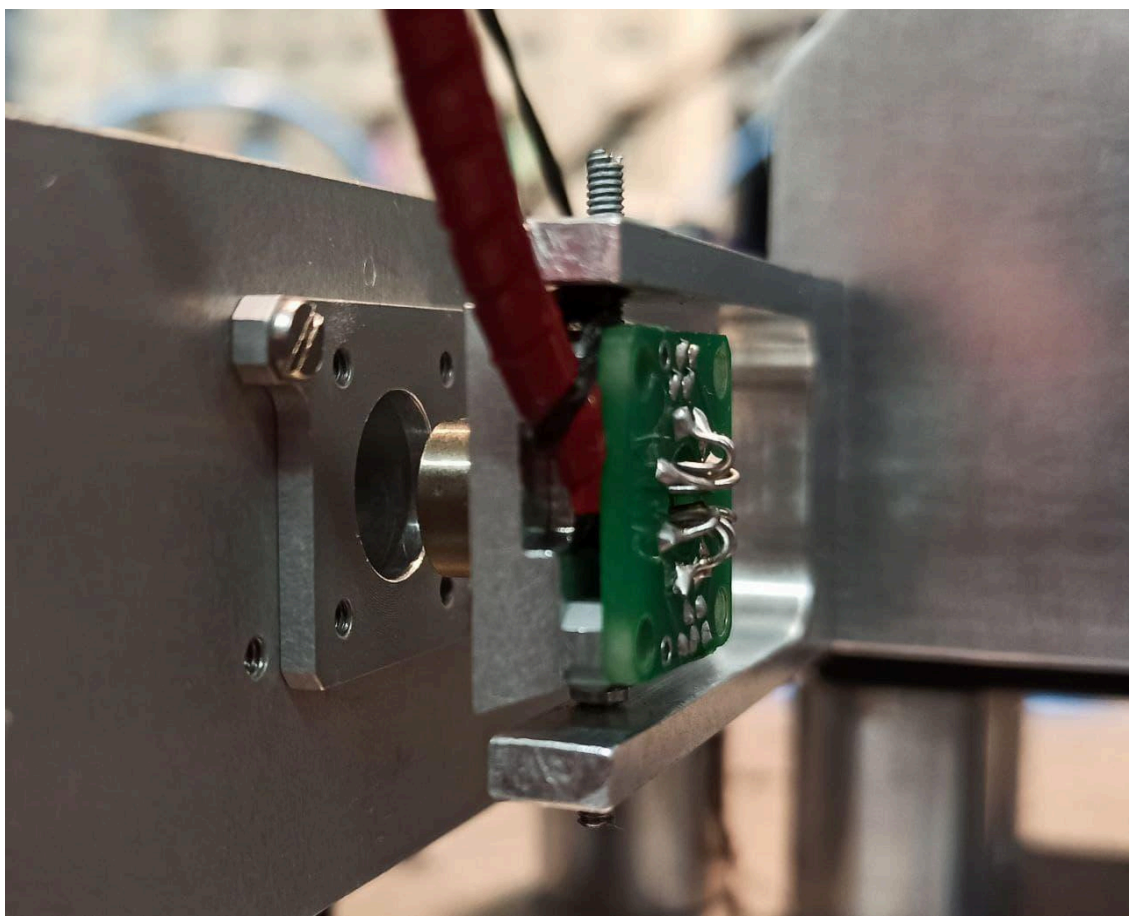

Рисунок 2.22 – Процесс юстировки лазерного модуля.

Качество спектральных данных, получаемых при помощи оптических методов анализа среды, напрямую зависит от итоговой юстировки системы. Жесткие требования, принятые в космическом приборостроении, необходимость проведения сложного цикла испытаний прибора не предполагают возможности внесения изменений в сборку оптической системы после закрепления всех элементов этой системы двухкомпонентным клеем К-400, представляющим собой субстанцию на основе эпоксидно-кремнийорганической смолы Т-111 и отвердителя – низкомолекулярного полиамида Л-20.

По этим причинам для более качественной настройки оптической системы ДЛС-Л, основанной на ИК-лазерах, был использован матричный фотоприемник производства компании

Sofradir (Франция), работающий в диапазоне 2-3 мкм. С его помощью стало возможным провести юстировку ИК-излучения лазерных диодов с точностью ~0.001 рад. Это в свою очередь позволило снизить число возможных отражающих поверхностей, на которых могла бы возникнуть оптическая интерференция. Узкий лазерный пучок проходит сквозь аналитическую кювету с диаметром полости 3 мм, не касаясь ее внутренней поверхности. Пример изображения съюстированного лазерного пучка на матричном фотоприемнике, находящемся на расстоянии 1 м от лазера представлен на рисунке 2.23.

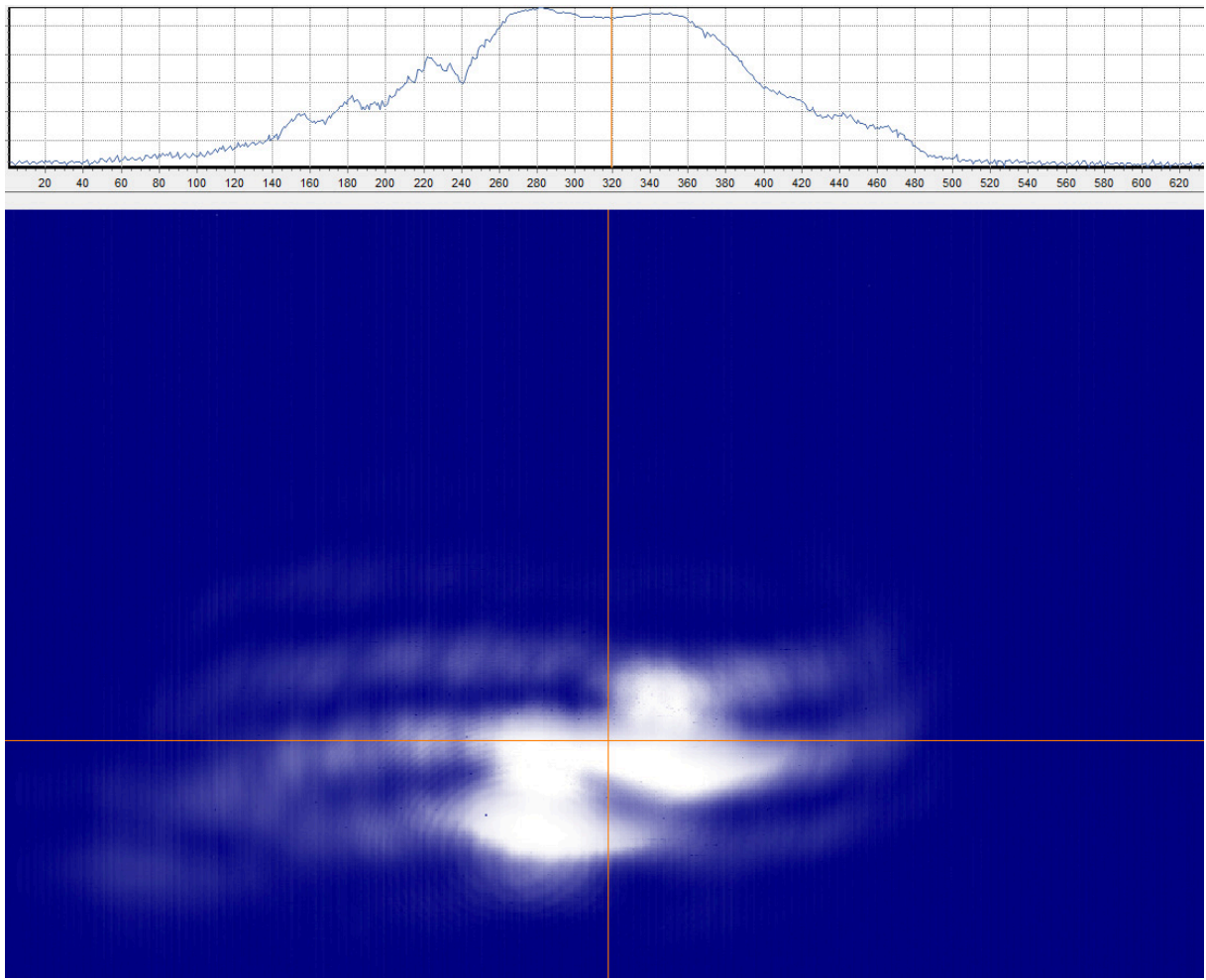

Рисунок 2.23 – Пример изображения съюстированного лазерного пучка на матричном фотосенсоре, находящемся на расстоянии 1 м от лазера.

Однако, как показала серия проведенных экспериментов по изучению газовых смесей в аналитическом объеме, полностью избавиться от вклада оптической интерференции для данной сборки прибора ДЛС-Л не удалось. Более того, по результатам многомесячного цикла измерений изотопии проб воды и $CO_2$ был сделан вывод о фактической невозможности полного подавления интерференции для выбранной конфигурации ДЛС-Л.

### 2.2.3. Электронная система управления ДЛС-Л

Микроконтроллер STM32F437ZIT6 (МК) выполняет все операции с данными, обмен данными и управление периферийными устройствами. Производительность МК обеспечивает частоту дискретизации 128 кГц для настройки тока лазера. 16-разрядный цифро-аналоговый преобразователь (ЦАП) управляет драйвером тока накачки диодных лазеров в диапазоне 0-200 мА. Прибор имеет три трансимпедансных усилителя (ТИУ) для преобразования фототока фотодиода аналитического канала, фотодиода опорного канала и фотодиода монитора в напряжение выходного сигнала. Коэффициент усиления ТИУ регулируется цифровым кодом в диапазоне от 5 кОм до 2 МОм. Далее три 16-разрядных аналого-цифровых преобразователя (АЦП) оцифровывают сигналы ТИУ.

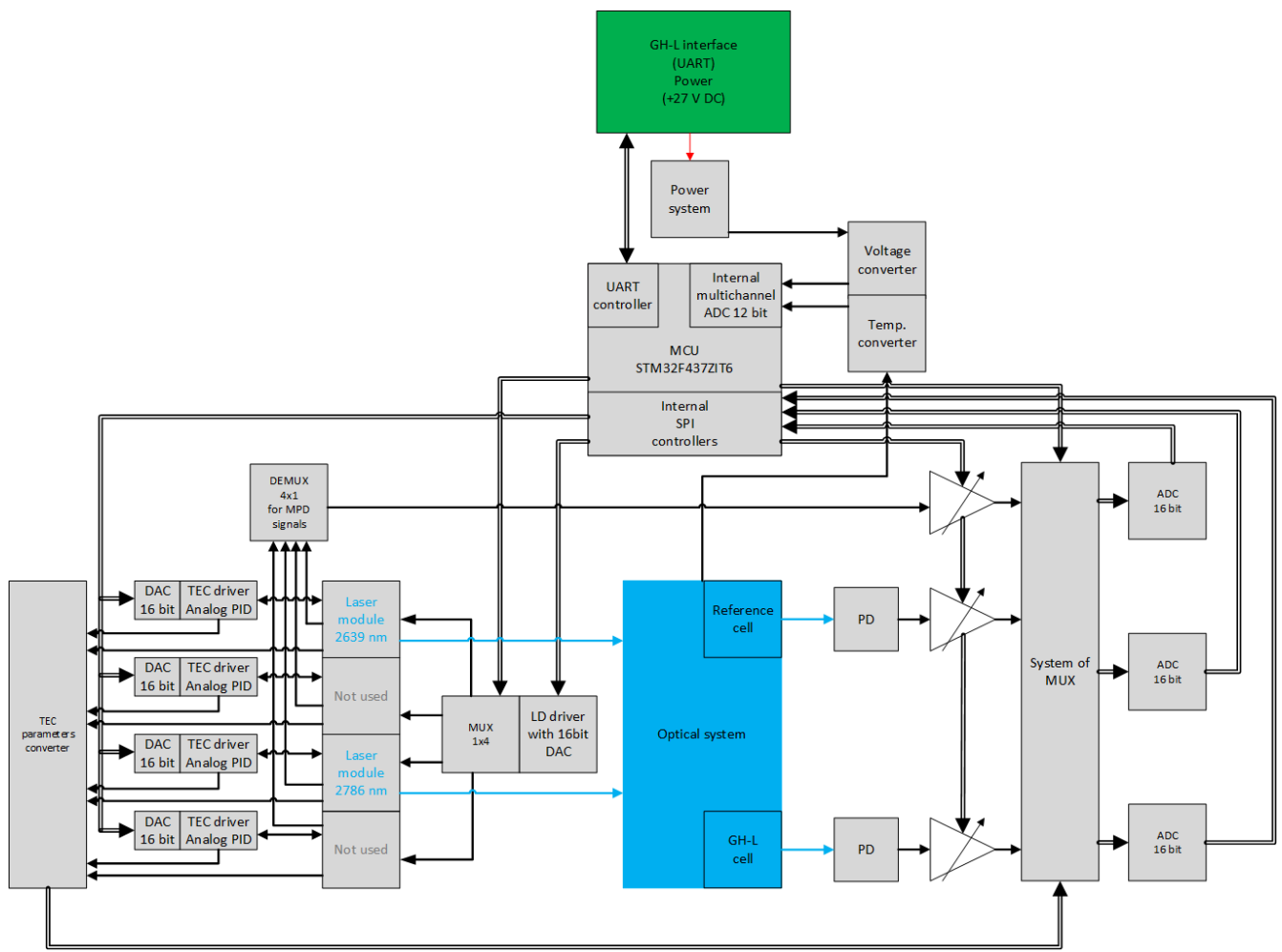

Рисунок 2.24 – Блок-схема управления ДЛС-Л.

Температура лазерных модулей контролируется схемой термоэлектрического охлаждения с аналоговым ПИД-регулятором. Температура устанавливается с помощью 16-разрядного ЦАП в диапазоне от 0 до 40°С. Температурная стабильность составляет 1 мК. Встроенные в MCU 12-разрядные АЦП оцифровывают напряжение питания платы и температуру с термисторов, расположенных на реперной кювете, корпусе прибора, фотодиодах и корпусах лазерных модулей. Средняя потребляемая мощность прибора в рабочем режиме составляет менее 5.5 Вт. Блок-схема управления ДЛС-Л представлена на рисунке 2.24.

Лазерные модули прибора ДЛС-Л работают по очереди, в течение нескольких десятков минут полного сеанса работы на цикл работы каждого лазера отводится 30 сек: 17 сек на точное установление его температурного режима, последующие 3 сек на запись спектральных данных и 10 сек на передачу 34.5 кБ данных текущего спектрального канала измерений по 250 Кбит/с каналу в прибор ГХ-Л. За час работы DLS-L измеряет 120 спектров объемом 4 Мб. Пример циклограммы управления лазерными модулями ДЛС-Л представлен на рисунке 2.25. Периодичность переключения и опроса спектральных каналов ДЛС-Л хорошо соответствует характерным временам динамики молекулярных фазовых переходов и химических реакций, происходящих в процессе контролируемого ГХ-Л пиролиза грунта, составляющим несколько десятков секунд.

Для улучшения соотношения сигнал/шум аналитического канала предусмотрено приборное усреднение сигналов по 32 последовательным записям спектра оптического поглощения газовой пробы в течение интервала времени 0.32 сек. Каждая запись содержит 64 точки нулевого тока, 192 точки начального значения и 1024 точки пилообразной перестройки. Длительность одного спектрального скана – 10 мс.

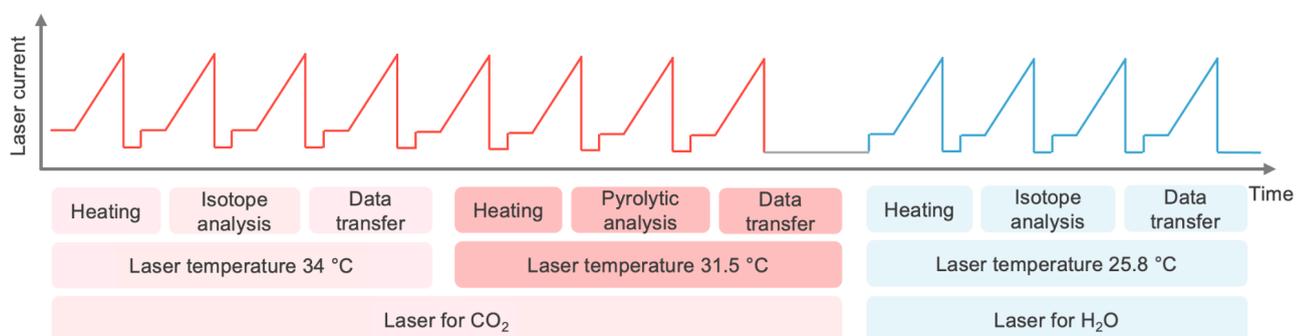

Рисунок 2.25 – Пример циклограммы управления лазерными модулями ДЛС-Л.

Строго синхронная регистрация спектров опорного канала осуществляется на основе компактной реперной кюветы, заполненной газовой смесью – 25% $H_2O$, 25% $CO_2$, и 50% $N_2$ с давлением 50 мбар при комнатной температуре с хорошо локализованными линиями спектрального поглощения. Это позволит выполнить привязку и совместную обработку всей совокупности данных сеанса работы ДЛС-Л для каждого спектрального канала, что обеспечит дополнительное улучшение соотношения сигнал/шум, а также получение прямых интегральных оценок содержания измеряемых газов.

Гибкость системы управления ДЛС-Л под управлением микроконтроллера STM32F437 позволяет как выполнять циклограмму работы ДЛС-Л с параметрами, определенными наземными калибровками, так и корректировать индивидуальные режимы работы каждого лазера по передаваемым с Земли командам управления, что может потребоваться после выполнения перелета космического аппарата «Луна-27» и его посадки на лунную поверхность, а также и по мере его долговременного нахождения и функционирования в условиях больших перепадов окружающей температуры и прочих факторов воздействия внешней среды на аппаратуру лунного автоматического посадочного зонда [143].

## 2.2.4. Циклограмма работы ГАК

Газоаналитический комплекс работает эпизодически, выполняя заданную циклограмму операций. В наборе циклограмм имеются как циклограммы анализа пробы грунта, так и технологические циклограммы, которые обеспечивают приведение систем комплекса в рабочее состояние для проведения анализа.

Перед началом работы все приборы комплекса приводятся в рабочее состояние. У прибора ТА-Л приводы переводят в состояние готовности приема грунта. Проводится прожиг соответствующей пиролитической ячейки перед ее загрузкой для очистки от возможных накопившихся загрязнений. Прибор ГХ-Л также заблаговременно проводит продувку гелием нагретых капиллярных колонок и адсорбционных накопителей для очистки от накопившихся загрязнений [143].

После подготовки комплекса к анализу проводятся операции по загрузке грунта в заданную ПЯ или в многоразовое десорбционное устройство (МДУ). Выгрузка образца реголита в приемные отверстия прибора ТА-Л проводится из грунтозаборного устройства (ГЗУ) манипулятора, куда образец перегружается из бурильного устройства. Затем грунт загружается в заданную ячейку, которая герметизируется для проведения анализа.

В начале стандартной циклограммы анализа проводится прогрев систем приборов для выхода в рабочее состояние. После этого запускается режим термического анализа, и ПЯ отрабатывает заданный профиль нагрева пробы грунта, снимается термограмма пробы. В процессе нагрева выделяющиеся газы переносятся потоком газа-носителя из ПЯ в прибор ГХ-Л. Количество выделяемых в ПЯ газов измеряется микрокатарометром $MK_3$ (рисунок 2.13). Далее газы переносятся потоком газа-носителя через трубку – аналитическую кювету – ДЛС (рисунок 2.26) в последовательно расположенные накопители $AH_2$ и $AH_1$. ДЛС проводит анализ смеси на содержание молекул $H_2O$ и $CO_2$ и измеряет их концентрацию [143].

При необходимости длительных измерений трубка ДЛС с интересующими газами может быть заперта с целью накопления сигнала для повышения точности. В этом случае может быть использован адсорбционный накопитель, представляющий из себя две адсорбционные ловушки – капилярные трубки, заполненные адсорбентом Tenax или Carbosieve SIII. Адсорбент задерживает выделенные при пиролизе газы при охлаждении и выделяет их при нагревании. Эта система аналогична используемой в приборе Sample Analysis at Mars на марсоходе Curiosity [150]. Таким образом, появляется возможность анализа смеси интересующих газов без влияния газа-носителя He. На выходе из трубки ДЛС летучие компоненты ($CO_2$, вода, органические соединения и др.) захватываются накопителем $AH_2$, а прошедшие через него постоянные газы ($CH_4$, CO, благородные газы, др.) собираются в накопителе $AH_1$.

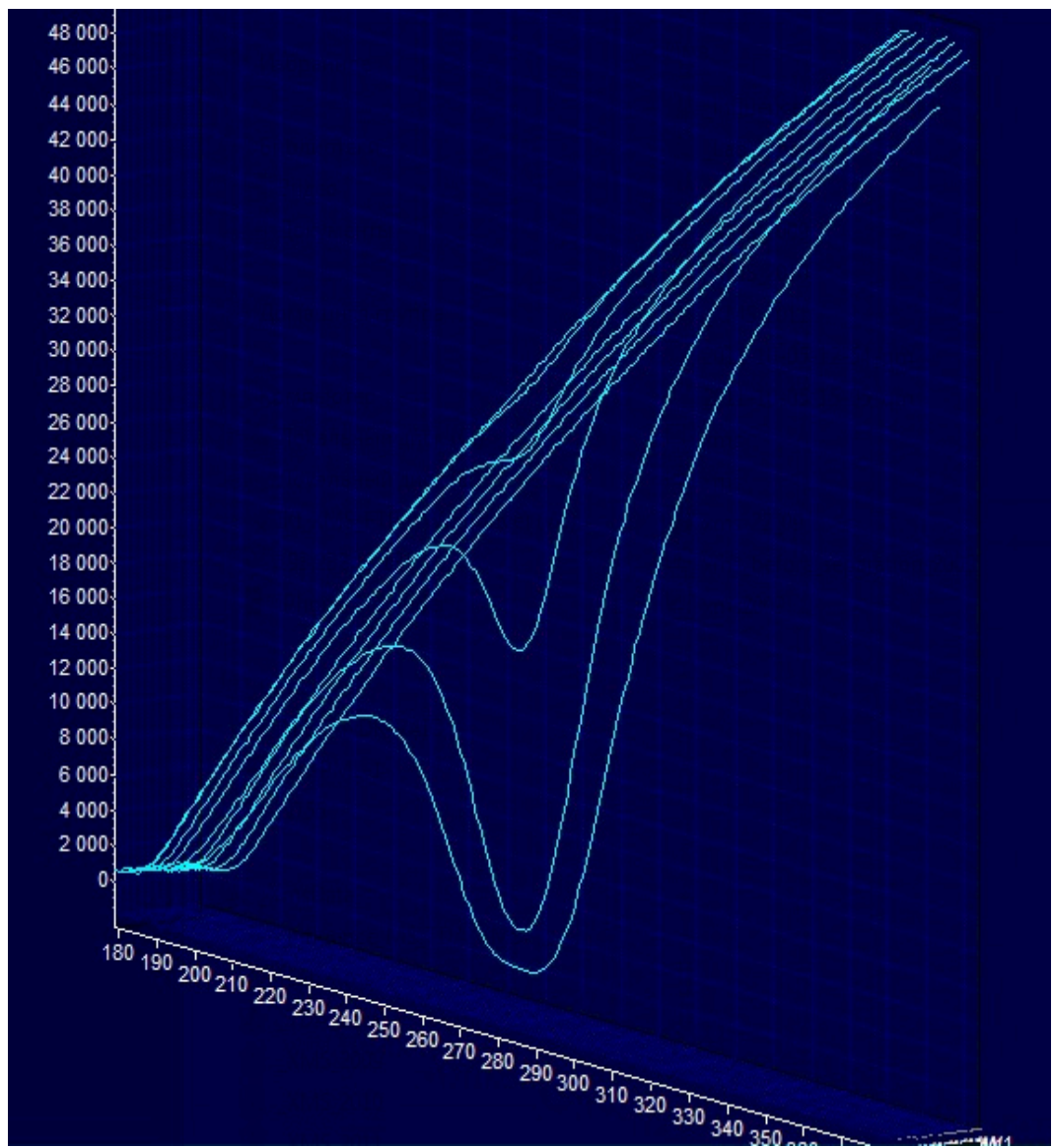

Рисунок 2.26 – Моделирование измерения последовательности спектров за 18 секунд в «проточном» режиме работы путем мгновенного ввода небольшой порции чистого $CO_2$ через технологический вход ГХ-Л и транспортировки ее с газом-носителем He по капиллярам и через трубку ДЛС.

Далее последовательно проводится хроматографический анализ газов из накопителей $AH_1$ и $AH_2$ на соответствующих колонках $KK_1$ и $KK_2$. После завершения аналитической циклограммы комплекс автоматически отключается, а при выборе следующего образца реголита для анализа весь цикл от подготовки, загрузки до анализа повторяется вновь. Время прохождения аналитической циклограммы может длиться от ~40 мин до полутора часов и зависит от поставленной аналитической задачи и энергетических возможностей КА [143].

Основной поток данных комплекса формирует модуль ДЛС, – примерно 4.5 Кбайт/с при общем максимальном потоке 5 Кбайт/с. Суммарный объем данных, генерируемых ГХ-Л и ТА-Л при выполнении типичной циклограммы измерений, составляет около 12 Мбайт за 110 мин работы.

Количество аналитических циклов с использованием высокотемпературных ПЯ ограничено их количеством, поэтому стратегия высокотемпературных измерений рассчитана на восемь таких циклов анализа. Получение проб летучих веществ с помощью МДУ прибора ТА-Л ограничено возможностями взятия образцов буровым устройством и ГЗУ ЛМК.

## 2.3. Определение рабочего спектрального диапазона прибора

Для решения научных задач эксперимента ДЛС-Л выбран метод абсорбционной диодно-лазерной спектроскопии, обеспечивающий высокую чувствительность к детектируемым газам за счет точной регистрации спектральных контуров линий молекулярного поглощения, соответствующих выбранным колебательно-вращательным переходам [101,147,148].

Предусмотрены два режима работы спектрометра ДЛС-Л в составе прибора ГХ-Л. В первом режиме работы прибора достигается субпроцентная погрешность определения содержания составляющих газовой смеси, что в сочетании с высокой скоростью выполнения рабочего алгоритма позволяет эффективно проводить измерения непосредственно в течение пиролитического процесса, сопоставляя интенсивность выхода измеряемых газов с температурой нагрева пробы грунта.

В другом режиме работы продолжительные по времени измерения заранее сформированной в аналитической кювете газовой пробы позволят с высокой точностью измерить интегральные параметры и изотопные соотношения D/H, $^{18}O/^{17}O/^{16}O$, $^{13}C/^{12}C$ для молекул $H_2O$ и $CO_2$.

КДО прибора ДЛС-Л работает в двух диапазонах длин волн, обусловленных используемыми на текущем этапе работы по созданию прибора лазерными диодами: 2639 нм и 2786 нм. В свою очередь используемые ДЛ подбирались по результатам анализа спектральных диапазонов, представляющих наибольший интерес для изучения сравнительно слабых линий поглощения изотопологов молекул $H_2O$ и $CO_2$, а также сильных линий исследуемых газов для удобства работы в режиме измерения концентрации в реальном времени без накопления, среди коммерчески доступных на момент финансирования закупок.

Спектральные линии поглощения молекул $H_2O$ и $CO_2$ в указанном диапазоне длин волн имеют заметно большую интенсивность, чем в значительно более комфортном для работы из-за высокого уровня развития телекоммуникационных технологий ближнем ИК-диапазоне. Так на момент написания диссертации наиболее удобные для применения в диодно-лазерной спектроскопии лазерные диоды с волоконным выводом излучения в силу оптических свойств используемого кварцевого волокна существуют для максимальной длины волны ~2.33 мкм. Однако относительно ближнего ИК-диапазона (1.35 мкм для $H_2O$ и 1.58 мкм для $CO_2$) интенсивность линий $H_2O$ в используемом в спектрометре ДЛС-Л диапазоне 2.64 мкм выше на порядок величины, а интенсивность линий $CO_2$ в диапазоне 2.79 мкм на 3 порядка выше.

При помощи ДЛС-Л содержание двуокиси углерода и водяного пара измеряется в окрестности длины волны 2786.5 нм по наиболее сильным линиям поглощения в охватываемом диапазоне длин волн, изотопологов воды HDO, $H_2^{17}O$ и $H_2^{18}O$ – около 2639 нм, $^{13}CO_2$, $^{17}OC^{16}O$ и $^{18}OC^{16}O$ – около 2786 нм.

Опционально возможны также измерения содержания ацетилена $C_2H_4$ в окрестности 1533 нм или 3077 нм, либо измерения содержания иного характерного углеводорода при наличии соответствующего лазера к моменту сборки, юстировки и окончательной настройки прибора.

Подробная таблица выбранных спектральных линий поглощения перечисленных изотопологов $H_2O$ и $CO_2$ и характеристик соответствующих колебательно-вращательных переходов приведена ниже. Табличные значения параметров соответствующих колебательно-вращательных переходов взяты из базы данных HITRAN [151]. Модельные зависимости коэффициента поглощения от волнового числа приведены для значений давления и температуры, соответствующих параметрам, при которых проводились функциональные испытания конструкторско-доводочного образца ДЛС-Л в составе газового хроматографа ГХ-Л.

При давлении газовой смеси ~30 мбар и поддерживаемой в объеме трубки температуре около 70ºС ожидаемая расчетная глубина линий поглощения может достигать нескольких десятков процентов для $H_2O$ и $CO_2$ и нескольких процентов или долей процента для изотопологов $H_2O$ и $CO_2$.

Таблица 2.2 – Список спектральных линий, вносящих наибольший вклад в поглощение в выбранных спектральных окнах ДЛС-Л.

| № линии | Изотополог | Волновое число, см$^{-1}$ | Интенсивность линии, см/мол | Энергия нижнего уровня, см$^{-1}$ | Ссылка |
|---------|------------|---------------------------|------------------------------|-----------------------------------|--------|
| \multicolumn| | | | | |

| № линии | Изотополог | Волновое число, см$^{-1}$ | Интенсивность линии, см/мол | Энергия нижнего уровня, см$^{-1}$ | Ссылка |
|---------|------------|---------------------------|------------------------------|-----------------------------------|--------|
| **Спектральное окно 2639 нм (3788.2–3789.8 см$^{-1}$)** | | | | | |
| 1 | HD$^{16}$O | 3788.336600 | 1.385e-23 | 221.9462 | [152,155] |
| 2 | H$_2^{16}$O | 3788.755950 | 2.770e-25 | 1718.7188 | [153,154] |
|   | H$_2^{17}$O | 3788.785180 | 2.094e-23 | 36.9311 | [152,154] |
| 3 | H$_2^{16}$O | 3788.809610 | 9.462e-24 | 1843.0296 | [153,154] |
| 4 | H$_2^{18}$O | 3788.912540 | 4.170e-23 | 223.8285 | [153,154] |
| 5 | H$_2^{16}$O | 3789.243660 | 1.092e-22 | 888.6327 | [154] |
| 6 | H$_2^{16}$O | 3789.277420 | 3.264e-22 | 888.5988 | [154] |
| 7 | H$_2^{16}$O | 3789.634760 | 5.473e-22 | 275.4970 | [154] |

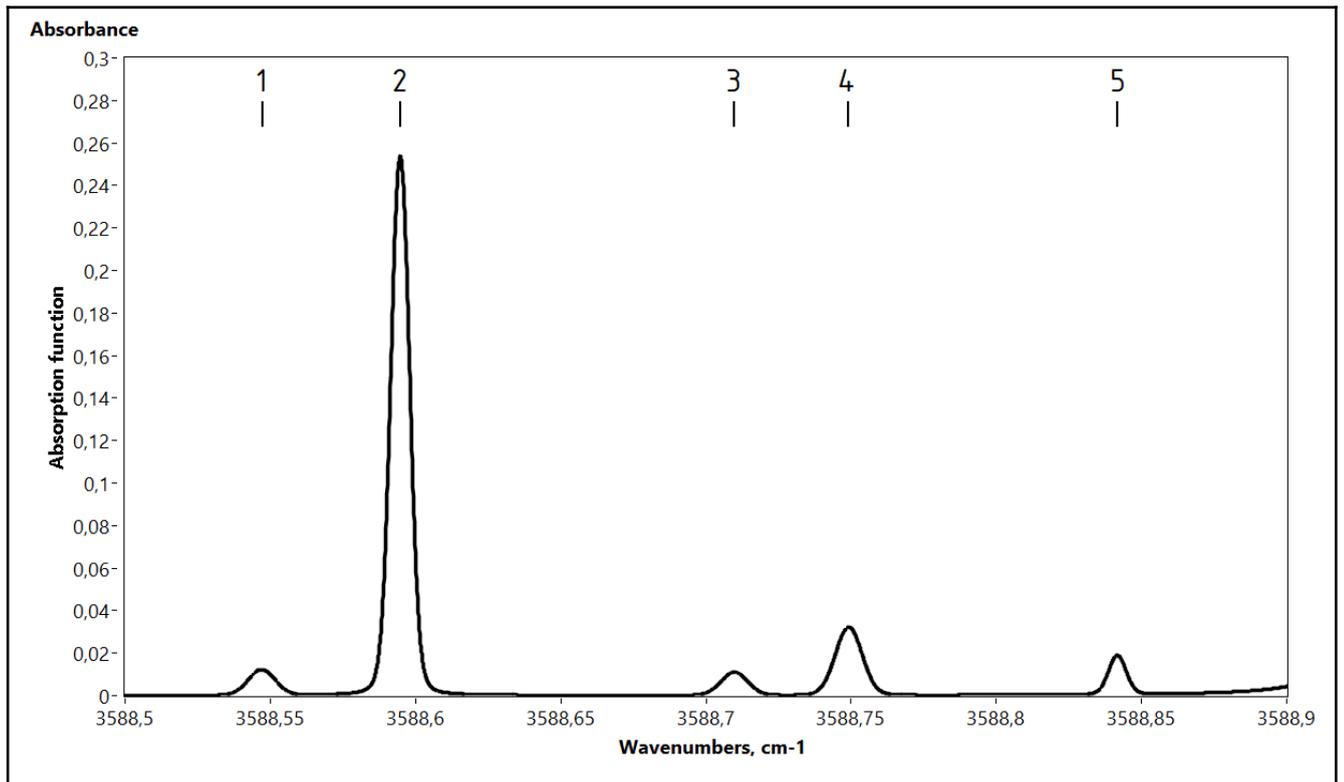

| | Спектральное окно 2786 нм (3588.5–3588.9 см⁻¹) | | | | |
|---|---|---|---|---|---|
| 1 | $H_2{}^{16}O$ | 3588.547150 | 1.328e-20 | 756.7248 | [154] |
| 2 | $^{12}C^{16}O_2$ | 3588.594784 | 1.268e-21 | 710.4163 | [156,157] |
| 3 | $H_2{}^{16}O$ | 3588.710130 | 1.184e-20 | 744.0637 | [154] |
| 4 | $H_2{}^{16}O$ | 3588.749660 | 3.514e-20 | 744.1627 | [154] |
| 5 | $^{12}C^{16}O_2$ | 3588.841584 | 7.491e-23 | 1602.3763 | [156,157] |

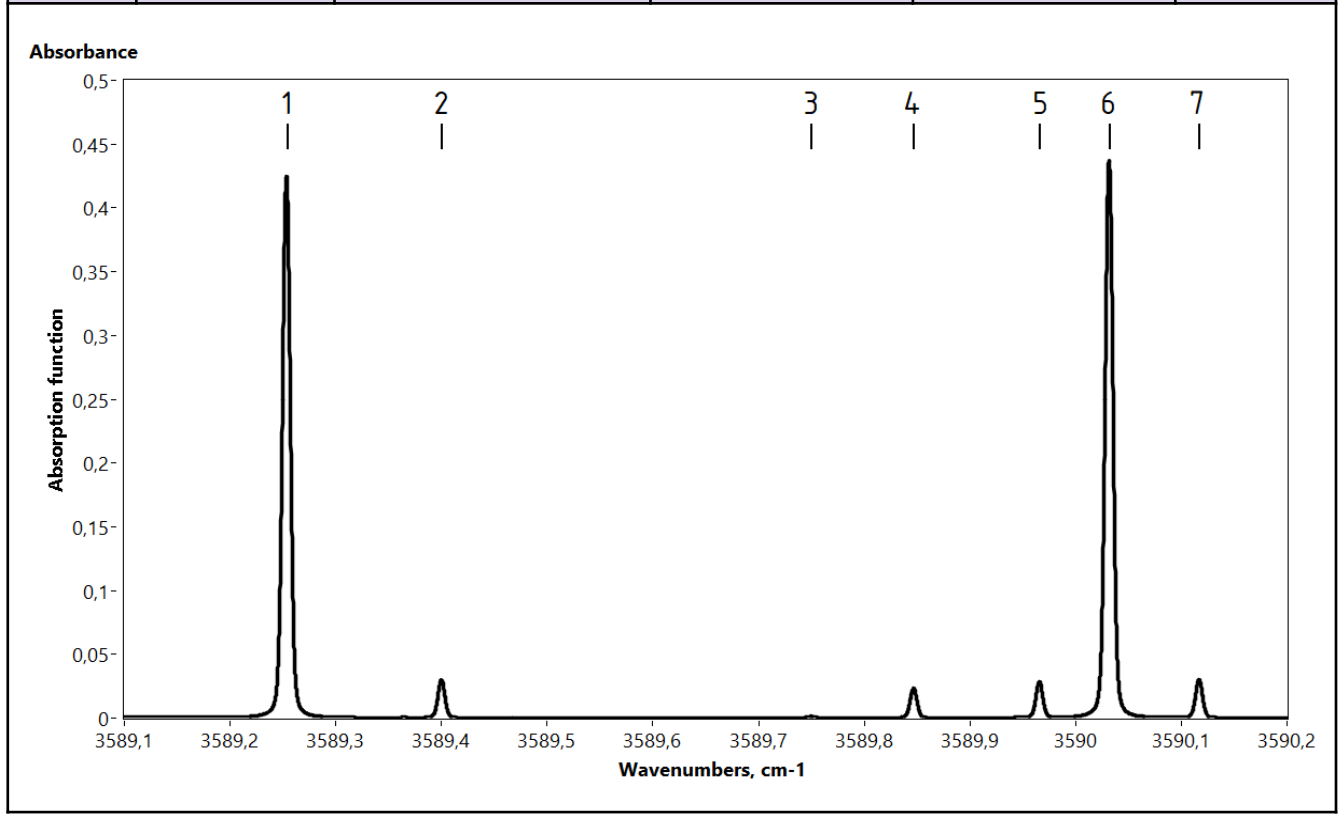

| Спектральное окно 2786 нм (3589.2–3590.2 см⁻¹) | | | | | |
|---|---|---|---|---|---|
| 1 | $^{12}C^{16}O_2$ | 3589.253315 | 1.330e-21 | 718.9419 | [156,157] |
| 2 | $^{16}O^{12}C^{18}O$ | 3589.399934 | 7.113e-23 | 278.2797 | [156,157] |
| 3 | $^{16}O^{12}C^{17}O$ | 3589.749561 | 3.152e-24 | 2.2717 | [156,157] |
| 4 | $^{13}C^{16}O_2$ | 3589.846641 | 5.520e-23 | 843.0701 | [156,157] |
| 5 | $^{16}O^{12}C^{18}O$ | 3589.966458 | 6.582e-23 | 298.8876 | [156,157] |
| 6 | $^{12}C^{16}O_2$ | 3590.031647 | 1.379e-21 | 728.4123 | [156,157] |
| 7 | $^{12}C^{16}O_2$ | 3590.115999 | 6.429e-23 | 1648.4210 | [156,157] |

Как можно видеть из приведенных модельных спектров, наибольшую точность восстановления изотопных отношений в ходе проведения функциональных испытаний прибора стоило ожидать для изотопологов воды $H_2^{17}O$ и $H_2^{18}O$, а также для изотополога двуокиси углерода $^{13}CO_2$ в силу наличия для указанных изотопологов сравнительно интенсивных уединенных линий поглощения, что положительно сказывается на качестве обработки экспериментальных данных.

## 2.4. Методика обработки аналитических данных

Методика диодно-лазерной спектроскопии базируется на существовании зависимости частоты *v* генерации диодного лазера от тока накачки активной области и температуры полупроводникового лазерного кристалла. Полагая температуру активной среды неизменной, будем рассматривать зависимость интенсивности излучения *I* лазерного диода от тока накачки. Генерируемое лазером излучение, проходя сквозь исследуемую газовую среду, поглощается на определенных частотах, соответствующих молекулярным переходам, и попадает на фотоприемное устройство. Напряжение на выходе фотоприемника *U* для случая полупроводникового фотодиода (ФД) с круглой светочувствительной площадкой можно представить в виде:

$$U = RAI\pi r^2 S_p(\nu) = RAe^{-k(\nu)}I_0\pi r^2 S_p(\nu),\qquad(2.2)$$

где *A* – доля излучения ДЛ, попадающая на фотоприемник, $k(\nu)$ – коэффициент поглощения исследуемой среды, *r* – радиус фоточувствительной площадки ФД [см], $S_p(\nu)$ – характеристика фоточувствительности выбранного ФД [А/Вт], *R* – эквивалентное сопротивление трансимпедансного усилителя фотоприемного устройства [Ом]. Определение концентрации исследуемых молекулярных газов в газовой смеси возможно при использовании закона Бугера-Ламберта-Бера [158]. Для узкого пучка излучения в изотропной поглощающей среде закон имеет вид:

$$I(\nu) = I_0(\nu)e^{-k(\nu)} = I_0(\nu)e^{-\alpha(\nu)L},\qquad(2.3)$$

где $\alpha(\nu)$ – объемный коэффициент поглощения [см$^{-1}$], *L* – длина оптического пути [см], $I_0(\nu)$ – интенсивность падающего излучения, $I(\nu)$ – интенсивность излучения, прошедшего поглощающую среду. Величина $I_0(\nu)$ называется спектральным континуумом или базовой линией, поскольку совпадает с уровнем оптического сигнала в отсутствие поглощения, относительного которого и измеряется спектральная функция пропускания среды [13]. В первом приближении базовая линия повторяет по форме характер зависимости токовой накачки лазера от времени, однако в силу нелинейной зависимости частоты генерации диодного лазера от тока накачки она также обладает нелинейностью, в большей степени заметной при широкой полосе частотной перестройки излучения. Также на нелинейные свойства базовой линии может влиять точность температурной стабилизации лазерного кристалла, которая реализуется посредством управления встроенным элементом Пельтье при помощи PID-регулятора, что позволяет добиться стабилизации температуры на уровне $10^{-2}$-$10^{-3}$ К. В случае необходимости более высокой точности стабилизации частоты лазерного излучения применяется, например, методика стабилизации частоты лазерного излучения по реперной спектральной линии [159]. Корректность учета этой нелинейности является основным источником ошибки в диодно-лазерной спектроскопии.

Тонкая структура базовой линии, связанная с флуктуацией тока накачки диодного лазера, попаданием рассеянного излучения внутрь активной области лазерного кристалла, интерференцией на оптических элементах устройства [160], может заметно искажать контуры спектральных линий и препятствовать идентификации линий небольшой интенсивности. Перечисленные причины возникновения тонкой структуры базовой линии могут быть устранены посредством высокоточной стабилизации частоты излучения, а также путем синхронного детектирования $I_o(\nu)$ и $I(\nu)$. Проблема дрейфа и тонкой структуры базовой линии может быть, в частности, решена с помощью детектирования молекулярного поглощения на удвоенной частоте модуляции тока накачки [13,161,162]. Также существуют методы компенсации тонкой структуры базовой линии при обработке сигнала [163,164], основанные на Фурье-анализе базовой линии.

Пренебрегая взаимодействием спектральных линий, коэффициент поглощения в формуле (2.2) можно представить следующим образом:

$$\alpha(\nu) = \sum_{i,j} \sigma_{i,j}(\nu) N_j, \qquad (2.4)$$

где $\sigma_{i,j}(\nu)$ – сечение поглощения i-й линии j-го вещества на частоте $\nu$ [см$^2$/молек], $N_j$ – объемная концентрация j-го вещества [молек/см$^3$]. В случае изолированной линии детектируемого молекулярного газа можно опустить подстрочные индексы. Тогда выражение (2.3) примет вид:

$$I(\nu) = I_0(\nu) e^{-\sigma(\nu) N L}. \qquad (2.5)$$

Сечение поглощения можно представить в виде:

$$\sigma(\nu) = S(T) f(\nu), \qquad (2.6)$$

где $S(T)$ – интегральная интенсивность линии [см$^{-1}$/молек/см$^{-2}$], $T$ – температура [К], $f(\nu)$ – нормированный контур спектральной линии [см].

Рассмотрим детальнее получение сырых данных и решение обратной задачи по восстановлению концентраций отдельных газов и их изотопологов в эксперименте ДЛС-Л. В ходе выполнения рабочей циклограммы прибор ДЛС-Л выдает 16-битный исходный код, содержащий служебную информацию, показания датчиков температуры и спектральные данные (рисунок 2.27). Для дальнейшей работы с экспериментальными данными требуется их конвертация из формата 16-битного кода в десятичные значения АЦП, которые разделяются на массив спектральных данных, показания датчиков температуры и т.д.

```
A3 28 58 00 00 4C 9F A0 40 7D BD 7F D9 72 20 72 20 5E 9F 6E 19 67 8F 00 00 80 00 80 00 C0 18
E3 00 00 10 80 8D A7 1B 44 6D 9C B9 E2 0B 2B 54 80 A1 C8 EE 17 3E 64 8E B8 D9 02 2A 48 70 9B C3
A 1E 3C 61 83 99 B9 D7 F8 1D 3F 66 8C AA CB EE 0A 2A 4B 6A 8A A1 B7 D5 EC 07 26 3C 4F 62 69 73 7F
B2 C7 E1 F4 0E 2A 3B 54 6B 7B 92 A8 BC D5 E3 F4 04 0F 24 3F 57 75 8E A1 B9 CC E7 FF 14 20 34 42 E1
4 45 67 83 9F B8 D5 00 00 00 2A 00 00 00 26 00 00 00 6C 00 00 00 4C B7 D2 5A 5D 44 55 10 F3 53 AC
A3 28 58 00 00 4C 9F A0 40 7D BD 7F D9 72 20 72 20 5E 9F 6E 19 67 8F 00 00 80 00 80 00 C0 18
E3 00 00 10 80 00 3A 3A 3A 3B 3B 3B 3B 3B 3B 3C 3C 3C 3C 3C 3D 3D 3D 3D 3D 3D 3E 3E 3E 3E 3E 3F 3F 3F 3F
8 68 68 69 69 69 69 69 69 69 6A 6A 6A 6A 6A 6A 6B 6B 6B 6B 6B 6B 6C 6C 6C 6C 6C 6C 6D 6D 6D 6D 6D 6D
6F 90 90 90 90 90 90 90 90 90 90 91 91 91 91 91 91 91 92 92 92 92 92 92 93 93 93 93 93 93 93
F AF AF AF AF AF 00 00 00 00 00 00 02 00 00 00 8E 00 00 00 40 06 05 03 03 03 03 08 F5 00 00 00 00 00
A3 28 58 00 00 4C 9F A0 40 7D BD 7F D9 72 20 72 20 5E 9F 6E 19 67 8F 00 00 80 00 80 00 C0 18
E3 00 00 10 80 D9 01 30 4F 7B A7 C7 F5 1E 46 71 98 C2 C6 0B 36 61 9D D3 F5 1E 46 70 9E BB E4 0D 2D 54 7F A0
E E2 FD 22 43 5D 80 9C BE E3 FC 1D 3E 52 6F 90 A8 C8 E7 09 30 4A 6B 92 A9 CA EE 09 28 46 60 83 9E BD DD F0 06 1A 18 1A
EE 0C 2C 40 5B 78 89 A3 B7 C9 E5 FB 19 3E 5A 7B 9A B0 CF ED FD 1C 2F 48 62 77 8D 9D B0 00 00 00 2A 00 00 00 6C 00 00 00 4C B7 D2 5A 5D 44 55 10 F7 53 AC 8E 71 3B C4 4C 9F 00
2 49 62 77 8D 9D B0 00 00 00 2A 00 00 00 26 00 00 00 6C 00 00 00 4C B7 D2 5A 5D 44 55 10 F7 53 AC 8E 71 3B C4 4C 9F 00
A3 28 58 00 00 4C 9F A0 40 7D BD 7F D9 72 20 72 20 5E 9F 6E 19 67 8F 00 00 80 00 80 00 C0 18
```

<div align="center">Рисунок 2.27 – Пример выдаваемого прибором ДЛС-Л 16-битного кода.</div>

Для анализа спектральных данных требуется их предварительная обработка. Исходные сигналы в аналитическом и реперном каналах, примеры которых представлены на рисунке 2.28, являются наборами последовательно записанных 32 раза результатов сканирования в одном спектральном диапазоне, усредненных за 0.32 сек. Усреднение полученного набора также позволяет улучшить отношение сигнал/шум, поскольку на используемых временах усреднения вклад возможного дрейфа электроники еще не сказывается.

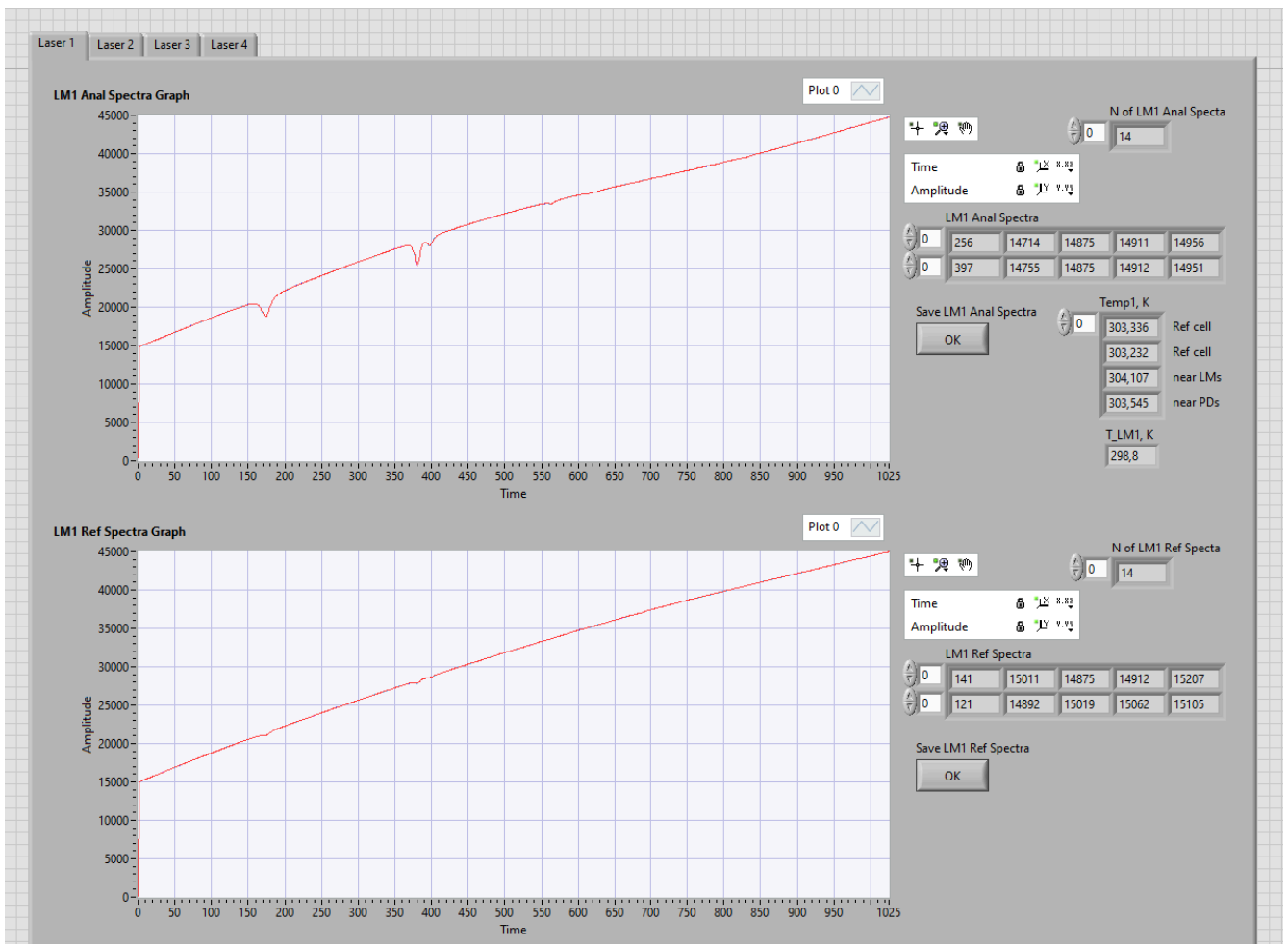

<div align="center">Рисунок 2.28 – Фрагмент интерфейса программы для преобразования 16-битного кода ДЛС-Л, в десятичные значения массива спектральных данных, показания датчиков температуры и пр.</div>

Для дальнейшей работы с полученными спектральными данными необходимо определить базовую линию принятого сигнала, являющуюся, как было сказано выше, уровнем оптического сигнала в отсутствие поглощения, относительного которого измеряется спектральная функция пропускания. Базовая линия не является постоянной величиной и изменяется во времени, в частности, из-за вклада оптической интерференции между отражающими поверхностями и ненулевой ширины полосы лазерной генерации.

С этой целью при первичной обработке данных спектрометра ДЛС-Л, использовалось два подхода. В рамках общепринятого классического подхода можно использовать полиномиальное приближение базовой линии сигнала. Возможен и иной подход к нормировке сигнала, основанный на аппроксимации базовой линии при помощи ортогональных полиномов [165,166]. В случае его применения меняется как алгоритм обработки экспериментальных данных, так и последовательность построения модельных синтетических спектров. Рассмотрим по отдельности каждый подход подробнее.

## 2.4.1. Базовый подход к обработке экспериментальных спектральных данных

При использовании полиномиального приближения базовой линии аналитического сигнала для его вычисления исключаются фрагменты, содержащие спектральные линии поглощения анализируемой газовой смеси. По оставшимся фрагментам сырого сигнала методом наименьших квадратов вычисляется полиномиальный фитинг $n$-той степени.

При этом для вычисления полиномиального приближения заданной степени, наилучшим образом описывающего базовую линию исходного сигнала, использовался алгоритм на основе поворота Гивенса из стандартной библиотеки алгоритмов функции полиномиального приближения, входящей в состав среды разработки LabView. В этой среде была выполнена разработка большей части программного обеспечения обработки экспериментальных данных и их дальнейшего анализа в силу простоты создания графического пользовательского интерфейса для дальнейшей работы. Также при построении полиномиального приближения производился выбор спектрального диапазона сырых спектральных данных, по которому это приближение строилось, что позволяло отсечь начальные и/или конечные участки, не содержащие значимой информации. Отображение сигнала прибора ДЛС-Л и построенного полиномиального приближения базовой линии представлено на рисунке 2.29. Здесь и далее будут приводиться шаги по работе со спектром поглощения водяного пара, однако все то же справедливо и для углекислого газа.

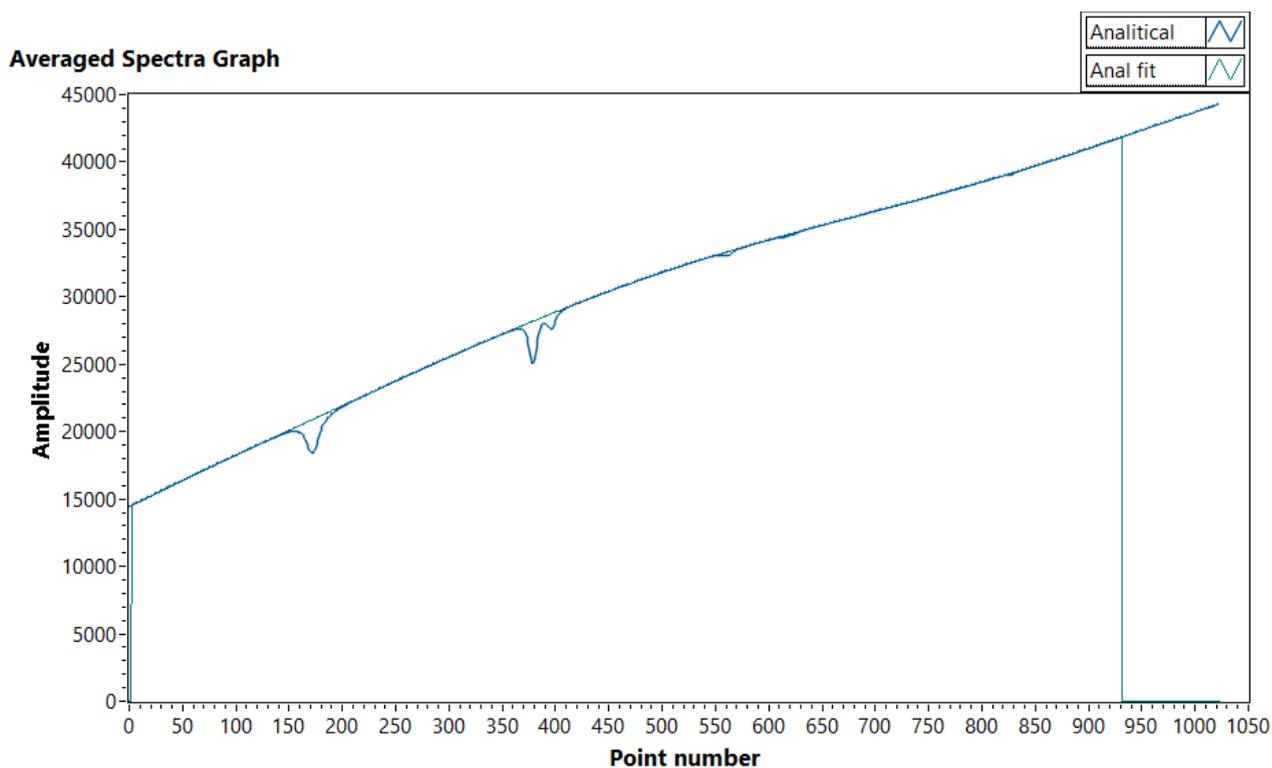

Рисунок 2.29 – Исходный результат сканирования диапазона 3788-3790 см$^{-1}$ по длине волны лазерного излучения в аналитическом канале прибора ДЛС-Л (синяя кривая) и полиномиальное приближение экспериментального результата (бирюзовая) в значениях АЦП.

Нормирование исходного сигнала на полиномиальное приближение базовой линии дает промежуточный результат, имеющий физический смысл функции пропускания газовой смеси (рисунок 2.30).

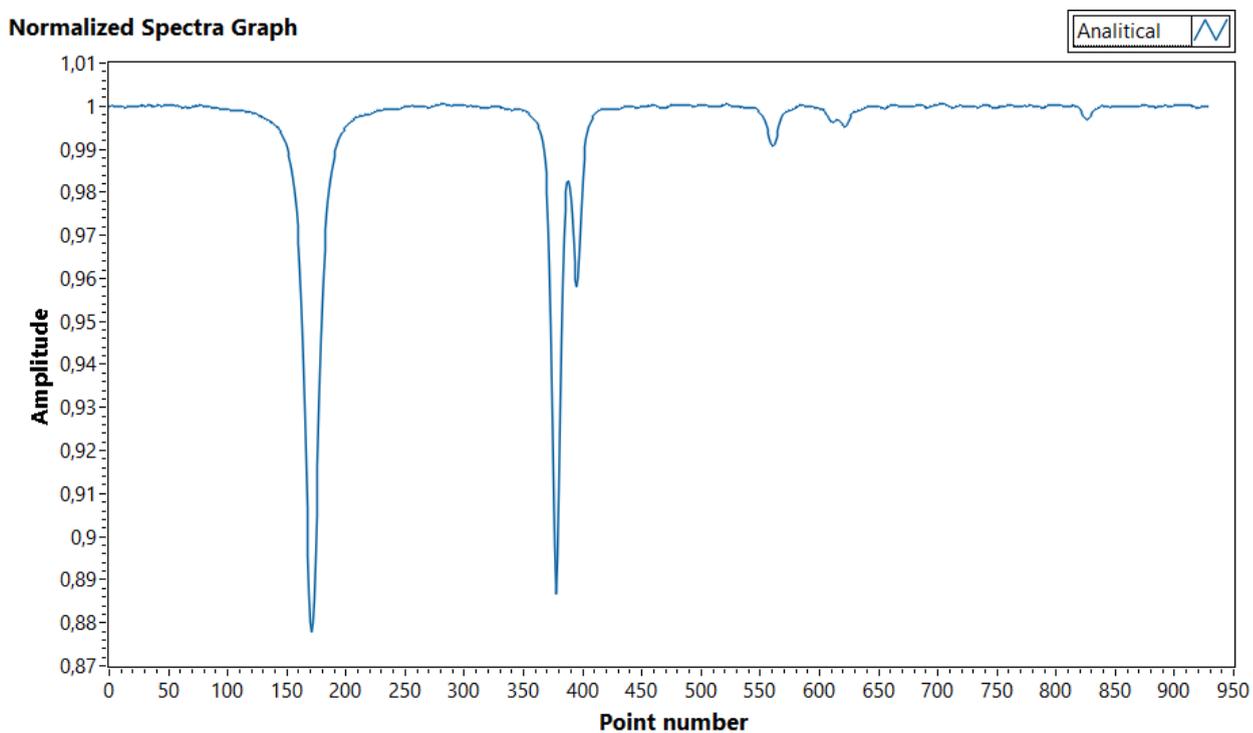

Рисунок 2.30 – Результат нормирования исходных данных на полиномиальное приближение базовой линии.

Как было отмечено выше, линейная токовая перестройка лазерного диода в широком диапазоне значений тока накачки, вариант которой для одного из лазерных модулей представлен на рисунке 2.31, при стабильной температуре кристаллической структуры лазера приводит к нелинейной перестройке по длине волны.

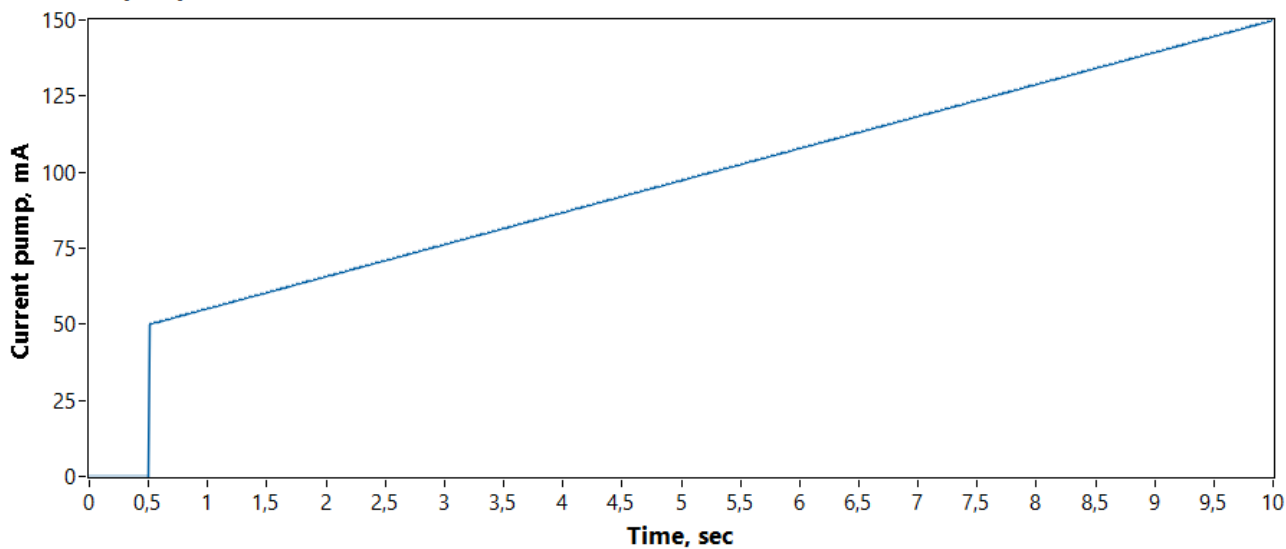

Рисунок 2.31 – Пример характера токовой накачки лазерного диода в приборе ДЛС-Л.

Поэтому следующая необходимая операция – линеаризация горизонтальной развертки обрабатываемых данных, которую можно реализовать с использованием резонатора Фабри-Перо (рисунок 2.32) и данных о положении центров линий поглощения на шкале волновых чисел по базе данных HITRAN (рисунок 2.33).

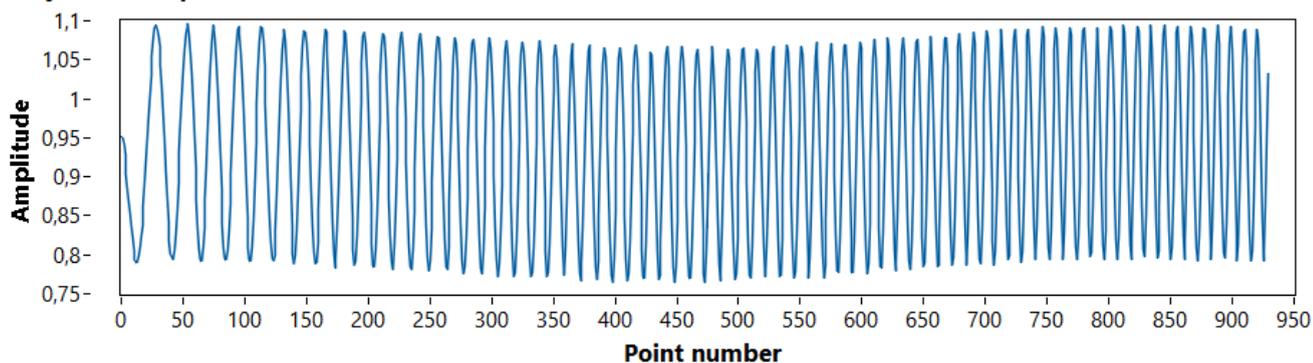

Рисунок 2.32 – Результат перестройки лазерного диода при прохождении излучения через резонатор Фабри-Перо.

Как видно из рисунка 2.32, интерференционные пики резонатора Фабри-Перо неравномерно распределены по области перестройки лазерного излучения при линейном росте тока накачки лазерного диода. В таком случае, зная междомодовое расстояние резонатора, равное в данном случае $2.59 \times 10^{-2}$ см$^{-1}$, можно получить кривую, связывающую номер точки в исходном сигнале с величиной перестройки по волновому числу.

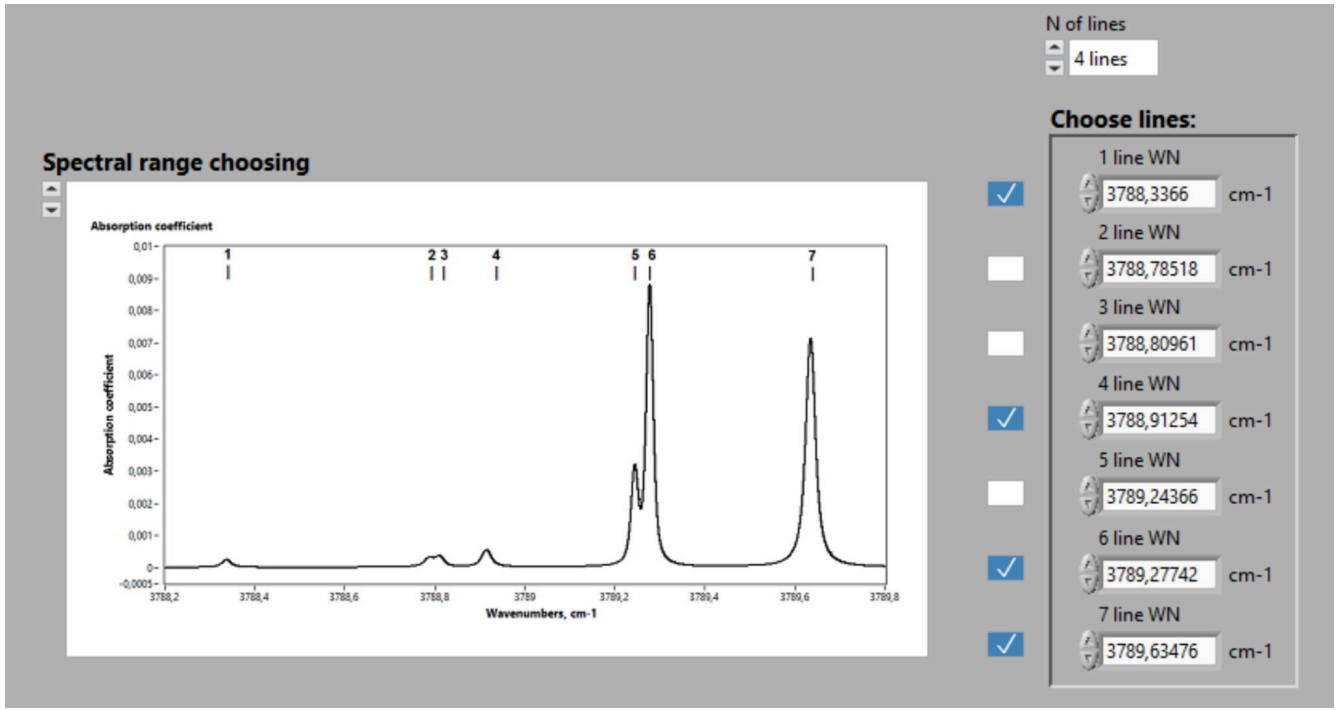

Рисунок 2.33 – Фрагмент интерфейса программы для приведения исходных данных к виду зависимости коэффициента поглощения от волнового числа.

Эту зависимость можно уточнить при помощи системы $N$ алгебраических уравнений, где $N$ – число используемых точных положений пиков спектральных линий поглощения, значения волновых чисел которых взяты из базы данных HITRAN (рисунок 2.34).

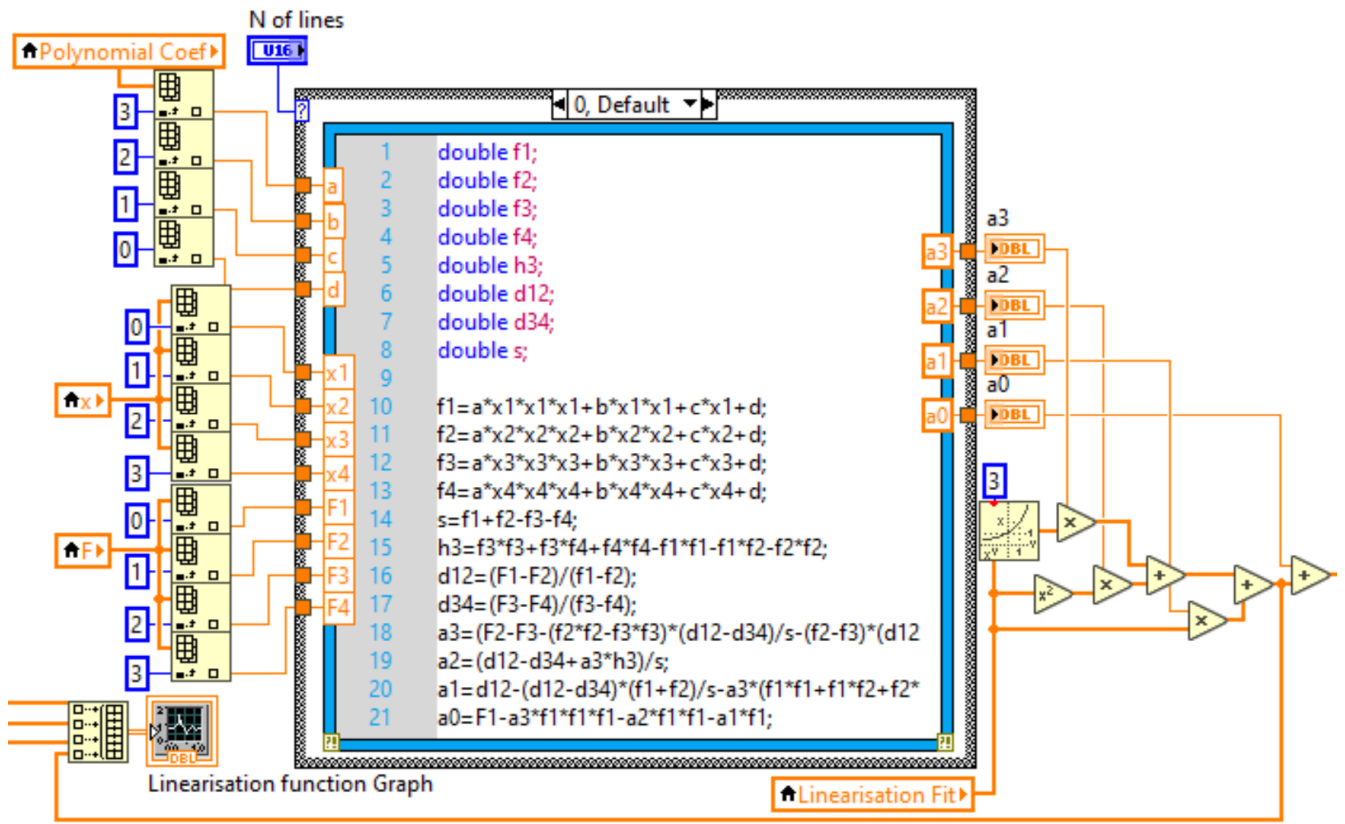

Рисунок 2.34 – Фрагмент блок-диаграммы программы для приведения исходных данных к виду зависимости коэффициента поглощения от волнового числа.

Для этого на этапе получения нормированного сигнала, представленного на рисунке 2.30, определяются положения пиков линий поглощения. Затем выбранному числу из них присваиваются значения волнового числа положения центра соответствующей линии согласно базе данных HITRAN, как показано на рисунке 2.33. В результате решения системы $N$ полиномиальных уравнений $N$-ой степени от функции относительной перестройки, полученной по интерференционным пикам резонатора Фабри-Перо, может быть получена уточненная по положениям пиков спектральных линий поглощения перестроечная кривая, представленная на рисунке 2.35 без учета свободного члена для удобства сопоставления.

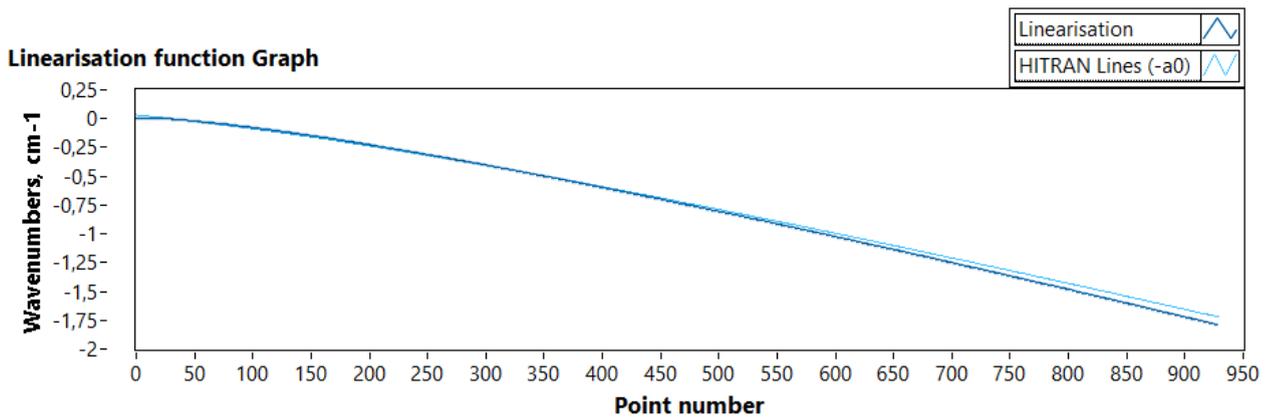

Рисунок 2.35 – Перестроечная кривая, полученная из интерференционных пиков резонатора Фабри-Перо (синяя), и уточненная по положениям пиков спектральных линий поглощения без учета свободного члена (голубая).

Используя полученную перестроечную кривую для проведения процедуры линеаризации горизонтальной шкалы результата нормировки (рисунок 2.30), получим зависимость функции пропускания анализируемой газовой среды от волнового числа (рисунок 2.36).

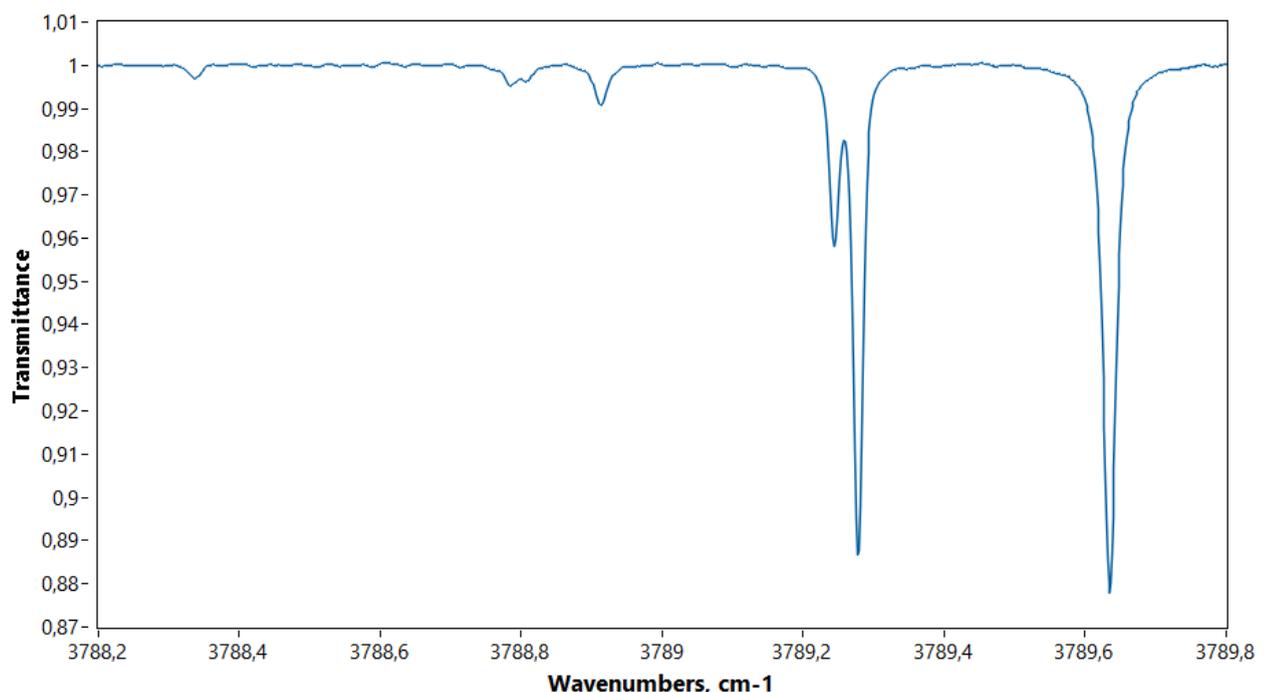

Рисунок 2.36 – Функция пропускания, полученная из экспериментальных данных.

Как видно из рисунка 2.36, такая процедура обработки экспериментальных данных позволяет свести к минимуму вклад медленной оптической интерференции, возникающей на коротких расстояниях. Избавиться от вклада быстрой интерференции, соответствующей дистанции в десятки см, оказывается значительно сложнее, поскольку ее частота не постоянна на всем интервале перестройки лазерного излучения, что отличает ее от возможных электрических наводок постоянной частоты, в случае которых может быть достаточно эффективна Фурье-фильтрация.

Однако, как будет показано далее, эффективность алгоритмического подавления вклада высокочастотной интерференции возрастает при проведении долговременного цикла измерений на временах порядка часа. В таком случае из-за постепенного нагрева прибора, работающего в вакуумированной камере, видимая на регистрируемом сигнале быстрая интерференция будет перемещаться вдоль шкалы волновых чисел. При достаточно длительном времени проведения эксперимента интерференция сдвинется более чем на период, тем самым нивелируя значение своего вклада в неточность определения усредненной за это время интенсивности линий поглощения.

Для дальнейшей работы с экспериментальным спектром стоит преобразовать функцию пропускания в коэффициент поглощения анализируемой газовой среды, показанный на рисунке 2.37.

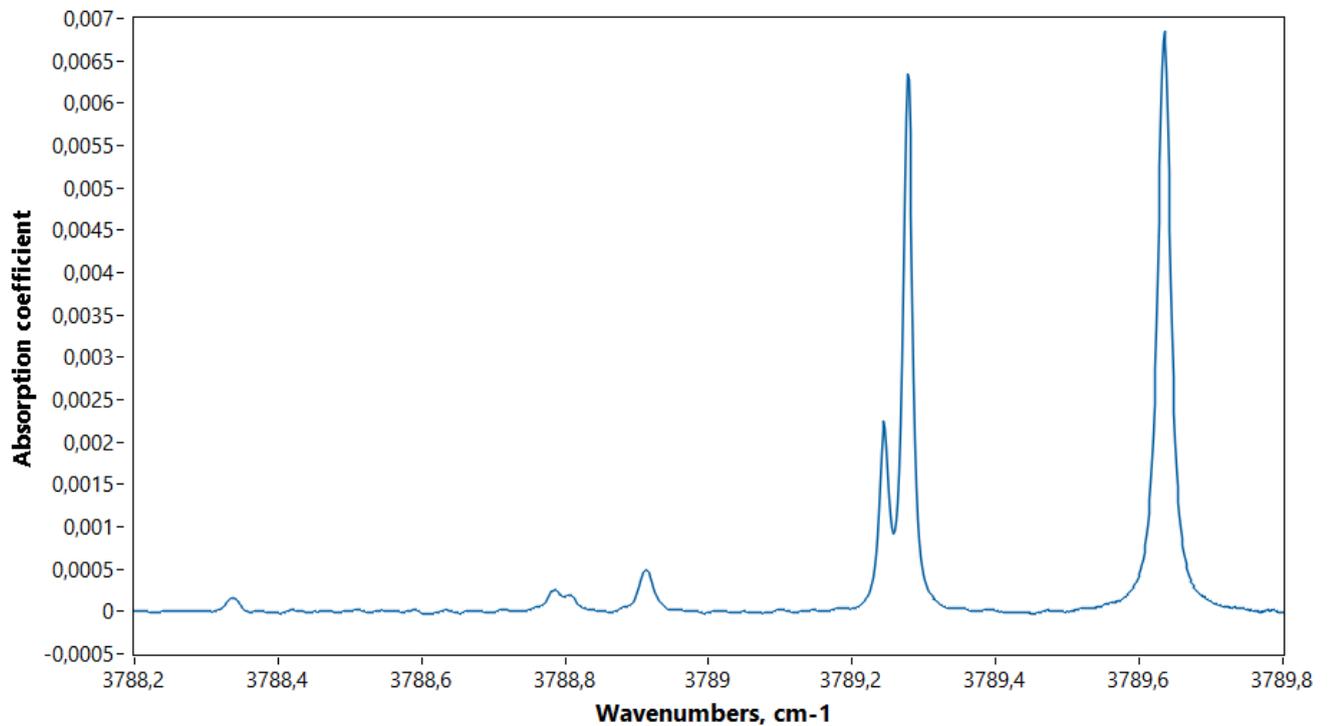

Рисунок 2.37 – Коэффициент поглощения анализируемой газовой среды.

Из закона Бугера-Ламберта-Бера и формулы 2.3 для коэффициента поглощения получим:

$$\alpha = -\frac{ln(T_f)}{L} = -\frac{ln(I/I_0)}{L},$$ (2.7)

где $T_f$ – функция пропускания, $L$ – длина аналитической кюветы.

Для удобства дальнейшей работы с полученным коэффициентом поглощения осуществлялась операция интерполяции с заданным шагом по волновому числу $10^{-5}$-$10^{-4}$ см$^{-1}$, совпадающему с шагом, используемым в модели, и применить фильтрацию Савицкого-Голая.

## 2.4.2. Подход к обработке экспериментальных спектральных данных на основе применения ортогональных полиномов

Метод преобразования интерференционной картины и вычитания фона [165,166] представляет собой итеративный процесс, включающий преобразование базовой линии и вычитание фона. Рассмотрим последовательность преобразований исходного сигнала в рамках этого подхода.

Спектральное поглощение целевого газа $\alpha(\nu)$ из формулы (2.7) можно выразить следующим образом для отдельного энергетического перехода:

$$\alpha(\nu) = S(T) \cdot P \cdot X_i \cdot \phi(\nu, T, P, X_i),$$ (2.8)

где $S$ [см$^{-2}$·атм$^{-1}$] – сила линии, $P$ [атм] – полное давление газа, $X_i$ – мольная доля частиц $i$, а $\varphi$ [см] – функция контура линии.

Как упоминалось выше, на точность методики ДЛС могут влиять многие факторы [167-172], такие как характеристики интерференционного вклада, флуктуации пропускания и эффекты оптического эталона. Принимая во внимание эти факторы, уравнение (2.8) можно расширить до:

$$\alpha_M(\nu) = \alpha_{MT}(\nu) + \alpha_{MBG}(\nu) =$$
$$= \alpha_{MT}(\nu) + \alpha_{MI}(\nu) + \alpha_{MF}(\nu) + \alpha_{ME}(\nu),$$ (2.9)

где $\alpha_M(\nu)$ – полное измеренное поглощение, $\alpha_{MT}(\nu)$ – поглощение целевого газа, а $\alpha_{MBG}(\nu)$ – фоновое поглощение, включающее $\alpha_{MI}(\nu)$, $\alpha_{MF}(\nu)$ и $\alpha_{ME}(\nu)$ – вызванные оптической интерференцией, флуктуациями пропускания и эффектами оптического эталона соответственно.

Аналогично полное модельное поглощение может быть выражено уравнением (2.10), когда поглощение целевого газа перекрывается соседними интерференционными элементами:

$$\alpha_S(\nu) = \alpha_{ST}(\nu) + \alpha_{SBG}(\nu) = \alpha_{ST}(\nu) + \alpha_{SI}(\nu),$$ (2.10)

где $\alpha_S(\nu)$ – полное синтетическое поглощение, $\alpha_{ST}(\nu)$ – синтетическое поглощение целевого газа, а $\alpha_{SBG}(\nu)$ – синтетическое фоновое поглощение, которое в данном случае состоит только из вклада интерференции $\alpha_{SI}(\nu)$.

Для подгонки измеренного профиля поглощения расчетным обычно используется контур Фойгта [173], температуру и концентрацию целевого газа можно получить из достижения наилучшего соответствия между измеренным и смоделированным профилем поглощения. Таким образом, определяя $\varepsilon(\nu)$ как невязку – разницу между измеренным и смоделированным коэффициентом поглощения, имеем

$$\alpha_M(\nu) = \alpha_S(\nu) + \varepsilon(\nu)$$ (2.11)

Когда интересующий частотный диапазон соответствует $N$ точкам $\nu_i$ ($0 \leqslant i \leqslant N - 1$), измеренное и смоделированное поглощение $\alpha_M(\nu_i)$ и $\alpha_S(\nu_i)$ на частоте $\nu_i$ могут быть выражены как:

$$\alpha_M(\nu_i) = \alpha_S(\nu_i) + \varepsilon(\nu_i)$$ (2.12)

Этот шаг преобразует решение в задачу минимизации значения $\varepsilon(\nu_i)$ для наилучшего соответствия измеренного поглощения модельному, что позволит найти оптимальные значения концентрации:

$$\alpha_{MT}(\nu_i) + \alpha_{MBG}(\nu_i) = \alpha_{ST}(\nu_i) + \alpha_{SBG}(\nu_i) + \varepsilon(\nu_i)$$ (2.13)

Более точные результаты могут быть получены при фитинге контуром Фойгта, если фоновое поглощение $\alpha_{MBG}(\nu_i)$ и $\alpha_{SBG}(\nu_i)$ можно определить и вычесть из необработанных измеренных и смоделированных значений поглощения. Для случаев слабого поглощения, когда справедливо линейное расширение формулы Ламберта-Бера, измеренное и смоделированное фоновое поглощение могут быть аппроксимированы полиномами. А чтобы избежать проблемы плохой обусловленности нормальных матриц при обычной полиномиальной аппроксимации высокого порядка, применяется метод ортогональной полиномиальной аппроксимации [174,175] с полиномиальным порядком обычно равным двум или меньше.

Таким образом, идея данного метода состоит в разложении экспериментального и расчетного профилей поглощения $\alpha_M(\nu)$ и $\alpha_S(\nu)$ в ряд ортогональных полиномов с последующим вычитанием первых $k_0$ компонентов разложения, учитывающих неизвестный фон, и анализа модифицированного указанным образом профиля поглощения. Такой метод имеет очевидное преимущество перед стандартным подходом, заключающееся в том, что он позволяет избежать эмпирического процесса выделения непоглощающих участков спектра при разметке исходных данных для процедуры определения базовой линии при помощи полиномиального приближения. Кроме того, количество переменных также уменьшается, что повышает надежность алгоритма подбора.

Проиллюстрируем этот подход формально. Сначала применяется ортогонализация Грама–Шмидта для формирования множества ортогональных многочленов [176]:

$$\alpha_M(\nu_i) = \sum_{k=0}^{K-1} \alpha_M^k P_M^k(\nu_i), \tag{2.14}$$

где $\alpha_M^k$ – коэффициент полинома $P_M^k(\nu_i)$, определяемый формулой

$$\alpha_M^k = \frac{\sum_{i=0}^{N} \alpha_M(\nu_i) P_M^k(\nu_i)}{\sum_{i=0}^{N} \left\lfloor P_M^k(\nu_i) \right\rfloor^2} \tag{2.15}$$

Набор многочленов $P_M^k(\nu_i)$ определяется как ортогональный базис и удовлетворяет следующему условию ортогональности:

$$\sum_{i=1}^{N} P_M^{k_1}(\nu_i) P_M^{k_2}(\nu_i) = \delta_{k_1 k_2}. \tag{2.16}$$

Этот базис можно построить с помощью рекуррентного уравнения:

$$P_M^{k+1}(\nu) = (\nu - \alpha_k) P_M^k(\nu) - \beta_k P_M^{k-1}(\nu), \tag{2.17}$$

где коэффициенты $\alpha_k$ и $\beta_k$ определяются как:

$$\begin{cases} \alpha_k = \dfrac{\sum_{i=0}^{N} \nu_i \left\lfloor P_M^k(\nu_i) \right\rfloor^2}{\sum_{i=0}^{N} \left\lfloor P_M^k(\nu_i) \right\rfloor^2} \\ \beta_k = \dfrac{\sum_{i=0}^{N} \left\lfloor P_M^k(\nu_i) \right\rfloor^2}{\sum_{i=0}^{N} \left\lfloor P_M^{k-1}(\nu_i) \right\rfloor^2} \end{cases} \tag{2.18}$$

И исходные ортогональные полиномы выбраны как:

$$P_M^0(\nu) = 1, \tag{2.19}$$

$$P_M^1(\nu) = \nu - \alpha_0. \tag{2.20}$$

Затем как измеренное, так и смоделированное поглощение разлагаются на множество ортогональных полиномов:

$$\alpha_M(\nu_i) = \sum_{k=0}^{K-1} \alpha_M^k P_M^k(\nu_i), \tag{2.21}$$

$$\alpha_S(\nu_i) = \sum_{k=0}^{K-1} \alpha_S^k P_S^k(\nu_i),$$

$$\tag{2.22}$$

После вычитания первых $k_0$ компонентов разложения модифицированные профили поглощения принимают вид:

$$\alpha_{M,mod}(\nu_i) = \alpha_M(\nu_i) - \sum_{k=0}^{k_0} \alpha_M^k P_M^k(\nu_i), \tag{2.23}$$

$$\alpha_{S,mod}(\nu_i) = \alpha_S(\nu_i) - \sum_{k=0}^{k_0} \alpha_S^k P_S^k(\nu_i). \tag{2.24}$$

Графически исходный сигнал в виде $\alpha_M(\nu)$ из формулы (2.7), разложение на ортогональные полиномы $\alpha_M(\nu_i)$ и базовая линия, вычисленная как $[\alpha_M(\nu_i) - \alpha_{M,mod}]$, на примере того же спектрального диапазона 3788-3790 см$^{-1}$ представлены на рисунке 2.38.

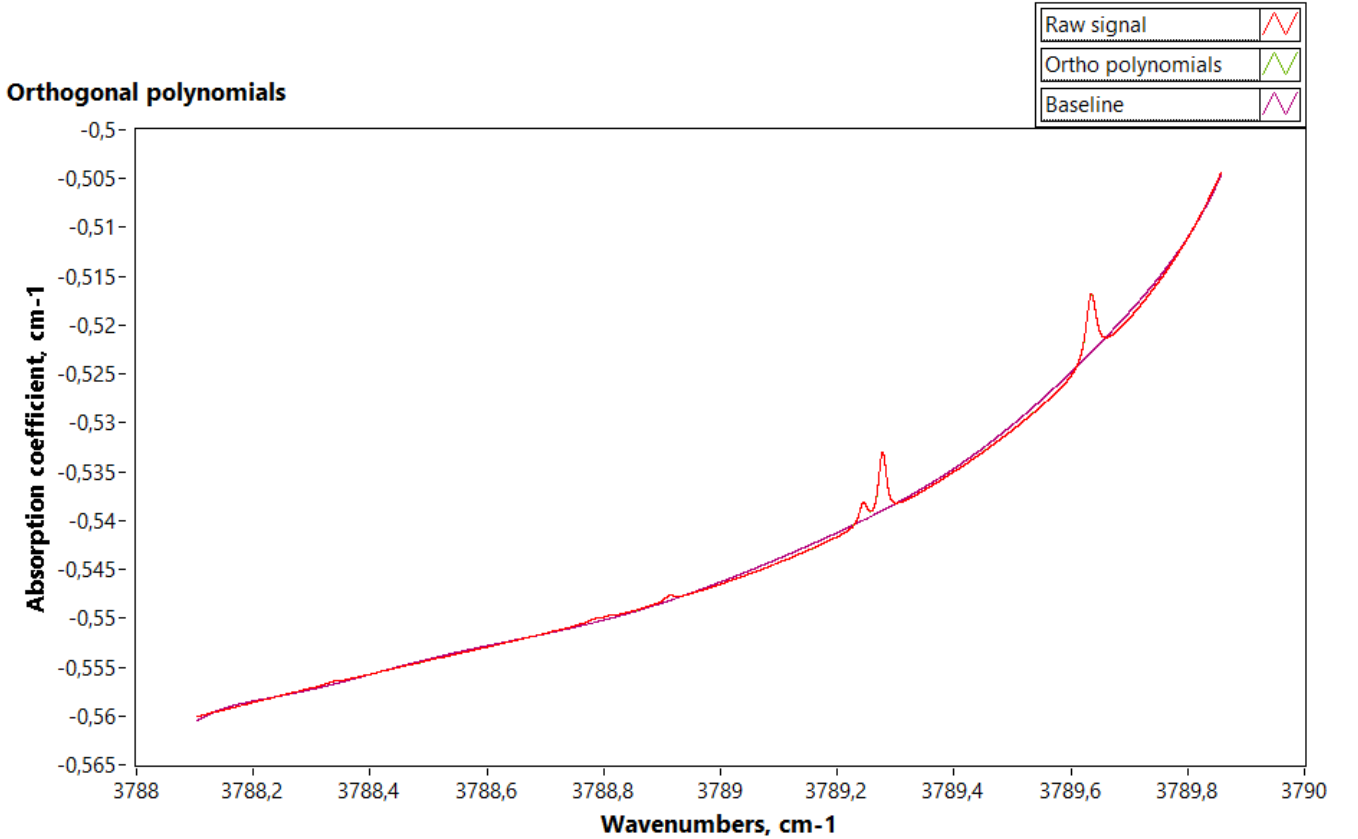

Рисунок 2.38 – Исходный сигнал (красный) в виде $\alpha_M(\nu)$ из формулы (2.7), разложение на ортогональные полиномы $\alpha_M(\nu_i)$ (зеленое) и базовая линия (фиолетовая), вычисленная как $[\alpha_M(\nu_i) - \alpha_{M,mod}(\nu_i)]$, на примере спектрального диапазона 3788-3790 см$^{-1}$.

Для удобства дальнейшей работы с полученным профилем поглощения $\alpha_{M,mod}$ осуществлялась операция интерполяции с заданным шагом по волновому числу $10^{-5}$–$10^{-4}$ см$^{-1}$, совпадающему с шагом, используемым в модели. Наконец, полученный профиль поглощения $\alpha_{M,mod}$ фиттируется $\alpha_{S,mod}$ с использованием критериев минимальной невязки, что будет подробнее описано далее:

$$\sum_{i=0}^{N} \left(\alpha_{M,mod}(\nu_i) - \alpha_{S,mod}(\nu_i)\right)^2 =$$
$$= \sum_{i=0}^{N} \left(\sum_{k=k_0}^{K} \alpha_M^k P_M^k(\nu_i) - \sum_{k=k_0}^{K} \alpha_S^k P_S^k(\nu_i)\right)^2 \to min. \tag{2.25}$$

Результаты обработки исходного сигнала двумя предложенными способами представлено на рисунке 2.39. В качестве результата предложенного в данном разделе метода обработки представлена величина $\alpha_{M,mod}$, которая используется далее при сравнении с модельным спектром поглощения.

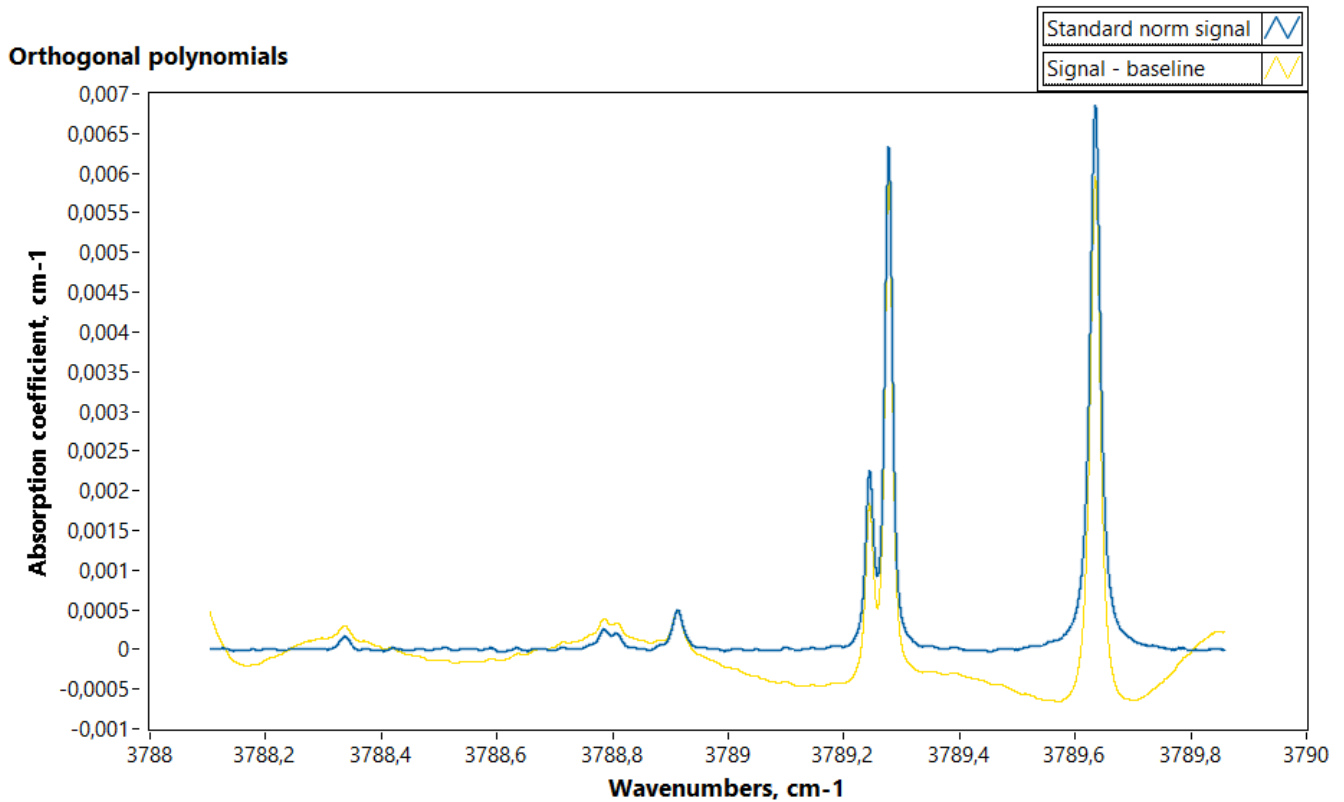

Рисунок 2.39 – Результаты обработки исходного сигнала при помощи базового подхода (синий) и методики на основе применения ортогональных полиномов (желтый).

Целесообразность применения методики ортогональных полиномов, которая может вызвать сомнения при изучении рисунка 2.39, будет показана в разделе 2.4.5 при сопоставлении с результатами моделирования, обработанными согласно этой методике.

### 2.4.3. Методы подавления паразитных составляющих сигнала

Важно отметить, что на временах в единицы секунд, соответствующих времени выполнения циклограммы работы одного лазера в одном спектральном диапазоне, дрейф сигнала незначителен, что позволяет им пренебречь и работать далее с усредненным за время одного цикла измерений результатом. Однако при проведении серии экспериментов с записью результатов сканирования одного спектрального диапазона с характерными временами в десятки минут от начала выполнения первой циклограммы до завершения последней пренебрегать наличием дрейфа сигнала уже не представляется возможным, как видно из рисунка 2.40.

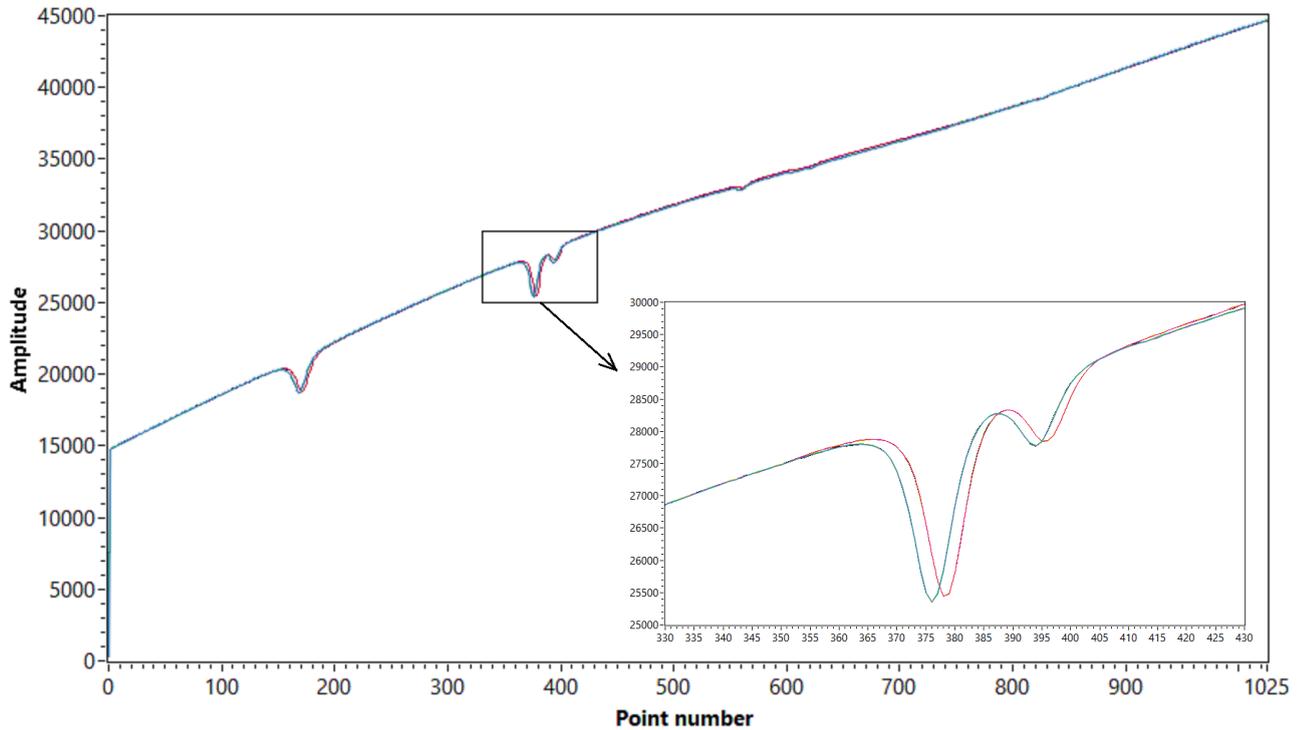

Рисунок 2.40 – Влияние электрического дрейфа на смещение спектральных линий по номерам соответствующих им пикселей для спектрального диапазона 3788-3790 см$^{-1}$ с временным интервалом между приведенными соседними сериями ~10 минут. В увеличенном виде представлен интервал вблизи линий поглощения 3789.24-3789.28 см$^{-1}$.

При этом возникает еще один фактор неопределенности, а именно отсутствие долговременной стабилизации параметров газовой смеси: температуры и давления. Опыт проведения функциональных испытаний показал, что для актуальной на время проведения функциональных испытаний в марте 2023 года сборки КДО ДЛС-Л оба параметра заметно меняются за время проведения эксперимента, что отражено на рисунках 2.41 и 2.42.

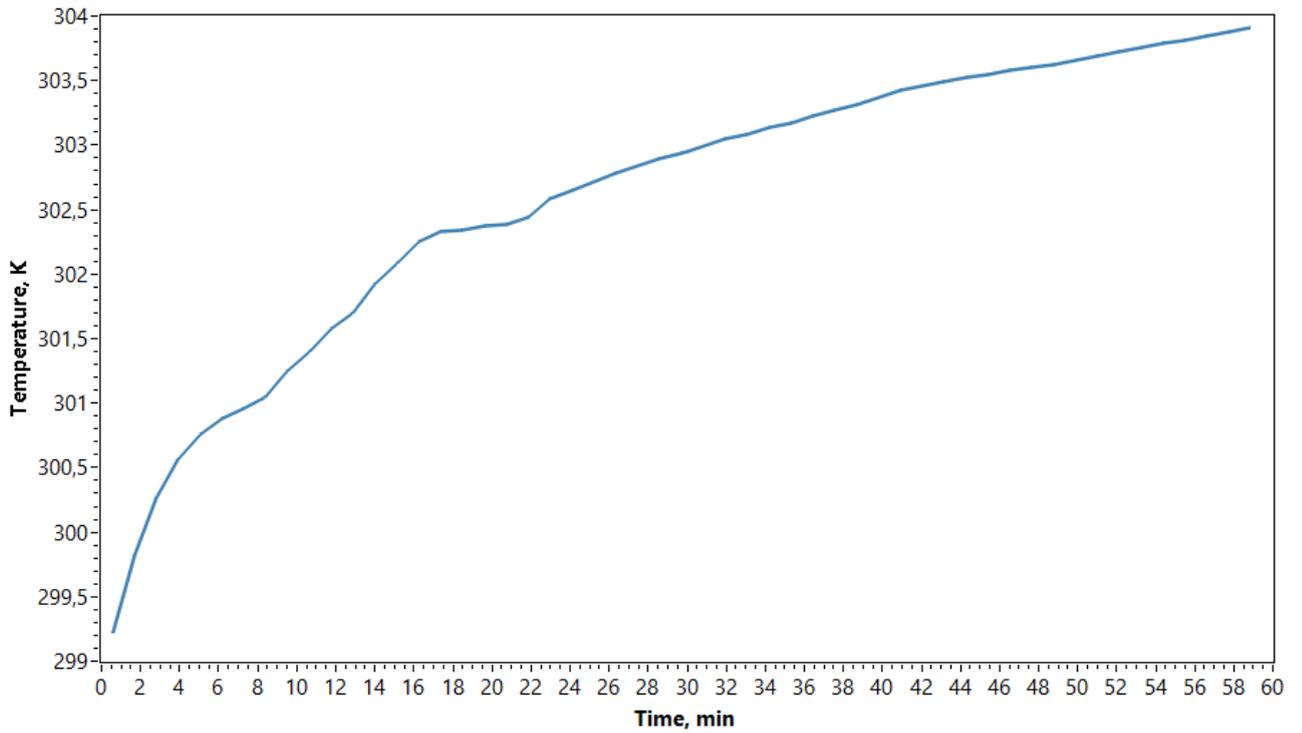

Рисунок 2.41 – Дрейф температуры исследуемой газовой смеси во время проведения эксперимента.

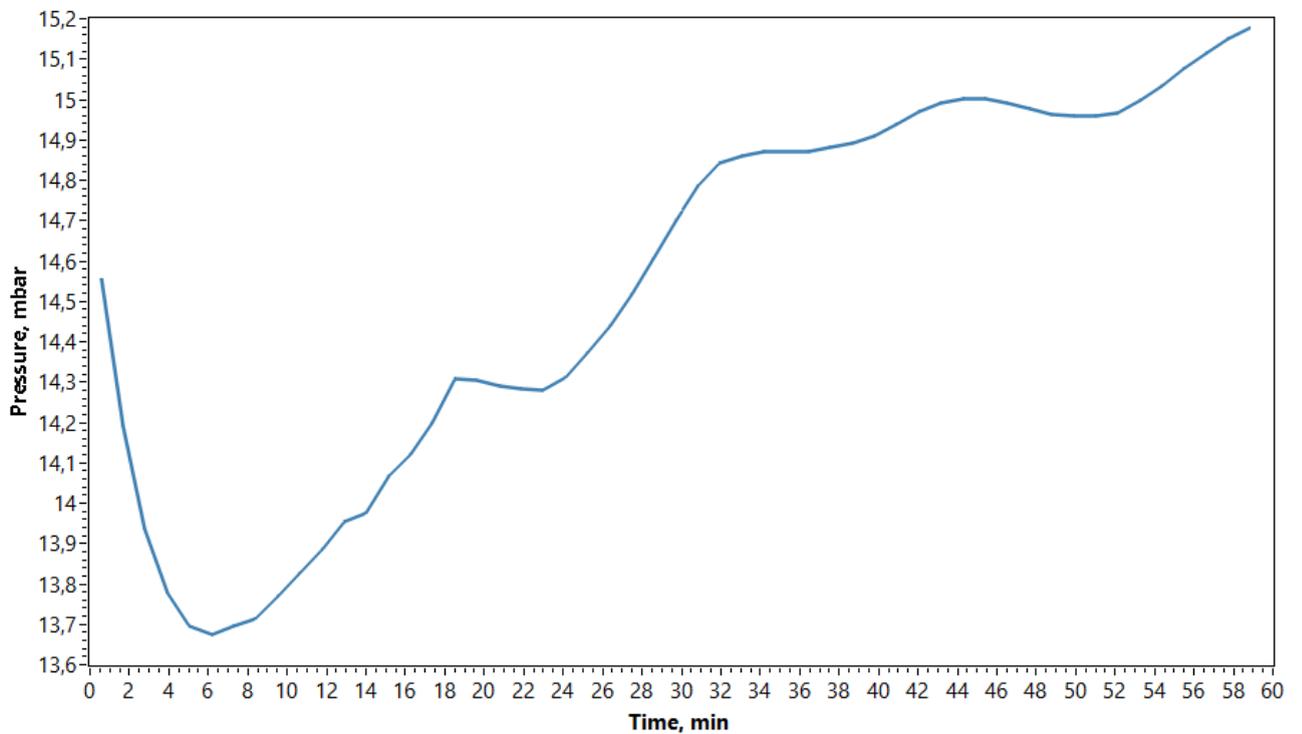

Рисунок 2.42 – Изменение давления исследуемой газовой смеси во время проведения эксперимента.

Это позволяет сделать вывод о необходимости отдельной обработки экспериментальных данных для каждой последующей циклограммы записи пропускания исследуемой газовой среды. При этом последующее усреднение результатов обработки выбранного спектрального

диапазона – коэффициента поглощения – для каждой серии позволяет уменьшить влияние оптической интерференции, за счет чего улучшить качество восстановления изотопных отношений. Это связано с движением высокочастотной интерференционной картины вдоль оси волновых чисел, связанным с изменением температуры системы. Было установлено, что при выходе температуры на относительно постоянную величину после 60 минут работы прибора в вакуумной камере скорость смещения интерференционной картины замедляется почти до нуля.

Таким образом, величина ошибки среднего для $\delta$ из формулы (2.1) после процедуры усреднения спектров в течение 60 минут падает сильнее, чем в $\sqrt{N}$ раз, что связано в том числе с усреднением вклада движущейся вдоль неподвижного спектра поглощения высокочастотной интерференции в связи с естественным термоциклированием системы прибора.

Что касается вклада низкочастотной интерференции, помимо уже упомянутого факта о том, что она довольно хорошо поддается вычитанию при помощи полиномиального приближения базовой линии зарегистрированного сигнала, стоит отметить, что она также может до определенной степени подавляться механической настройкой различных элементов блока ДЛС-Л. То же касается и шумов системы электроники, частично подавляемых путем минимизации наводок извне, улавливаемых сигнальными кабелями и кабелями питания прибора. В качестве примера можно привести исходный регистрируемый сигнал, полученный в 2021 году, и его качественное улучшение к марту 2023 года, приведенный на рисунке 2.43.

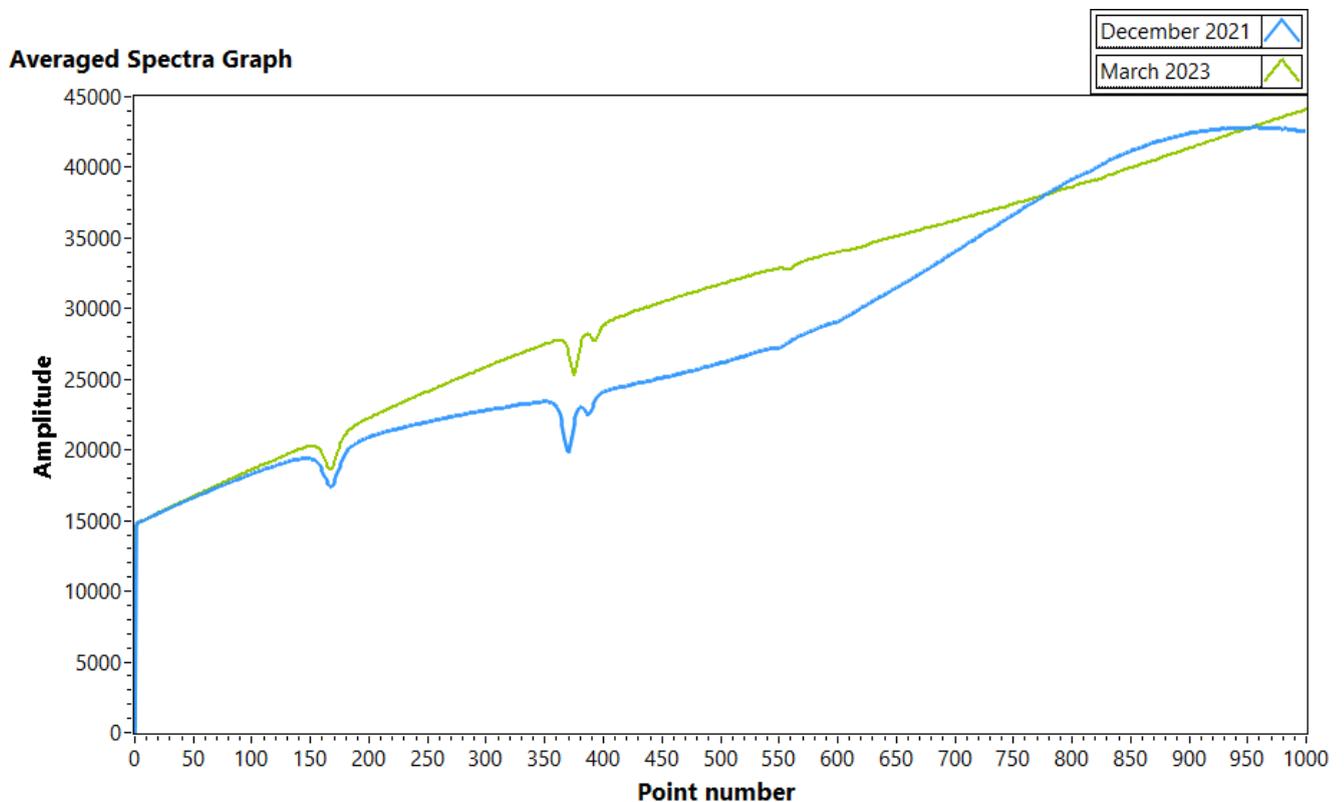

Рисунок 2.43 – Регистрируемый сигнал, соответствующий спектральному диапазону 3788-3790 см$^{-1}$, полученный в 2021 году, и его качественное изменение к марту 2023 года.

## 2.4.4. Сравнение различных профилей спектральных линий поглощения

Результат моделирования по данным базы HITRAN выбранного спектрального диапазона для конкретного изотополога определенной молекулы газа будет напрямую зависеть от используемого контура уширения спектральных линий поглощения. Для получения модельных спектров, соответствующих результатам экспериментов, проведенных в рамках испытаний прибора ДЛС-Л, учитывались основные механизмы уширения спектральных линий поглощения. Естественное уширение, связанное с принципом неопределенности Гейзенберга и характеризующееся временем жизни возбужденного состояния, – теоретический предел ширины спектральной линии, составляющий в ближнем и среднем ИК-диапазонах для полуширины линии на полувысоте порядка 0.1 МГц. Столкновительное уширение, обусловленное флуктуациями амплитуды, частоты и фазы колебательно-вращательных процессов в молекулах при столкновении с другими молекулами [13]. Наконец допплеровское уширение связано с тепловым движением молекул газа. Первые два механизма вносят вклад в лоренцеву составляющую контура уширения, допплеровский механизм уширения определяет вид контура уширения, имеющего форму гауссиана.

Спектральная линия уширяется около волнового числа энергетического перехода $\nu_0$, причем разброс представлен нормированной функцией формы линии $f$. Применительно к механизму уширения Допплера профиль Гаусса, описывающий форму линии поглощения при низком давлении в среде, можно представить в виде:

$$f_G = \sqrt{\frac{ln2}{\pi \cdot \gamma_D^2}} exp\left[-\frac{ln2 \cdot (\nu - \nu_0)^2}{\gamma_D^2}\right], \tag{2.26}$$

где $\gamma_D$ – полуширина спектральной линии на полувысоте (ПШПВ) [см$^{-1}$], $\nu$ – частота сканирующего излучения [см$^{-1}$], $\nu_0$ – центральное волновое число спектральной линии в вакууме [см$^{-1}$]. Величина полуширины спектральной линии на полувысоте $\gamma_D$ определяется как

$$\gamma_D = \sqrt{\frac{2N_a \cdot k_B T \cdot ln2}{M}} \cdot \frac{\nu_0}{c} = 3.581 \times 10^{-7} \cdot \sqrt{\frac{T}{M}}\nu_0, \tag{2.27}$$

где $N_a$ – число Авогадро, $k_B$ – постоянная Больцмана, $c$ – скорость света, $T$ – температура среды [К], $M$ – молярная масса выбранного изотополога [г/моль].

В условиях высокого давления, что соответствует, например, нижним слоям земной атмосферы, преобладает столкновительное уширение спектральных линий, описывающий этот механизм уширения профиль Лоренца выражается следующим образом:

$$f_L = \frac{1}{\pi} \cdot \frac{\gamma_L}{\gamma_L^2 + [\nu - (\nu_0 + \delta_L)]^2}, \tag{2.28}$$

где $\gamma_L$ – лоренцева полуширина спектральной линии на полувысоте [см$^{-1}$], $(\nu_0 + \delta_L)$ – смещенное положение центра линии, обусловленное значительным влиянием давления $p$, $\delta_L$ – параметр сдвига [см$^{-1}$], определяемый как:

$$\delta_L = \delta_{air} \cdot (p - p_{self}) + \delta_{self} \cdot p_{self}. \tag{2.29}$$

Здесь $\delta_{air}$ – параметр сдвига, связанный с давлением воздуха, $\delta_{self}$ – параметр сдвига, связанный с собственным давлением исследуемого газа. Параметр сдвига должен также включать температурную зависимость, но этот эффект до сих пор не учитывается в HITRAN.

Лоренцева полуширина на полувысоте $\gamma_L(p, T)$ для газа при давлении $p$, температуре $T$ и парциальном давлении выбранного газа $p_{self}$ рассчитывается как:

$$\gamma_L = \left(\frac{T}{T_{ref}}\right)^{n_{air}} [\gamma_{air} \cdot (p - p_{self})] + \left(\frac{T}{T_{ref}}\right)^{n_{self}} \gamma_{self} \cdot p_{self} + $$
$$+ 2.5 \times 10^{-10} \tag{2.30}$$

где $\gamma_{air}$ — столкновительная полуширина на полувысоте уширения воздухом [см$^{-1}$/атм] при $T_{ref}$ = 296 К и эталонном давлении $p_{ref}$ = 1 атм, $\gamma_{self}$ — столкновительная полуширина на полувысоте самоуширения [см$^{-1}$/атм] при тех же параметрах. В отсутствие других данных коэффициент температурной зависимости полуширины самоуширения $n_{self}$ принимался равным таковому полуширины уширения воздухом $n_{air}$. Свободный член обусловлен вкладом естественного уширения, также описываемого функцией Лоренца.

Для учета и столкновительного и доплеровского уширений спектральных линий используется контур Фойгта [173], который обеспечивает плавный переход от доплеровского контура линии при низком давлении к лоренцевому контуру при высоком давлении, поскольку является сверткой двух описанных ранее контуров:

$$f_V = \frac{1}{\gamma_D} \sqrt{\frac{ln2}{\pi}} \frac{y}{\pi} \int_{-\infty}^{\infty} \frac{e^{-t^2}}{y^2 + (x - t)^2} \, dt, \tag{2.31}$$
$$x = \sqrt{ln2} \frac{\nu - \nu_0}{\gamma_D},$$
$$y = \sqrt{ln2} \frac{\gamma_L}{\gamma_D}.$$

Интенсивность (сила осциллятора) спектральной линии определяется для отдельной молекулы на единицу объема, ее можно представить в виде области под контуром линии на рисунке 2.44. Интенсивность спектральной линии перехода $S$ [см$^{-1}$/молек/см$^{-2}$] между двумя колебательно-вращательными состояниями при температурах, отличных от эталонной температуры 296 К задается как

$$S(T) = S(T_{ref}) \cdot \frac{Q(T_{ref})}{Q(T)} \cdot \exp\left( c_2 E^{"} \left( \frac{1}{T_{ref}} - \frac{1}{T} \right) \right) \cdot \frac{\left[ 1 - \exp\left( -\frac{c_2 \cdot \nu_0}{T} \right) \right]}{\left[ 1 - \exp\left( -\frac{c_2 \cdot \nu_0}{T_{ref}} \right) \right]}, \qquad (2.32)$$

где $Q(T)$ – полная статистическая сумма, $c_2$ – вторая радиационная постоянная ($c_2 = 1.4387769$ см·К), $E^{"}$ – энергия нижнего уровня перехода [см$^{-1}$].

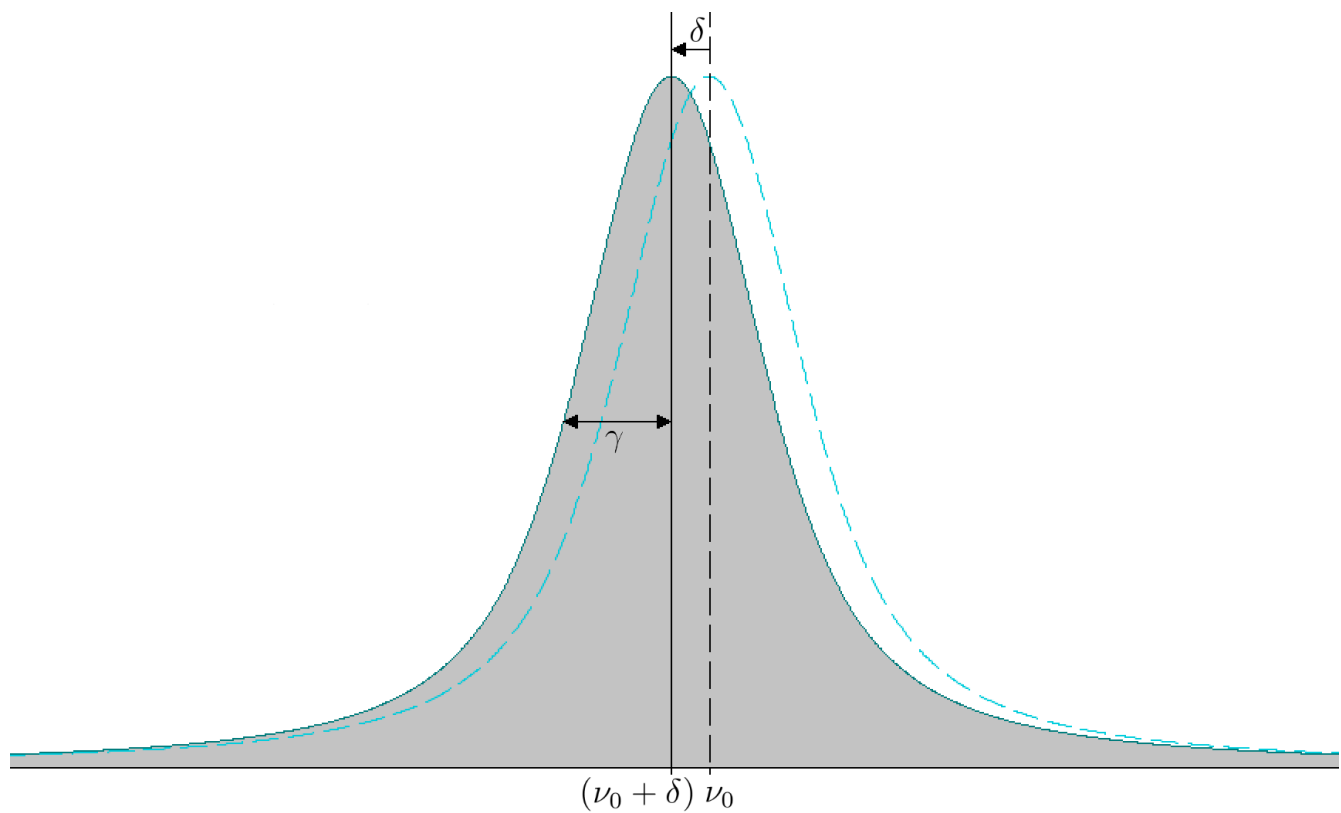

Рисунок 2.44 – Фундаментальные спектроскопические параметры перехода по базе данных HITRAN. Пунктирный профиль соответствует переходу при нулевом сдвиге, обусловленным давлением. Сплошной профиль с полушириной на полувысоте $\gamma$ сдвинут относительно центрального волнового числа $\nu_0$ на величину $\delta$. Заштрихованная область под профилем соответствует интенсивности спектральной линии $S$.

Концентрацию молекул выбранного газа можно определить как

$$N = N_L \left( \frac{T_0}{T} \frac{p_{self}}{p_0} \right), \qquad (2.33)$$

где $N_L$ – число Лошмидта ($N_L = 2.686781 \times 10^{19}$ молек/см$^3$), $T_0 = 273.15$ K и $p_0 = 1$ атм.

Параметры $\nu_0$, $S$, $\gamma_{air}$, $\gamma_{self}$, $E^{"}$, $n_{air}$, $n_{self}$, $\delta_{air}$, $\delta_{self}$, $Q(T_{ref})$ содержатся в базе данных HITRAN. Таким образом, качество получаемых модельных спектров для выбранного диапазона давлений в единицы-десятки мбар во многом определяется точностью приведенных в базе данных HITRAN вышеописанных параметров.

Сравнение форм контуров уширения, описываемых функцией Лоренца, Гаусса, и их сверткой – контуром Фойгта, представлено на рисунке 2.45. Как можно видеть, гауссова составляющая дает преимущественно вклад в амплитуду, тогда как крылья используемого для построения модельных спектров контура Фойгта определяются вкладом контура Лоренца. При этом также видно, что форма и амплитуда контура Фойгта заметно отличается от двух исходных контуров.

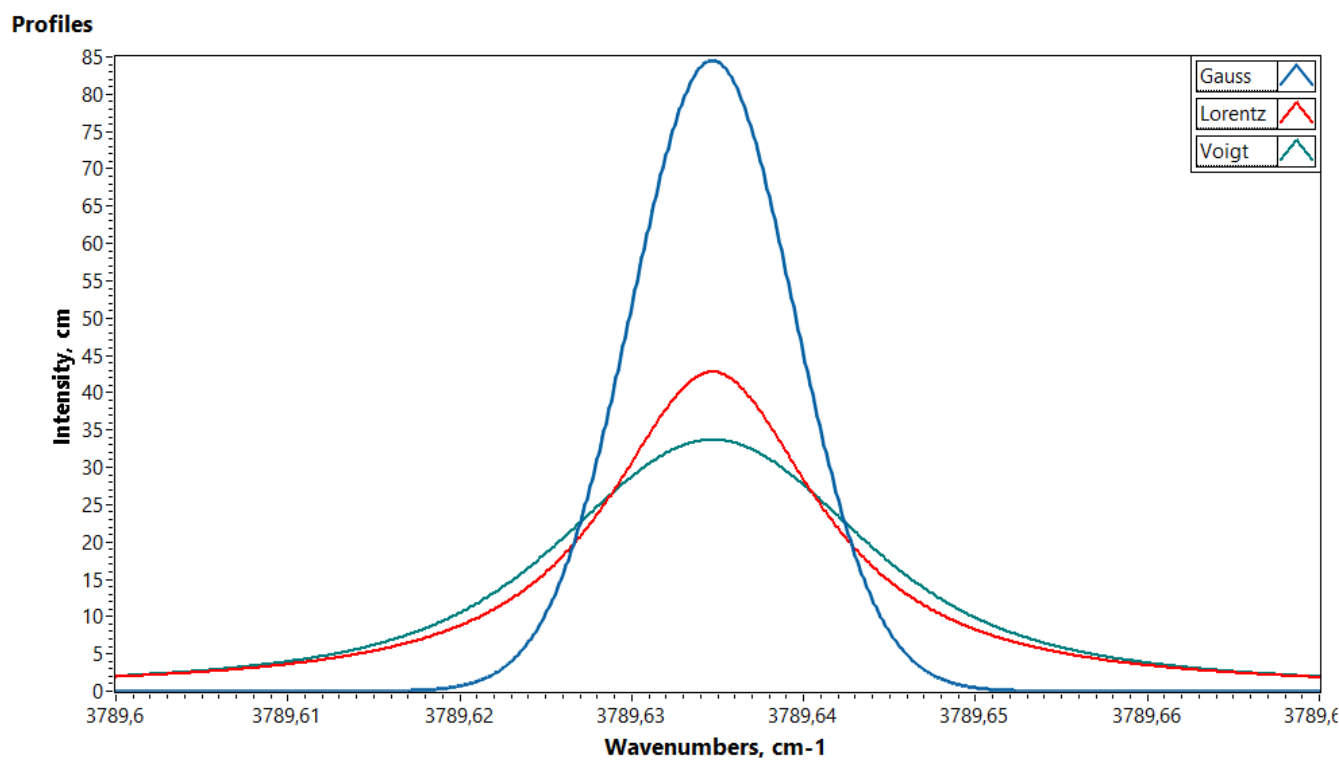

Рисунок 2.45 – Сравнение контуров уширения Гаусса (синий), Лоренца (красный) и Фойгта (зеленый) для линии основного изотополога $H_2O$ с $\nu_0 = 3789.63476$ см$^{-1}$.

Применимые для областей высокого давления контур уширения Лоренца, учитывающий столкновительное уширение, и для областей низкого давления контур уширения Гаусса, учитывающий термическое уширение, для диапазона рабочих давлений прибора ДЛС-Л плохо описывают экспериментальные результаты. Однако свертка этих двух контуров уширения – контур уширения Фойгта значительно лучше справляется с этой задачей.

Для увеличения скорости численного расчета контура Фойгта относительно скорости вычисления свертки двух контуров для каждой спектральной линии поглощения используется метод, основанный на применении функции Фаддеевой – комплексной функции ошибок от комплексного аргумента [177]:

$$w(z) = e^{-z^2} erfc(-iz) = e^{-z^2} \left( 1 + \frac{2i}{\sqrt{\pi}} \int_0^z e^{t^2}\, dt \right), \qquad (2.34)$$

$$z = x + iy. \qquad (2.35)$$

Точное, но эффективное с точки зрения затраченного компьютерного времени вычисление функции Фаддеевой на протяжении десятилетий является проблемой в разных областях физики. Подход к ее вычислению на основе рациональных аппроксимаций привлек значительное внимание и используется во многих программах, часто в сочетании с другими методами.

Была разработана программа, вычисляющая функцию Фаддеевой в верхней полуплоскости $z = x + iy$, т.е. для $y \geqslant 0$ (рисунок 2.46). Верхняя полуплоскость может быть подразделена на три области, в каждой из которых используется наиболее корректный для каждого случая программный код. Код Гумличека «CPF12» [178] обеспечивает точность от пяти до шести значащих цифр во всей комплексной плоскости [179]. Рациональная аппроксимация Вайдмана [180] может быть использована для малых $|x| + y < 15$, а рациональная аппроксимация Гумличека [181] наиболее эффективна с точки зрения необходимого вычислительного времени для всех $x$ и $y$, как было показано Шрайером [182], с максимальной относительной погрешностью как действительной, так и мнимой частей $< 10^{-4}$.

Контур Фойгта может быть выражен через вещественную часть функции Фаддеевой при $y > 0$ [182], что в нашем случае будет справедливо:

$$f_V = \sqrt{\frac{ln2}{\pi}} \frac{1}{\gamma_D} Re\{w(z)\},\qquad(2.36)$$

где значения вещественных $x$ и $y$ из (2.35) будут вычисляться как

$$x = \frac{\sqrt{ln2}}{\gamma_D}[\nu - (\nu_0 + \delta_L)],\qquad(2.37)$$

$$y = \sqrt{ln2}\frac{\gamma_L}{\gamma_D}.\qquad(2.38)$$

Здесь мы дополнительно учитываем фактор сдвига частоты перехода $\delta_L$ под действием внешнего и собственного давления газа.

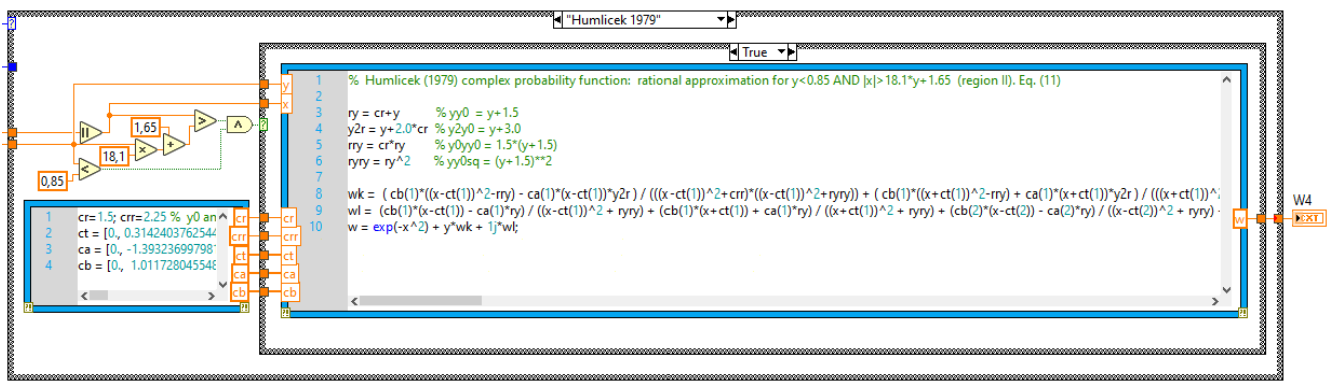

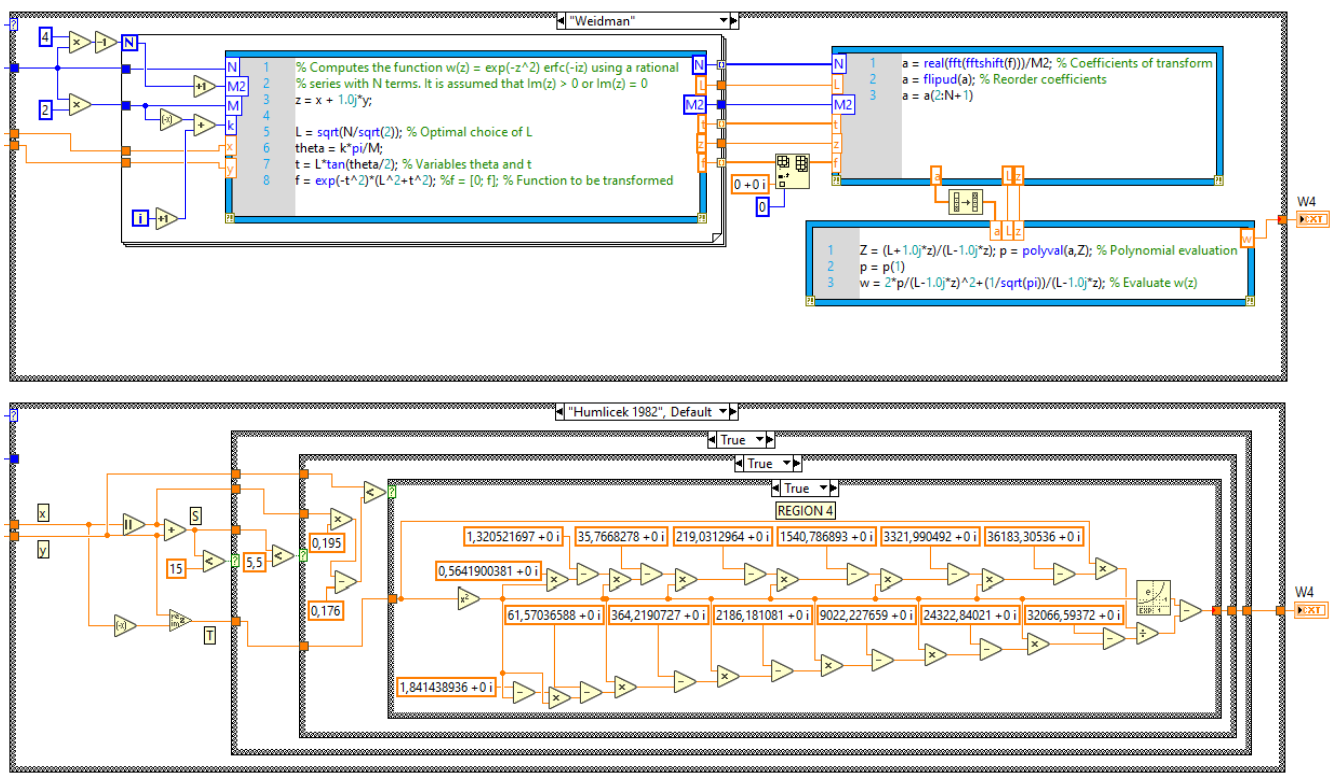

Рисунок 2.46 – Фрагмент блок-диаграммы программы для вычисления функции Фаддеевой.

Известно, что профиль Фойгта не дает полностью точного представления о форме спектральной линии [183] и его использование может привести, например, к систематическому занижению экспериментальных интенсивностей линий [184-186]. Использование профиля Фойгта для моделирования спектров молекулярных газов, регистрируемых как в условиях лаборатории, так и в атмосферном воздухе с высоким разрешением, приводит к характерным М-образным невязкам [187-192] между экспериментальными и модельными спектрами.

Недостатки профиля Фойгта, почти повсеместно используемого базами данных и кодами переноса излучения для учета эффектов давления и доплеровского уширения в изолированных колебательно-вращательных и чисто вращательных переходах, привели к разработке множества альтернативных моделей контуров спектральных линий. Эти модели отражают большее число физических эффектов, влияющих на форму линий, как правило, ценой их большей сложности.

При рассмотрении различных моделей контуров линий поглощения важно помнить, что по мере их усложнения они будут содержать больше свободных параметров, что приводит к лучшему фитингу измеренных спектров. Однако хорошее совпадение модельного набора спектральных линий с экспериментально измеренными не является гарантией достоверности используемой модели формы линии. В таблице 2.3 перечислены ключевые модели профилей линий, упорядоченные по количеству параметров, необходимых для описания одного спектрального перехода.

Таблица 2.3 – Модели контуров линий поглощения. N – количество параметров, необходимых для характеристики формы линии одного изолированного перехода при данной температуре для данной молекулы. ЗС – зависимый от скорости.

| Название | N | Параметры | Зависимость от скорости | Механизм соударений | Корреляция |
|---|---|---|---|---|---|
| Допплер | 1 | $\gamma_D$ | Нет | Нет | Нет |
| Лоренц | 2 | $\gamma_L$, $\delta_L$ | Нет | Нет | Нет |
| Фойгт | 3 | $\gamma_D$, $\gamma_L$, $\delta_L$ | Нет | Нет | Нет |
| Галатри | 4 | $\gamma_D$, $\gamma_L$, $\delta_L$, $\nu_{VC}$ | Нет | Мягкие | Нет |
| Раутиан | 4 | $\gamma_D$, $\gamma_L$, $\delta_L$, $\nu_{VC}$ | Нет | Жесткие | Нет |
| ЗС-Фойгт | 5 | $\gamma_D$, $\gamma_L$, $\delta_L$, $\gamma_2$, $\delta_2$ | Да | Нет | Нет |
| ЗС-Галатри | 6 | $\gamma_D$, $\gamma_L$, $\delta_L$, $\gamma_2$, $\delta_2$, $\nu_{VC}$ | Да | Мягкие | Нет |
| ЗС-Раутиан | 6 | $\gamma_D$, $\gamma_L$, $\delta_L$, $\gamma_2$, $\delta_2$, $\nu_{VC}$ | Да | Жесткие | Нет |
| Хартманн-Тран | 7 | $\gamma_D$, $\gamma_L$, $\delta_L$, $\gamma_2$, $\delta_2$, $\nu_{VC}$, $\eta$ | Да | Жесткие | Да |

Влияние изменения скорости, вызванного столкновением, на форму спектральной линии называют сужением Дике [193]. В этом случае важную роль приобретает сила столкновений, то есть механизм влияния на изменение скорости. Модели жестких столкновений предполагают, что скорости молекул до и после каждого столкновения полностью декоррелированы – каждое столкновение настолько сильное, что молекула полностью «теряет память» о своей скорости до столкновения, а скорость после столкновения определяется распределением Максвелла. Соответствующий профиль линии называется профилем Раутиана [194]. Гипотеза мягких столкновений, при которой для существенного изменения скорости молекул необходимо множество столкновений, определяет профиль Галатри [195]. Как модели жестких, так и мягких столкновений вводят один дополнительный параметр $\nu_{VC}$ для количественной оценки частоты столкновений с изменением скорости [см$^{-1}$].

Профиль Галатри можно выразить как [196]:

$$f_G = \sqrt{\frac{ln2}{\pi}} \frac{1}{\gamma_D} Re \left[ \frac{1}{\frac{1}{2z} + y - ix} M \left( 1; 1 + \frac{1}{2z^2} + \frac{y - ix}{z}; \frac{1}{2z^2} \right) \right], \qquad (2.39)$$

где значения вещественных $x$ и $y$ вычисляются согласно (2.37) и (2.38) соответственно, $M$ – вырожденная гипергеометрическая функция [197].

Профиль Раутиана можно выразить через функцию Фаддеевой [196] следующим образом:

$$f_R = \sqrt{\frac{ln2}{\pi}} \frac{1}{\gamma_D} Re\left[\frac{w(x, y + \zeta)}{1 - \sqrt{\pi}\zeta w(x, y + \zeta)}\right],$$ (2.40)

где значения вещественных *x* и *y* также вычисляются согласно (2.37) и (2.38) соответственно, а $\zeta$ определяется как

$$\zeta = \frac{\nu_{VC}}{\gamma_D}.$$ (2.41)

Зависимость времен релаксации от скорости, рассматриваемая как единственный источник сужения линии, приводит к зависимому от скорости профилю Фойгта (ЗС-Фойгта) [198,199]. Алгоритм Буна-Уокера-Берната [192] позволяет рассчитать контур ЗС-Фойгта из двух профилей Фойгта.

Приписывание сужения линии исключительно столкновениям с изменением скорости (СИС) часто приводит к ошибочным значениям параметра $\nu_{VC}$. Такое поведение означает, что необходимы другие методы учета сужения линий. Для этого в модель мягких столкновений вводится зависимость от скорости через параметр зависимости от скорости ширины линии $\gamma_2$ [см$^{-1}$] и параметр зависимости от скорости сдвига линии $\delta_2$ [см$^{-1}$], что приводит к ЗС-Галатри [200]. ЗС-Галатри сводится к профилю Галатри при отсутствии зависимости от скорости ($\gamma_2 = 0$) и к ЗС-Фойгту при отсутствии столкновений, изменяющих скорость ($\nu_{VC} = 0$). Аналогичным образом, зависимость от скорости, введенная в профиль Раутиана, определяет зависящий от скорости профиль Раутиана [201].

Модели ЗС-Галатри и ЗС-Раутиана предполагают, что аспекты столкновения, связанные с изменением скорости и вращательным состоянием молекул, независимы. На практике это не так: изменение скоростей уравновешивается изменением внутреннего состояния молекул, согласно закону сохранения энергии. Таким образом, механизмы СИС и ЗС могут работать одновременно, а соответствующие параметры их моделей коррелируют. Модели профиля, учитывающие корреляцию этих двух столкновительных эффектов, описываются функциями, включающими дополнительный безразмерный параметр корреляции $\eta$.

Один из методов введения корреляции заключается в использовании частично коррелированной модели жестких столкновений для столкновений с изменением скорости и состояния [202-204]. Наиболее простая в применении модель учитывает зависимость скорости только квадратично, что дает частично коррелированный квадратично-зависимый от скорости профиль жестких столкновений [205-207]. Эта модель является гибкой и имеет главное преимущество: ее можно представить в относительно простой форме, что позволяет проводить быструю вычислительную оценку.

ИЮПАК рекомендует принять частично коррелированный профиль жестких столкновений, квадратично зависящий от скорости, в качестве наиболее оптимальной модели для спектроскопии высокого разрешения [208]. Для простоты его можно называть профилем Хартмана-Трана. Этот профиль достаточно сложен, чтобы уловить различные вклады столкновений в форму изолированной линии, но может быть достаточно быстро вычислен и сводится к более простым профилям, включая профиль Фойгта, при определенных упрощающих предположениях.

Профиль Хартмана-Трана можно выразить как [205-207]:

$$f_{HT} = \frac{1}{\pi} Re \left[ \frac{A(\nu)}{1 - [\nu_{VC} - \eta(C_0 - \frac{3}{2}C_2)]A(\nu) + \eta C_2 B(\nu)} \right], \qquad (2.42)$$

где введены параметры

$$C_0 = \gamma_L + i\delta_L, \qquad (2.43)$$

$$C_2 = \gamma_2 + i\delta_2, \qquad (2.44)$$

а значения $A(\nu)$ и $B(\nu)$ вычисляются как

$$A(\nu) = \frac{c}{2\nu_0} \sqrt{\frac{m}{\pi k_B T}} [w(iz_1) - w(iz_2)], \qquad (2.45)$$

$$B(\nu) = \frac{2k_B T}{m(1-\eta)C_2} \left[ -1 + \frac{1}{2}\sqrt{\frac{\pi}{y}}(1 - z_1^2)w(iz_1) - \frac{1}{2}\sqrt{\frac{\pi}{y}}(1 - z_2^2)w(iz_2) \right], \quad (2.46)$$

где $m$ — масса молекулы, а $z_1$ и $z_2$ определены как

$$z_1 = \sqrt{x+y} - \sqrt{y}, \qquad (2.47)$$

$$z_2 = \sqrt{x+y} + \sqrt{y}, \qquad (2.48)$$

а $x$ и $y$ в свою очередь как

$$x = \frac{i(\nu - \nu_0) + \left(C_0 - \frac{3}{2}C_2\right)(1-\eta) + \nu_{VC}}{C_2(1-\eta)}, \qquad (2.49)$$

$$y = \frac{1}{\left[ 2\frac{\sqrt{ln2}}{\gamma_D} C_2(1-\eta) \right]^2}. \qquad (2.50)$$

На рисунке 2.47. показано сравнение профилей Фойгта, Галатри, Раутиана и Хартмана-Трана с параметрами, подобранными для наглядности этого сравнения.

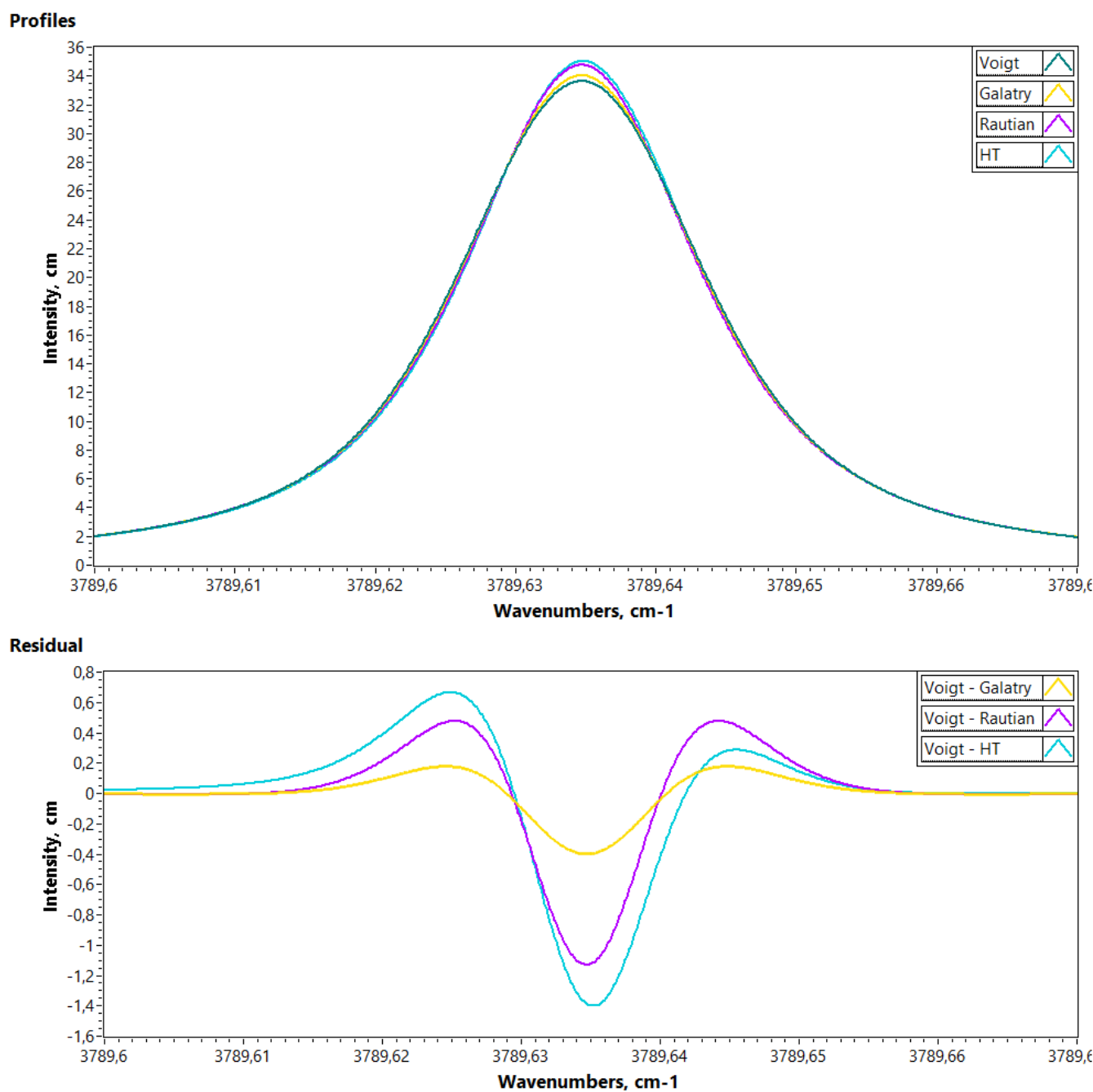

Рисунок 2.47 – Сверху: сравнение профилей Фойгта (зеленый), Галатри (желтый), Раутиана (фиолетовый) и Хартмана-Трана (голубой) для линии основного изотополога $H_2O$ с $v_0 =$ 3789.63476 см$^{-1}$. Снизу: невязка между профилем Фойгта и профилем Галатри (желтый), Раутиана (фиолетовый) и Хартмана-Трана (голубой) для той же спектральной линии.

Стоит отметить, что работа с перечисленными профилями, учитывающими более тонкие эффекты, влияющие на форму спектральных линий, чрезвычайно осложняется тем, что большая часть табличных параметров помимо используемых для профиля Фойгта, необходимых для их расчета, отсутствует в базе данных HITRAN 2020 года.

Несмотря на то, что разработчики этой базы в своей библиотеке HAPI для языка Python [209] закладывают возможность применения контура Хартмана-Трана и всех контуров, к которым он сводится при поочередном занулении его параметров, перечисленных в таблице 2.3, указанные параметры можно найти лишь для некоторых переходов отдельных молекулярных газов. Так для выбранных спектральных диапазонов энергетических переходов молекул $H_2O$ и $CO_2$ спектрометра ДЛС-Л необходимые параметры отсутствуют.

Поскольку подбор нескольких параметров, влияющих на форму линии, при обработке экспериментальных данных, полученных при помощи спектрометра ДЛС-Л, не являющегося сертифицированным средством измерения, не кажется убедительным методом анализа, предоставляющим заведомо физически корректные результаты, после безуспешных попыток найти необходимые параметры для выбранных при подборе спектральных диапазонов энергетических переходов молекул $H_2O$ и $CO_2$ в опубликованных на момент выполнения данной работы научных статьях было принято решение вернуться к использованию контура Фойгта. В данном случае теряется возможная точность методики, что проявляется в образовании М-образных невязок между экспериментальными и модельными спектрами. Однако при этом не вводятся дополнительные параметры, которые могут неконтролируемо искажать результаты анализа в большей степени.

## 2.4.5. Моделирование синтетических спектров поглощения

После выбора подходящего профиля спектральных линий для дальнейшего использования при моделировании синтетических спектров поглощения следующим шагом нужно вычислить значения коэффициента поглощения выбранного изотополога для каждого волнового числа $\nu$ с заданным шагом:

$$\alpha_S(\nu) = f_X^i(\nu) \cdot S_i \cdot N, \tag{2.51}$$

где $f_X^i$ – профиль выбранной модели для $i$-той линии поглощения, $S_i$ – интенсивность $i$-той линии поглощения, $N$ – концентрация молекул выбранного газа в смеси.

Для корректного вычисления синтетического спектра коэффициента поглощения необходимо сперва вернуться к предварительно обработанным экспериментальным данным. Нужно определить интегральную величину под контуром уединенной спектральной линии поглощения основного изотополога выбранной молекулы, которая будет пропорциональна интенсивности линии $S_i$ и концентрации молекул в смеси $N$.

Эта процедура проводится для наиболее подходящей спектральной линии полученного в ходе обработки исходного сигнала коэффициента поглощения с учетом выбранного для процедуры интерполяции шага по волновому числу (рисунок 2.48).

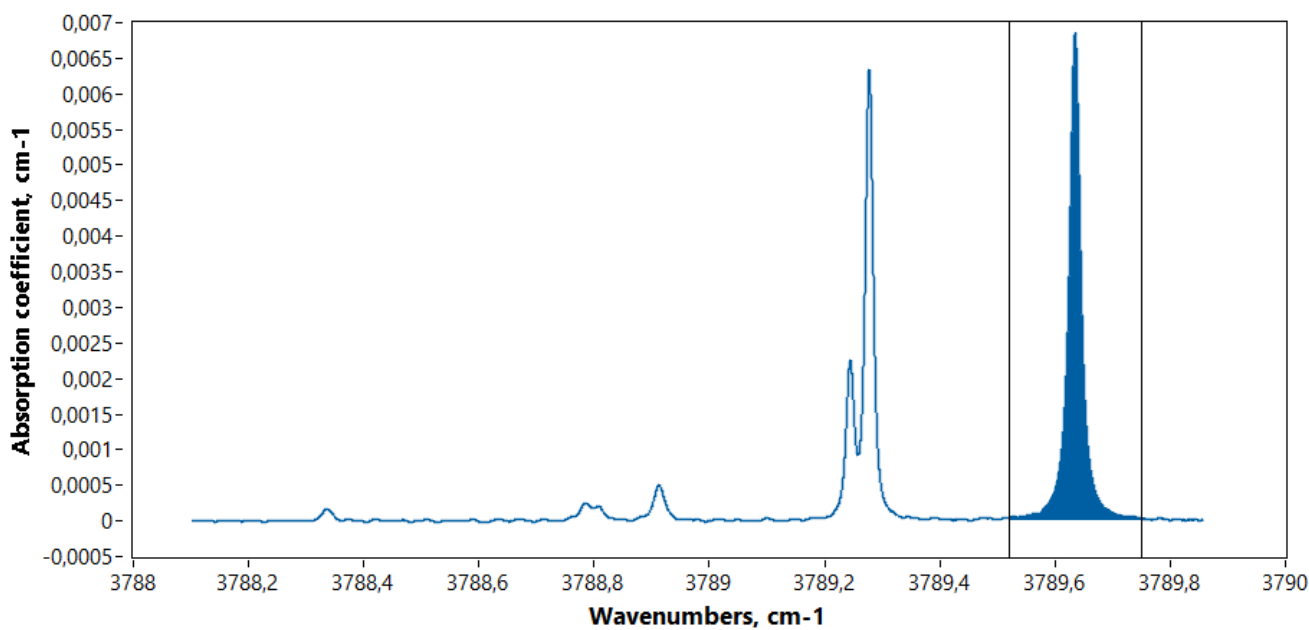

Рисунок 2.48 – Полученный коэффициент поглощения водяного пара при давлении 15 мбар. Вертикальными линиями ограничена область интегрирования линии с $v_0 = 3789.63476$ см$^{-1}$.

Далее, получив целевое значение интегральной величины заштрихованной на рисунке 2.48 области под контуром измеренного коэффициента поглощения, переходим к подбору таких параметров синтетического спектра коэффициента поглощения основного изотополога выбранного молекулярного газа, чтобы интеграл ограниченной тем же диапазоном волновых чисел области под контуром уже синтетического спектра (рисунок 2.49) совпал с целевым значением.

Варьируемыми параметрами выступают парциальное давление исследуемого газа, общее давление смеси и температура. В случае корректной работы датчиков температуры и общего давления прибора полученные в результате этой операции параметры модельного спектра будут с хорошей точностью совпадать с их показаниями.

Затем проводится вычисление коэффициентов поглощения для других представляющих интерес изотопологов выбранного молекулярного газа при тех же значениях парциального давления, общего давления смеси и температуры.

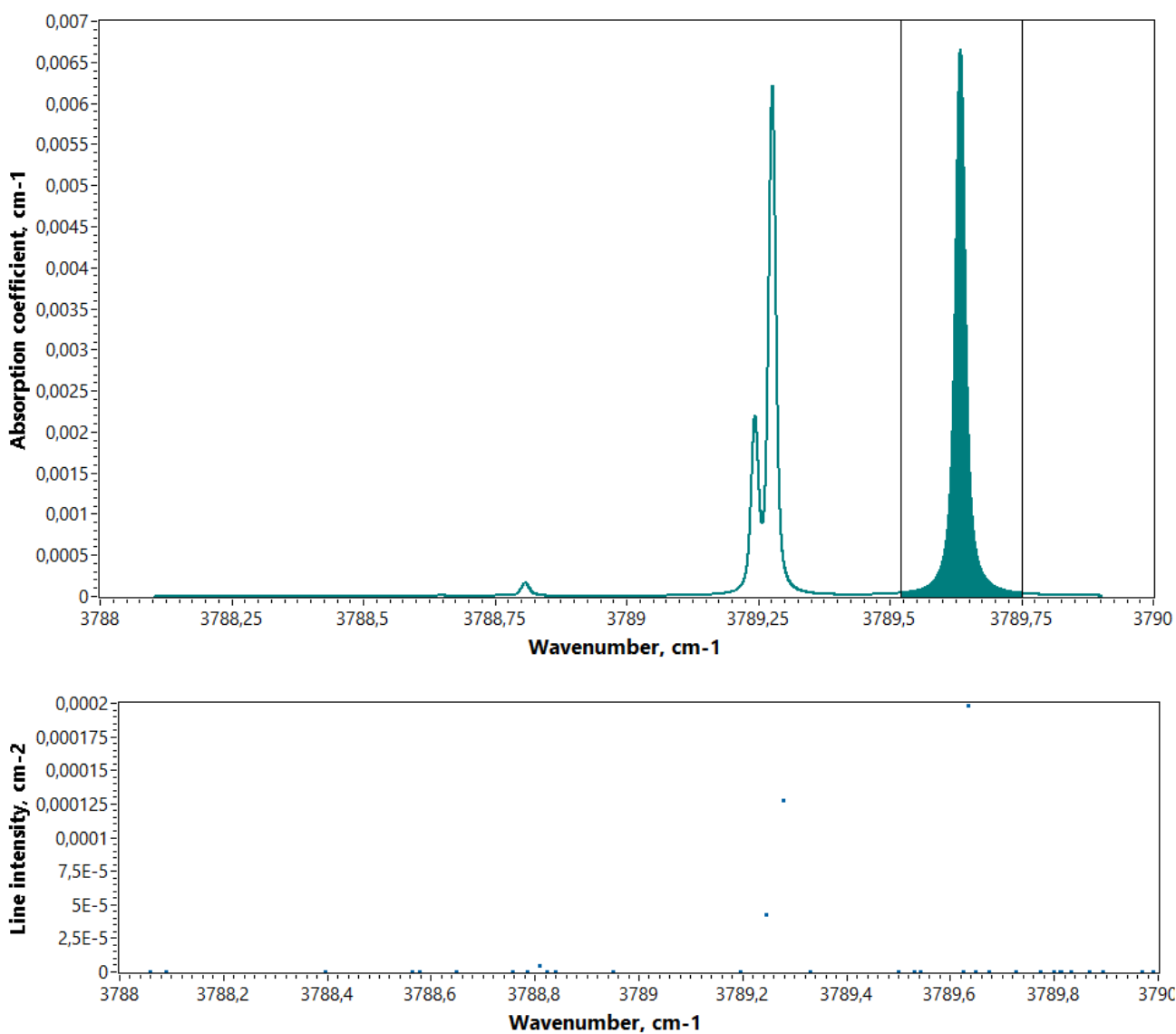

Рисунок 2.49 – Сверху: результат расчета модельного спектра коэффициента поглощения по данным базы HITRAN с использованием контура Фойгта основного изотополога $H_2O$. Вертикальными линиями ограничена область интегрирования линии поглощения с $v_0 = 3789.63476$ см$^{-1}$. Снизу: значения произведения интенсивности $S_i$ спектральных линий поглощения на величину концентрации выбранного газа $N$ для выбранного диапазона.

Для дальнейшей работы требуется провести нормировку вычисленных спектров коэффициента поглощения на величины их распространенности, приведенные в базе данных HITRAN в разделе метаданных изотопологов молекулярных газов (рисунок 2.50).

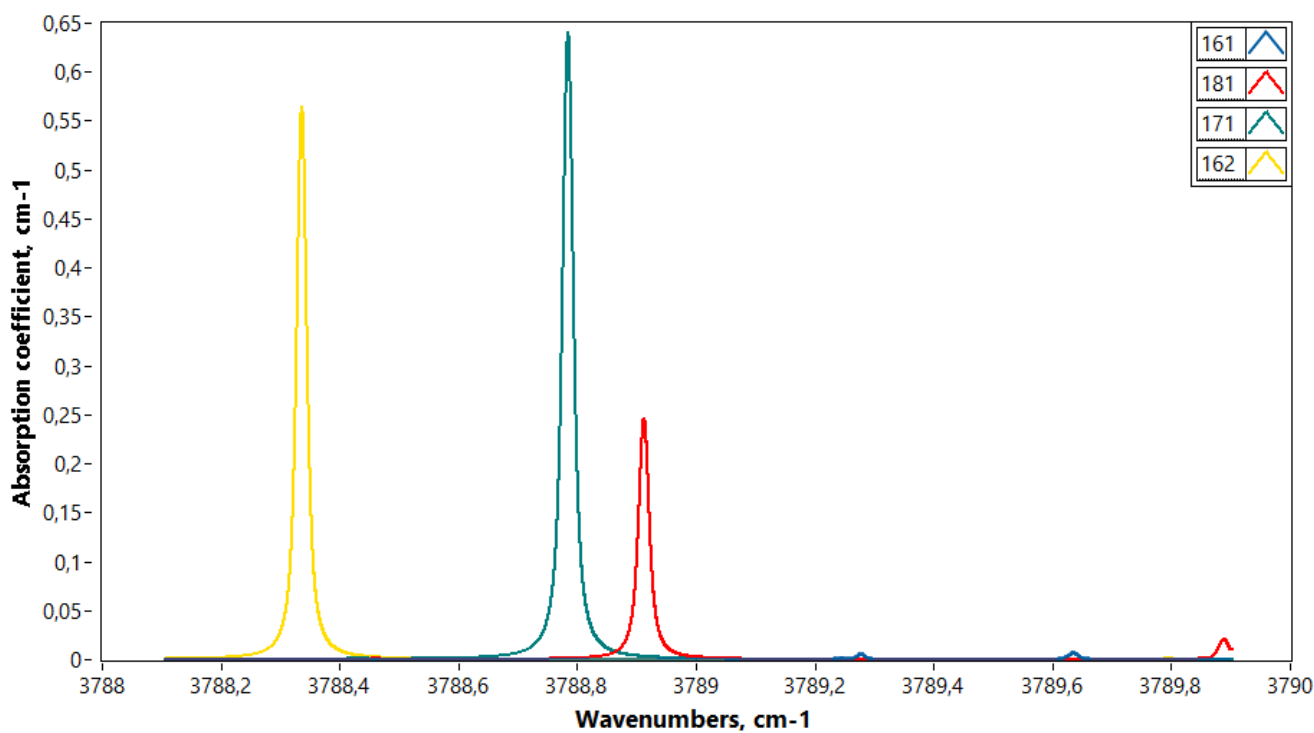

Рисунок 2.50 – Спектры коэффициента поглощения четырех изотопологов $H_2O$, нормированные на значения их распространенности по HITRAN. Синий – $H_2^{16}O$, красный – $H_2^{18}O$, зеленый – $H_2^{17}O$, желтый – $HD^{16}O$.

Работать с изотопологами $H_2O$ и $CO_2$ с распространенностью менее $10^{-4}$ для прибора ДЛС-Л не представляется возможным, таким образом возникает ограничение на лист доступных для исследования изотопологов. Естественная распространенность первых четырех изотопологов водяного пара и углекислого газа по HITRAN приведена в таблице 2.4.

Таблица 2.4 – Естественная распространенность ряда изотопологов $H_2O$ и $CO_2$.

| Формула | Код AFGL | Распространенность | Формула | Код AFGL | Распространенность |
|---|---|---|---|---|---|
| $H_2^{16}O$ | 161 | $9.97317 \times 10^{-1}$ | $^{12}C^{16}O_2$ | 626 | $9.84204 \times 10^{-1}$ |
| $H_2^{18}O$ | 181 | $1.99983 \times 10^{-3}$ | $^{13}C^{16}O_2$ | 636 | $1.10574 \times 10^{-2}$ |
| $H_2^{17}O$ | 171 | $3.71884 \times 10^{-4}$ | $^{16}O^{12}C^{18}O$ | 628 | $3.94707 \times 10^{-3}$ |
| $HD^{16}O$ | 162 | $3.10693 \times 10^{-4}$ | $^{16}O^{12}C^{17}O$ | 627 | $7.33989 \times 10^{-4}$ |

До сих пор в данном разделе рассматривался случай базового подхода к обработке спектральных данных, однако, как упоминалось в разделе 2.4.2, в работе по обработке данных прибора ДЛС-Л использовался также подход, основанный на применении ортогональных полиномов. Для этого случая синтетический спектр коэффициента поглощения каждого изотополога подвергался обработке, аналогичной той, что была описана в разделе 2.4.2. Тогда спектры коэффициента поглощения рассматриваемых изотопологов будут иметь вид, представленный на рисунке 2.51.

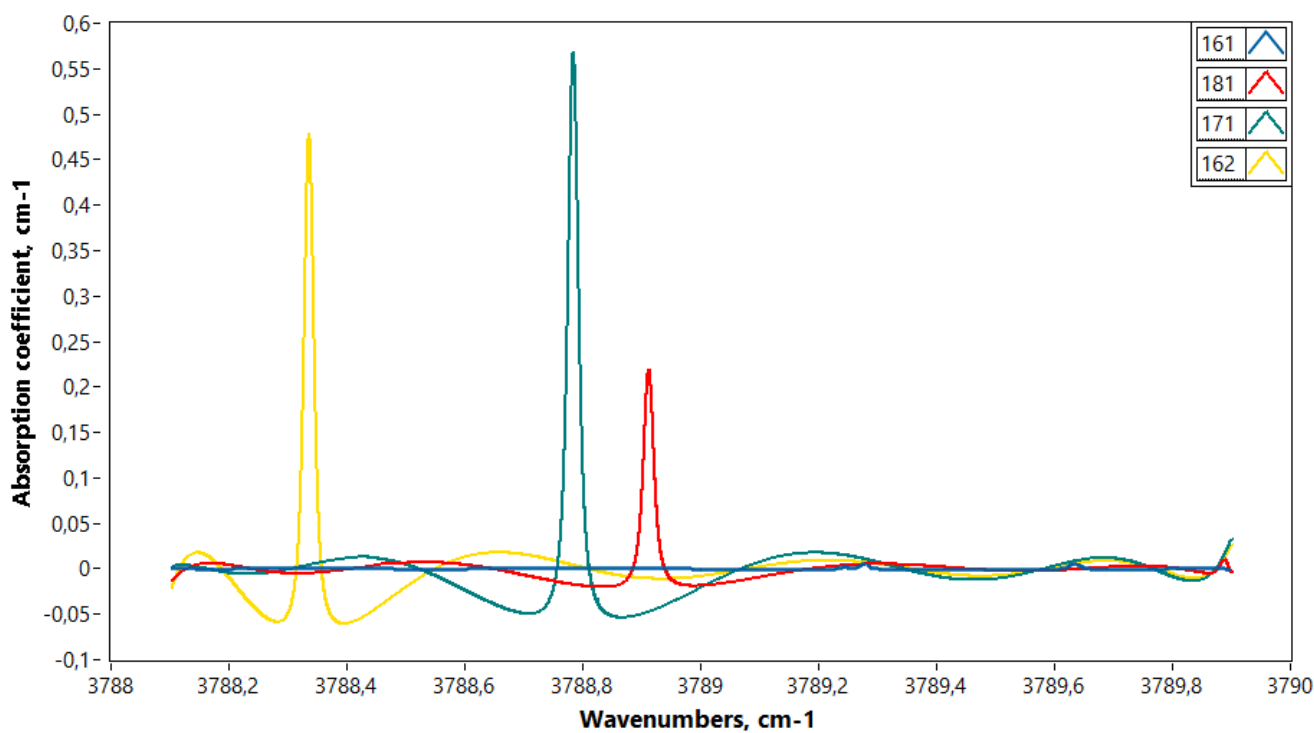

Рисунок 2.51 – Спектры коэффициента поглощения, обработанные согласно методике ортогональных полиномов, четырех изотопологов $H_2O$, нормированные на значения их распространенности по HITRAN. Синий – $H_2^{16}O$, красный – $H_2^{18}O$, зеленый – $H_2^{17}O$, желтый – $HD^{16}O$.

### 2.4.6. Определение изотопных отношений

Следующим шагом после вычисления набора синтетических спектров коэффициента поглощения для каждого изотополога выбранного молекулярного газа проводится вычисление финального синтетического спектра коэффициента поглощения путем суммирования приведенных на рисунке 2.49 спектров изотопологов с определенными весами:

$$\alpha_S^{H_2O}(\nu) = g_1 \cdot \alpha_S^{161}(\nu) + g_2 \cdot \alpha_S^{181}(\nu) + g_3 \cdot \alpha_S^{171}(\nu) + g_4 \cdot \alpha_S^{162}(\nu), \qquad (2.52)$$

где $g_i$ – веса для синтетических коэффициентов поглощения соответствующих изотопологов, определяемые при помощи фитинга $k$-мерной линейной функцией (в данном случае $k = 4$) методом наименьших квадратов на основе алгоритма сингулярного разложения измеренного коэффициента поглощения $\alpha_M$. Тогда полученные веса будут иметь физический смысл распространенности соответствующего изотополога исследуемого газа в аналитическом объеме кюветы ДЛС-Л. Получив таким образом значения $\alpha_M(\nu)$ и $\alpha_S(\nu)$ можем сравнить их графически и получить графическое отображение невязки экспериментального и синтетического спектров $\varepsilon(\nu)$ на рисунке 2.52. Отметим, что невязка, приведенная здесь и далее на рисунках, вычислялась как -$\varepsilon(\nu)$ из формулы (2.11).

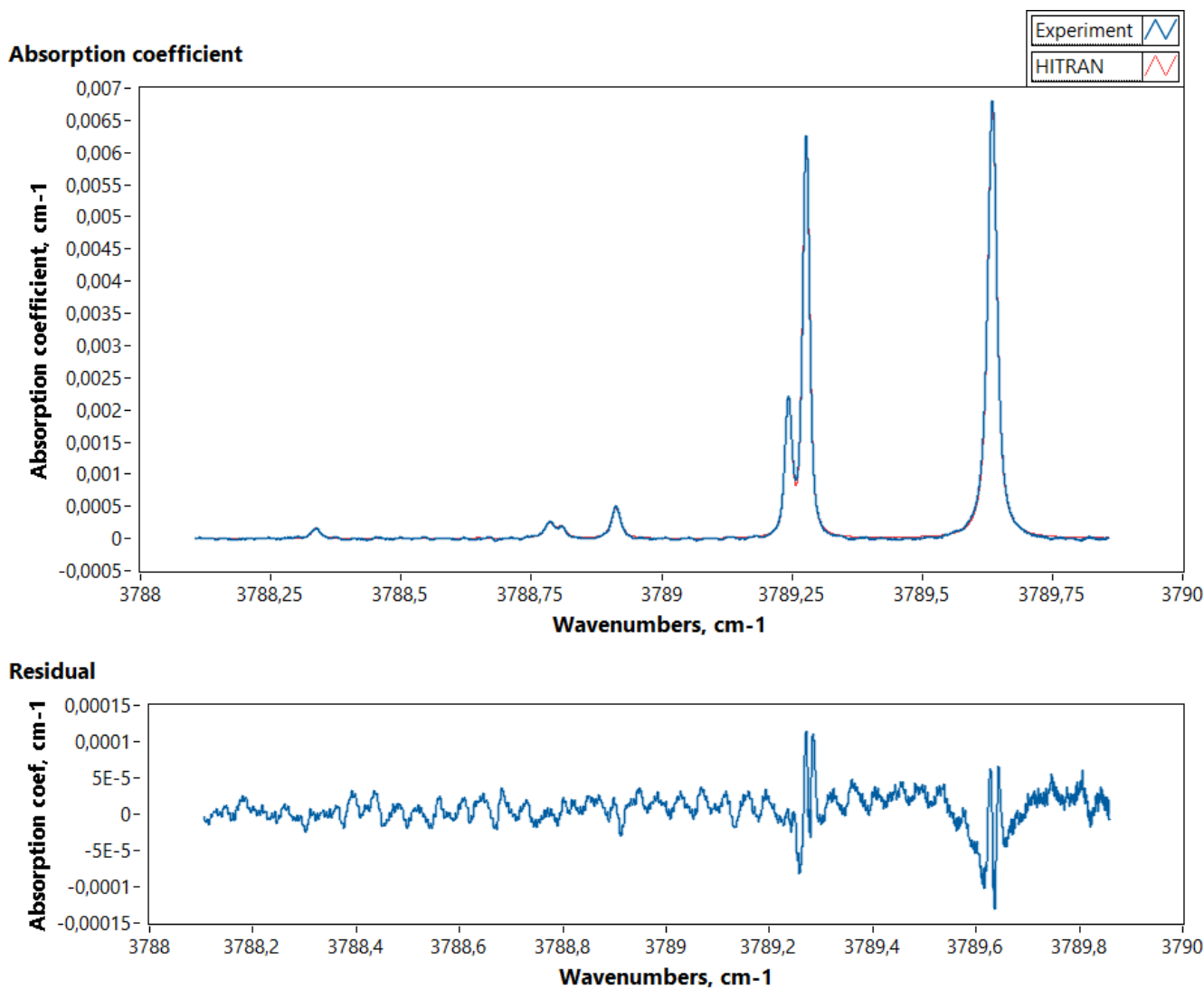

Рисунок 2.52 – Сопоставление экспериментально полученного и синтетического спектров (сверху) и их невязка (снизу).

На этом этапе для каждого обрабатываемого спектра вычислялись значения $R_{sample}$ из формулы (2.1) как отношение веса соответствующего изотополога $g_i$ из формулы (2.52) к весу основного изотополога:

$$R^i_{sample} = \eta \cdot \frac{g_i}{g_1}, \qquad (2.53)$$

где $\eta$ = ½ для HDO, $^{18}OCO$ и $^{17}OCO$ из-за наличия двух изотопов одного элемента в изучаемой молекуле и $\eta$ = 1 для $H_2^{18}O$ и $^{13}CO_2$. Следовательно можно вычислить значения изотопных сигнатур для каждого исследуемого изотополога – $\delta_{VSMOW}D$, $\delta_{VSMOW}^{18}O$, $\delta_{VSMOW}^{17}O$ для $H_2O$ и $\delta_{VPDB}^{13}C$, $\delta_{VSMOW}^{18}O$, $\delta_{VSMOW}^{17}O$ для $CO_2$.

В случае обработки согласно методики ортогональных полиномов сопоставление экспериментально полученного и синтетического спектров будут выглядеть иначе, тогда как их невязка будет мало отличаться от аналогичного результата при стандартной обработке (рисунок 2.53).

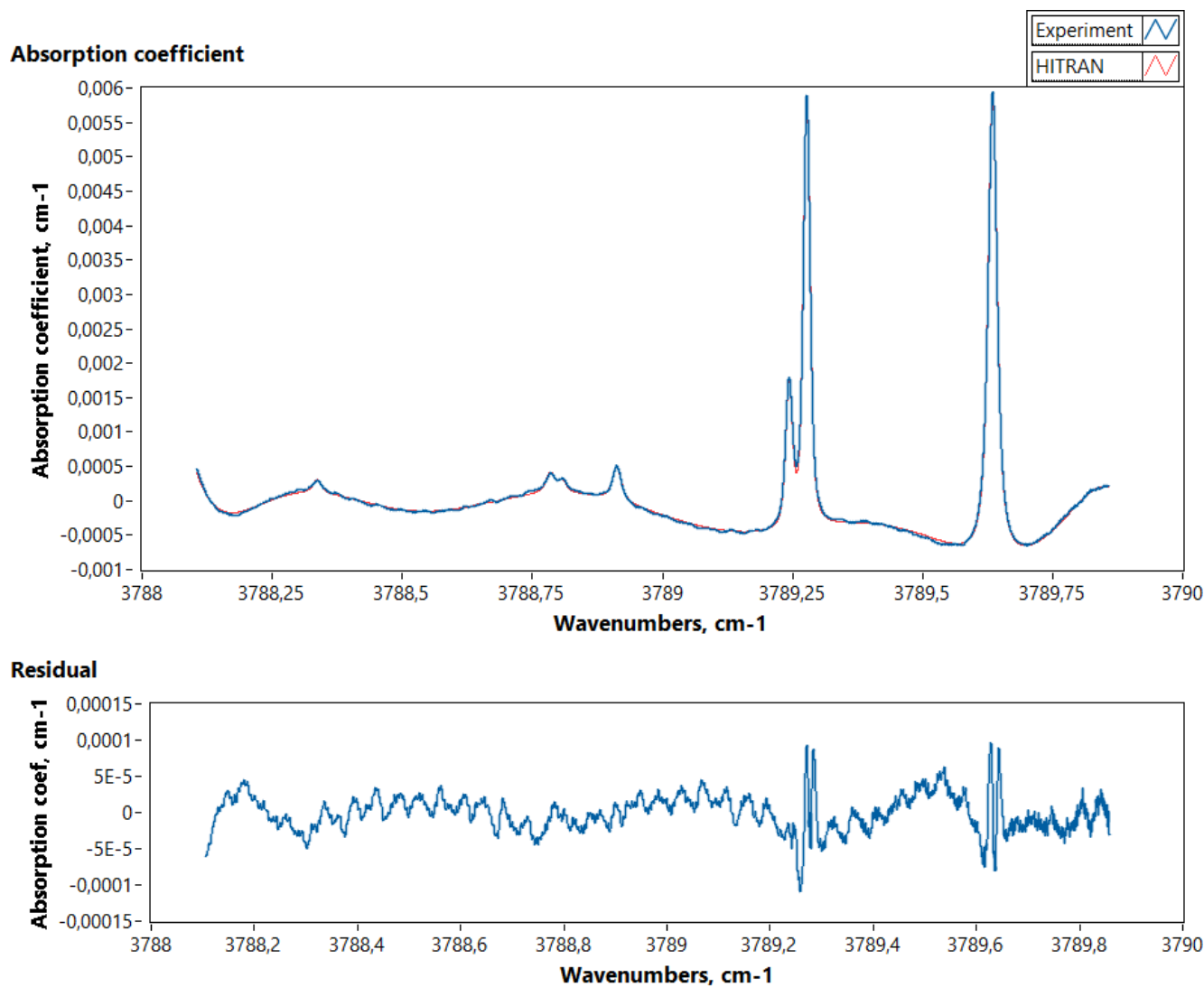

Рисунок 2.53 – Сопоставление экспериментально полученного и синтетического спектров (сверху) и их невязка (снизу) в случае обработки согласно методики ортогональных полиномов.

При выборе в пользу той или иной методики краеугольным критерием будет являться вид невязки измеренного и синтетического спектров (рисунок 2.54). Нужно отметить, что при очевидной разнице в процессах, заложенных в обработку согласно двум изложенным подходам, результаты – то есть формы невязки в этих двух случаях – очень мало отличаются. Оба подхода позволяют довольно эффективно справляться с вкладом низкочастотной интерференции, тогда как вклад высокочастотной интерференции во многом сохраняется так же в обоих случаях, что для методики ортогональных полиномов может быть связано в описываемом случае с разницей интенсивности спектральных линий и амплитуды высокочастотной интерференции на 2-3 порядка. Нельзя, впрочем, исключать возможность того, что для плохо знакомого со спецификой обработки спектральных данных оператора методика ортогональных полиномов позволит получать лучшие результаты в силу меньшей вовлеченности оператора в процесс.

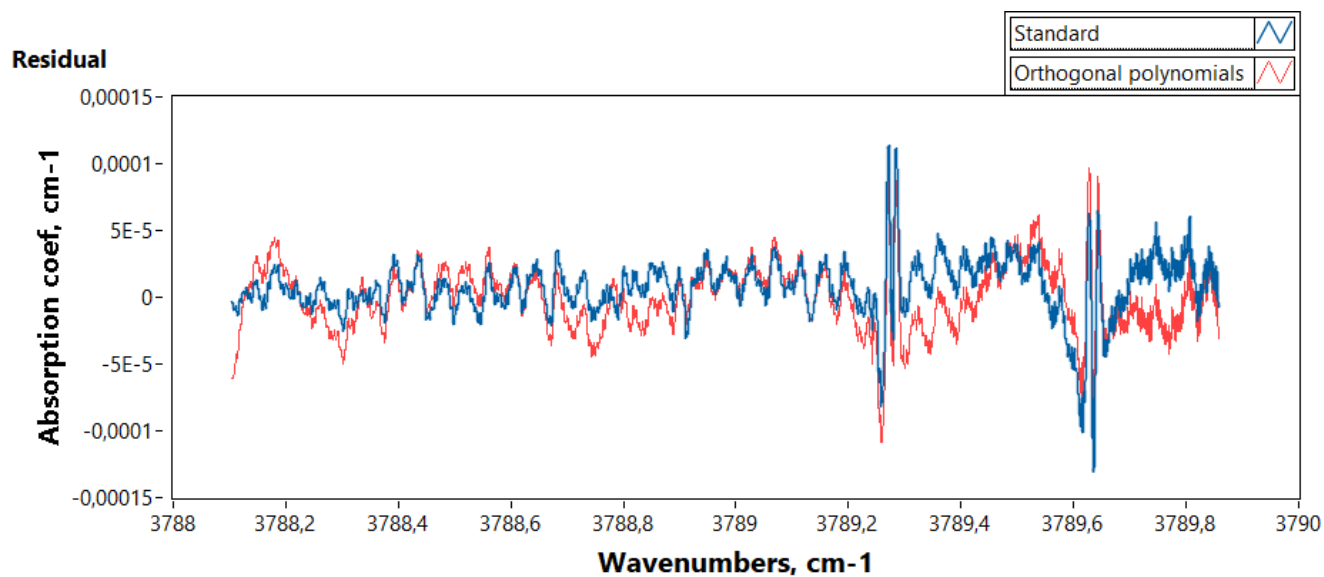

Рисунок 2.54 – Сопоставление невязки измеренного и синтетического спектров в случае стандартной методики обработки (синяя) и методики ортогональных полиномов (красная).

Однако совмещая обе предложенные методики можно добиться подавления интерференции на средних частотах. В качестве опыта, не описанного ранее в литературе, подход к обработке сигнала согласно методики ортогональных полиномов применялся к невязке измеренного и синтетического спектров, полученной при стандартной обработке (рисунок 2.55).

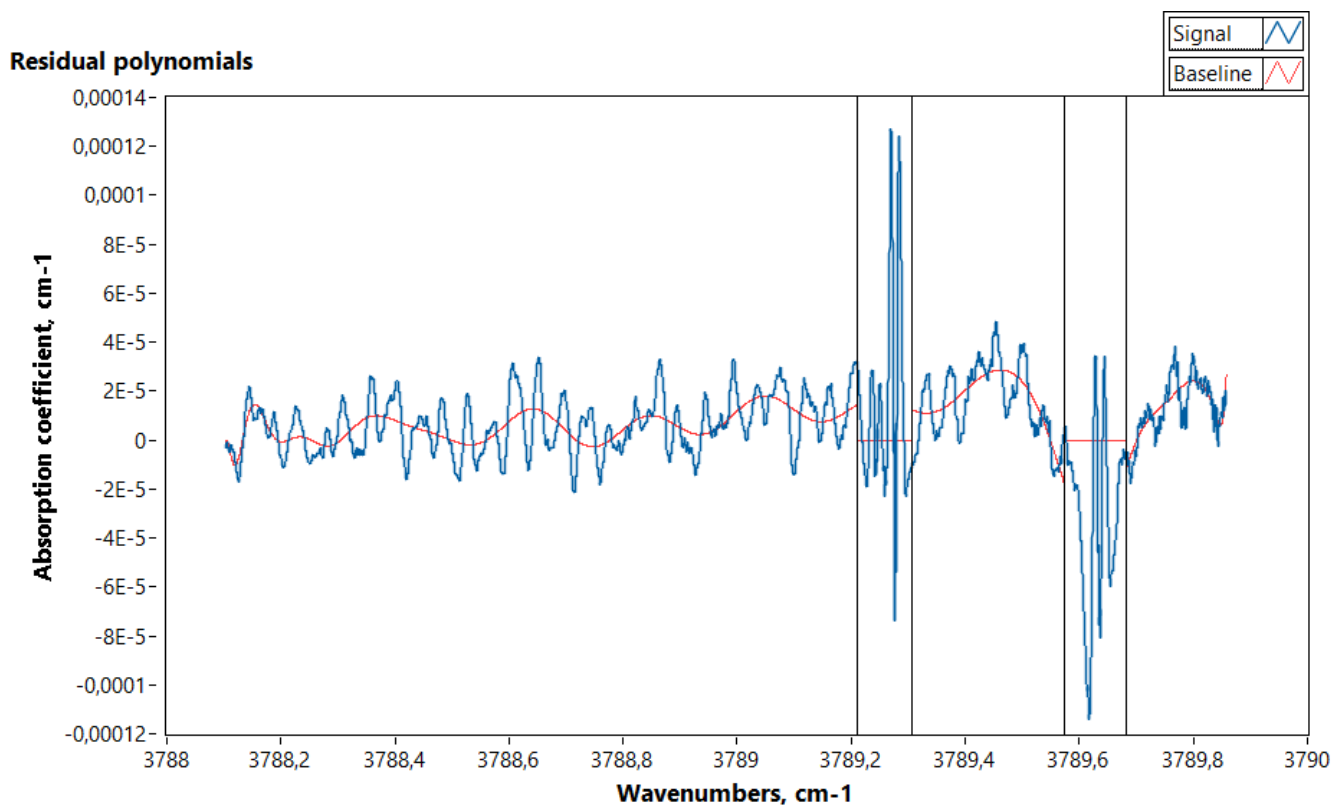

Рисунок 2.55 – Применение методики ортогональных полиномов для подавления интерференционного вклада на средних частотах. Синяя линия – невязка, полученная при стандартной обработке. Красная – «базовая линия» невязки, полученная при обработке по методике ортогональных полиномов.

Вычитая полученную таким образом «базовую линию» невязки, построенную по диапазонам спектра, не содержащим сильных линий поглощения, из измеренного спектра коэффициента поглощения, получим менее зашумленный результат, содержащий только высокочастотный вклад интерференции, соответствующий биениям на расстояниях в десятки сантиметров (рисунок 2.56). «Базовая линия», представляющая вклад среднечастотной интерференции, соответствует биениям на оптическом пути порядка единиц сантиметров.

Аналогично можно работать и не выходя за рамки методики ортогональных полиномов, применяя ее в два этапа и к обработке исходного сигнала и к невязке с синтетическим спектром, поскольку, как было отмечено выше и видно из рисунка 2.54, характер высокочастотной интерференции не меняется в зависимости от применяемой методики, а имеющую отличия среднечастотную интерференцию второй этап применения методики ортогональных полиномов позволяет подавить.

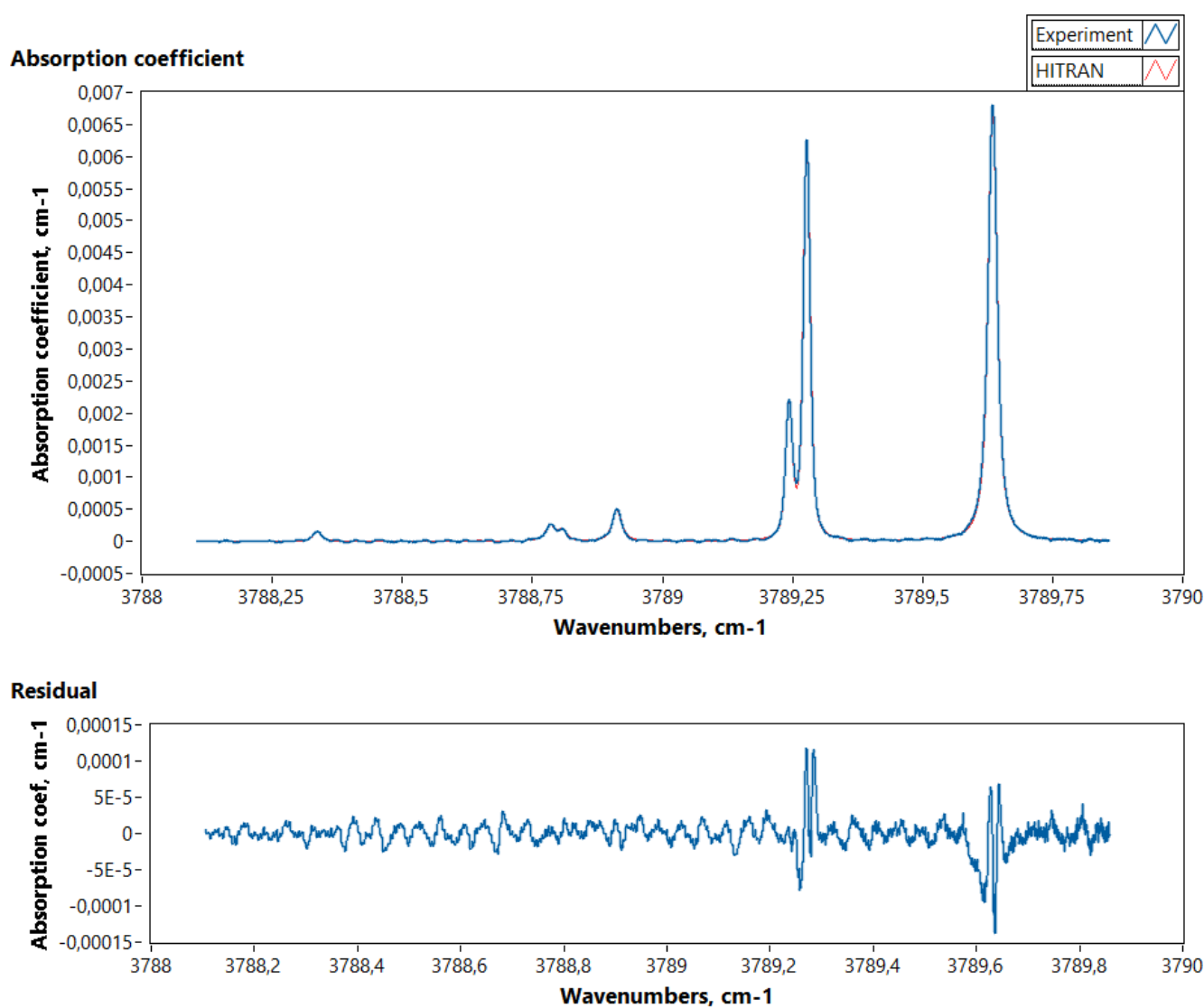

Рисунок 2.56 – Сопоставление экспериментально полученного и синтетического спектров (сверху) и их невязка (снизу) при подавлении вклада интерференции на средних частотах.

Поскольку в ходе физических испытаний прибора ДЛС-Л данные снимались в течение одного часа для каждого газа, была набрана достаточно большая статистика. Это позволило с хорошей точностью рассчитать величины ошибки среднего для каждой изотопной сигнатуры, а также обработать усредненный по пулу из нескольких десятков измеренный спектр коэффициента поглощения. Сопоставление такого усредненного измеренного спектра коэффициента поглощения с синтетическим спектром для водяного пара приведено на рисунке 2.57.

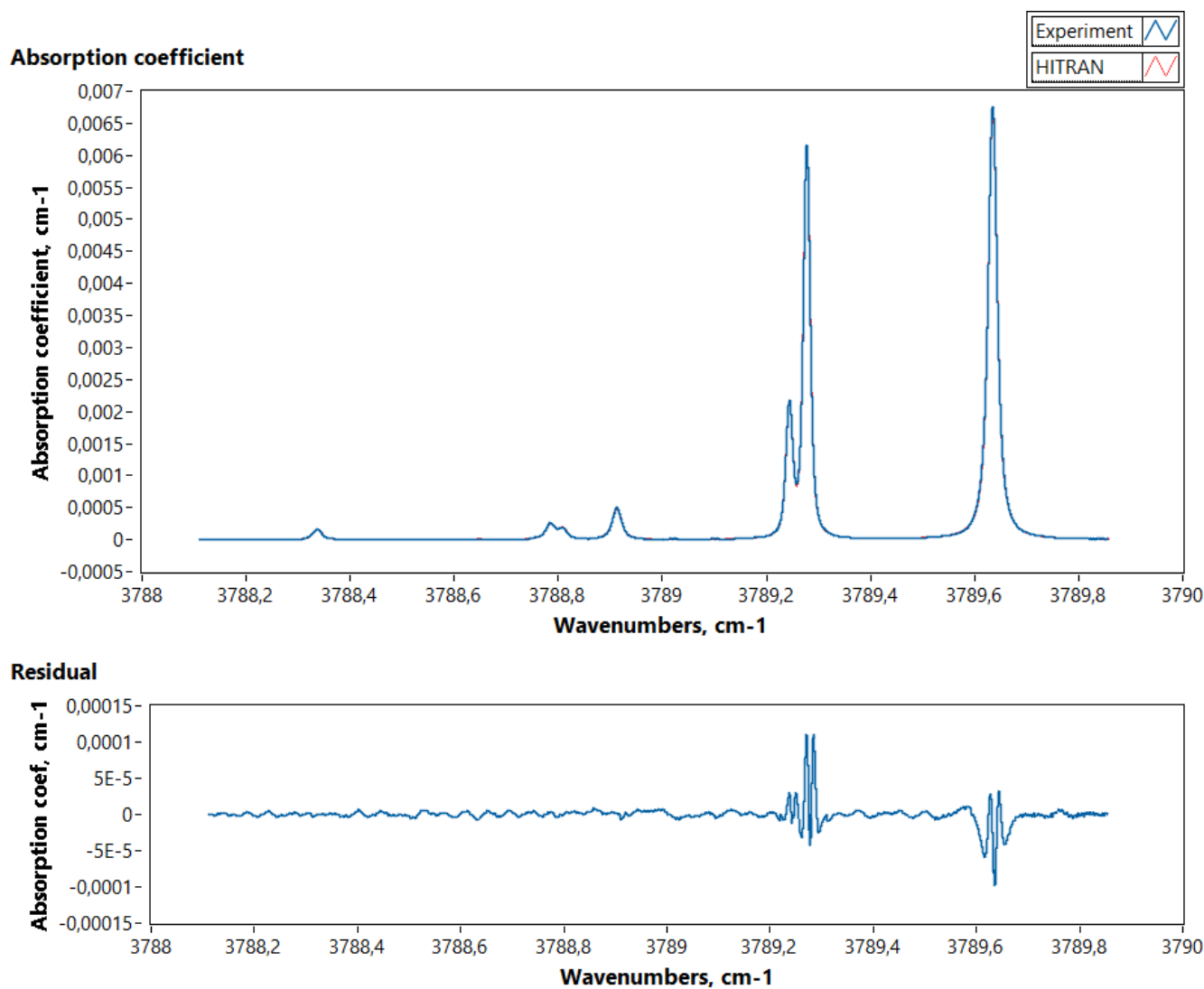

Рисунок 2.57 – Сопоставление усредненного по пулу измеренных и синтетического спектров коэффициента поглощения $H_2O$ (сверху) и их невязка (снизу).

Для представленного на рисунке 2.57 усредненного по пулу измеренных спектров коэффициента поглощения $H_2O$ размах шумовой дорожки от пика до пика составляет $\sim 7 \div 9 \times 10^{-6}$ см$^{-1}$, таким образом отношение сигнал/шум для сильных линий основного изотополога составило $\sim 800$-$2500$, отношение сигнал/шум для HDO – $\sim 93$, для $H_2^{18}O$ – $\sim 198$ и для $H_2^{17}O$ – $\sim 120$.

Также без промежуточных шагов, которые полностью аналогичны работе со спектрами водяного пара, приведем сопоставление усредненного измеренного спектра коэффициента поглощения с синтетическим спектром для $CO_2$ на рисунке 2.58.

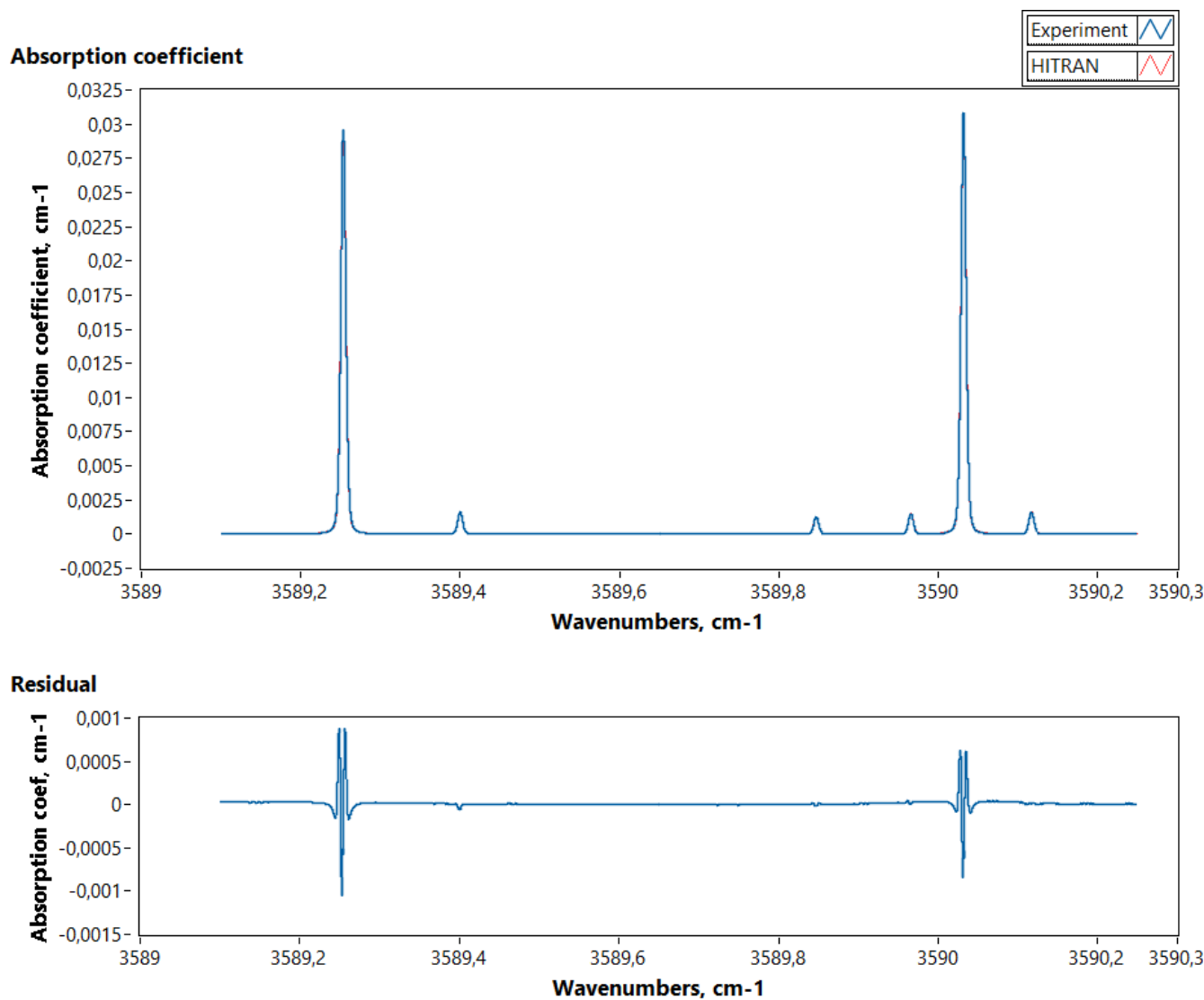

Рисунок 2.58 – Сопоставление усредненного по пулу измеренных и синтетического спектров коэффициента поглощения $CO_2$ (сверху) и их невязка (снизу).

Для представленного на рисунке 2.58 усредненного по пулу измеренных спектров коэффициента поглощения $CO_2$ размах шумовой дорожки от пика до пика составляет $\sim 3 \div 4 \times 10^{-6}$ см$^{-1}$, таким образом отношение сигнал/шум для сильных линий основного изотополога составило $\sim 2 \times 10^4$, отношение сигнал/шум для $^{13}CO_2$ – $\sim 1253$, для $^{18}OC^{16}O$ – $\sim 1027$ и для $^{17}OC^{16}O$ – $\sim 66$.

## 2.5. Анализ полученных в ходе наземных испытаний прибора ДЛС-Л аналитических данных

Продолжительные физические испытания спектрометра ДЛС-Л, стартовавшие после его сборки и окончательной юстировки оптической системы, заняли в общей сложности около двух лет. В это время прорабатывалась, адаптировалась под специфику работы прибора и совершенствовалась методика обработки данных.

Параллельно с этим совершенствовались и алгоритмы проведения экспериментов. В ходе финальных испытаний были значительно снижены возможные вклады десорбции в газовой системе используемого стенда с вакуумной камерой, в которой помещался прибор ДЛС-Л, благодаря длительной подготовке к каждому эксперименту, сводящейся к нескольким откачкам стенда до высокого вакуума ~$10^{-3}$ мбар по показаниям лабораторного датчика давления Atovac ACM200, чередовавшимся с напусками в систему чистого азота, на протяжении 2-3 дней перед проведением эксперимента. Используемая для проведения испытаний прибора ДЛС-Л вакуумная камера представлена на рисунке 2.59.

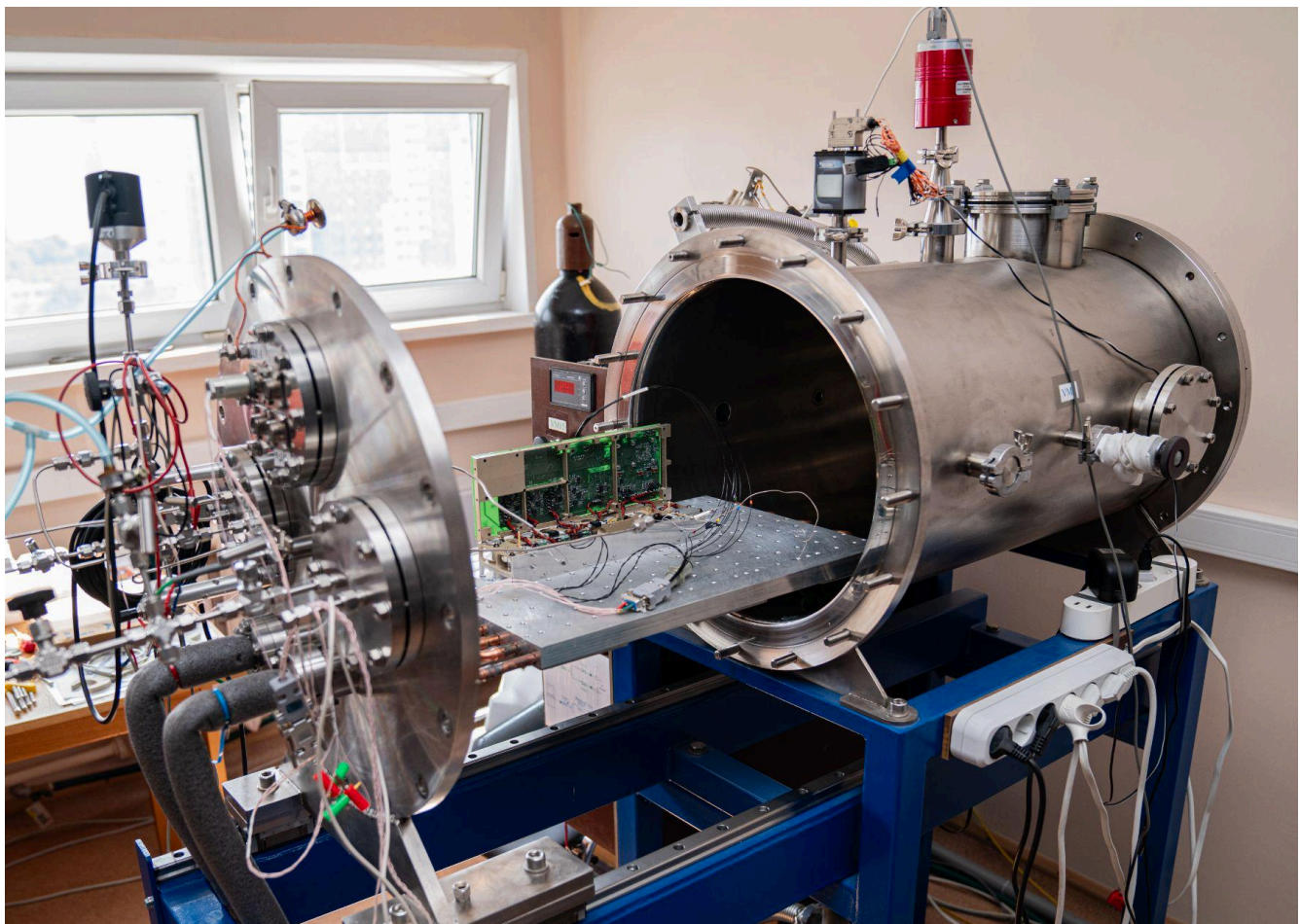

Рисунок 2.59 – Используемая для проведения испытаний прибора ДЛС-Л вакуумная камера в открытом состоянии.

Также были приложены усилия для подавления шумов электроники прибора и электрических наводок, игравших на начальных этапах работы роль, не меньшую, чем вклад интерференции. При этом сохранились очевидные ограничения, связанные с невозможностью помещения газовой системы стенда в печь для гарантированного избавления от возможных обменных процессов исследуемой пробы с адсорбированной водой в стенках системы.

Для возможности анализа качества результатов, получаемых в ходе наземных испытаний прибора ДЛС-Л, требовалось получить независимую оценку. Для этого были предприняты разные стратегии для двух исследуемых газов. Стоит отдельно отметить, что из-за ограниченного финансирования проекта возможности приобретения откалиброванных относительно стандартов VSMOW и VPDB воды и углекислого газа не было. По этой причине использовались способы получения независимой оценки используемых газовых проб, не требующие больших финансовых затрат.

В случае с водяным паром выход был найден в использовании в качестве пробы природной минеральной воды, прошедшей при подготовке эксперимента предварительную процедуру дегазации, с тем условием, что вода выбранного бренда имеет выходные данные об используемой скважине, то есть не должна претерпевать значимых изменений в своем составе ввиду большого объема подземного источника, а также была исследована в лабораторных условиях на современном оборудовании. Выбранная минеральная вода Perrier, добываемая к юго-востоку от французской коммуны Вержез, этим требованиям удовлетворяет. Она входит в список бутилированной минеральной воды, изученной на оборудовании лабораторного класса – масс-спектрометрах для изотопного анализа [210,211]. В подобных публикациях приводятся значения изотопных сигнатур для HDO и для $H_2^{18}O$, тогда как из-за сложности пробоподготовки значение изотопной сигнатуры для $H_2^{17}O$ отсутствует. По этой причине для оценки значения $\delta_{VSMOW}{}^{17}O$ использовалась зависимость $\delta^{17}O_{VSMOW} = 0.52 \times \delta^{18}O_{VSMOW}$ [129]. Сравнение результатов измерения изотопных отношений $H_2O$ при помощи прибора ДЛС-Л и данных, приведенных в литературе, представлено в таблице 2.5.

Таблица 2.5 – Сравнение результатов измерения изотопных отношений $H_2O$ при помощи прибора ДЛС-Л и данных, приведенных в литературе.

| Величина | Результаты из литературы [210,211] | Результаты измерения ДЛС-Л | Отклонение |
|---|---|---|---|
| $\delta_{VSMOW}{}^{18}O$ | -6.33±0.02‰ | -13.23±1.71‰ | -6.90‰ |
| $\delta_{VSMOW}D$ | -42.00±0.20‰ | -54.34±4.11‰ | -12.34‰ |
| $\delta_{VSMOW}{}^{17}O$ [129] | -3.28‰ | -4.20±8.69‰ | -0.92‰ |

Приведенные в таблице стандартные ошибки среднего для изотопных сигнатур, полученных по данным спектрометра ДЛС-Л, вычислялась для каждого изотополога по массиву из нескольких десятков значений, полученных для каждого измеренного спектра коэффициента поглощения. Как видно, все полученные при помощи прибора ДЛС-Л результаты занижены относительно данных, известных из литературы. При этом точность методики такова, что для единичного спектра вклад высокочастотной интерференции может выливаться в ошибку определения $\delta_{VSMOW}^{18}O$ до ~10‰, для $\delta_{VSMOW}^{17}O$ в ошибку до ~20‰ и для $\delta_{VSMOW}D$ в ошибку до ~30‰. Однако благодаря температурному дрейфу в ходе эксперимента, представленному на рисунке 2.41, высокочастотная интерференция двигалась вдоль анализируемого участка спектра со средней скоростью ~0.4 см$^{-1}$/час. Поскольку период высокочастотной интерференции составляет ~0.04 см$^{-1}$, за длительность эксперимента около часа она успевает преодолеть ~10 периодов, тем самым сильно уменьшая свой вклад в неточность определения изотопных отношений.

Одной из причин более существенного занижения результата изотопной сигнатуры $\delta_{VSMOW}D$ может быть то, что ввиду сравнительно большой площади поверхности стенок относительно используемого аналитического объема кюветы, а также большей скорости адсорбции HDO относительно $H_2O$ за длительное время подготовки эксперимента, занимающее несколько часов, и время самого эксперимента, составляющее около часа, тяжелая вода могла осесть на стенки капиллярной газовой системы в большей степени чем основной изотополог.

В случае с $CO_2$ небольшая порция, напущенная из используемого 40-литрового баллона с некалиброванным газом, была передана для изучения на масс-спектрометре для изотопного анализа Thermo Fisher Scientific Delta XP с пиролитической приставкой TC/EA в ГЕОХИ РАН. Сравнение результатов измерения изотопных отношений $CO_2$ при помощи прибора ДЛС-Л и данных, полученных в ГЕОХИ РАН, приведено в таблице 2.6.

Таблица 2.6 – Сравнение результатов измерения изотопных отношений $CO_2$ при помощи прибора ДЛС-Л и данных, полученных в ГЕОХИ РАН.

| Величина | Результаты ГЕОХИ РАН | Результаты измерения ДЛС-Л | Отклонение |
|---|---|---|---|
| $\delta_{VPDB}^{13}C$ | -47.55±0.06‰ | -46.64±1.12‰ | 0.91‰ |
| $\delta_{VSMOW}^{18}O$ | 20.17±0.27‰ | 37.52±0.82‰ | 17.35‰ |

В данном случае точность методики такова, что для единичного спектра вклад высокочастотной интерференции может выливаться в ошибку определения $\delta_{VPDB}^{13}C$ до ~2‰, а для $\delta_{VSMOW}^{18}O$ в ошибку до ~3‰. Однако благодаря температурному дрейфу в ходе эксперимента высокочастотная интерференция двигалась вдоль анализируемого участка спектра со средней скоростью ~0.3 см⁻¹/час. Поскольку период высокочастотной интерференции составляет ~0.04 см⁻¹, за длительность эксперимента около часа она успевает преодолеть ~8 периодов, тем самым заметно снижая свой вклад в неточность определения изотопных отношений.

Сравнение полученных для измеренных в ходе эксперимента спектров значений изотопных сигнатур с независимыми оценками представлено на рисунке 2.60. Хорошо видно заметное влияние движущейся высокочастотной интерференции на получаемые результаты.

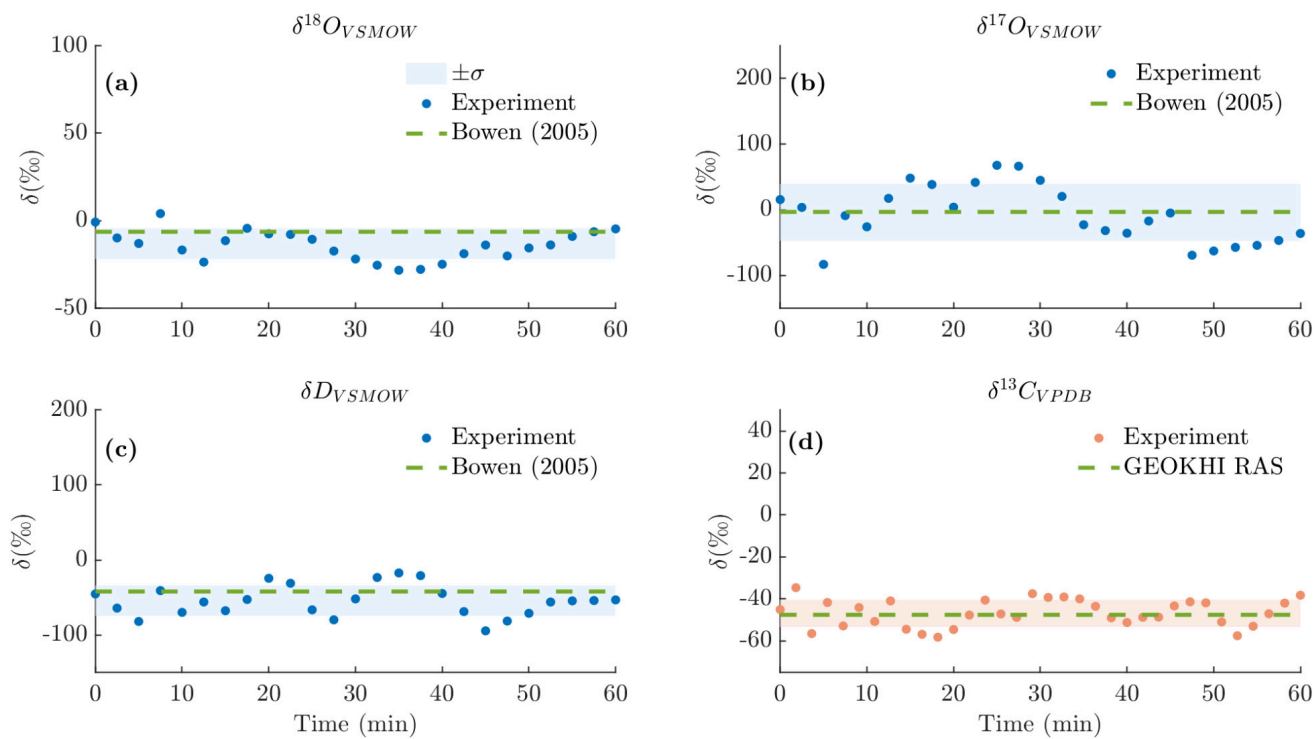

Рисунок 2.60 – Вариации измеряемых величин изотопных сигнатур в процессе эксперимента: точки – значения δ; зеленая пунктирная линия – независимая оценка; цветная область – доверительный интервал ±σ. **(a)** измеренное значение $\delta_{VSMOW}{}^{18}O$ в сравнении с обзорами [219,211]. **(b)** измеренное значение $\delta_{VSMOW}{}^{17}O$ в сравнении с обзорами [210,211,129]. **(c)** измеренное значение $\delta_{VSMOW}D$ в сравнении с обзорами [210,211]. **(d)** измеренное значение $\delta_{VPDB}{}^{13}C$ в сравнении с результатами ГЕОХИ РАН.

Для каждого изотополога было рассчитано отношение сигнал/шум и предел обнаружения (LOD), приведенные в таблице 2.7. Параметр $LOD_{iso}$ определялся для каждого анализируемого изотополога как минимальная концентрация этого изотополога $N_{iso}$, которую можно выявить на фоне шума $\alpha_{noise\ spectrum}$ без спектральных особенностей:

$$LOD_{iso} = 3 \cdot min(N_{iso}) = 3 \cdot \frac{STD(\alpha_{noise\_spectrum})}{\sigma_{iso}}. \tag{2.54}$$

Таблица 2.7 – Отношение сигнал/шум и предел обнаружения изотопологов водяного пара и углекислого газа прибором ДЛС-Л.

| Изотополог | Отношение сигнал/шум | Предел обнаружения |
|---|---|---|
| $H_2O$ | 800-2500 | $1.95 \times 10^{14}$ mol/cm³ |
| $H_2^{18}O$ | 200 | $2.5 \times 10^{15}$ mol/cm³ |
| $H_2^{17}O$ | 120 | $5.1 \times 10^{15}$ mol/cm³ |
| HDO | 93 | $8.2 \times 10^{15}$ mol/cm³ |
| $CO_2$ | 20 000 | $3.4 \times 10^{13}$ mol/cm³ |
| $^{13}CO_2$ | 1250 | $8.2 \times 10^{14}$ mol/cm³ |

Подводя итог, можно констатировать, что несмотря на крайне скромные габариты и электропотребление, а также, что наиболее важно, внутренний объем аналитической кюветы чуть больше 1 мл ввиду использования кюветы капиллярного типа из-за специфики газовой системы всего газоаналитического комплекса и некоторые огрехи при проектировании оптической системы спектрометра ДЛС-Л, результаты физических испытаний показали, что при актуальной на момент проведения работы наполненности базы данных HITRAN можно получить значения изотопных сигнатур для изотопологов молекулы воды HDO, $H_2^{18}O$ и $H_2^{17}O$, а также для изотополога молекулы углекислого газа $^{13}CO_2$ с точностью, достаточной для сопоставления с результатами, полученными в ходе советских миссий «Луна-16», «Луна-20» и «Луна-24», а также серии миссий NASA «Аполлон».

Однако до объявленной даты старта в 2028 году планируется проведение некоторых доработок прибора. В программном обеспечении будет увеличена частота дискретизации и реализован алгоритм стабилизации длины волны излучения лазерных модулей по спектральным линиям поглощения с использованием реперной кюветы. При обработке данных возможен переход к применению профиля спектральных линий поглощения Хартмана-Трана, – поскольку база данных HITRAN обновляется каждые 4 года, дополнительные параметры линий для выбранных спектральных областей могут быть доступны в версии 2024 или 2028 года. Наконец, планируется проведение дополнительных испытаний прибора и его калибровка по сертифицированным эталонам.

При этом проведение *in situ* исследований лунного грунта вблизи места посадки на поверхность Луны избавило бы от возможности загрязнения исследуемых образцов, связанной с многолетним хранением и изучением в земных лабораториях. Стоит отметить, что число подобных устройств для изучения поверхностного грунта космических тел на сегодняшний день крайне мало. Из успешных миссий можно привести в пример прибор Sample Analysis at Mars (SAM) в составе Mars Science Laboratory марсохода Curiosity [150]. Также недавно были опубликованы принципы работы прибора для изотопного анализа $^{15}N/^{14}N$ в лунном грунте *in situ* на аппарате «Чанъэ-7», планируемом к запуску в 2026 году [212].

Изучение лунного грунта *in situ* могло бы приблизить получение стройного знания о фундаментальных процессах формирования Луны и источниках летучих веществ на ней и помогло бы расставить точки над *i* в прикладных вопросах о поставках летучих веществ для будущих миссий по исследованию Луны.

## Выводы к главе 2

По результатам, изложенным в главе 2, можно сформулировать следующие основные выводы:

1. Использование метода диодно-лазерной абсорбционной спектроскопии дает возможность создавать газоаналитические сенсоры для изотопного анализа, удовлетворяющие требованиям космического приборостроения – критически малой массы и низкого энергопотребления, а также электромагнитной совместимости с оборудованием всего научного комплекса и радиационной стойкости;

2. Представленный спектрометр ДЛС-Л работет с бо́льшим числом изотопологов молекул воды и углекислого газа, чем могут позволить стандартные решения для масс-спектрометров лабораторного класса. Спектрометр для исследования *in situ* изотопных отношений содержащихся в лунном грунте $H_2O$ и $CO_2$, разработан впервые;

3. Отсутствие режима стабилизации по положению пиков линий поглощения в основном или реперном каналах прибора приводит к дрейфу сигнала между соседними сериями измерений, что значительно увеличивает время обработки ввиду последовательной работы с каждой серией полученных спектральных данных;

4. Разработанный программный комплекс для анализа спектральных данных позволяет успешно обрабатывать и анализировать экспериментальные данные прибора ДЛС-Л с учетом специфических особенностей эксперимента – дрейф сигнала, оптическая интерференция, шумы электроники и проч.;

5. Спектрометр ДЛС-Л вкупе с разработанным программным обеспечением для обработки полученных спектральных данных позволяет проводить изотопный анализ для представляющих наибольший фундаментальный интерес в вопросах о процессах формирования Луны и источниках летучих веществ на ней изотопологов молекул воды и углекислого газа с точностью от 1‰ до 1%;

6. В разработанном программном комплексе для анализа спектральных данных предусмотрена возможность применения более сложных контуров спектральных линий, что позволит в будущем дополнительно улучшить качество обработки при добавлении в грядущие версии базы данных HITRAN необходимых для этого, но пока отсутствующих табличных значений;

7. В результате сборки, настройки и юстировки спектрометра ДЛС-Л был создан значительный задел, позволяющий в дальнейшем разрабатывать компактные и чувствительные спектрометры для локальных измерений малых газовых составляющих в условиях земной атмосферы – для задач экологического мониторинга, в частности, мониторинга парниковых газов на территориях естественных эмиссий (болота, районы тающей многолетней мерзлоты) или антропогенных загрязнений, коммерческого применения для отслеживания качества воздуха в индустриальных зонах и т.д.

# ГЛАВА 3. КОНЦЕПЦИЯ ИНФРАКРАСНОГО ДИСТАНЦИОННОГО ГАЗОАНАЛИЗАТОРА ЛИДАРНОГО ТИПА ДЛЯ МОНИТОРИНГА ЭМИССИЙ МЕТАНА

Описанный в предыдущей главе метод абсорбционной лазерной спектроскопии в качестве одного из основных алгоритмических ограничений достижимой точности определения концентрации отдельных составляющих анализируемой газовой смеси или изотопологов выбранного газа имеет сложность корректного учета базовой линии (спектрального континуума) исходного сигнала. Помимо двух описанных ранее подходов к решению вопроса учета базовой линии и периодических шумов исходного сигнала существует сравнительно распространенный подход, основанный на вейвлет-преобразовании [213-216]. Однако нельзя с уверенностью говорить о том, что сочетание абсорбционной спектроскопии на основе непрерывно перестраиваемых полупроводниковых лазеров с каким-либо из перечисленных методов обработки исходного сигнала позволяет получить неискаженный спектр поглощения анализируемой газовой смеси. По этой причине во многих приложениях используется метод модуляционной спектроскопии, позволяющий уйти от решения проблемы корректного описания базовой линии [161,217-219].

Примером практической реализации метода модуляционной спектроскопии в сочетании с квадратурным детектированием анализируемого сигнала может служить разработанный в МФТИ Газоанализатор для Измерения Метана Лидарный Инфракрасный (ГИМЛИ).

В качестве целевого газа для разработанного газоанализатора был выбран метан, поскольку, во-первых, молекула этого газа имеет подходящую для спектрального анализа энергетическую структуру, во-вторых, метан является важным фактором парникового эффекта в климатической системе Земли, распространенным продуктом микробиологической активности и широко используемым энергоносителем. Таким образом, изучение содержания метана в приземном слое атмосферного воздуха представляет большой научный интерес для экологии и климатологии, но может быть адаптировано к коммерческому применению – поиску утечек из магистральных газопроводов и в районах промышленных объектов, связанных с добычей, транспортировкой, переработкой, хранением метана или использованием его в качестве энергоресурса.

Данная глава посвящена обоснованию и постановке задачи мониторинга содержания метана в приземных слоях атмосферного воздуха, подробному описанию разработки лабораторного макета прибора ГИМЛИ, описанию его структуры и методики измерений. Экспериментальные результаты, полученные в ходе отработки предложенной методики на созданном лабораторном стенде, приведены в конце главы. Разработке версии прибора для полевых измерений посвящена следующая глава.

Личный вклад автора заключается в сборке и юстировке лабораторного стенда для отработки предложенной методики, разработке ПО для управления стендом, проведении серии экспериментов для отработки методики, анализе режимов работы стенда, разработке методики обработки экспериментальных данных, проведении анализа спектральных данных при помощи разработанных автором алгоритмов, получении итоговых результатов по оценке точности определения концентрации метана при помощи созданного лабораторного стенда. В настоящем разделе используются материалы публикации автора [162].

## 3.1. Значимость проведения мониторинга содержания метана в атмосфере Земли

Составляющие атмосферу газы с разной степенью эффективности поглощают уходящее тепловое излучение поверхности Земли, благодаря чему баланс между приходящим солнечным и уходящим тепловым излучением формируется около тропопаузы. В силу ряда процессов, среди которых доминирует адиабатическое перемешивание, температура приземного слоя воздуха превышает эффективную температуру планеты на величину, приблизительно равную $\Delta T = \frac{gH_t}{C_p}$, где $g$ – ускорение свободного падения, $C_p$ – теплоемкость воздуха при постоянном давлении, $H_t$ – высота тропопаузы. Для Земли эта величина довольно существенная, в среднем около 38℃. Это явление, известное как парниковый эффект, наблюдается на всех планетах, обладающей атмосферой, и обеспечивает постоянную среднюю температуру у поверхности Земли около 15℃ [220]. Однако в последние годы, говоря о парниковом эффекте, акцент делается не на естественном процессе, а на усугублении эффекта, которое, как считается, в значительной степени вызвано деятельностью человека и ответственно за повышение средней температуры у поверхности Земли.

Парниковые газы (ПГ) прозрачны в видимом диапазоне спектра, но интенсивно поглощают и переизлучают энергию в ИК-диапазоне, вызывая парниковый эффект. Основными парниковыми газами являются водяной пар, углекислый газ, метан и закись азота. Их общий объем менее 1% от объема атмосферы. Степень воздействия каждого ПГ на глобальное потепление климата зависит от трех ключевых факторов: его содержания в атмосфере, времени жизни в атмосфере и потенциала глобального потепления, который выражает общую энергию, которая может быть поглощена данной массой ПГ за определенный период времени, по сравнению с той же массой $CO_2$ за тот же период времени [221].

Среди перечисленных газов $CH_4$ является вторым по распространенности антропогенным ПГ после $CO_2$ – на его долю приходится 17% мировых выбросов парниковых газов в результате деятельности человека [222], как показано на рисунке 3.1.

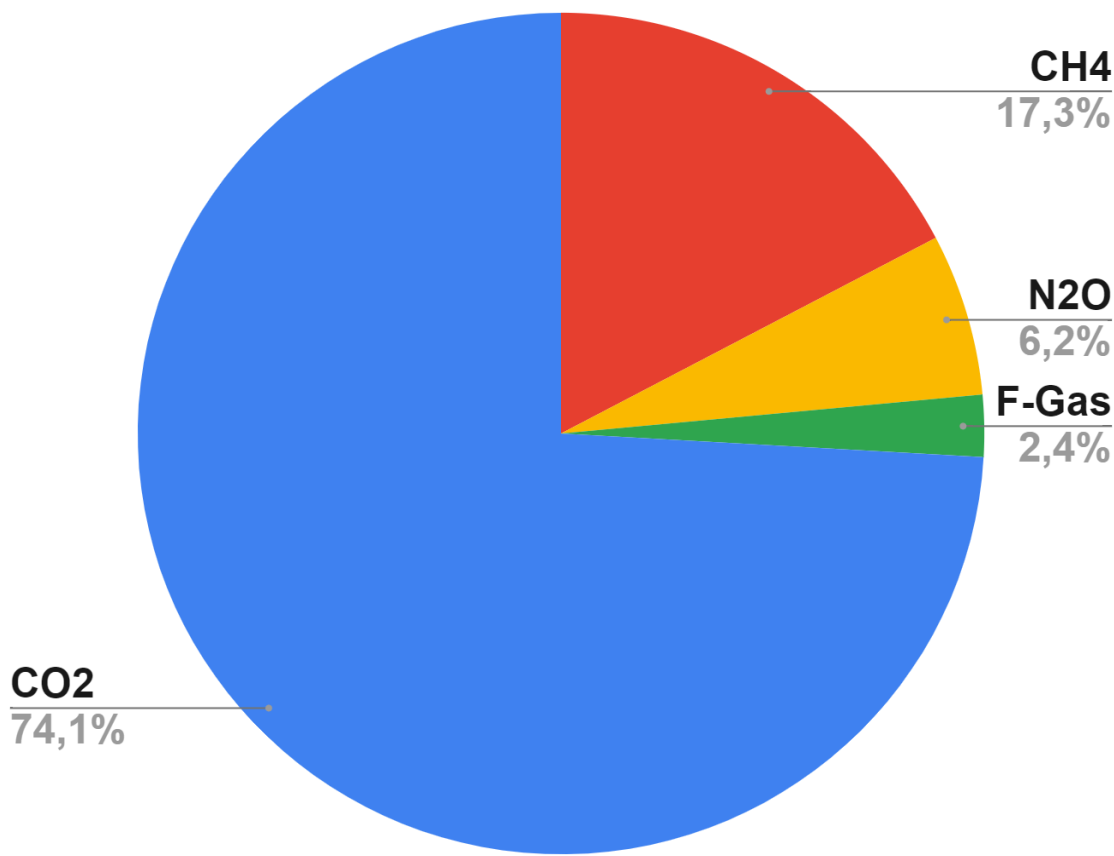

Рисунок 3.1 – Распределение выбросов парниковых газов согласно отчету Межправительственной группы экспертов по изменению климата (МГЭИК) за 2019 год.

Наряду с природным газом источниками метана в атмосферном воздухе могут быть биогенные и эндогенные (геологические) процессы. Например, болотистая местность является важным естественным источником метана. Помимо естественных способов попадания в атмосферу 50-65% глобальных выбросов $CH_4$ связано с антропогенным фактором [223]. К ним относятся:

- добыча и транспортировка угля, природного газа и нефти,
- животноводство,
- сельское хозяйство (особенно рисоводство),
- землепользование (в том числе сведение лесов),
- разложение органических отходов на полигонах захоронения отходов и в сточных водах,
- утечки из газовых систем и горнодобывающих районов.

Таким образом, значительное сокращение выбросов $CH_4$ может быть достигнуто за счет устранения утечек в трубопроводах и промышленных установках в районах добычи нефти и газа [223]. На рисунке 3.2 представлены глобально усредненные данные Национального управления океанических и атмосферных исследований США по среднемесячному содержанию метана в атмосфере Земли за разные периоды времени [224]. Можно видеть, что за период с 1983 по 2023 год содержание метана в атмосферном воздухе выросло на ~20%.

Вышесказанное вызывает серьезную озабоченность, поскольку метан по сравнению с $CO_2$ гораздо эффективнее в качестве парникового газа – его кумулятивный потенциал глобального потепления превышает аналогичный показатель для $CO_2$ в 28 раз за период в 100 лет и в 84 раза за период в 20 лет [221,225,226].

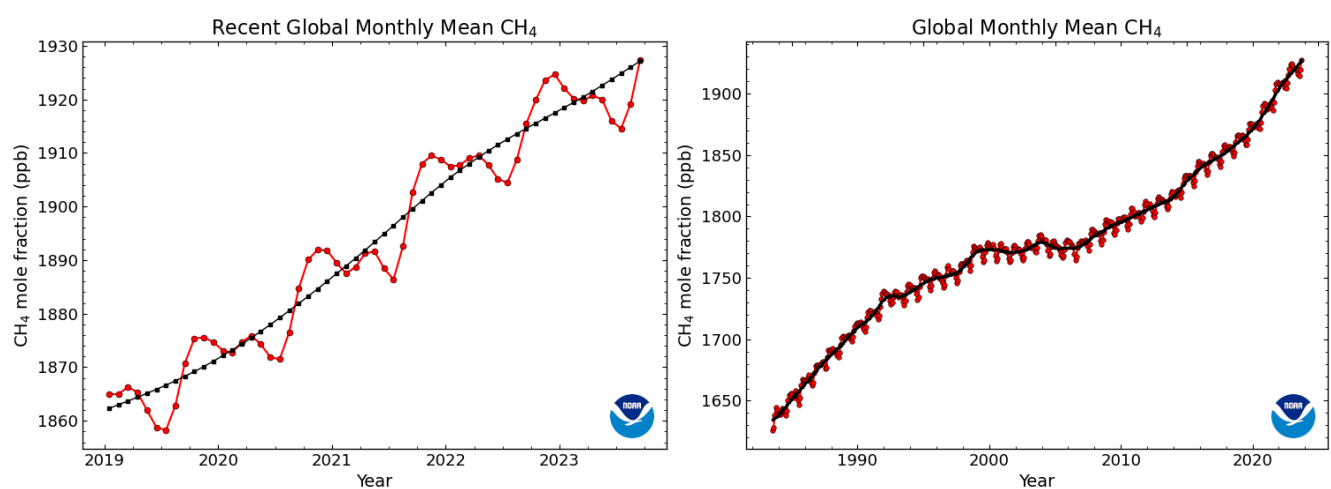

Рисунок 3.2 – Глобально усредненные данные Национального управления океанических и атмосферных исследований США по среднемесячному содержанию метана в атмосфере Земли за 5 лет с 2019 года (слева) и за 40 лет с 1983 года (справа).

Еще одним фактором неопределенности является тающая многолетняя мерзлота в Арктическом регионе. Вблизи поверхности арктической многолетней мерзлоты в почве содержится большое количество органических соединений углерода и остатков растительной и животной органики, которые не могут разлагаться и гнить, в то время как слои многолетней мерзлоты, расположенные глубже, содержат минеральные грунты. При таянии многолетней мерзлоты происходит высвобождение метана и углекислого газа в атмосферный воздух.

В докладе Межправительственной группы экспертов по изменению климата (МГЭИК) «Океан и криосфера в условиях меняющегося климата» говорится, что средняя температура многолетней мерзлоты с 2007 по 2016 год повысилась на рекордную величину 0.39℃ ± 0.15℃ [227]. В условиях потепления Арктики более длинные и теплые вегетационные периоды могут усилить биогенное разрушение органических соединений, захороненных под слоем многолетней мерзлоты и внутри нее, увеличивая масштабы высвобождения $CO_2$ и $CH_4$ в атмосферу.

Кроме того, быстрое таяние многолетней мерзлоты может происходить по всей Арктике, изменяя гидрологию поверхности, что может способствовать дальнейшему таянию. Из-за этих локальных обратных связей деградация многолетней мерзлоты может происходить гораздо быстрее, чем это можно было бы предсказать на основе одних только изменений температуры воздуха. Это потенциально может привести к необратимому ускорению эмиссии метана в атмосферу, что в свою очередь может оказать значительное влияние на климатическую систему [227]. Стоит отметить, что экспедиции на Восточно-Сибирском арктическом шельфе действительно выявили значительный поток метана из тающей мерзлоты на дне в атмосферу [228].

Несмотря на определенный скепсис, существующий в научном сообществе по отношению к концепции «метановой бомбы» [229], влияние эмиссии CH$_4$ на текущие климатические изменения подчеркивается МГЭИК, Европейским космическим агентством (ESA) и Национальным управлением по аэронавтике и исследованию космического пространства (NASA), которые выступили с совместными инициативами по его мониторингу [223]. Карты, созданные в рамках Инициативы ESA по изменению климата, дают представление о таянии многолетней мерзлоты в Северном полушарии. На рисунке 3.3 показана площадь многолетней мерзлоты в 2017 году по сравнению с 2003 годом.

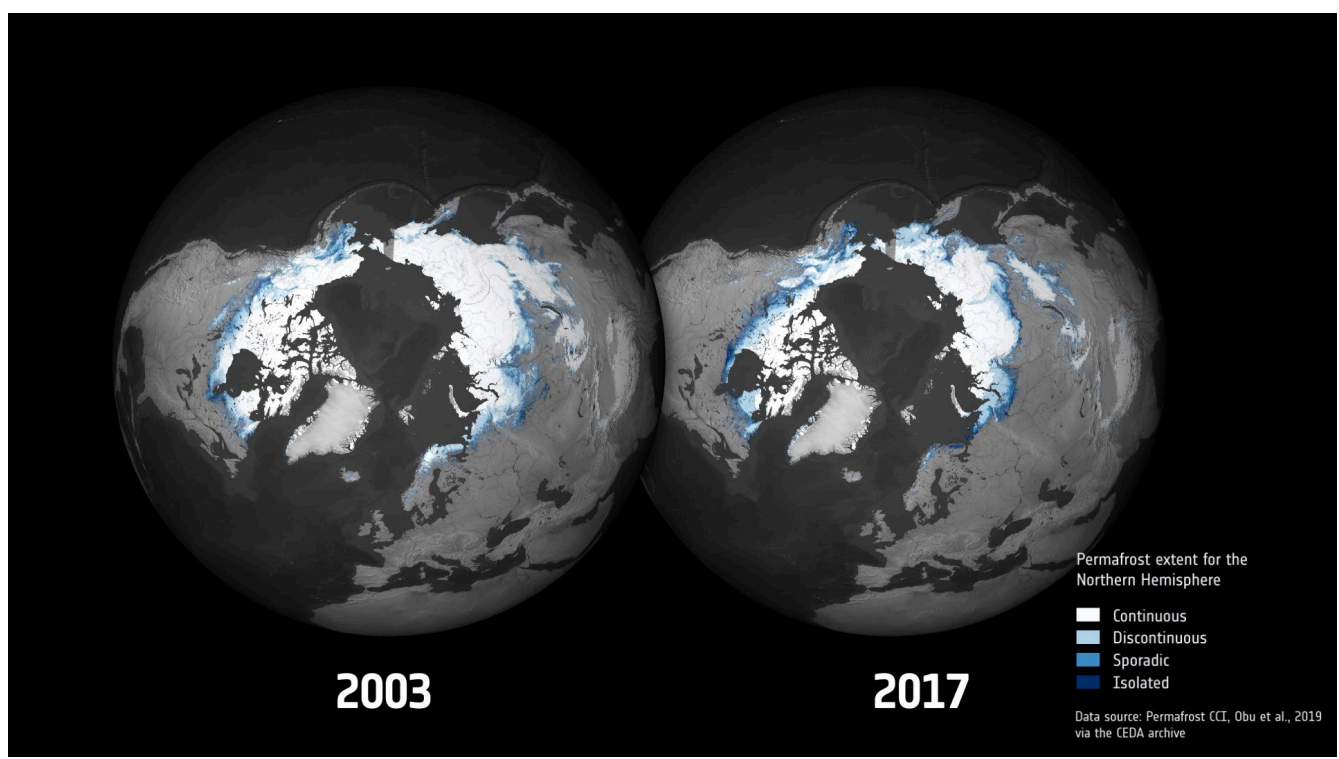

Рисунок 3.3 – Распространенность многолетней мерзлоты в 2017 году по сравнению с 2003 годом по данным ESA [230].

При этом, говоря о России, проблема таяния многолетней мерзлоты с последующей эмиссией парниковых газов в атмосферу является, возможно, куда более довлеющей, чем для иных государств, в силу географических особенностей, что видно из рисунка 3.4.

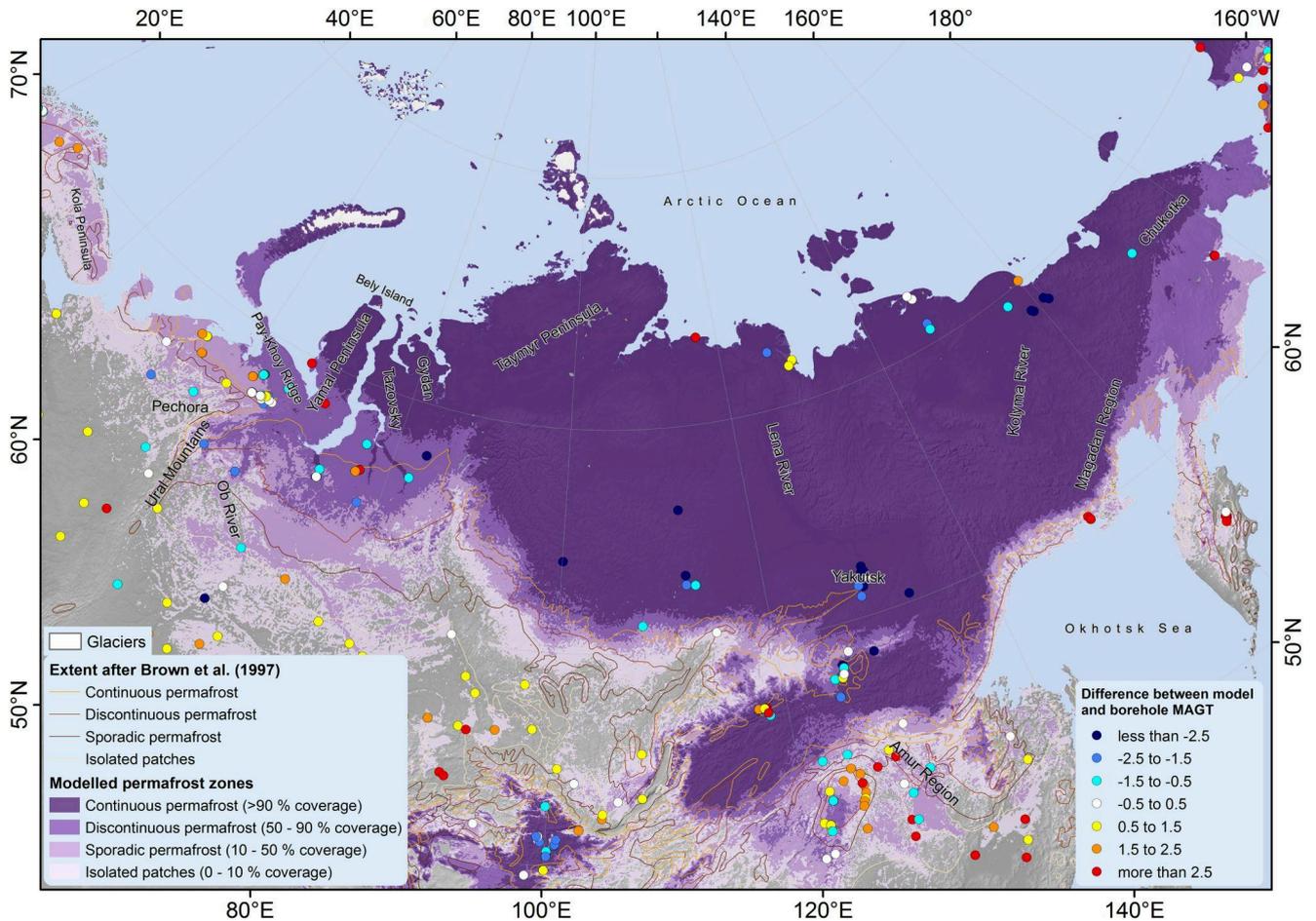

Рисунок 3.4 – Площадь многолетней мерзлоты России [230,231].

Районы многолетней мерзлоты составляют более 60% территории страны [232], к тому же стоит помнить о мерзлоте, находящейся на дне арктических морей, оценки объемов которой представляются методологически заметно более сложными. Динамика многолетней мерзлоты на территории России требует пристального внимания как научного сообщества, так и государства, поскольку помимо медленных эффектов может приводить и к заметным последствиям в короткой перспективе, связанными с выходом из строя инфраструктуры жизнеобеспечения арктического региона постройки советского периода, как это случилось в 2020 году с одним из резервуаров дизельного топлива ТЭЦ-3 в Норильске.

Заинтересованность отдельных организаций и государств в мониторинге антропогенной эмиссии метана связана помимо необходимости в минимизации их последствий с неравномерностью выбросов в разных частях планеты, что видно на рисунке 3.5, полученным по данным спутника ESA Copernicus Sentinel-5P с пространственным разрешением 7×7 км.

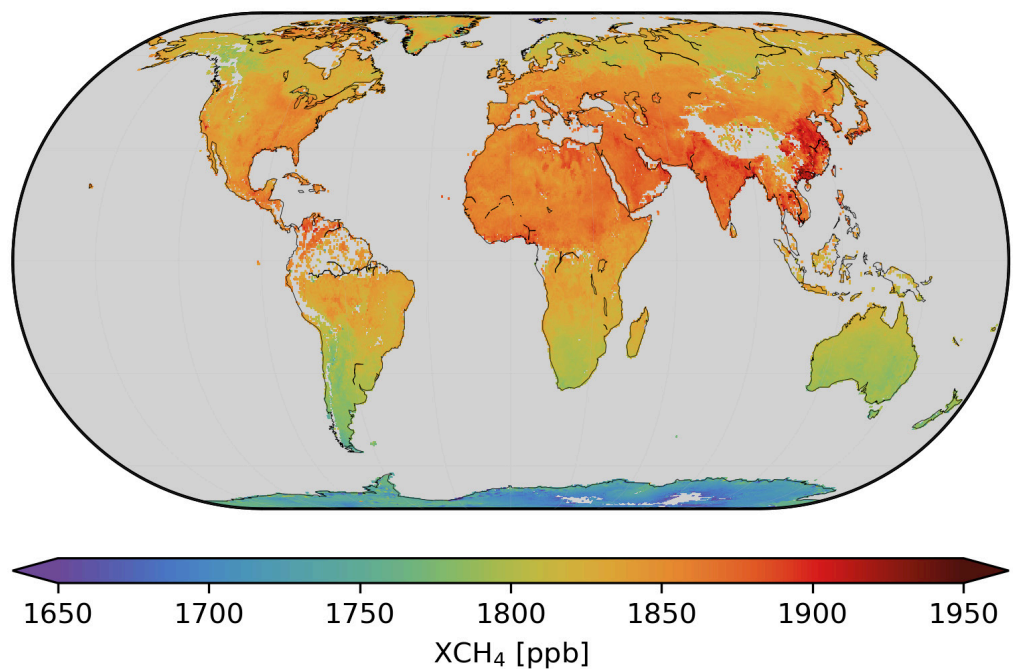

Рисунок 3.5 – Распределение метана в атмосфере Земли по данным спутника Copernicus Sentinel-5P.

Такая ситуация привела к возникновению системы торговли квотами (СТК) на выбросы парниковых газов – рыночного инструмента сокращения выбросов парниковых газов. На государственном уровне правительства устанавливают пороговое значение для общего объема выбросов в различных секторах экономики, после чего компании в выбранных секторах обязывают получать разрешения на каждую единицу своих выбросов. Такие разрешения покупаются у государства и компаний, участвующих в системе.

На межгосударственном уровне в 1997 году в качестве дополнения к Конвенции ООН об изменении климата был разработан Киотский протокол, содержащий конкретные обязательства государств по сокращению выбросов парниковых газов. Протокол вступил в силу в феврале 2005 года. Его участниками являются 157 государств.

Международный рынок торговли квотами находится на этапе становления. Однако на государственном и региональном уровнях продажа квот на выбросы парниковых газов давно и успешно функционирует. Первая полноценная СТК появилась в 2005 году в Европейском Союзе. К 2022 году объем торгов рынка эмиссионных квот ЕС превысил 40 млрд долларов США. В 2011 году правительство Китая объявило о планах разработки рынка торговли квотами на выбросы $CO_2$, целиком система должна запуститься до 2025 года и охватить более 5 Гт годовых выбросов $CO_2$ по данным департамента многостороннего экономического сотрудничества Минэкономразвития России. Система торговли квотами КНР покроет 1/7 часть глобальных выбросов $CO_2$ от сжигания ископаемого топлива, что делает ее самой крупной в мире.

Существуют перспективы торговли квотами с другими странами, которая на данном этапе может проходить в рамках Парижского соглашения в рамках Рамочной конвенции ООН об изменении климата. Статья 6 Парижского соглашения предусматривает участие в торговле квотами на уровне государств. Страны, которые являются покупателями или продавцами таких квот, должны делать соответствующие корректировки в рамках своих обязательств по Парижскому соглашению. Также существует механизм торговли квотами между компаниями. Спрос и предложение на эти инструменты определяются строгостью критериев климатических проектов и качеством верификации реальных выбросов.

Очевидно, что для глобального мониторинга эмиссии парниковых газов и, в частности, для верификации выбросов парниковых газов участниками рынка необходимы системы и устройства, подходящие для этих задач. С 2021 года в рамках пилотного проекта Минобрнауки России запустило 17 «карбоновых полигонов» – тестовых площадок на неурбанизированных территориях, на которых разрабатывают и испытывают технологии измерения, мониторинга и контроля парниковых газов. Разработанный в МФТИ анализатор метана может быть использован в составе комплекса газоаналитического научного оборудования таких площадок, а также для проведения дистанционного мониторинга утечек в магистральных трубопроводах и промышленных установках в районах добычи углеводородов, для мониторинга эмиссии свалочного газа с мусорных полигонов.

## 3.2. Обоснование выбора методики и рабочего спектрального диапазона

Хорошую возможность поиска и наблюдения утечек природного газа или свалочного газа, более чем наполовину состоящего из метана, дает современное развитие беспилотной летательной техники. Как уже говорилось, именно эта газовая составляющая атмосферного воздуха была выбрана в качестве целевого газа при апробации предлагаемой методики, лежащей в основе лидара для мониторинга индустриальных загрязнений, подходящего для установки на беспилотный летательный аппарат (БПЛА).

Возможность установки на БПЛА в свою очередь является важным экономическим фактором для потенциального применения устройства в целях экологического мониторинга или контроля технологических процессов, например, для обнаружения утечек из магистральных газопроводов. Использование легких БПЛА является экономически более целесообразным в сравнении с применением для тех же задач пилотируемой авиации. Пример БПЛА с максимальной полезной нагрузкой ~8 кг, подходящего для таких задач, представлен на рисунке 3.6.

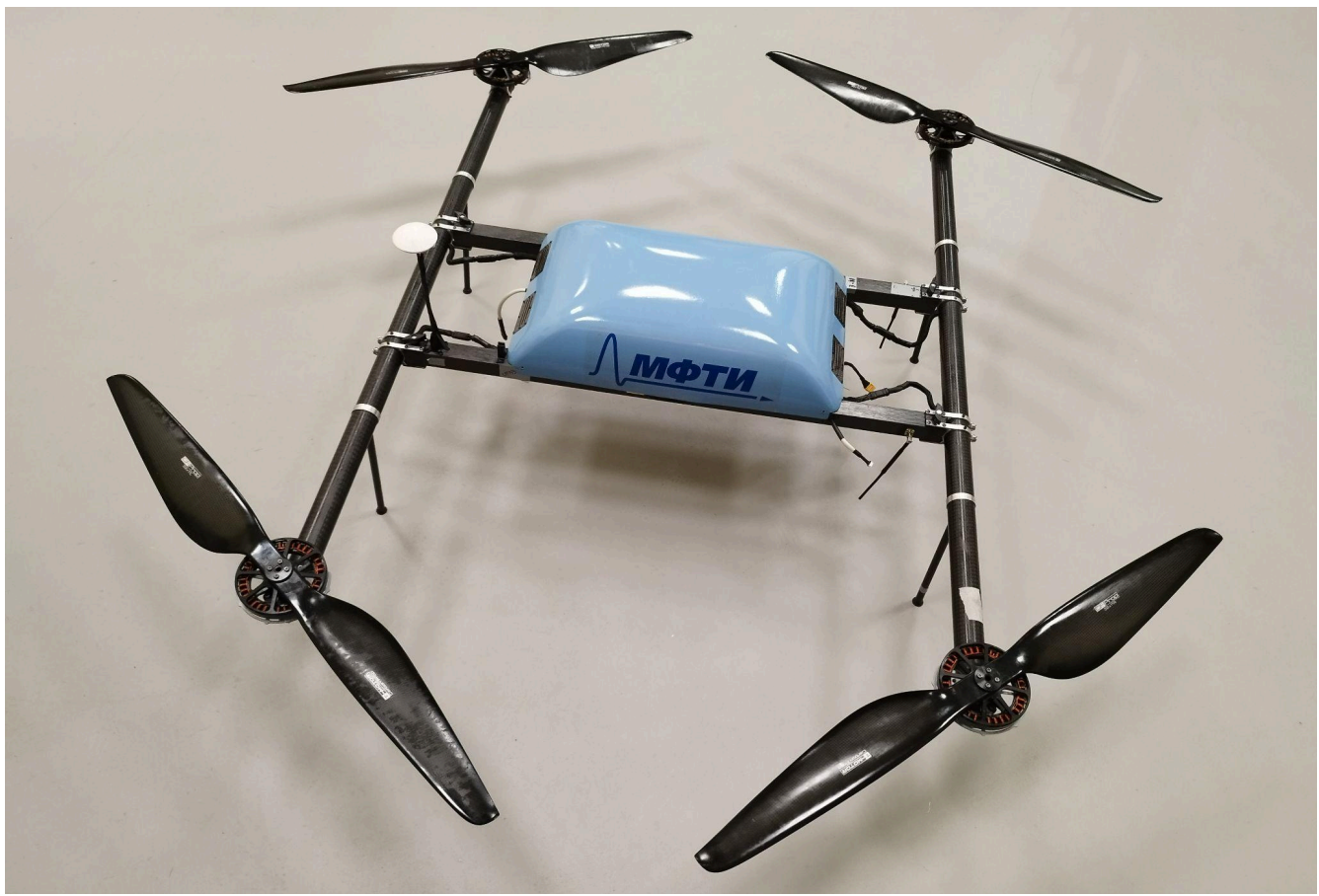

Рисунок 3.6 – Квадрокоптер «Ирбис-432» с максимальной полезной нагрузкой ~8 кг, эксплуатируемый в МФТИ.

Поскольку стандартная скорость перемещения БПЛА мультироторного типа составляет 15-20 м/с, частота поперечного сканирования может составлять 45°/с при высоте полета ~50 м, а пространственное разрешение зондирования должно быть не хуже характерных размеров утечки вблизи места истечения – единиц метров, для решения поставленной задачи необходима методика, допускающая частоту повторения полного цикла спектральных измерений на уровне единиц-десятков Гц. При этом чувствительность такой методики должна обеспечивать определение превышения естественного фона метана на единицы-десятки процентов. Типичное содержание метана в атмосферном воздухе в слабо урбанизированной местности составляет ~1.8-2.2 ppm, что на дистанции в 100 м, соответствующей удвоенной высоте БПЛА над земной поверхностью 50 м, дает ~200 ppm·m. Подобным качествам, как будет показано далее, соответствует предлагаемая методика, основанная на применении модуляционной лазерной абсорбционной спектроскопии и квадратурного детектирования рассеянного от поверхности излучения.

Линии поглощения метана вблизи 3.24 мкм (3086 см⁻¹), принадлежащие фундаментальной моде колебаний связи C-H, примерно в 40 раз превосходят по интенсивности линии обертона этой моды вблизи 1.65 мкм (6057 см⁻¹), что видно на рисунке 3.7.

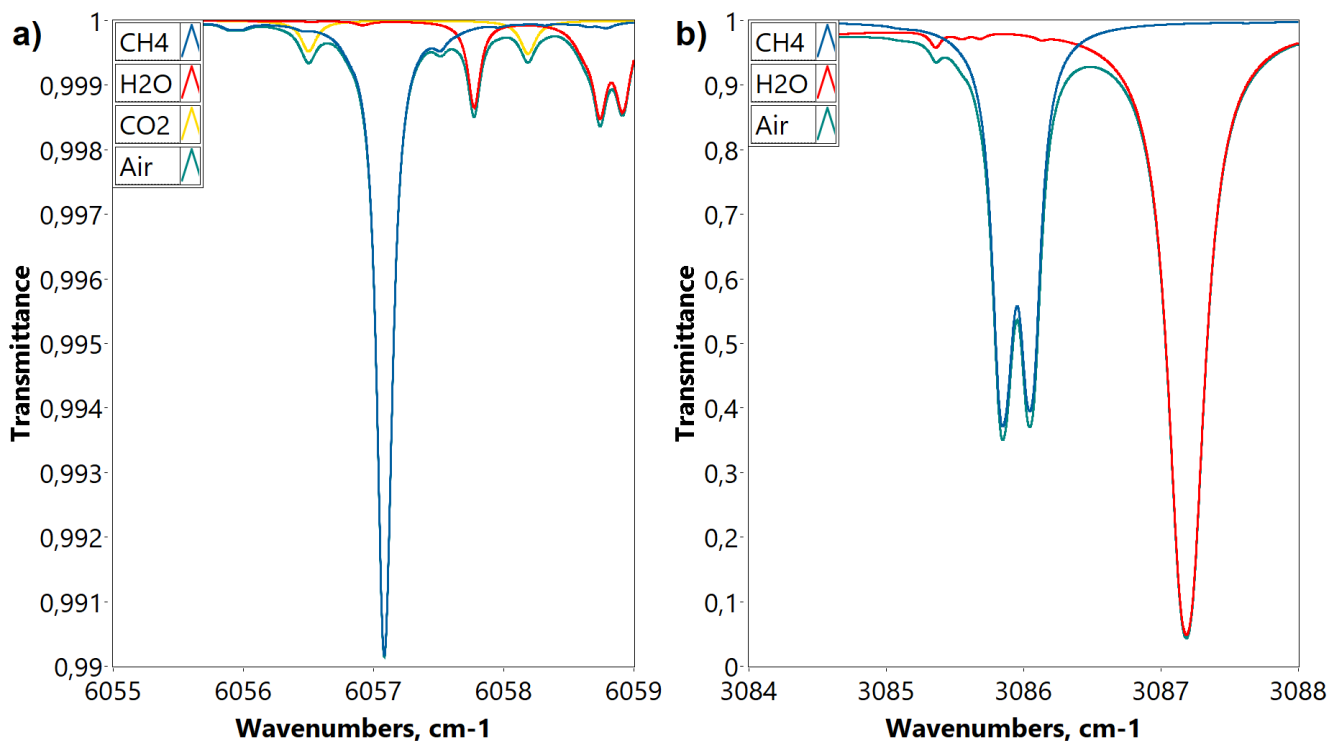

Рисунок 3.7 – Спектральные линии поглощения: **(a)** функция пропускания для 2 ppm метана в атмосферном воздухе при 298 К на дистанции 100 м в диапазоне 6055-6059 см⁻¹, полученная по данным базы HITRAN; **(b)** функция пропускания для 2 ppm метана в атмосферном воздухе при 298 К на дистанции 100 м в диапазоне 3084-3088 см⁻¹, полученная по данным базы HITRAN.

Это компенсируется возможностью применения в ближнем ИК-диапазоне недорогих и компактных полупроводниковых фотодиодов на основе InGaAs, удельная обнаружительная способность $D^*$ которых в своем рабочем диапазоне ($\sim 5 \times 10^{12}$ $\frac{\text{см}\cdot\sqrt{\text{Гц}}}{\text{Вт}}$ для размера чувствительной площадки 1 мм) на 3 порядка выше, чем у неохлаждаемых InAs-фотодиодов ($\sim 5 \times 10^{9}$ $\frac{\text{см}\cdot\sqrt{\text{Гц}}}{\text{Вт}}$ для 1 мм), работающих в области 3 мкм, и на 2 порядка превосходит удельную обнаружительную способность неохлаждаемых HgCdTe-фотодиодов ($\sim 3 \times 10^{10}$ $\frac{\text{см}\cdot\sqrt{\text{Гц}}}{\text{Вт}}$ для 1 мм), рассчитанных на работу в диапазоне 2-4 мкм.

Фотодиоды с несколькими каскадами термоэлектрического охлаждения, обладающие лучшей чувствительностью, рассматриваемой задачи создания компактного маломассивного устройства с низким энергопотреблением, подходящего для установки на БПЛА, не вполне соответствуют из-за необходимости в случае их применения использования массивных радиаторов или активной системы охлаждения для отведения избыточного выделяемого тепла. InGaAs-фотодиоды указанными недостатками не обладают, к тому же такие фотодиоды необходимого для спектроскопии качества в разы дешевле благодаря развитию телекоммуникационных технологий.

Также стоит отметить, что существуют коммерчески доступные диодные лазеры с центральной длиной волны излучения 1651 нм с мощностью излучения до 100 мВт, тогда как для длины волны 3240 нм доступны мощности излучения на порядок меньше, но стоимость такого лазера будет значительно выше. К тому же в ближнем ИК-диапазоне зачастую не требуется проведения сложных юстировочных работ, поскольку фотонные элементы могут использоваться с волоконным вводом/выводом излучения, так как стандартные кварцевые оптические волокна применяются до длины волны ~2.3 мкм. Приведенные доводы подтолкнули к работе над апробацией методики и дальнейшей разработкой прибора в ближнем ИК-диапазоне.

Заявленный способ дистанционного измерения концентрации атмосферных газовых составляющих на примере метана возможен посредством непрерывного измерения глубины линии поглощения этого газа. Поскольку предложенная методика корректно работает для изолированной спектральной линии, в качестве рабочего диапазона был выбран мультиплет $R_4$ первого обертона полосы $2\nu_3$ на длине волны 1.651 мкм, приведенный на рисунке 3.7а. Как видно из рисунка, ширина на полувысоте линии поглощения $R_4$ при атмосферном давлении составляет $\Delta\nu \approx 0.15$ см$^{-1}$. Не имея перекрытия с линиями поглощения $H_2O$, эта линия может наблюдаться в условиях средней и высокой влажности.

Таким образом, для реализации предложенного способа оптимальным вариантом источника лазерного излучения является полупроводниковый диодный лазер с длиной волны излучения в области 1.65 мкм. Такой лазер может быстро перестраиваться по длине волны в окрестности пика выбранной линии поглощения путем модуляции тока инжекции, при этом температура кристалла может быть стабилизирована на уровне $10^{-3}$ К, что в выбранном диапазоне соответствует точности управления частотой излучения ~$10^{-3}$ см$^{-1}$.

## 3.3. Формирование сканирующего излучения в рамках методики модуляционной спектроскопии в сочетании с квадратурным приемом излучения

Для достижения более высокой чувствительности, чем при использовании классической абсорбционной спектроскопии, может быть использована техника модуляции. Она дает два основных преимущества: во-первых, измеряется разностный сигнал, который прямо пропорционален концентрации анализируемой газовой составляющей, и, во-вторых, измеряемый сигнал сдвигается в более высокочастотную область, тем самым обеспечивая большее отношение сигнал/шум и, следовательно, более высокую чувствительность.

На практике эта техника делится на два подхода. В случае спектроскопии модуляции длины волны излучения (Wavelength modulation spectroscopy, WMS) [233,234] частота модуляции $f$ значительно меньше ширины спектральной линии поглощения выбранного газа, и проводится анализ сигнала на частоте модуляции $f$ или следующих гармоник $2f$ и т.д. Обычно используются частоты модуляции от нескольких кГц до нескольких МГц. Этот метод применяется с начала 1970-х годов с использованием перестраиваемых диодных лазеров [234]. С другой стороны, частотно-модуляционная спектроскопия (Frequency Modulation Spectroscopy, FMS) применяется для частот модуляции, которые сравнимы или превышают ширину интересующей спектральной особенности [235-237]. В данной работе применялся метод модуляции длины волны излучения, поэтому в дальнейшем речь пойдет о нем.

Этот метод предполагает высокочастотную модуляцию тока инжекции лазера, что приводит к гармоническому изменению длины волны излучения лазера во времени. Такой подход позволяет устранить проблемы, связанные с базовой линией, поскольку принимаемый фотоприемником сигнал имеет горизонтальную базовую линию благодаря форме модуляции, слабо влияющую на конечный результат, к тому же, используя синхронный детектор в качестве приемного устройства, можно более чем на порядок увеличить отношение сигнала к шуму относительно результатов, полученных с помощью методики прямого детектирования [238].

Для корректной работы этой методики необходима точность динамической стабилизации центральной длины волны, относительно которой модулируется лазерное излучение, на уровне не хуже $10^{-3}$ см$^{-1}$. Такой уровень стабилизации достижим в случае использования в качестве реперной длины волны, соответствующей пику линии поглощения исследуемого газа. Однако в большинстве случаев при использовании модуляции длины волны излучения ограничиваются использованием регулирования температуры лазерного кристалла при помощи элемента Пельтье без обратной связи по длине волны генерируемого излучения относительно положения реперной линии поглощения, что может гарантированно обеспечить точность стабилизации лишь на уровне $10^{-2}$ см$^{-1}$ [239-242].

В качестве формы модуляции используется гармоническая функция. Принимаемый синхронным детектором сигнал на частоте модуляции $f$ пропорционален изменению мощности падающего лазерного излучения, которая зависит от длины волны. Демодулированный сигнал на частоте модуляции $f$ или, другими словами, первая гармоника детектируемого сигнала, полученная сканированием линии поглощения при помощи модулированного излучения, имеет форму, напоминающую производную контура линии поглощения. Первая гармоника сигнала имеет физический смысл общей интенсивности принимаемого излучения, также по сдвигу ее фазы можно определить расстояние до рассеивающей излучение лазера поверхности. Также вычисляются вторая и третья гармоники сигнала на удвоенной и утроенной соответственно частоте модуляции, наличие которых связано с нелинейностью, вызванной присутствием линии поглощения в анализируемом сигнале. В отсутствии же спектральных особенностей в сканируемой области 2*f*- и 3*f*-сигнал будут равны нулю. Вторая гармоника сигнала имеет физический смысл интенсивности выбранной спектральной линии поглощения. Третья гармоника дает возможность стабилизации лазерного излучения по спектральной линии поглощения, поскольку в отличие от первой гармоники не имеет подставки, а точка ее симметрии совпадает по частоте с центром линии поглощения. Иллюстрация поведения первых трех гармонических составляющих принимаемого сигнала приведена на рисунке 3.8.

Анализ составляющей сигнала на удвоенной частоте модуляции 2*f*, имеющей форму второй производной контура линии [161] позволяет уменьшить влияние изменения мощности лазерного излучения, соответствующего модулированному току инжекции. Такой подход позволяет снизить фликкер-шум 1/*f* лазерного излучения и чувствительность измерений к тепловым флуктуациям. Эти преимущества позволяют проводить измерения поглощения на уровне $10^{-6}$-$10^{-7}$ при высокочастотной модуляции длины волны в диапазоне 10 МГц [243]. Недостатком такой методики является необходимость калибровки по известным концентрациям измеряемого газа или при помощи поверенных газоаналитических приборов из-за сложности решения обратной задачи.

Рассмотрим подробнее принципы формирования сканирующего выбранную спектральную область излучения согласно методу абсорбционной диодно-лазерной спектроскопии с токовой гармонической модуляцией лазерного излучения при стабилизированной температуре лазерного кристалла и применение квадратурного детектирования принимаемого сигнала.

### 3.3.1. Формирование сканирующего сигнала в рамках методики модуляционной лазерной спектроскопии

Согласно закону Бугера-Ламберта-Бера [158] в случае однородной газовой среды при концентрации метана $N$ [молек/см³] и длине оптического пути $L$ [см] функция пропускания $T(v)$ монохроматического лазерного пучка, прошедшего сквозь слой этого газа, согласно формуле (2.5) может быть выражена как

$$T(\nu) = \frac{I(\nu)}{I_0(\nu)} = e^{-\sigma(\nu)NL} \tag{3.1}$$

где $\sigma(v)$ – сечение поглощения метана на частоте $v$ [см$^2$/молек], зависимость которого от частоты определяется контуром линии. В нижних слоях атмосферы доминирует столкновительное уширение спектральных линий. Рассмотрим описывающий этот механизм уширения профиль Лоренца. Согласно (2.38):

$$f_L = \frac{1}{\pi} \cdot \frac{\gamma_L}{\gamma_L^2 + [\nu - (\nu_0 + \delta_L)]^2}, \tag{3.2}$$

где $\gamma_L$ – лоренцева ширина спектральной линии на полувысоте [см$^{-1}$], $(\nu_0 + \delta_L)$ – смещенное положение центра линии, обусловленное влиянием давления, $\delta_L$ – параметр сдвига [см$^{-1}$], определенный в (2.39), $v$ – частота излучения [см$^{-1}$], $v_0$ – центральная частота линии [см$^{-1}$]. Тогда сечение поглощения согласно (2.6) можно представить в виде

$$\sigma(\nu) = \frac{1}{\gamma_L^2 + [\nu - (\nu_0 + \delta_L)]^2} \cdot \sigma_0, \tag{3.3}$$

где интегральное сечение $\sigma_0 = \frac{S \cdot \gamma_L}{\pi}$ пропорционально интенсивности линии $S$ [см$^{-1}$/молек/см$^{-2}$].

Для оптически тонкой среды $\sigma(v) \cdot N \cdot L \ll 1$ (при $L = 100$ м для естественного содержания метана в атмосферном воздухе $\sigma(v_0) \cdot N \cdot L \approx 10^{-2}$), и коэффициент пропускания можно выразить как

$$T(\nu) = 1 - \sigma(\nu) \cdot N \cdot L = 1 - \frac{1}{\gamma_L^2 + [\nu - (\nu_0 + \delta_L)]^2} \cdot \sigma_0 \cdot N \cdot L. \tag{3.4}$$

В качестве источника света используется полупроводниковый лазер, частота излучения которого $v$ зависит от температуры и тока накачки. В случае синусоидальной модуляции тока инжекции с частотой $\omega = 2\pi f$ при постоянной температуре, мгновенную частоту $v$ [см$^{-1}$] и интенсивность $I$ [Вт/см$^2$] лазерного излучения можно выразить следующими уравнениями

$$\nu(t) = \nu_0 + \Delta\nu cos(\omega t), \tag{3.5}$$

$$I(t) = I_0 + \Delta I cos(\omega t + \varphi), \tag{3.6}$$

где $\varphi$ – сдвиг между фазой частотной модуляции и фазой модуляции интенсивности. Амплитуду модуляции интенсивности излучения лазера будем считать пропорциональной амплитуде модуляции тока инжекции, что справедливо для большинства практических целей [244]. Коэффициент пропускания (3.4) может быть разложен в ряд Тейлора в точке $v_{laser}$, тогда его выражение до второго порядка $\Delta v$ можно представить в виде

$$T(\nu) = T_0 + T_1 \Delta\nu cos(\omega t) + T_2 \frac{\Delta\nu^2}{4} \left[1 + cos(2\omega t)\right], \tag{3.7}$$

где

$$T_0 = 1 - \frac{1}{\gamma_L^2 + [\nu_{laser} - (\nu_0 + \delta_L)]^2} \cdot \sigma_0 \cdot N \cdot L, \qquad (3.8)$$

$$T_1 = \frac{2\left[\nu_{laser} - (\nu_0 + \delta_L)\right]}{[\gamma_L^2 + [\nu_{laser} - (\nu_0 + \delta_L)]^2]^2} \cdot \sigma_0 \cdot N \cdot L, \qquad (3.9)$$

$$T_2 = \frac{2\left[\gamma_L^2 - 3\left[\nu_{laser} - (\nu_0 + \delta_L)\right]^2\right]}{[\gamma_L^2 + [\nu_{laser} - (\nu_0 + \delta_L)]^2]^3} \cdot \sigma_0 \cdot N \cdot L. \qquad (3.10)$$

При фазочувствительном синхронном детектирвоании излучения можно выделить сигналы, пропорциональные $T_1$ и $T_2$. Как видно из рисунка 3.8, при совпадении центральной частоты лазерного излучения $\nu_{laser}$ со смещенным положением центра линии поглощения метана ($\nu_0 + \delta_L$) значение члена разложения $T_0$, повторяющего форму коэффициента пропускания (рисунок 3.8a), становится минимальным. Член $T_1$, соответствующий сигналу на частоте модуляции $f$ (рисунок 3.8b), в этом случае становится равным уровню базовой линии, а абсолютное значение члена $T_2$, соответствующего сигналу на удвоенной частоте $2f$ (рисунок 3.8c), становится максимальным.

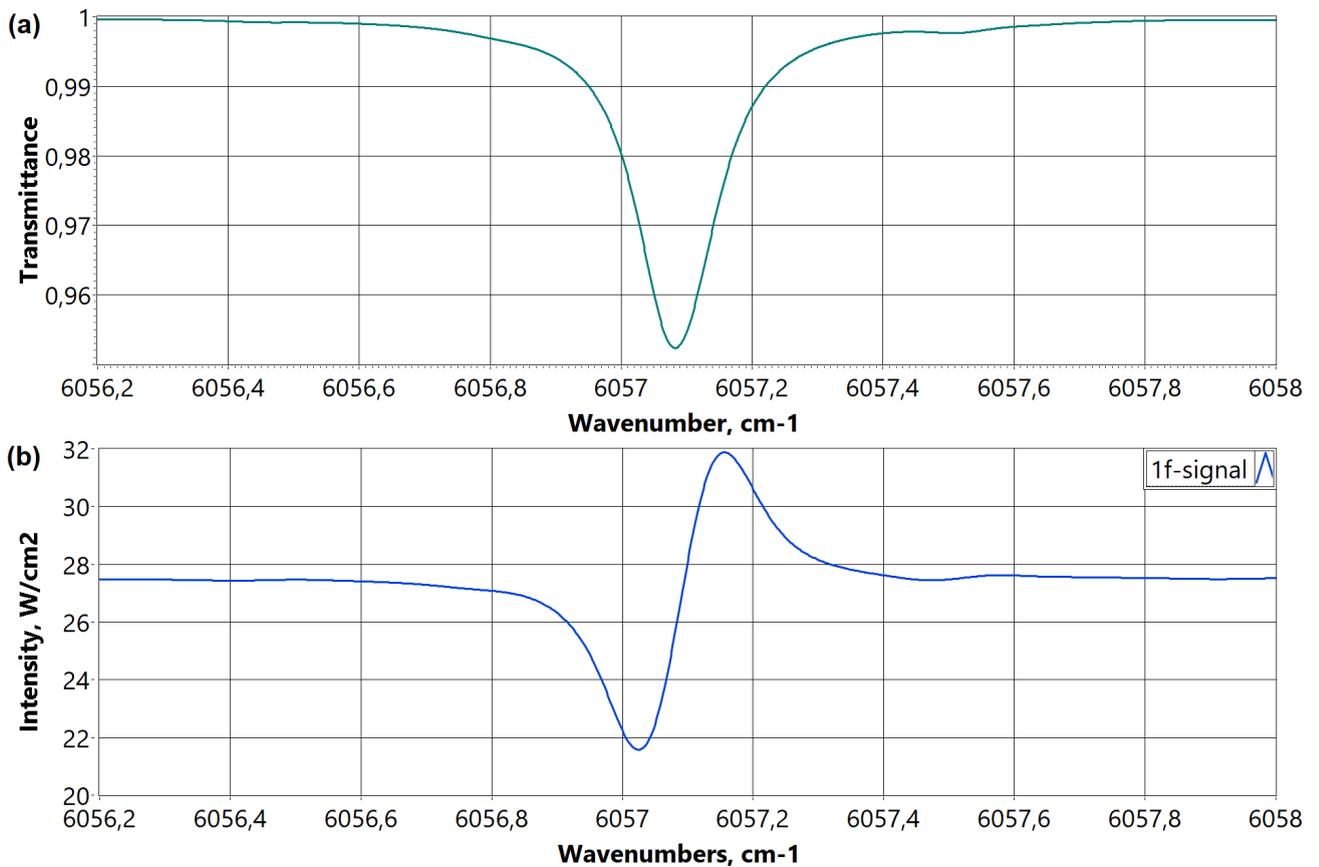

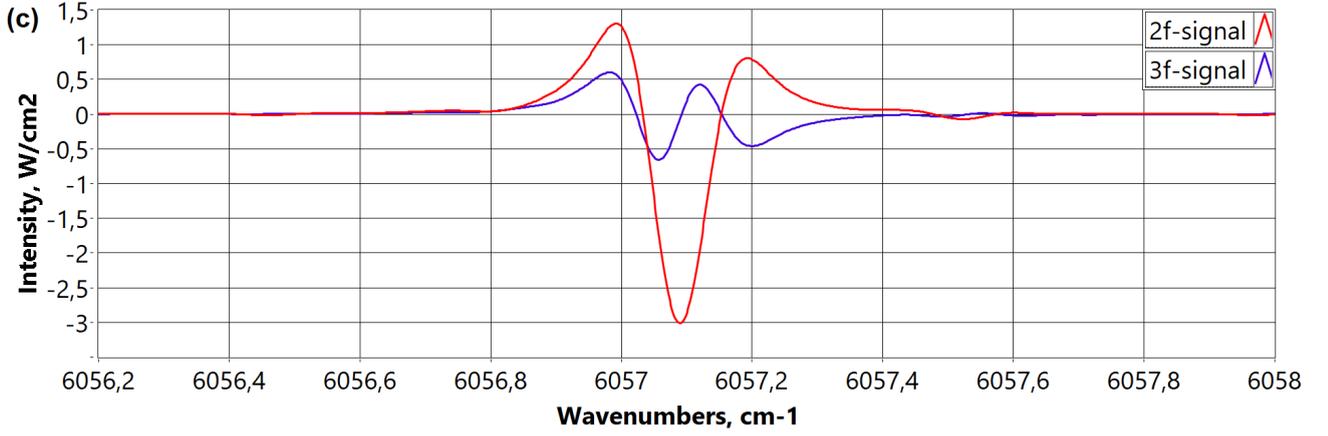

Рисунок 3.8 – **(a)** Функция пропускания 1000 ppm·m метана в области мультиплета R₄ Q-ветви полосы 2ν₃ (F₂), соответствующего длине волны 1.651 мкм, полученный по данным базы HITRAN [151,245]; **(b)** результат моделирования формы гармонической составляющей принимаемого сигнала прошедшего сквозь облако метана в атмосферном воздухе с содержанием 1000 ppm·m излучения на частоте модуляции *f*; **(c)** формы гармонических составляющих на удвоенной *2f* и утроенной *3f* частотах.

Если коэффициент потерь излучения на всей длине оптического пути от источника излучения до фотоприемника равен *K*, интенсивность излучения P [Вт/см²], попавшего на фотоприемник, пренебрегая возможными искажениями, связанными с нелинейными оптическими эффектами, можно выразить как

$$P(t) = K \cdot I(t) \cdot T(t) =$$

$$= K \left[ I_0 + \Delta I cos(\omega t + \varphi) \right] \cdot \left[ T_{02} + T_1 \Delta \nu cos(\omega t) + T_2 \frac{\Delta \nu^2}{4} cos(2\omega t) \right], \quad (3.11)$$

где $T_{02} = T_0 + T_2 \frac{\Delta \nu^2}{4}$. Выразим компоненту $P(\omega)$, меняющуюся с частотой $\omega$:

$$P(\omega = 2\pi f) = K \left[ T_{02} \Delta I cos(\omega t + \varphi) + T_1 I_0 \Delta \nu cos(\omega t) \right]. \quad (3.12)$$

Составляющие интенсивности сигнала *P(ω)* на частоте модуляции имеют разные фазы, как видно из (3.12). При совпадении центральной частоты лазерного излучения $v_{laser}$ с центром $(v_0 + \delta_L)$ линии $\sigma(v) = \frac{\sigma_0}{\gamma_L^2}$, тогда из условия $\sigma(v) \cdot N \cdot L \ll 1$ для (3.8) и (3.9) получим $T_{02} = 1$ и

$$P(\omega) = K \left[ \Delta I cos(\omega t + \varphi) \right]. \quad (3.13)$$

Тогда максимальный вклад интенсивности сигнала на частоте модуляции будет составлять

$$P(\omega)_{max} = K \Delta I. \quad (3.14)$$

В случае синхронного приема сигнала на удвоенной частоте *2ω* интенсивность сигнала можно записать как

$$P(2\omega = 4\pi f) = K\left[T_2 I_0 \frac{\Delta\nu^2}{4} cos(2\omega t)\right],$$ (3.15)

что при условии $\nu_{laser} = (\nu_0 + \delta_L)$ даст

$$P(2\omega) = K I_0 \frac{\Delta\nu^2}{2\gamma_L^4}\sigma_0 \cdot N \cdot L.$$ (3.16)

Тогда из уравнений (3.14) и (3.16) получим

$$\frac{P(2f)}{P(f)} = \frac{P(2\omega)}{P(\omega)} = \frac{\Delta\nu^2 I_0}{2\gamma_L^4 \Delta I}\sigma_0 \cdot N \cdot L,$$ (3.17)

таким образом, по отношению сигналов на частоте модуляции *f* и удвоенной частоте *2f* можно определить концентрацию детектируемого газа:

$$N = \frac{2\gamma_L^4 \Delta I}{\Delta\nu^2 I_0}\frac{1}{\sigma_0 L}\frac{P(2f)}{P(f)} = \frac{2\gamma_L^3 \Delta I}{\Delta\nu^2 I_0}\frac{\pi}{SL}\frac{P(2f)}{P(f)},$$ (3.18)

где интенсивность спектральной линии $S$ для линии $R_4$ равна $1.52\times10^{-21}$ см/мол. Более подробно принципы модуляционной абсорбционной спектроскопии уже рассматривались в соответствующих книгах и обзорах [6,161,215,246,247].

### 3.3.2. Квадратурное детектирование принимаемого сигнала

Для избежания влияния на уровень принимаемого сигнала флуктуаций фазы модуляции, что может приводить к ошибке определения концентрации на порядок величины и более, в качестве методики обработки анализируемого излучения был выбран принцип квадратурного детектирования [248-250].

Падающее на фотодиод излучение индуцирует фототок. Ток фотодиода аналитического канала, усиленный и прошедший частотный фильтр с полосой пропускания $f - 2f$, преобразуется в напряжение $V_{in}$, которое регистрируется АЦП:

$$V_{in}(t) = P(t) \cdot S_{PD} \cdot \eta \cdot R_{eq},$$ (3.19)

где $S_{PD}$ – площадь светочувствительного элемента фотодиода [см$^2$], $\eta$ – чувствительность фотодиода [А/Вт], $R_{eq}$ – эквивалентное сопротивление предусилителя [Ом].

Поскольку буфер памяти рассчитан на $N'$ периодов синусоиды, на первом шаге обработки происходит усреднение по $N'$ периодам:

$$V_{in}^{av} = \frac{\sum_{i=1}^{N'} V_{in}^i}{N'}, \tag{3.20}$$

где $V_{in}^i$ – сигнал в $i$-том периоде синусоиды. Для восстановления модулированного сигнала необходимо избавиться от базовой линии принимаемого сигнала:

$$V_{in}^f = V_{in}^{av} - \frac{\sum_{k=1}^{K'} V_k^{av}}{K'}, \tag{3.21}$$

где $V_k^{av}$ – $k$-ая точка массива значений принимаемого сигнала, усредненного по $N'$ периодам, $K'$ – число точек на период синусоиды. Процедура вычитания базовой линии, связанной с постоянной составляющей сигнала, также может быть реализована при помощи полосового частотного фильтра. Определить косинусную и синусную компоненты гармонической составляющей сигнала на частоте $nf$ при мгновенном значении $v$ частоты излучения лазера можно как

$$S_{nf}^{cos} = \sum_{k=1}^{K'} V_k^f cos(n\omega t_k), \tag{3.22}$$

$$S_{nf}^{sin} = \sum_{k=1}^{K'} V_k^f sin(n\omega t_k), \tag{3.23}$$

где $V_k^f$ – $k$-ая точка массива $V_{in}^f$. Тогда для абсолютного значения $S_{nf}$ гармонической составляющей сигнала на частоте $nf$ при мгновенном значении $v$ частоты излучения лазера справедливо:

$$S_{nf} = \sqrt{\left(S_{nf}^{cos}\right)^2 + \left(S_{nf}^{sin}\right)^2}. \tag{3.24}$$

Таким образом можно определить гармонические составляющие принимаемого сигнала на частоте $f$, $2f$ и $3f$. При этом в отсутствие спектральных особенностей в сканируемой спектральной области, а также в отсутствие нелинейных искажений, вносимых электроникой, интенсивность $2f$- и $3f$-сигнала будет равна нулю. Составляющая сигнала на частоте $f$ позволяет оценить общую интенсивность принимаемого излучения, $2f$-сигнал – интенсивность выбранной линии. Поэтому, как было показано ранее, по отношению составляющих сигнала на частоте $2f$ и $f$ можно определить концентрацию детектируемого газа при совпадении центральной частоты $v_{laser}$ излучения лазера и положения пика линии поглощения ($v_0 + \delta_L$).

Также по отношению синусной и косинусной компоненты сигнала на частоте модуляции $f$ можно определить изменение фазы $\Delta\varphi$ на длине оптического пути $L$, что в свою очередь позволяет определить дистанцию до рассеивающей излучение поверхности. Рассмотрим подробнее демодуляцию сигнала на частоте *f*. При совпадении центральной частоты лазерного излучения $v_{laser}$ с центром $(v_0 + \delta_L)$ линии или же в отсутствии спектральных особенностей в области сканирования можно представить величину $V_k^f$ как

$$V_k^f = V sin(\omega t_k + \Delta\varphi), \tag{3.25}$$

где *V* – амплитуда гармонической составляющей сигнала на частоте модуляции *f*. Тогда косинусную и синусную компоненты гармонической составляющей сигнала на частоте модуляции *f* можно представить в виде

$$S_f^{cos} = \sum_{k=1}^{K'} V_k^f cos(\omega t_k) = \sum_{k=1}^{K'} \frac{V}{2}[sin(\Delta\varphi) + sin(2\omega t_k + \Delta\varphi)] \sim sin(\Delta\varphi), \tag{3.26}$$

$$S_f^{sin} = \sum_{k=1}^{K'} V_k^f sin(\omega t_k) = \sum_{k=1}^{K'} \frac{V}{2}[cos(\Delta\varphi) - cos(2\omega t_k + \Delta\varphi)] \sim cos(\Delta\varphi). \tag{3.27}$$

Тогда изменение фазы $\Delta\varphi$ на длине оптического пути *L* можно выразить как

$$\Delta\varphi = arctan\left(\frac{S_f^{cos}}{S_f^{sin}}\right), \tag{3.28}$$

откуда можно найти дистанцию до рассеивающей излучение поверхности

$$H = \frac{L}{2} = \frac{1}{2}\frac{c}{f}\frac{\Delta\varphi}{2\pi} = \frac{c}{4\pi f}arctan\left(\frac{S_f^{cos}}{S_f^{sin}}\right), \tag{3.29}$$

где *c* – скорость света. Однако данная процедура на практике тоже может требовать калибровки, поскольку изменение фазы $\Delta\varphi$ будет складываться не только из задержки, связанной с распространением излучения, но и с влиянием на сигнал аналоговой электроники – предусилителя и полосового фильтра.

Первая гармоническая составляющая сигнала смещена относительно нуля, что вызвано остаточной амплитудной модуляцией лазерного излучения. Также с наличием амплитудной модуляции связана асимметрия *2f-* и *3f-*сигнала [219]. Это смещение отсутствует в гармонической составляющей на частоте *3f*, что делает ее более подходящей для реализации алгоритма температурной стабилизации лазерного диода. Такая стабилизация необходима в системах, работающих в широком динамическом диапазоне уровней возвращаемого сигнала [251].

В связи с этим аналогичной обработке подвергается сигнал реперного канала, благодаря чему по третьей гармонической составляющей непрерывно реализуется алгоритм стабилизации лазерного диода с точностью до $10^{-4}$ см$^{-1}$ по нулевому положению *3f*-сигнала, соответствующему пику спектральной линии поглощения.

## 3.4. Лабораторный макет газоанализатора для дистанционного мониторинга концентрации атмосферных газовых составляющих

Для отработки описанной выше методики в 2020 году в МФТИ был собран лабораторный макет прибора, принципиальная схема которого представлена на рисунке 3.9.

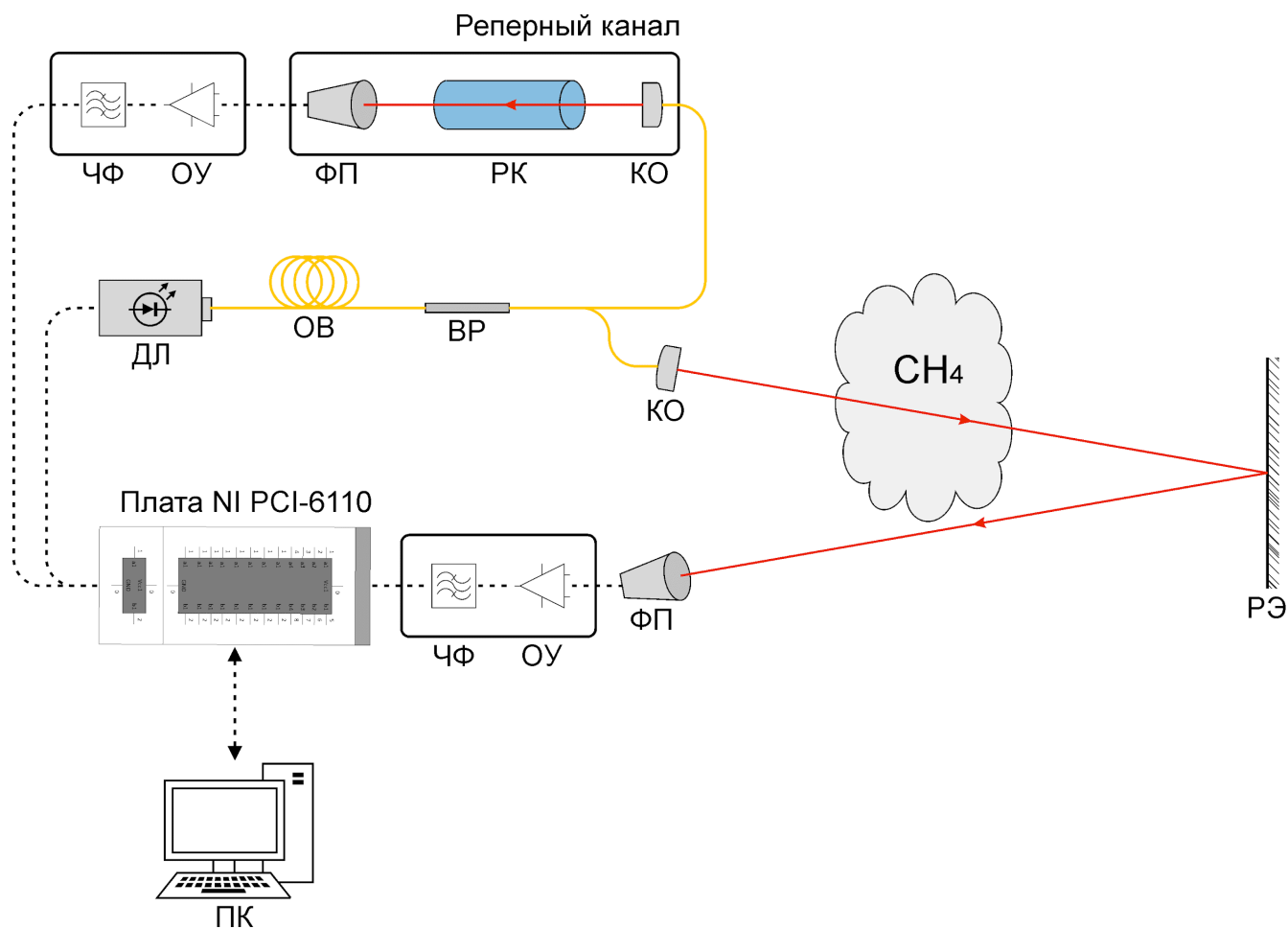

Рисунок 3.9 – Принципиальная схема экспериментальной установки, которая состоит из следующих основных элементов: ДЛ - диодный лазер, ОВ – оптическое волокно, ВР – волоконный разветвитель, КО – коллимирующая оптика, РЭ – рассеивающий экран, РК – реперная газовая кювета, ФП – фотоприемник, ОУ – операционный усилитель, ЧФ – П-образный частотный фильтр, ПК – персональный компьютер.

В качестве источника излучения использовался диодный РОС-лазер компании LasersCom (Беларусь) с центральной длиной волны излучения 1.65 мкм, шириной полосы генерируемого излучения 500 кГц и диапазоном температурной перестройки 50 см$^{-1}$. В корпус лазера установлен термоэлектрический элемент Пельтье, позволяющий прецизионно управлять температурой лазерного кристалла, термистор и мониторный фотодиод.

Ток накачки лазера был модулирован функцией синуса, при этом постоянная величина тока составляла 100 мА, амплитуда модуляции рассчитывалась по ширине на полувысоте используемой линии поглощения метана согласно индексу модуляции – отношению амплитуды модуляции к ширине линии газа, – который для максимизации полезного сигнала, восстанавливаемого из лоренцевского контура линии, выбирается равным 2.2 [252]. Частота модуляции $f$ составляла 50 кГц.

На волоконном выходе лазера был смонтирован оптический изолятор, в свою очередь присоединенный к волоконному разветвителю (ВР). Излучение разделялось в соотношении 1/9 и разводилось на два канала. Большая часть излучения попадала в аналитический канал, меньшая – в реперный. В аналитическом канале выходная коллимирующая оптика (КО) направляла лазерный пучок, который рассеивался от экрана (РЭ) и собирался входной оптикой блока фотоприемника (ФП), проходя при этом через область повышенной концентрации метана вблизи контролируемого источника, имитирующего утечку. Дистанция до рассеивающего экрана составляла $L = 6.3$ м.

Входная оптика представляла собой стандартную плосковыпуклую линзу производства ThorLabs диаметром $d = 51.2$ мм, просветленную в ближнем ИК-диапазоне. В фокусе линзы находился фотодиод Hamamatsu G8370-82 (Япония) с размером чувствительной площадки 2 мм, преобразовывавший падающее излучение в фототок. Фототок попадал в двухкаскадный предусилитель, состоявший из операционного усилителя (ОУ) и полосового частотного фильтра (ЧФ), с эквивалентным трансимпедансным коэффициентом передачи $R = 75$ МОм и полосой пропускания $B = 50$-150 кГц. Затем аналоговый сигнал считывался 12-битным АЦП платы National Instruments PCI-6110.

Реперный канал, необходимый для стабилизации длины волны лазерного излучения по пику выбранной линии поглощения метана, содержал коллиматор, направлявший лазерное излучение через кювету, наполненную метаном при давлении 10 мбар, на фотодиод Hamamatsu G8370-81 с размером чувствительной площадки 1 мм. Сигнал реперного канала также проходил через двухкаскадный предусилитель с полосой пропускания 50-150 кГц и считывался другим независимым 12-битным АЦП платы NI PCI-6110. Внешний вид оптической схемы смонтированного лабораторного стенда для отработки предложенной методики представлен на рисунке 3.10.

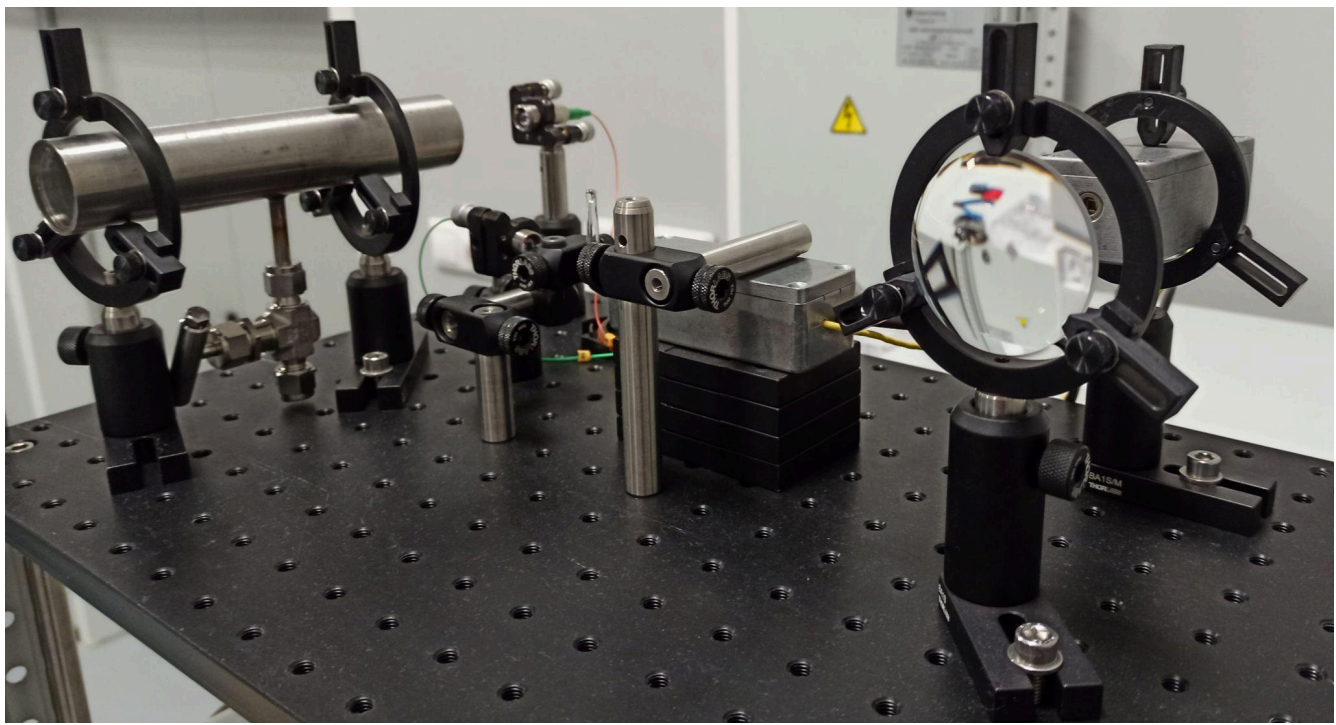

Рисунок 3.10 – Внешний вид смонтированного лабораторного стенда для отработки методики.

Управление лабораторным макетом лидара осуществлялось при помощи персонального компьютера (ПК) с программным обеспечением (ПО), разработанным в среде программирования LabVIEW. Указанное ПО позволяло управлять током накачки диодного лазера и термоэлектрическим элементом Пельтье при помощи 16-битного ЦАП платы NI PCI-6110, снимать показания с мониторного фотодиода и термистора лазера, фотоприемников аналитического и реперного каналов, а также проводить в режиме реального времени обработку получаемых данных.

## 3.5. Результаты лабораторной отработки предложенной методики измерения

Как было сказано, для отработки методики при помощи лабораторного макета был выбран мультиплет $R_4$ Q-ветви полосы $2\nu_3$ ($F_2$) метана на длине волны 1.651 мкм [151,245]. На рисунке 3.11 представлена функция пропускания этого газа, соответствующая содержанию метана в атмосферном воздухе на оптическом пути лазерного излучения на уровне 1000 ppm·м, рассчитанная на основе данных базы HITRAN.

Для выбранной спектральной линии было проведено моделирование прохождения лазерного излучения, генерируемого ДЛ с параметрами тока инжекции $100 \pm 9$ мА при температуре, стабилизированной по линии $R_4$, сквозь облако метана, последующего рассеяния излучения от экрана, находящегося на расстоянии $D = 6.3$ м от собирающей линзы с диаметром $d$ = 51.2 мм, и обработки принятого сигнала. Результатом обработки модельного сигнала являются рассчитанные гармонические составляющие сигнала на частоте модуляции $f$, удвоенной частоте $2f$ и утроенной частоте $3f$. Формы полученных $f$-, $2f$- и $3f$-составляющих сигнала представлены на рисунке 3.8b-c.

При помощи температурной перестройки длины волны лазерного излучения было осуществлено сканирование пропускания вблизи линии $R_4$ при тех же параметрах тока инжекции ДЛ, дистанции до рассеивающего экрана, апертуры приемного оптического тракта и содержания метана в атмосферном воздухе. Полученный в ходе такого сканирования сигнал был обработан согласно принципам, изложенным ранее.

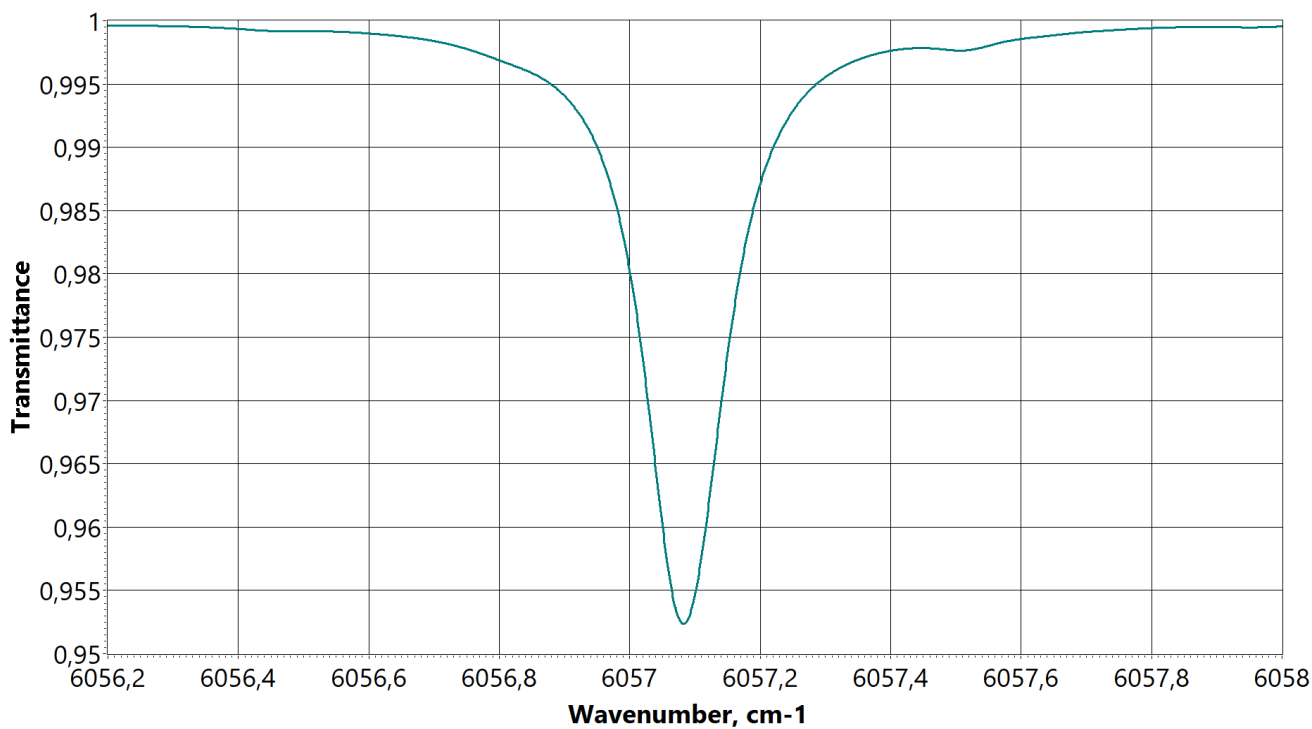

Рисунок 3.11 – Функция пропускания метана в атмосферном воздухе, соответствующая содержанию метана на уровне 1000 ppm·м.

Форма рассчитанного в результате проводимой в течение эксперимента обработки отношения гармонических составляющих сигнала на частоте модуляции $f$ и удвоенной частоте $2f$ представлена на рисунке 3.12. Также на рисунке 3.12 представлены результаты моделирования отношения сигналов $2f$ и $1f$ для содержания метана в атмосферном воздухе на уровне 924.8 ppm·м, 965.6 ppm·м и 1006,4 ppm·м для возможности визуальной интерпретации полученного экспериментально результата.

При моделировании не учитывался возможный вклад остаточной амплитудной модуляции сигнала, что привело к отклонению формы экспериментально полученного отношения $2f$-сигнала к $1f$-сигналу от результатов моделирования в области бóльших частот. Впрочем, как видно из рисунке 3.12, такой подход позволяет определить точность используемой модели, по шумовой дорожке экспериментально полученного сигнала, связанной с собственными шумами фотоприемника, цифровыми шумами платы NI PCI-6110 и шумами интенсивности ДЛ. Вклад характерного для лазерной спектроскопии фликкер-шума в данном случае отсутствует благодаря использованной методики и наличию полосового частотного фильтра в блоке приема и усиления сигнала.

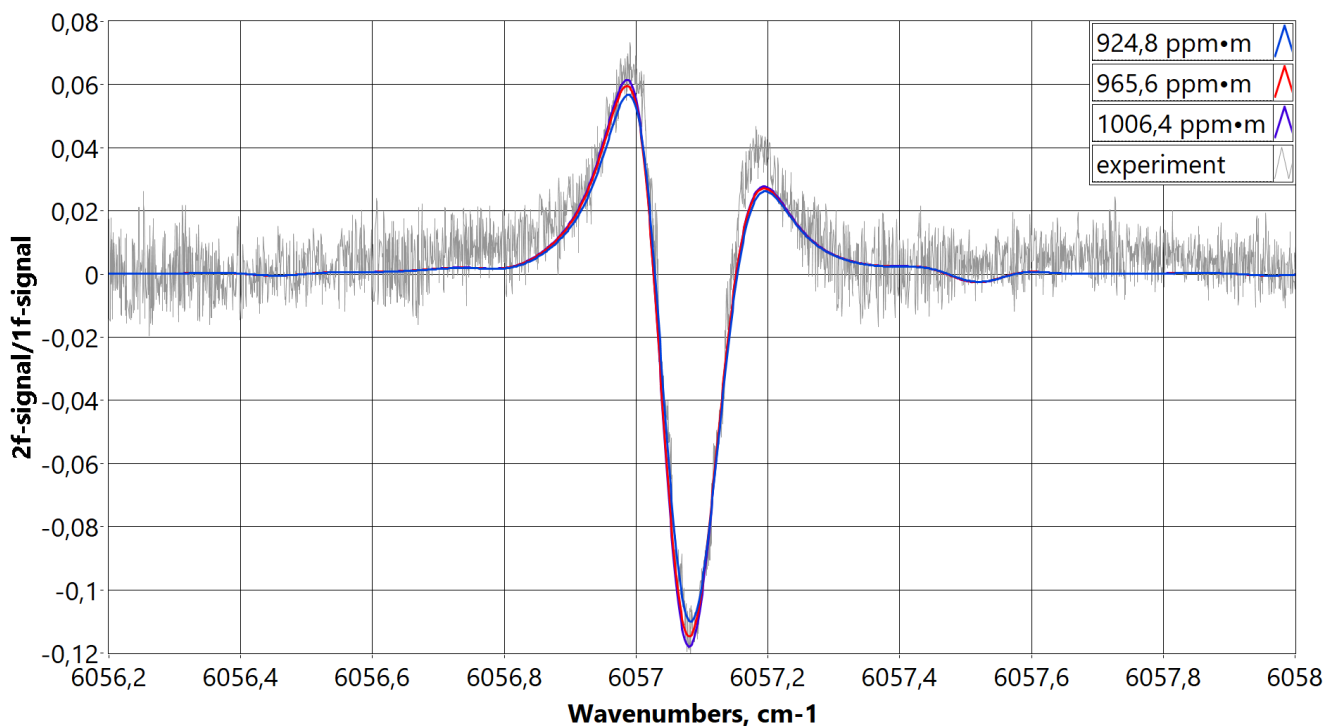

Рисунок 3.12 – Отношение гармонических составляющих сигнала на удвоенной частоте *2f* и частоте модуляции *f*. Экспериментальные данные и результаты моделирования для содержания метана на уровне 924.8 ppm·м, 965.6 ppm·м и 1006.4 ppm·м.

Решением обратной задачи является значение концентрации метана, на основе определенных *1f*- и *2f*-сигналов согласно уравнению (3.18). Тогда для экспериментально определенных гармонических составляющих, как видно из рисунка 3.12, можно получить величину концентрации детектируемого газа:

$$N_{exp} = (4.44 \pm 0.19) \times 10^{17} \frac{mol}{cm^3}, \qquad (3.30)$$

что при длине содержащего метан в атмосферном воздухе объема 5.44 см, температуре воздуха 25℃ и атмосферном давлении соответствует (965.6 ± 40.8) ppm·м. Результат определения содержания метана в измерительном тракте при помощи датчика давления Infcon DI200 составил $N = (4.6 \pm 0.1) \times 10^{17} \frac{\text{мол}}{\text{см}^3}$, что соответствует (1000.6 ± 23.2) ppm·м. Таким образом, отклонение среднего значения экспериментально определенной концентрации газа от содержания метана, измеренного при помощи датчика давления Infcon DI200, составила 3.5%.

Для определения чувствительности лабораторного макета прибора было выбрано содержание метана в атмосферном воздухе, также соответствующее 1000 ppm·м. При указанных параметрах установки и усреднении сигнала в течение 2 мс, то есть числе усреднений $N' = 100$, определялось стандартное отклонение и среднее значение сигнала по 50 последовательным значениям отношения *2f*- к *1f*-сигналу, что позволило определить чувствительность лабораторного макета прибора для общего времени работы 100 мс:

$$D_{LM} = N \cdot \frac{STD \left[\frac{S_{2f}}{S_{1f}}\right]_{N'}}{\frac{\sum_i^{N'} \frac{S_{2f}}{S_{1f}}}{N'}} = 42 ppm \cdot m, \tag{3.31}$$

что соответствует 21% от 200 ppm·м – фонового содержания метана ~2 ppm на длине оптического пути 100 м. Для сравнения в одном из серийно выпускаемых газоанализаторов [253], устанавливаемом на пилотируемую авиационную технику ввиду больших габаритов и массы, уровень чувствительности при усреднении сигнала за 500 мсек составляет 25 ppm·м.

## Выводы к главе 3

По результатам, изложенным в главе 3, можно сформулировать следующие основные выводы:

1. Метод абсорбционной диодно-лазерной спектроскопии с токовой гармонической модуляцией лазерного излучения при стабилизированной температуре лазерного кристалла в сочетании с применением квадратурного детектирования принимаемого сигнала позволяет решить задачу базовой линии с точностью, достаточной для дистанционного мониторинга выбранного газа в атмосферном воздухе;

2. Продемонстрированы результаты работы лабораторного макета прибора, основанного на применении методики модуляционной лазерной абсорбционной спектроскопии в комбинации с квадратурным приемом сигнала;

3. Оценочные масса и уровень энергопотребления прибора, основанного на предложенных принципах, соответствуют характеристикам полезной нагрузки существующих БПЛА, что значительно упростит и удешевит поиск утечек из магистральных газопроводов и мониторинг воздушной среды вблизи опасных производств, мусорных полигонов, в арктических регионах, районах с болотистой местностью и т.д;

4. Показано, что достижимый уровень чувствительности в десятки ppm·м при характерных дистанциях в десятки метров не уступает существующим лазерным газоанализаторам, применяемым для мониторинга метана в атмосферном воздухе;

5. Предложенная методика позволит осуществлять измерения интегральных значений концентрации выбранного газа с потенциально большей чувствительностью, чем у существующих лазерных спектрометров лидарного типа, за счет предложенного алгоритма высокоточной стабилизации длины волны генерируемого лазерным источником излучения по пику линии поглощения, детектируемому в реперном канале прибора;

6. Дополнительно данная методика позволяет независимо определять высоту полета, необходимую для корректного вычисления концентрации анализируемой газовой составляющей атмосферного воздуха в реальном времени или постфактум, что может быть полезно ввиду возможной несогласованности протоколов записи данных с бортом или же ошибки записи трека полета БПЛА по данным GPS/ГЛОНАСС;

7. В результате сборки, настройки и юстировки лабораторного макета газоанализатора метана был создан значительный задел для последующей разработки компактных полевых газоанализаторов лидарного типа.

# ГЛАВА 4. РАЗРАБОТКА ИНФРАКРАСНОГО ДИСТАНЦИОННОГО ГАЗОАНАЛИЗАТОРА ЛИДАРНОГО ТИПА ДЛЯ МОНИТОРИНГА ЭМИССИЙ МЕТАНА В ПОЛЕВЫХ УСЛОВИЯХ

С начала работ по адаптации предложенной методики модуляционной лазерной спектроскопии с квадратурным детектированием сигнала к задаче мониторинга метана в атмосферном воздухе, описанных в предыдущей главе, даже при наличии регулирования в сфере выбросов парниковых газов в развитых странах, программ по переходу на «зеленую» энергетику, контроля за промышленностью и иных мер признаваемые Программой ООН по окружающей среде (ЮНЕП) данные показывают, что содержание ПГ в атмосфере продолжает расти [254-258], что отражено на рисунке 4.1.

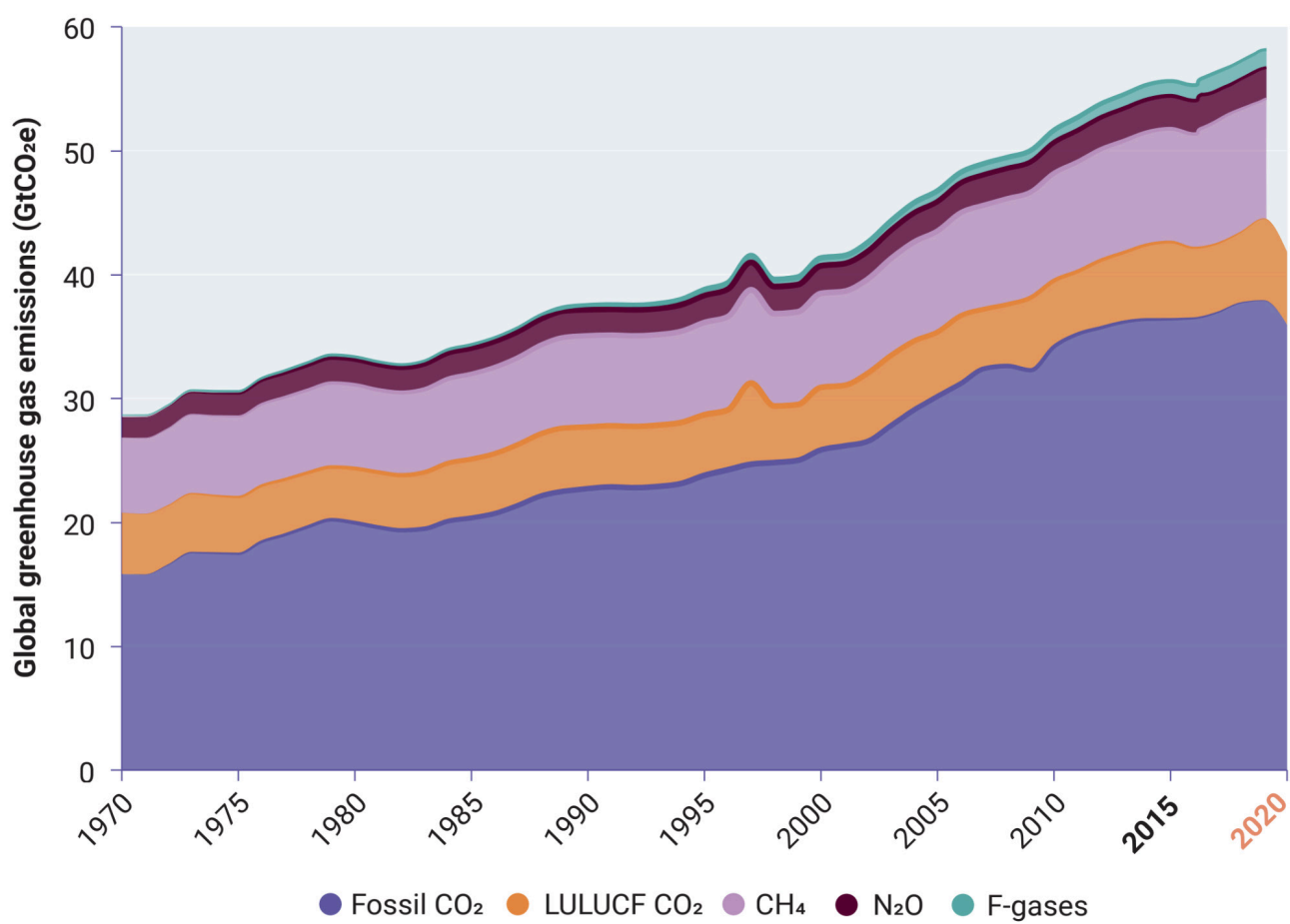

Рисунок 4.1 – Глобальные выбросы парниковых газов из всех источников в гигатоннах $CO_2$-эквивалента за 1970-2020 гг [254-258]. LULUCF – землепользование и лесное хозяйство.

В отчете ЮНЕП 2019 года говорится, что несмотря на десятилетие повышенного внимания политиков и общества к проблеме изменения климата и подписание Парижского соглашения, глобальные выбросы парниковых газов не удалось заметно сократить, а разрыв между оценочными значениями выбросов ПГ, соответствующих тренду на удержание повышения средней температуры в XXI веке в пределах 1.5-2℃, и оценками реальных выбросов велик как никогда [259].

Отчет ЮНЕП о разрыве в выбросах ПГ 2021 года показывает, что новые национальные обязательства в области климата в сочетании с другими мерами по смягчению последствий изменения климата ведут к тому, что к концу века глобальная температура повысится на 2.7℃ [257]. Это намного выше целей, намеченных Парижским соглашением по климату, и приведет к значительным изменениям климата Земли. Чтобы удержать повышение средней температуры ниже 1.5℃ в этом столетии, что является оптимальным значением согласно Парижскому соглашению, необходимо вдвое сократить ежегодные общемировые выбросы ПГ до 2030 года. На рисунке 4.2. представлены глобальные выбросы парниковых газов при различных сценариях и разрыв в выбросах в 2030 году между сценариями, предполагающими повышение средней температуры до конца века на 1.5-2℃, и вероятными сценариями при разной политике регулирования выбросов ПГ.

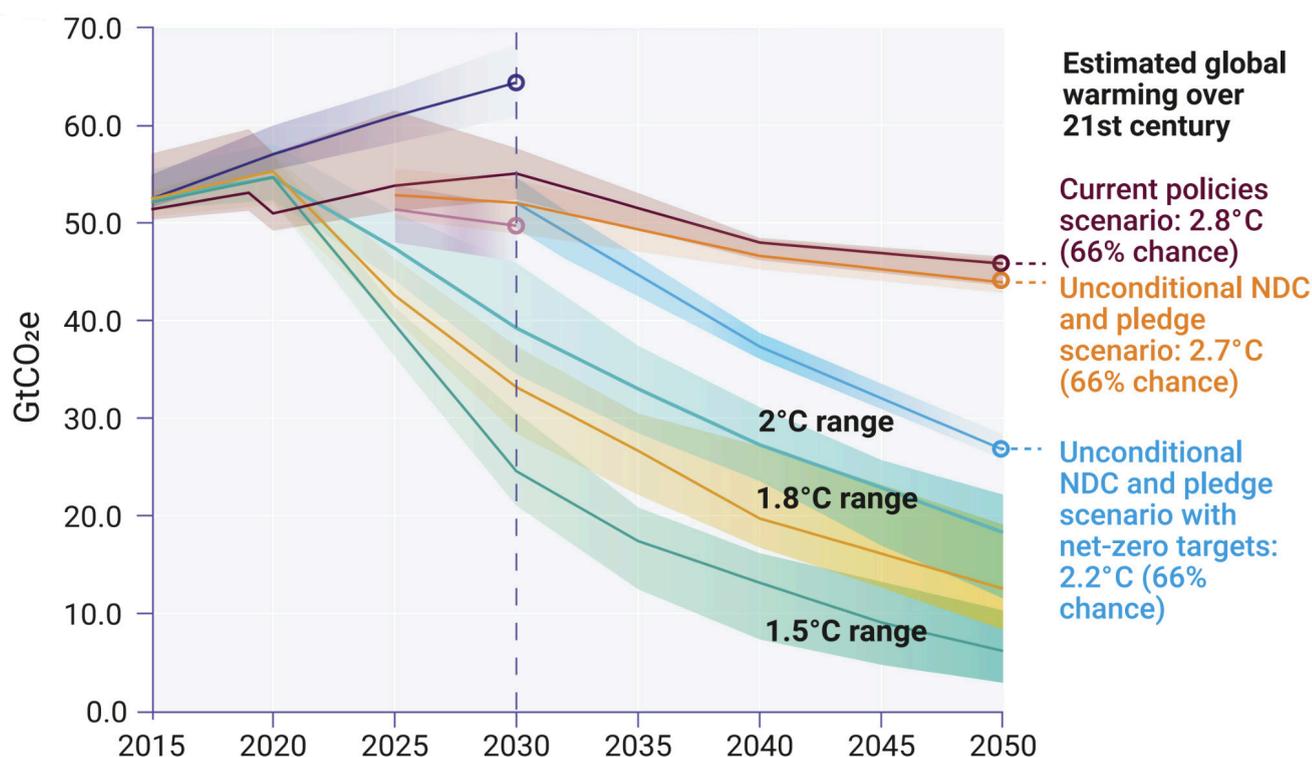

Рисунок 4.2 – Глобальные выбросы парниковых газов при различных сценариях и разрыв в выбросах в 2030 году между сценариями, предполагающими повышение средней температуры до конца века на 1.5-2℃, и вероятными сценариями при разной политике регулирования выбросов ПГ. По вертикали – эмиссия ПГ в гигатоннах $CO_2$-эквивалента, по горизонтали временной промежуток до 2050 года [257].

При эффективном выполнении обязательств по сокращению выбросов до нуля потепление может быть ограничено до 2.2℃, что ближе к цели Парижского соглашения, которая не превышает 2℃. Однако принятие многих национальных климатических планов откладывается на период после 2030 года. Сокращение выбросов метана из секторов ископаемого топлива, отходов и сельского хозяйства может помочь сократить разрыв в выбросах и снизить потепление в краткосрочной перспективе.

За последние два десятилетия основной причиной роста атмосферных выбросов метана является увеличение антропогенных выбросов, причем как от сельского хозяйства и отходов в Южной и Юго-Восточной Азии, Южной Америке и Африке, так и от ископаемого топлива в Китае, Российской Федерации и Соединенных Штатах Америки [260]. Локализация различных факторов эмиссии метана в атмосферу приведена на рисунке 4.3 [261].

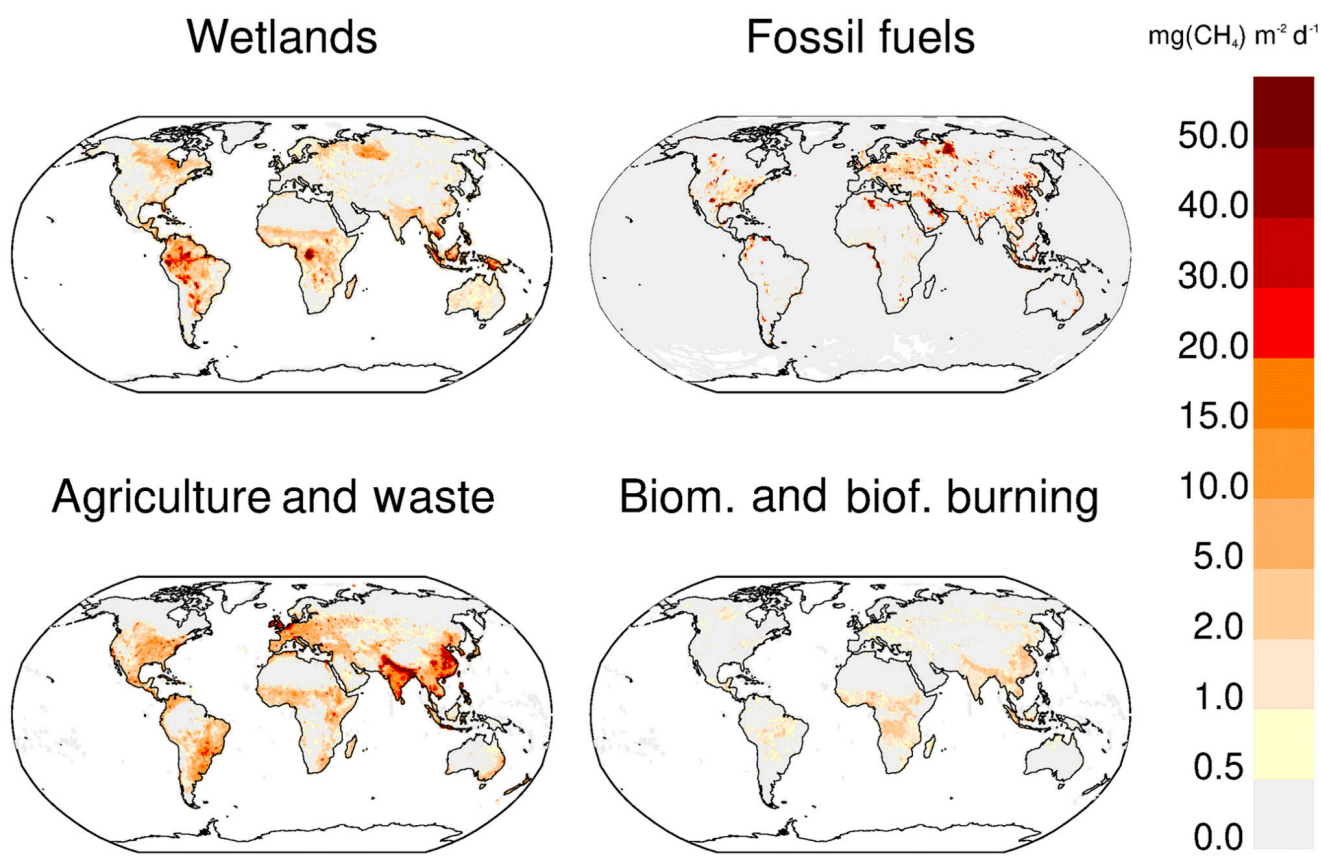

Рисунок 4.3 – Выбросы метана из четырех категорий источников (слева направо, сверху вниз): естественные водно-болотные угодья (за исключением озер, прудов и рек), ископаемое топливо, сельское хозяйство и отходы, сжигание биомассы и биотоплива за 2008-2017 гг [261].

Важную роль играют и естественные источники метана. Причем выбросы из природных источников могут увеличиваться по мере потепления водно-болотных угодий, увеличения количества тропических осадков и, как говорилось в предыдущей главе, значительно вырасти в связи с таянием многолетней мерзлоты. Карты ряда естественных источников метана, а также карта поглощения метана почвой представлены на рисунке 4.4 [261-264].

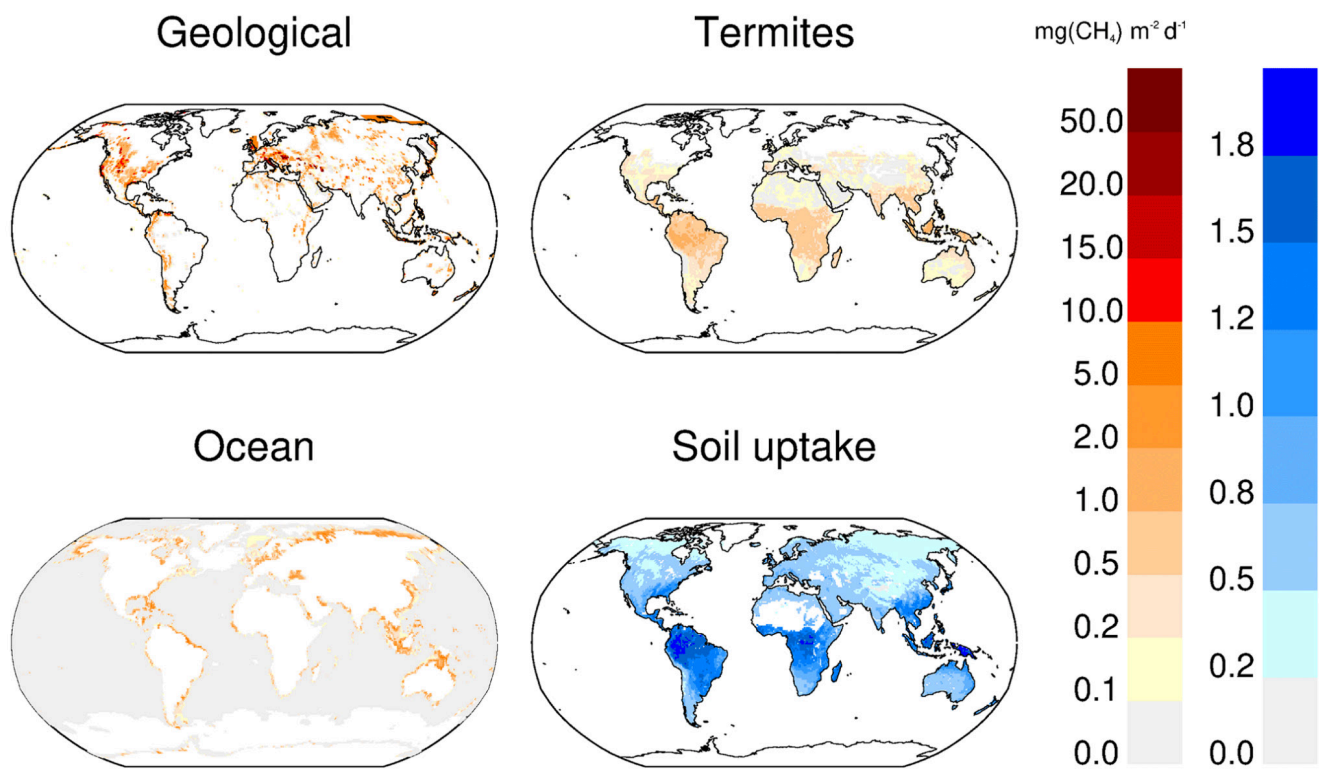

Рисунок 4.4 – Эмиссия метана из трех природных источников (левая цветовая шкала): геологических [262], термитов [261] и океанов [263]. Поглощение метана в почвах представлено в положительных единицах (правая цветовая шкала) [264].

Россия согласно своему национальному плану обязалась добиться к 2030 году снижения выбросов ПГ до 70% от уровня 1990 года. Однако планируемое уменьшение выбросов метана относительно 2015 года согласно национальному плану РФ к 2030 году должно составить 5%, тогда как для следования сценарию Парижского соглашения необходимо снижение на 35% [265].

Как уже упоминалось в предыдущей главе метан – это парниковый газ, кумулятивный потенциал глобального потепления которого превышает аналогичный показатель для $CO_2$ в 84 раза за период в 20 лет [221,225,226]. Снижение выбросов метана до уровней, предусмотренных Парижским соглашением, может помочь значительно замедлить динамику глобального потепления [266]. Но что мешает приблизиться к достижению плана Парижского соглашения помимо проблем регуляторного характера?

Важной проблемой является недостаточно высокое качество методов учета ПГ, попадающих в атмосферу. Следовательно необходимо улучшение методов измерения выбросов ПГ, поскольку известно, что применяемые на сегодняшний день методики оценок, например, в нефтегазовой отрасли могут недооценивать выбросы метана в атмосферу в 1.5-2 раза [267]. Обнаружение и корректная количественная оценка выбросов метана играют все более важную роль в снижении выбросов нефтегазовой промышленности. Так выбросы метана были измерены на 6650 объектах в шести основных нефтегазодобывающих регионах Канады для изучения региональных тенденций выбросов и составления инвентаризационной оценки для нефтегазового сектора Канады [268]. Ожидаемо, интенсивность выбросов при добыче природного газа сильно варьировалась: на старых месторождениях с низкой производительностью наблюдалась высокая интенсивность выбросов, а на новых месторождениях интенсивность выбросов была значительно ниже. В целом же, оказалось, что кадастр метана в нефтегазовой отрасли Канады недооценен в 1.5 раза.

Обнаружение крупных источников эмиссии метана, на долю которых приходится значительная часть выбросов нефтегазовых систем, реализуется при помощи разнообразных подходов. За последние несколько лет были разработаны передовые методы обнаружения утечек, например, основанные на оптической визуализации утечек, решения на базе пилотируемой авиации, а также мобильные анализаторы на основе спектроскопии затухания излучения в резонаторе (Cavity Ring Down Spectroscopy, CRDS). Были разработаны и более дешевые альтернативы на основе применения БПЛА, которые предлагают преимущества дистанционного зондирования без ограничений, связанных с необходимостью в наличии дорог, или пилотирования на малых высотах до 100 м, недоступных для пилотируемой авиации, что позволяет улучшить пространственное разрешение.

Методы обнаружения и количественного определения утечек, основанные на использовании малых беспилотных летательных аппаратов могут обеспечить точность, близкую к современным традиционным методам исследования, при одновременном увеличении расстояния и уменьшении времени отбора проб [269]. Преимущество использования беспилотных летательных аппаратов в некоторых случаях позволяет лучше локализовать источники выбросов и обеспечивает большую гибкость в работе.

Стоит отметить, что применение БПЛА в различных исследовательских проектах переживает стремительное развитие. Последние полтора десятилетия неуклонно растет число проектов, нацеленных на создание устройств, позволяющих проводить мониторинг состояния выбранной цели с борта БПЛА. В 2012 году анализатор атмосферных газов был интегрирован в БПЛА Национального управления по аэронавтике и исследованию космического пространства (NASA) Sensor Integrated Environmental Remote Research Aircraft (SIERRA) и использовался для высокоточных измерений метана, углекислого газа и водяного пара с частотой 1 Гц в Калифорнии и на Шпицбергене [270]. Был продемонстрировано устройство на основе VCSEL-лазера для измерения содержания ПГ в атмосфере, установленное на беспилотном вертолете [271]. Вулканические выбросы были зафиксированы над итальянским грязевым вулканом Ле Салинелле с помощью тепловизионной камеры, установленной на беспилотный гексакоптер [272]. Пространственно-временной анализ термокарстового озера проводился с помощью RGB-изображений, полученных с использованием самолета и БПЛА. Для изучения кипения метана рассматривались характеристики пузырьков на изображениях [273]. Определение местоположения и морфологии термитников с высокой пространственной детализацией в австралийском национальном парке Личфилд проводилось с помощью БПЛА [274]. Ограничившись перечисленными примерами, отметим обзорную работу по дистанционному зондированию с использованием БПЛА [275], описывающую многие приложения, включая экологический мониторинг извержений вулканов, эрозии почвы и т.д.

Активно применяются БПЛА и для изучения эмиссий ПГ из различных источников. Так, например, привычная методология обследования мусорных полигонов на предмет эмиссии свалочного газа основана на ручных газовых детекторах. Во время обхода отбираются пробы окружающего воздуха на высоте 50 мм от поверхности почвы с целью определения концентрации выбросов метана. Однако этот подход имеет ограничения, связанные с техническими характеристиками коммерчески доступных детекторов метана и особенностями самой среды полигона. Компания Terra Sana Consultants Pty Ltd в 2019 году разработала БПЛА, оснащенный лазерным датчиком, способным обнаруживать поверхностные выбросы свалочного газа с привязкой к геолокации [276]. В 2020 году камеры видимого, ближнего ИК- и теплового ИК-диапазонов использовались для картографирования рельефа местности и создания цифровых карт высот мусорного полигона для выявления проблемных зон, где присутствуют локализованные выбросы метана, с помощью статического прототипа полупроводникового датчика [277].

В 2017 году был предложен сенсор метана открытого типа для обнаружения утечек в газовой инфраструктуре, установленный на БПЛА с размахом крыла 2.3 м, работающий от аккумулятора и рассчитанный на 90 минут полета, представляющий собой диодно-лазерный абсорбционный спектрометр, установленный с приемопередатчиком на одном крыле и светоотражателем на другом крыле (длина пути = 4.6 м), и плат управления измерениями в основном фюзеляже [278]. В 2019 году был предложен прибор, основанный на перестраиваемой диодно-лазерной абсорбционной спектроскопии, установленный на БПЛА с лазерным дальномером, для реконструкции двумерных шлейфов в реальных условиях [279].

В 2020 году набор электрохимических датчиков AlphaSense был использован на беспилотном летательном аппарате DJI серии 100, который выполнял полеты по зигзагообразной и спиральной траектории вблизи стационарного источника $NO_2$ на высотах до 9 м [280]. Для обнаружения и количественной оценки утечек из трубопроводов США использовался БПЛА, оснащенный спектрометром, принимающим обратно рассеянное излучение, с перестраиваемым диодным лазером. Во время исследований БПЛА находился на расстоянии 4 м от трубопровода, а минимальный предел обнаружения истечения составлял 0.06 г/с [281].

Поскольку применение БПЛА в целях изучения эмиссий ПГ стало в последние годы активно применяться, перечислить все достижения в этой сфере в рамках данного текста не представляется возможным. С этой целью можно обратиться к актуальному на момент написания обзору малых БПЛА для химического зондирования [282]. Однако можно заметить, что в большинстве своем устройства для мониторинга выбросов метана, устанавливаемые на малые БПЛА, не рассчитаны на работу на высотах в десятки-сотни метров. При этом безопасной высотой полета БПЛА в слабо урбанизированной местности стоит считать не менее 30-50 м, соблюдение которой необходимо для избежания столкновения борта с деревьями, мачтами ЛЭП и другими высокими объектами, а также для возможности совершения маневра в нештатной ситуации для сохранения БПЛА и полезной нагрузки. Таким образом, число возможных аналогов разработанного в МФТИ Газоанализатора для Измерения Метана Лидарного Инфракрасного (ГИМЛИ) значительно сужается.

Данная глава посвящена подробному рассмотрению разработки прототипа ГИМЛИ для полевых измерений, описанию его структуры, полученных результатов в разных погодных условиях в различных регионах при нескольких форматах проведения экспериментов. Также дано описание разработки следующей версии полевого прибора.

Личный вклад автора заключается в проектировании, разработке рабочей конструкторской документации, сборке и юстировке двух версий прибора, разработке унифицированного управляющего ПО прибора, проведении серии экспериментов и анализе режимов работы прибора, проведении анализа спектроскопических данных при помощи разработанного автором комплекса ПО, получении результатов по оценке точности определения концентрации метана и калибровке прибора, подготовке научных публикаций по результатам работы и заявки на получение патента на изобретение.

На разработанный прибор был получен патент РФ на изобретение RU2736178C1 «Способ и устройство для автономного дистанционного определения концентрации атмосферных газовых составляющих». В настоящем разделе используются материалы публикаций [283,284], подготовленных автором.

## 4.1. Существующие аналоги разработанного прибора

Говоря об устанавливаемых на БПЛА устройствах для анализа содержания выбранного газа в атмосферном воздухе, нужно сразу определиться с задачей, которую эти устройства должны решать. Это нужно, поскольку такие устройства можно разделить на две большие группы: устройства для *in situ* анализа и устройства для дистанционного зондирования. К первой группе можно отнести контактные химические датчики, мультисенсорные устройства для обнаружения нескольких целевых газов, собранные из таких датчиков, и системы электронного носа – все они не способны работать в дистанционном режиме, имеют достаточно медленный отклик и не очень высокую чувствительность, к тому же страдают от кросс-чувствительности к нецелевым газам. Хотя при этом они очень дешевы, их энергопотребление не превышает нескольких мВт, а масса десятков грамм [285].

К этому же классу можно отнести устройства, анализирующих газовую смесь, закачанную при помощи насоса в аналитическую ячейку. Два основных подхода, реализуемых в лазерных спектрометрах такого типа – спектроскопия затухания излучения в резонаторе (CRDS) и спектроскопия внеосевого полного внутрирезонаторного выхода (Off-Axis Integrated Cavity Output Spectroscopy, OA-ICOS). Оба метода чрезвычайно чувствительны, но до сих пор приборы, разработанные на их основе, были слишком тяжелыми и энергоемкими для применения на БПЛА. Однако недавно была разработана легкая (< 4 кг) версия устройства с открытой CRDS-ячейкой с невысоким энергопотреблением (< 30 Вт), которая подходит для установки на БПЛА [286]. Этот прибор обладает временным откликом 1 с и чувствительностью 30 ppb, позволяя обнаруживать шлейф метана на расстоянии более 60 м по ветру от источника выбросов.

Такие газоанализаторы оказываются эффективны при построении вертикальных профилей распределения концентрации различных газовых составляющих [287]. Но в случае применения подобных устройств для детектирования утечек природного газа из трубопроводов их принципиальным недостатком являются низкая точность локализации утечки и высокая вероятность ложного детектирования, обусловленные тем, что небольшие превышения естественной концентрации метана в воздухе могут не быть связанными с утечкой из газопроводов. К тому же применение такого рода устройств для поиска утечек является зачастую довольно затруднительным из-за необходимости учитывать направление и скорость ветра для возможности движения БПЛА только с подветренной стороны от места истечения. Дополнительно работа с такими приборами требует крайне высокой квалификации оператора

БПЛА, поскольку полеты будут предполагать работу на небольших высотах либо же только на открытой местности, что резко сужает возможную географию их применения.

Решением этих проблем представляется применение сенсоров второй группы – способных к дистанционному мониторингу на больших дистанциях. Примером таких устройств являются камеры оптической визуализации газа, позволяющих получать видеопотоки утечек газов на основе теплового контраста между фоном и истекающим газом [288]. Это инфракрасные или тепловизионные камеры с оптическим фильтром, настроенным на полосу поглощения целевого газа. При благоприятных атмосферных условиях и расстоянии съемки менее 10 м такая камера может обнаружить ~80 % всех утечек на газодобывающем объекте [289]. Основными недостатками этих камер являются их высокая стоимость [290], сложность количественного определения объема утечки [291] и высокий предел обнаружения – 10000 ppm для $CH_4$ [289].

Но есть класс приборов, способных регистрировать утечки метана на заметно больших расстояниях. Оптические газоанализаторы используют характерные спектры поглощения газов при их возбуждении ИК- или УФ-излучением. Они могут селективно обнаруживать на уровне ppb молекулярные газы с непересекающимися спектральными областями, например $CO_2$ и $CH_4$ в среднем ИК-диапазоне [292]. Поскольку оптические анализаторы измеряют физические свойства газа, эти измерения могут быть более быстрыми и надежными, чем те, которые основаны на активном чувствительном слое, включающем физико-химические реакции. Как было отмечено, лазерные спектрометры со встроенной многопроходной аналитической ячейкой обеспечивают более высокое качество измерений, но необходимое для заполнения газовой ячейки время приводит к понижению частоты выдачи данных. Приборы с открытым контуром обычно менее точны, но легче и быстрее.

На этом принципе построены устройства, принимающие излучение, рассеянное от поверхности – излучатель и детектор располагаются на одном конце оптического пути. Лазерный пучок излучается в сторону удаленной поверхности, а рассеянный свет собирается фотодиодом. В отличие от описанных выше приборов, такие устройства измеряют интегральную концентрацию газа в единицах ppm·м вдоль всего лазерного луча, поэтому за одно измерение можно собрать информацию по большой площади, что в ином случае потребовало бы множества точечных измерений.

Иллюстрация физического смысла единицы ppm·м в описываемом приложении приведена на рисунке 4.5 – пусть БПЛА находится на высоте 10 м над поверхностью земли, тогда интегральное содержание выбранного газа, измеряемого установленным на БПЛА газоанализатором, составит 50 ppm·м и в случае зондирования облака с концентрацией выбранного газа 50 ppm размером вдоль зондирующего луча 1 м и в случае облака с концентрацией 5 ppm и размером 10 м.

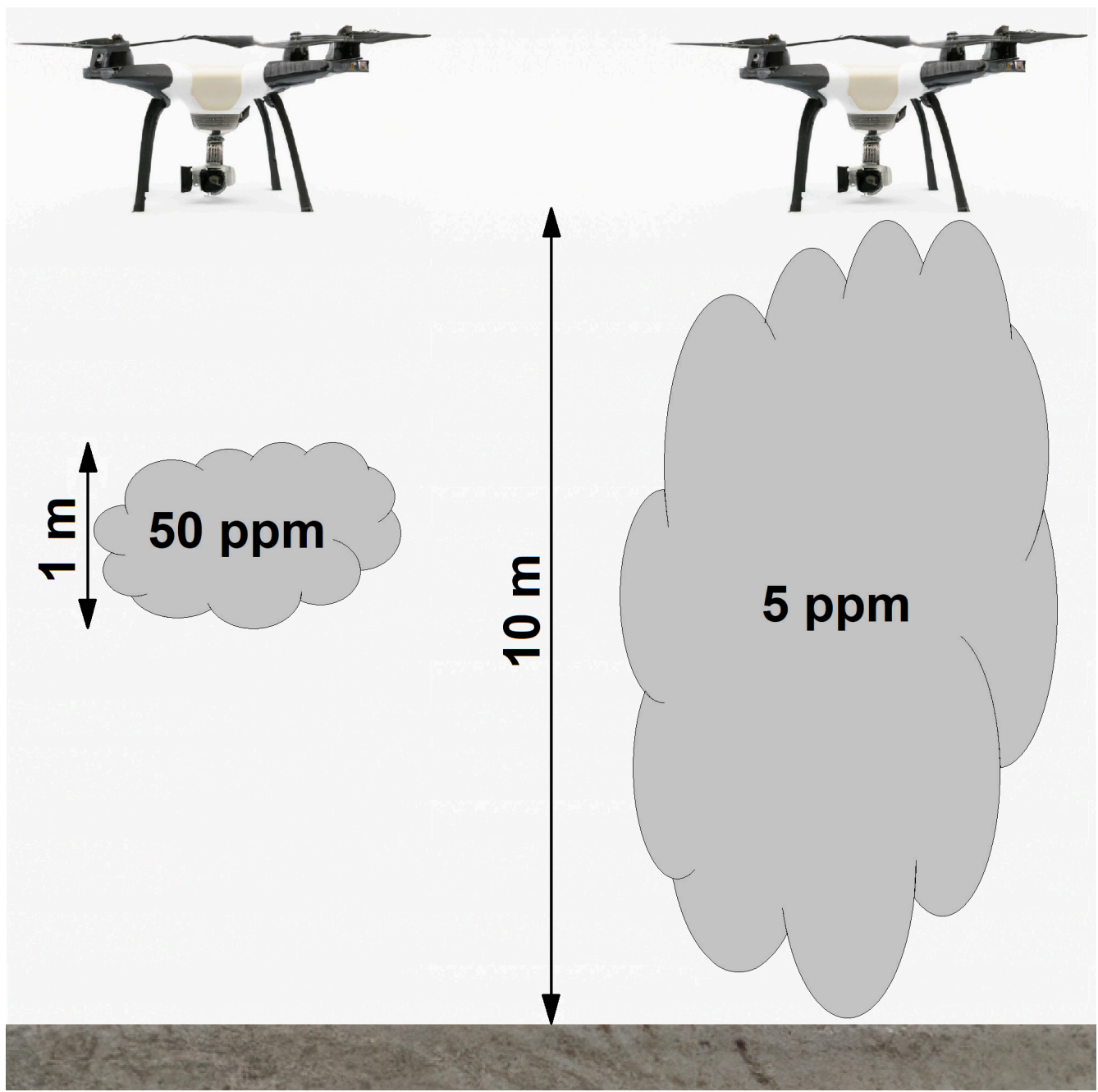

Рисунок 4.5 – Иллюстрация физического смысла единицы ppm · м.

Существует ряд патентов на устройства такого рода, основанных на применении разных спектроскопических методик. Одним из лидеров на рынке таких устройств является компания АО "Пергам-Инжиниринг" (Россия), являющаяся дистрибьютором широкого спектра устройств для поиска утечек метана и разработчиком ряда подобных устройств. Однако научных публикаций, посвященных разработкам такого оборудования, не слишком много. Одна из них посвящена прародителю линейки газоанализаторов, устанавливаемых на БПЛА, – детектору утечек метана ДЛС-Пергам ALMA G2 [253]. Данный прибор основан на применении принципа кросс-корреляционной спектроскопии. Реализация такого подхода позволяет добиться высокой чувствительности относительно выбранного газа на уровне 100 ppm·м для расстояния от устройства до рассеивающей генерируемое лазерное излучение подстилающей поверхности 100 м. Однако указанная реализация требует установки устройства на пилотируемые летательные аппараты из-за больших габаритов – 600×600×250 мм, массы 22 кг, высокого энергопотребления на уровне 100 Вт, а также необходимости присутствия на борту летательного аппарата оператора, контролирующего работу устройства при помощи мобильного компьютера, используемого для обработки данных. Учитывая значительную протяженность трасс магистральных газопроводов и высокую стоимость полетов пилотируемой авиационной техники, использование таких устройств является экономически нецелесообразным.

Что касается более современных разработок – газоанализаторов, устанавливаемых на БПЛА, среди наиболее часто используемых устройств – приборы производства Gastar Co. Ltd (Япония), особенно детектор Laser Methane mini (LMm), который был разработан в 1992 году компанией Tokyo Gas Engineering Solutions (Япония) как портативный переносной детектор для наземных инспекций [293]. В 2013 году он стал коммерчески доступным [294] и затем был интегрирован в различные системы дистанционного зондирования с борта БПЛА, включая Pergam Laser Methane Copter (LMC) компании АО "Пергам-Инжиниринг".

В последние годы компания Tokyo Gas Engineering Solutions разработала детектор Laser Falcon, предназначенный специально для БПЛА [295]. Можно найти работы, описывающие результаты их применения для задач, связанных с измерением интегрального содержания метана вблизи газовой инфраструктуры или над мусорными полигонами. Pergam Laser Falcon или Pergam Laser Methane Copter (LMC) – легкие приборы массой до 1 кг, обеспечивающие относительно высокую частоту сэмплирования ⩾2 Гц и точность в 100-1000 ppm·м, но требуют малых рабочих высот до 30 м и невысоких горизонтальных скоростей движения БПЛА для стабильной работы [296,297].

К классу легких устройств до 1 кг стоит отнести и выпущенный в 2019 году компанией AILF Instrument Co., Ltd. (Китай) в партнерстве с SZ DJI Technology Co., Ltd (Китай) детектор метана U10. Этот прибор применялся, например, для вертикального мониторинга интегральных концентраций $CH_4$ в закрытых и открытых объектах станций очистки сточных вод [298].

Более массивные устройства, требующие либо БПЛА с жестким крылом, как Pergam Laser Monitoring Fixed Wing (LMF), либо БПЛА с грузоподъемностью свыше 10 кг имеют более ограниченный спрос из-за большей стоимости и сложности эксплуатации, несмотря на лучшие показатели чувствительности к метану.

Основные характеристики коммерчески доступных на мировом рынке на момент написания газоанализаторов метана, подходящих для установки на борт БПЛА, приведены в таблице 4.1.

Таблица 4.1 – Основные характеристики коммерчески доступных на мировом рынке газоанализаторов метана, подходящих для установки на борт БПЛА.

| Название | Макс. высота | Масса | Частота сэмплирования | Ошибка | Динамический диапазон | Стоимость |
|---|---|---|---|---|---|---|
| NextTech MS-04 Methane Sensor | 30 м | | 10 Гц | ±10% | 0-5000 ppm·м | ~35 000 $ |
| DJI U10 Laser Methane Leakage Detector | 100 м | 0.53 кг | 40 Гц | ±5% | 0-50000 ppm·м | ~70 000 $ |
| Tokyo Gas Engineering Solutions Laser Methane Mini | 30 м | 1 кг | 2 Гц | ±10% | 0-50000 ppm·м | ~10 000 $ |
| Tokyo Gas Engineering Solutions Laser Falcon | 100 м | 0.3 кг | 2 Гц | ±10% | 0-50000 ppm·м | ~25 000 $ |
| Pergam LMF | 100 м | 1.7 кг | 10 Гц | | 0-25000 ppm·м | |
| Pergam ALMA G4 mini | 150 м | 12 кг | 25 Гц | ±1% | 0-20000 ppm·м | |

Зависимость чувствительности от дистанции до рассеивающей излучение поверхности коммерчески доступных на мировом рынке на момент написания газоанализаторов метана, подходящих для установки на борт БПЛА с указанием этих значений в спецификации производителя приведена на рисунке 4.6.

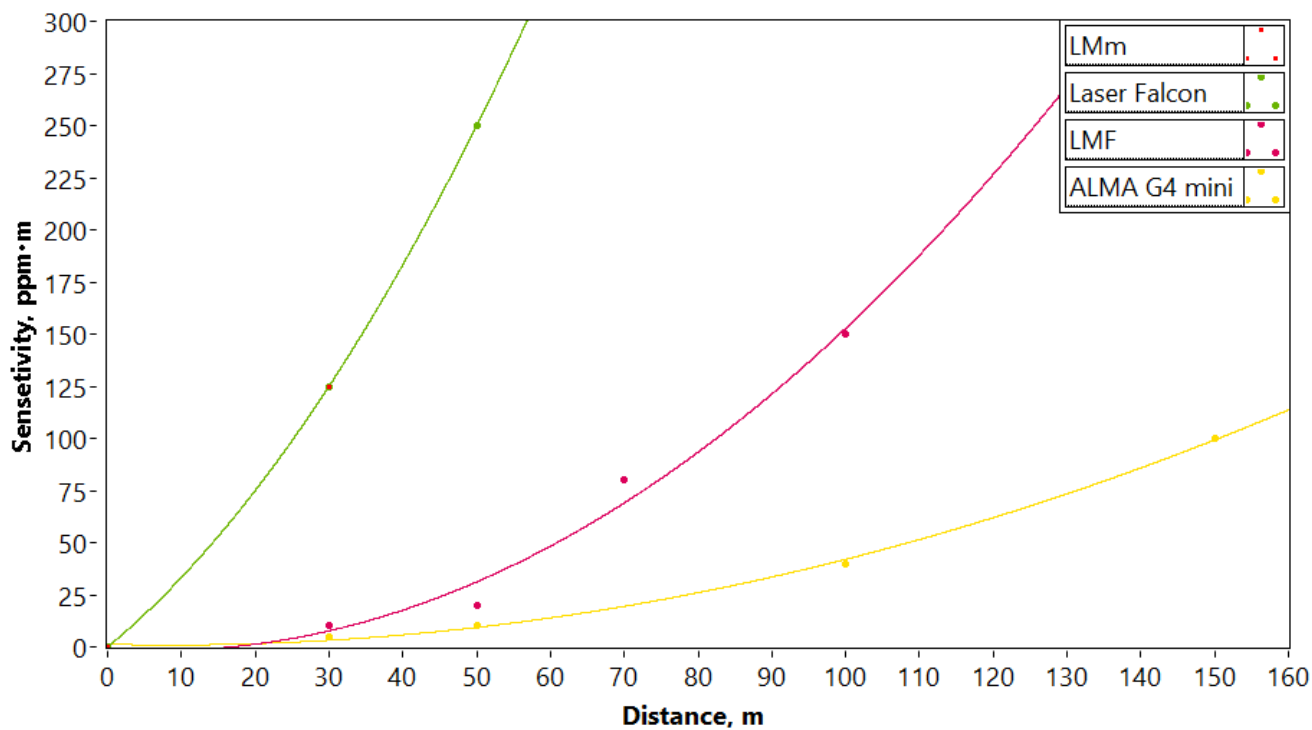

Рисунок 4.6 – Зависимость чувствительности от высоты полета БПЛА коммерчески доступных на мировом рынке газоанализаторов метана, подходящих для установки на борт БПЛА.

## 4.2. Разработка прототипа полевого прибора

После апробации предложенной методики следующим этапом работы стало создание полевого прибора, удовлетворяющего требованиям к установке на БПЛА, подходящего для работы более часа с массой полезной нагрузки 4-5 кг. Его разработка началась в МФТИ в 2021 году. После подбора фотоники, подготовки управляющей электроники прибора и его корпусирования начались полевые испытания, нацеленные на выявление ошибок в применяемых алгоритмах и возможных непредвиденных проблем с эксплуатацией устройства в реальных условиях. Ввиду многочисленных трудностей как в отладке прибора, так и в проведении полевых испытаний на борту БПЛА процесс доводки прототипа прибора занял около трех лет. По окончании отладки была проведена калибровка разработанного устройства по данным мобильного газоанализатора LI-COR LI-7810. Внешний вид разработанного прототипа полевого газоанализатора представлен на рисунке 4.7.

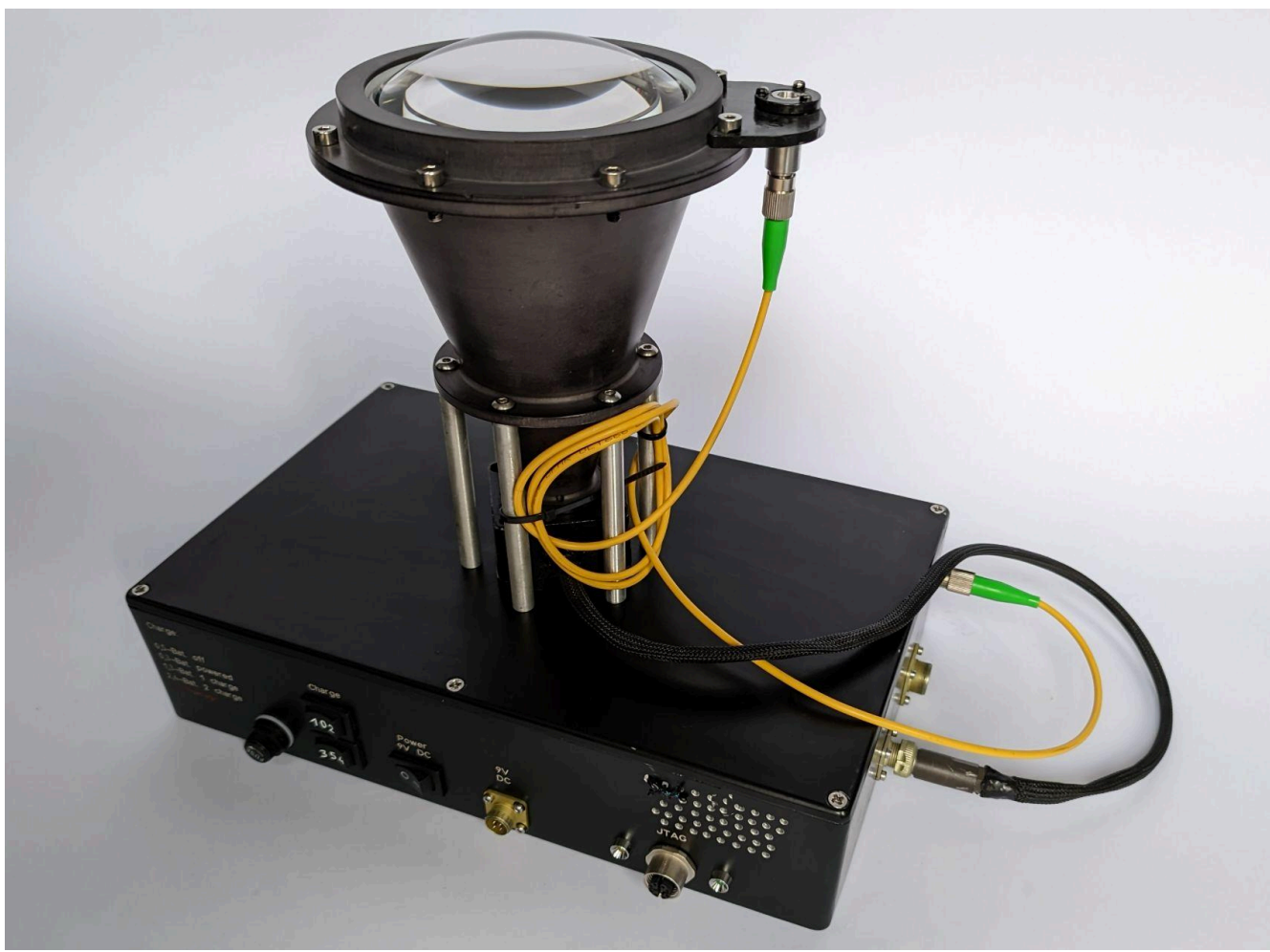

Рисунок 4.7 – Внешний вид разработанного прототипа полевого газоанализатора.

### 4.2.1. Корпус и оптическая схема прототипа газоанализатора

Как уже многократно упоминалось, разработанный прототип газоанализатора должен удовлетворять требованиям к полезной нагрузке легкого БПЛА. В таком случае для возможности проведения полетов длительностью не менее часа на одном заряде аккумуляторов БПЛА масса полезной нагрузки не должна превышать 4-5 кг. Также габаритами планируемого к использованию БПЛА будут ограничены максимальные размеры полезной нагрузки. Согласно этим требованиям было осуществлено корпусирование прототипа прибора. На рисунке 4.8 представлена модель корпуса в одном из возможных исполнений.

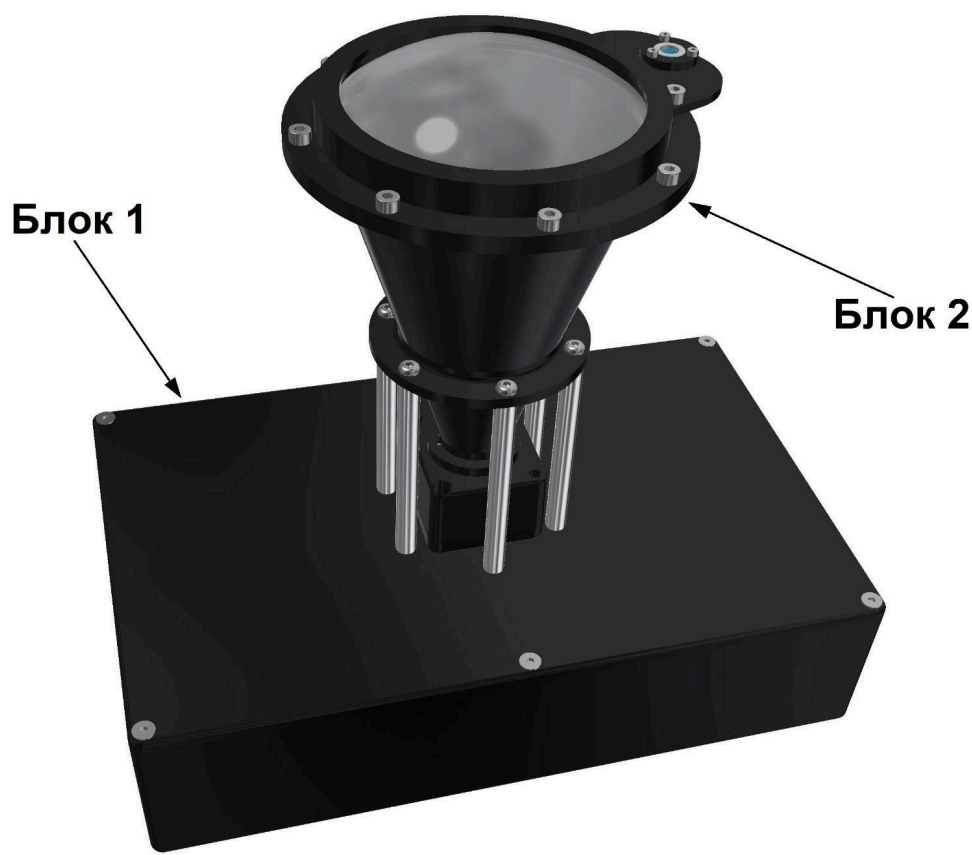

Рисунок 4.8 – Модель корпуса прототипа полевого газоанализатора.

Корпус прибора состоит из двух блоков. Блок 1 – блок управляющей электроники и реперного канала. Блок 2 – блок аналитического канала прибора. При необходимости расположение двух блоков друг относительно друга может меняться. Также для реализации возможности питания прибора как от встроенных аккумуляторов, так и от бортового питания предусмотрен блок вторичного источника питания (ВИП), преобразующий входное постоянное напряжение в диапазоне 9-36 В в напряжение питания прибора 9 В. Рассмотрим подробнее каждый блок прототипа газоанализатора.

Блок 1 содержит в себе основную плату управления, плату драйвера лазера, диодный лазер с волоконным выводом излучения и внешним оптическим изолятором, волоконный светоделитель, волоконный коллиматор, запаянную кювету из боросиликатного стекла, фотоприемник, аккумуляторы и кулер. Внутреннее устройство блока 1 показано на рисунке 4.9. Рассмотрение управляющей электроники будет далее, здесь же остановимся на описании оптической схемы.

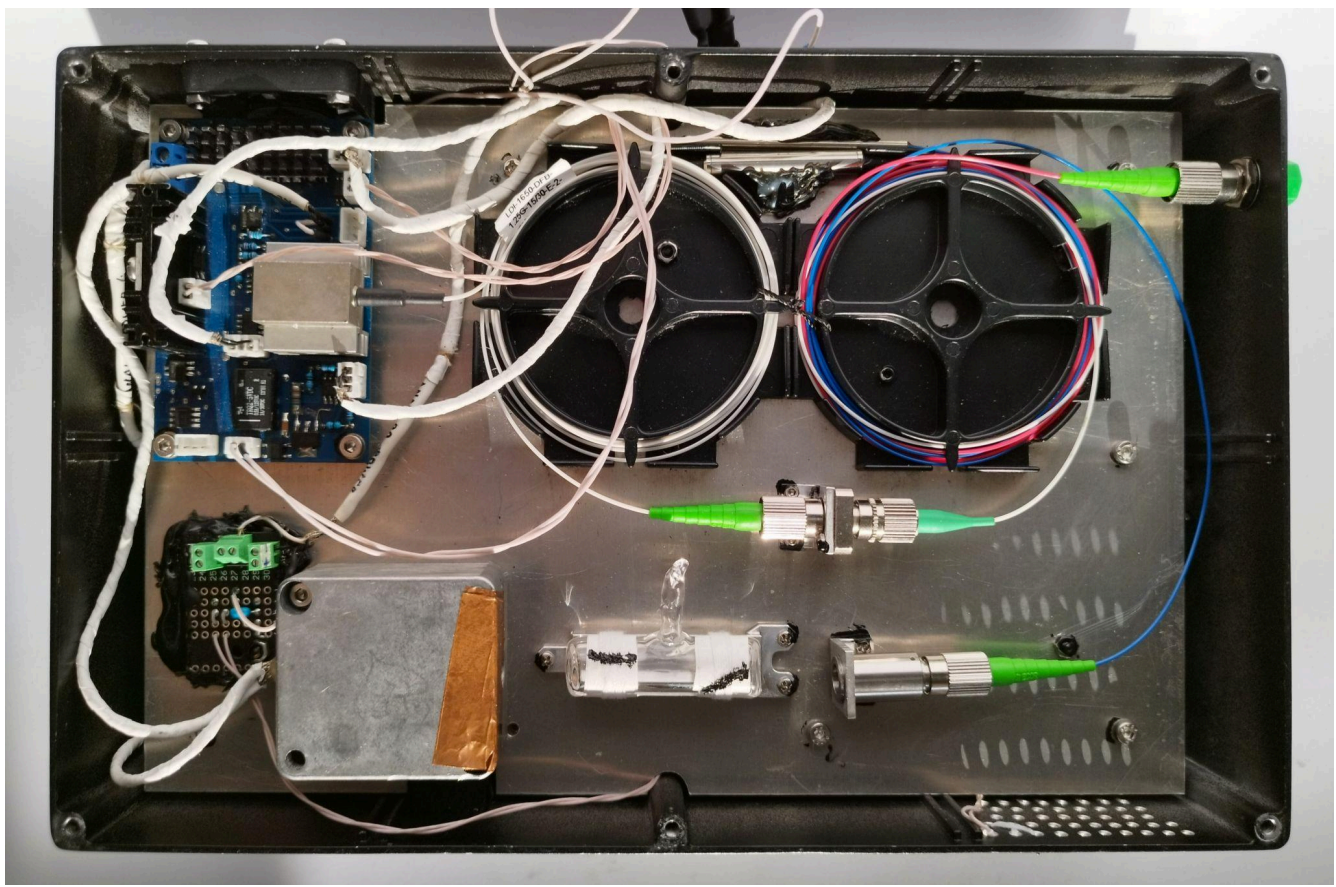

Рисунок 4.9 – Внутреннее устройство блока 1.

В качестве источника излучения был выбран такой же лазерный диод, что и для лабораторного макета, – РОС-лазер компании LasersCom (Беларусь) LDI-1650-DFB-1.25G-15/30 с центральной длиной волны излучения 1.65 мкм, шириной полосы генерируемого излучения 500 кГц и диапазоном температурной перестройки 50 см$^{-1}$. Выходная мощность излучения этого лазера в рабочем режиме составляет ~8 мВт.

Лазерное излучение подается в волоконный светоделитель с соотношением мощности в двух выходных каналах 97/3. При этом около 10% мощности теряется на соединителях волокна. Большая часть излучения направляется в аналитический канал, принадлежащий блоку 2, меньшая в реперный канал. Излучение в реперном канале коллимируется при помощи волоконного коллиматора ThorLabs F260APC-1550 и проходит сквозь запаянную кювету из боросиликатного стекла, изготовленную под заказ, длиной 46.5 мм, заполненную смесью метана и азота, после чего попадает на фотоприемное устройство. Фотоприемник состоит из

InGaAs-фотодиода FGA10 компании ThorLabs (США) с размером светочувствительной площадки 1 мм и платы предусилителя, рассчитанного на работу с третьей гармонической составляющей сигнала.

Блок 2, представленный на рисунке 4.10, состоит из плоско-выпуклой конденсорной линзы, конуса из алюминиевого сплава АМг6, повторяющего геометрию сходящегося пучка принимаемого излучения, InGaAs-фотодиода FGA21 компании ThorLabs (США) с размером светочувствительной площадки 2 мм с платой предусилителя, рассчитанного на работу в полосе частот, соответствующей первой и второй гармоническим составляющим принимаемого сигнала, в изолирующем от внешних наводок металлическом корпусе, а также коллимирующего лазерное излучение длиннофокусного волоконного коллиматора ThorLabs F280APC-1550.

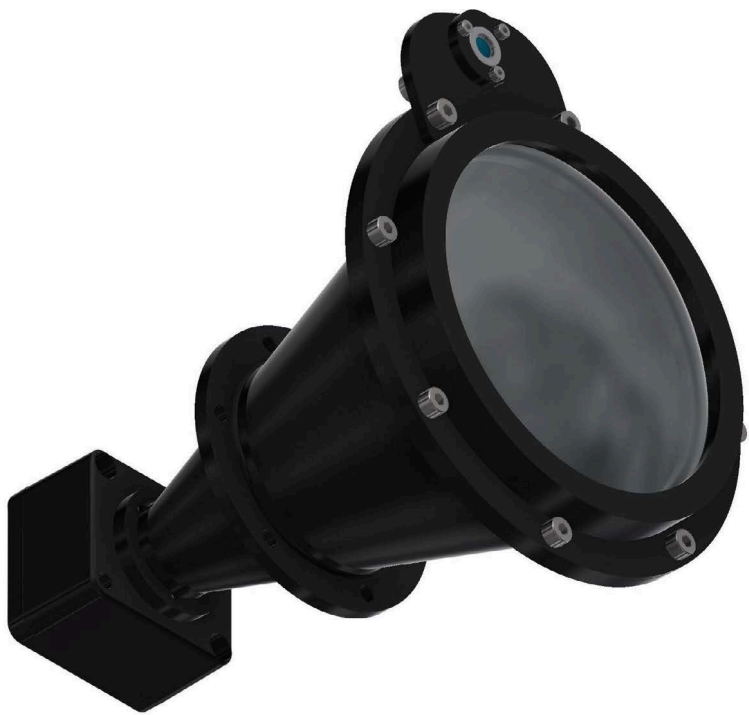

Рисунок 4.10 – Модель блока 2.

При проектировании этого блока в качестве принимающей оптики была выбрана плоско-выпуклая конденсорная линза из боросиликатного стекла диаметром 100 мм и эффективным фокусным расстоянием 136.2 мм. Это решение повлекло за собой использование несоосной схемы расположения волоконного коллиматора лазерного излучения и принимающей оптической схемы из-за сложности механической обработки оптических элементов из N-BK7. Это было учтено при проектировании следующей версии газоанализатора, как будет показано дальше. Однако на этапе работы с прототипом прибора такое решение оказалось удачным, поскольку позволяет при работе с установкой блока 2 на треногу и использования в качестве мишени вертикального экрана, рассеивающего излучение, наводиться на цель, заменяя источник излучения на внешний видимый лазер с волоконным выводом излучения. При этом из-за применения такой схемы возникла проблема минимальной рабочей дистанции до рассеивающей поверхности, поскольку схема была съюстирована для дистанции 50 м. Оптическая схема блока 2 представлена на рисунке 4.11. Полная оптическая схема прототипа прибора продемонстрирована на рисунке 4.13.

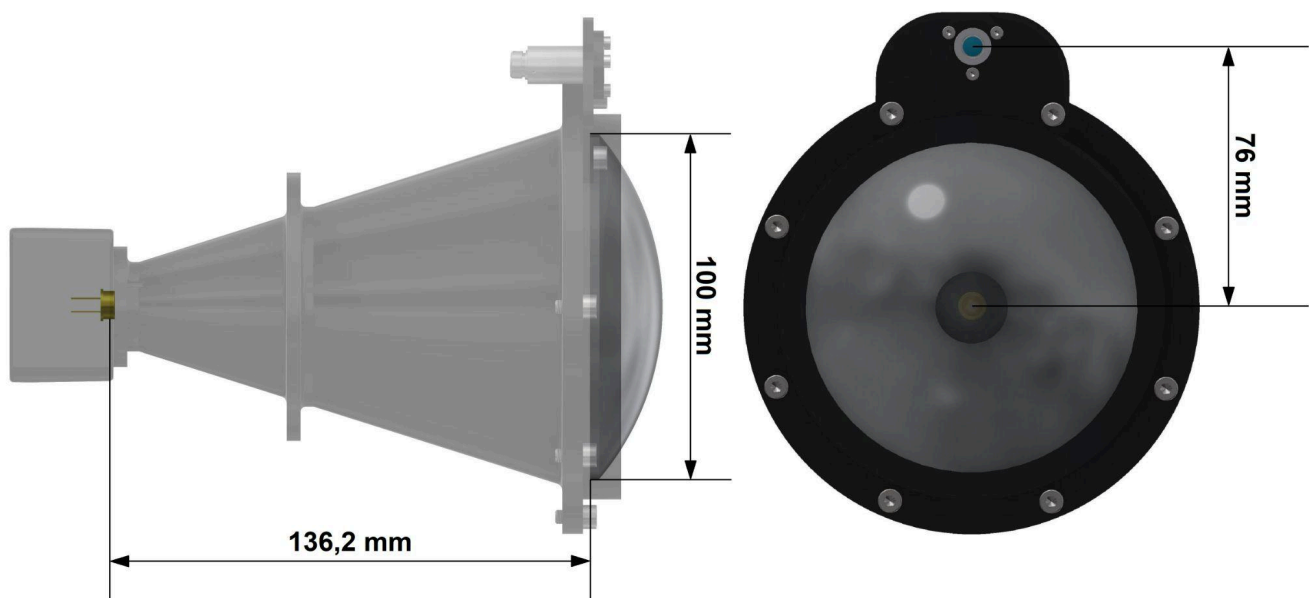

Рисунок 4.11 – Оптическая схема блока 2. Чувствительная площадка фотодиода находится в фокусе коллимирующей принимаемое излучение линзы. Расстояние между осью волоконного коллиматора и оптической осью принимающей рассеянное излучение линзы – 76 мм.

Поскольку блоки 1 и 2 соединены оптическим волокном, сигнальным кабелем и кабелем питания, их длина позволяет использовать разные конфигурации положения блоков друг относительно друга. В случае слишком малой высоты опорных стоек БПЛА, блоки могут быть размещены как на рисунке 4.12.

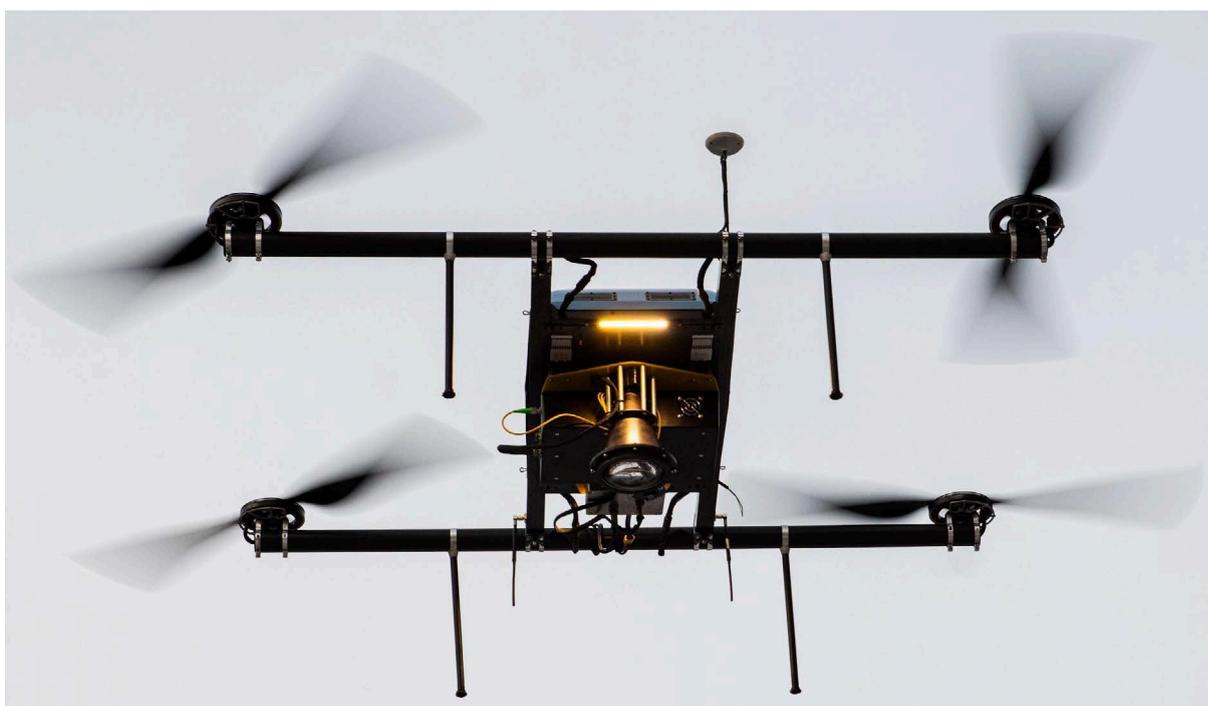

Рисунок 4.12 – Прототип газоанализатора ГИМЛИ на борту квадрокоптера «Ирбис-432».

Общие характеристики разработанного прототипа газоанализатора ГИМЛИ приведены в таблице 4.2. Характеристики, связанные с чувствительностью прибора на разных дистанциях, будут описаны далее.

Таблица 4.2 – Спецификация прототипа газоанализатора ГИМЛИ.

| Масса | ~4 кг |
|---|---|
| Габариты (ШВГ) | 275×255×175 или 275×190×400 |
| Апертура принимающей оптики | 100 мм |
| Длина волны излучения | 1651 нм |
| Мощность лазерного излучения | ~10 мВт |
| Частота сэмплирования | ~19 Гц |
| Потребление (стандартное/пиковое) | 12/35 Вт |
| Источник питания | встроенный аккумулятор/бортовой источник |
| Время работы от аккумулятора | ~5 ч |
| Рабочий диапазон температур | от -20℃ до +40℃ |

### 4.2.2. Электроника прототипа газоанализатора

Для прототипа газоанализатора ГИМЛИ была разработана собственная управляющая электроника, реализованная в шести печатных платах: цифровой плате управления, плате драйвера лазера, платах предусилителей ФД аналитического и реперного каналов, плате зарядного устройства аккумулятора, плате ВИП. Блок-схема электроники полевого прибора приведена на рисунке 4.13. Разнесение цифровой платы управления и платы драйвера лазера, которое могло приводить к наводкам из-за соединительных кабелей, было устранено в следующей версии прибора.

Блок управления, приема и обработки данных прототипа прибора выполнен в виде аналогового модуля управления током инжекции и температурой диодного лазера и усиления принимаемого аналогового сигнала и многофункциональной цифровой платы, содержащей АЦП и два ЦАП, программируемый МК и программируемую логическую интегральную схему (ПЛИС).

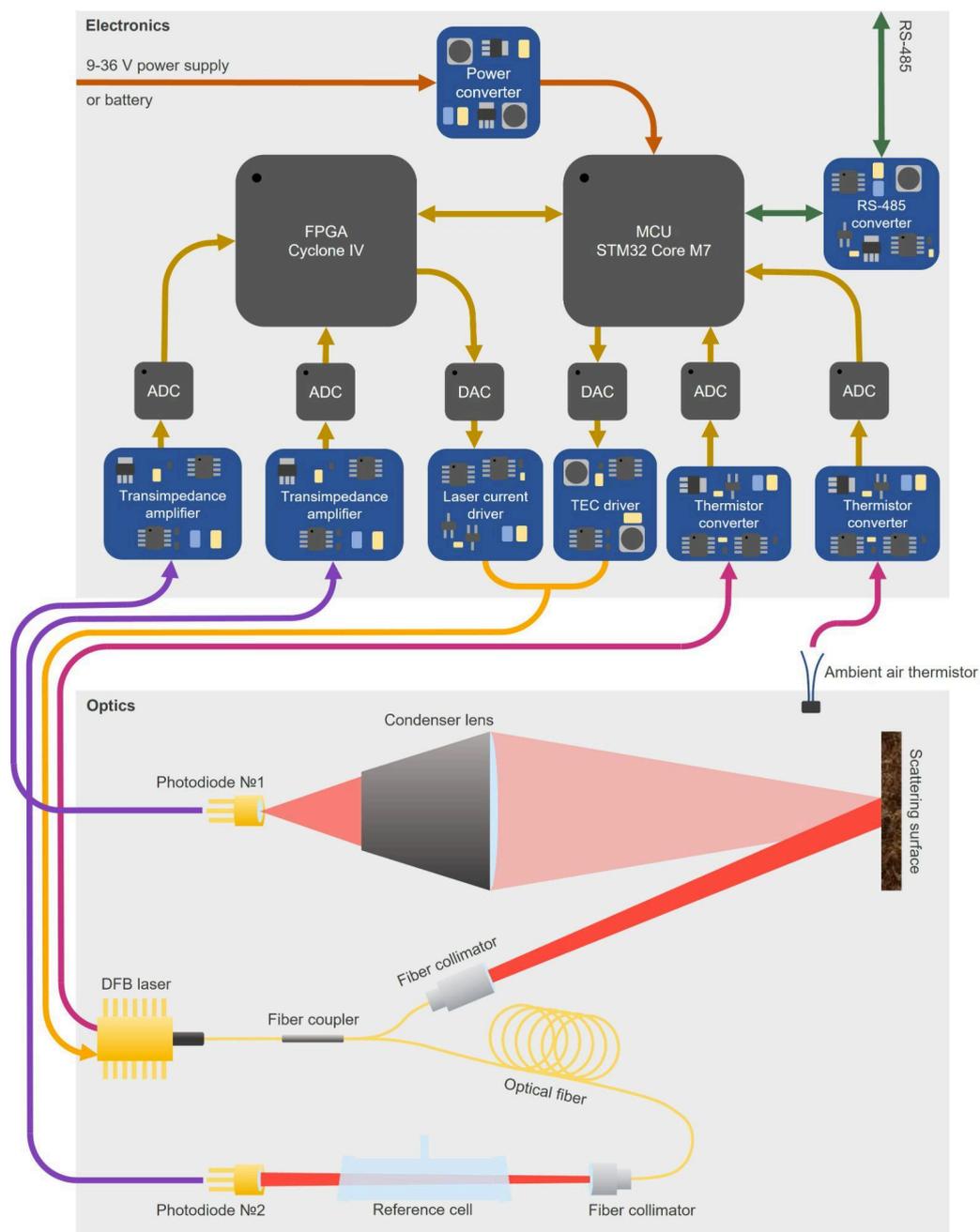

Рисунок 4.13 – Блок-схема оптики и электроники прототипа прибора.

Быстродействующий 16-битный ЦАП выдает ток накачки лазера, а также задает напряжение на элементе Пельтье, встроенном в корпус лазера. 16-битный АЦП принимает предварительно усиленные и прошедшие фильтрацию сигналы с фотодиодов аналитического и реперного каналов, а также сигнал с мониторного фотодиода диодного лазера, на 12-битный АЦП подается напряжение с термистора, по которому рассчитывается текущая температура лазерного кристалла.

Выбранная ПЛИС Cyclone IV в сочетании с используемым алгоритмом обработки данных позволяет достичь частоты выдачи предварительно обработанных данных, усредненных 1024 раз, при частоте модуляции $f$ = 26 кГц и периода синусоидальной модуляции, составляющего 192 точки, ~26 Гц. Однако частоту повторения полного цикла спектральных измерений лимитирует максимальная частота работы МК STM32 с ядром ARM Cortex-M7 ~ 18.5 Гц.

Предусилители ФД аналитического и реперного канала рассчитаны на работу в диапазоне *1f-2f* и вблизи *3f* соответственно. Поскольку для выбранного волоконного коллиматора F280APC-1550 диаметр выходного пучка *d* составляет ~3.2 мм, угол дифракционной расходимости для длины волны $\lambda = 1651$ нм составит

$$\theta_d = \frac{2\lambda}{\pi \cdot d} \cdot \frac{180}{\pi} = 0.019°, \tag{4.1}$$

что при дистанции до рассеивающей поверхности *h* = 100 м даст диаметр пучка на этом расстоянии ~66 мм. В случае косинусного рассеяния света от подстилающей поверхности [299] при диаметре принимающей излучение оптики D = 100 мм получим оценку для доли падающей на ФД мощности от рассеянной с единицы площади:

$$K \approx \frac{\pi \frac{D^2}{4}}{\pi h^2} = \left(\frac{D}{2h}\right)^2 = 10^{-6}. \tag{4.2}$$

При фокусном расстоянии используемой собирающей на ФД линзы 150 мм диаметр пятна попадающего на чувствительную площадку ФД на дистанции 100 м составит ~1.3 м, что много больше обусловленного дифракционной расходимостью диаметра лазерного пучка на этой дистанции. Таким образом, не учитывая атмосферное поглощение и рассеяние на аэрозольных взвесях в воздухе, получим оценку падающей на ФД мощности излучения ~6 нВт при излучаемой мощности 6.5 мВт. В таком случае при чувствительности фотодиода ~1 А/Вт эквивалентное сопротивление предусилителя аналитического канала на частоте модуляции лазерного излучения *f* и удвоенной частоте *2f* должно составлять в рабочей полосе частот сотни МОм, как видно на рисунке 4.14.

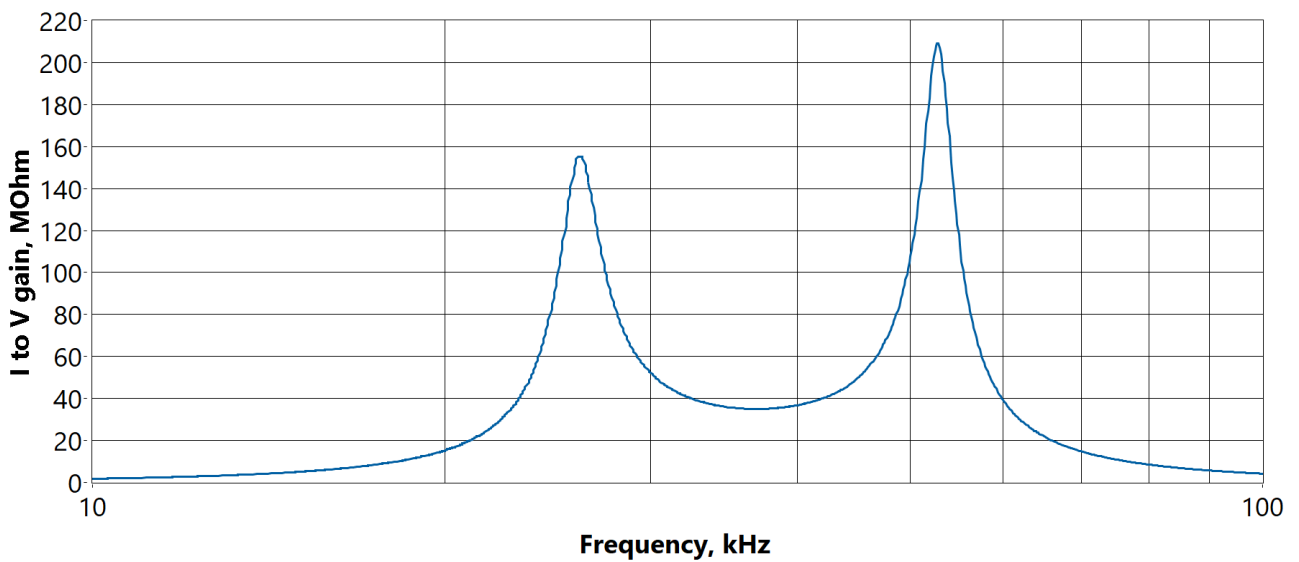

Рисунок 4.14 – Эквивалентное сопротивление предусилителя аналитического канала.

Амплитудно-частотные характеристики (АЧХ) и фазово-частотные характеристики (ФЧХ) предусилителя, рассчитанного для аналитического канала, представлены на рисунке 4.15.

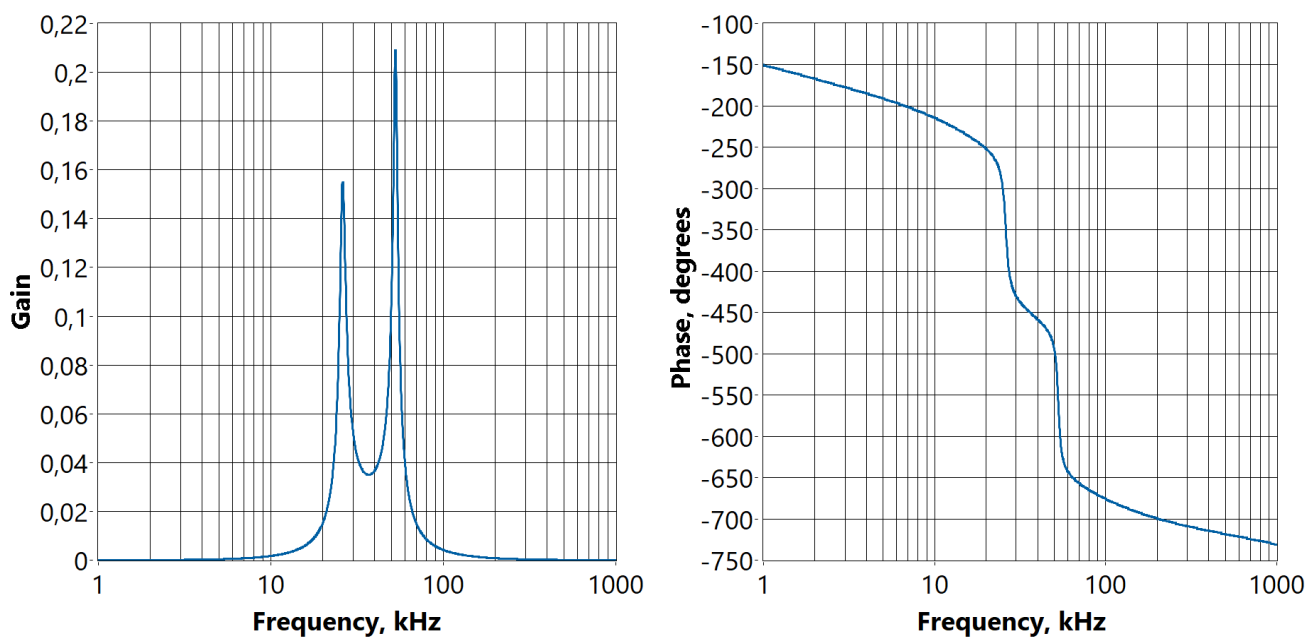

Рисунок 4.15 – АЧХ (слева) и ФЧХ (справа) аналитического канала.

Можно заметить, что у реального предусилителя фототока эквивалентное сопротивление на частоте модуляции лазерного излучения $f$ и удвоенной частоте $2f$ будет отличаться, тем самым усложнив формулу (3.19) необходимостью учета разных значений $R_{eq}$ для каждой гармонической составляющей сигнала. Но ввиду постоянства величин $R_{eq}$ для сигнала на частоте $f$ и $2f$ дальнейшие рассуждения, приведенные в предыдущей главе, остаются корректными.

Однако, как было обнаружено при полевых испытаниях прототипа прибора, чувствительность выбранного фотодиода может значительно отклоняться от значения ~1 А/Вт, актуального для 25℃, при изменении температуры. Для регистрации температуры окружающего воздуха, что необходимо для учета обнаруженной специфики работы используемых InGaAs-фотодиодов, подробно описанной в пункте 4.3.2, был использован NTC-термистор B57861-S.

В реперном канале на ФД падает ~10 мкВт, таким образом, эквивалентное сопротивление предусилителя в этом канале должно составлять в рабочей полосе частот, десятки кОм, что видно из рисунка 4.16. Максимальное усиление используемого предусилителя фототока в данном случае должно быть на утроенной частоте модуляции лазерного излучения $3f$, используемой для реализации алгоритма стабилизации частоты генерируемого лазерного излучения по выбранной спектральной линии поглощения.

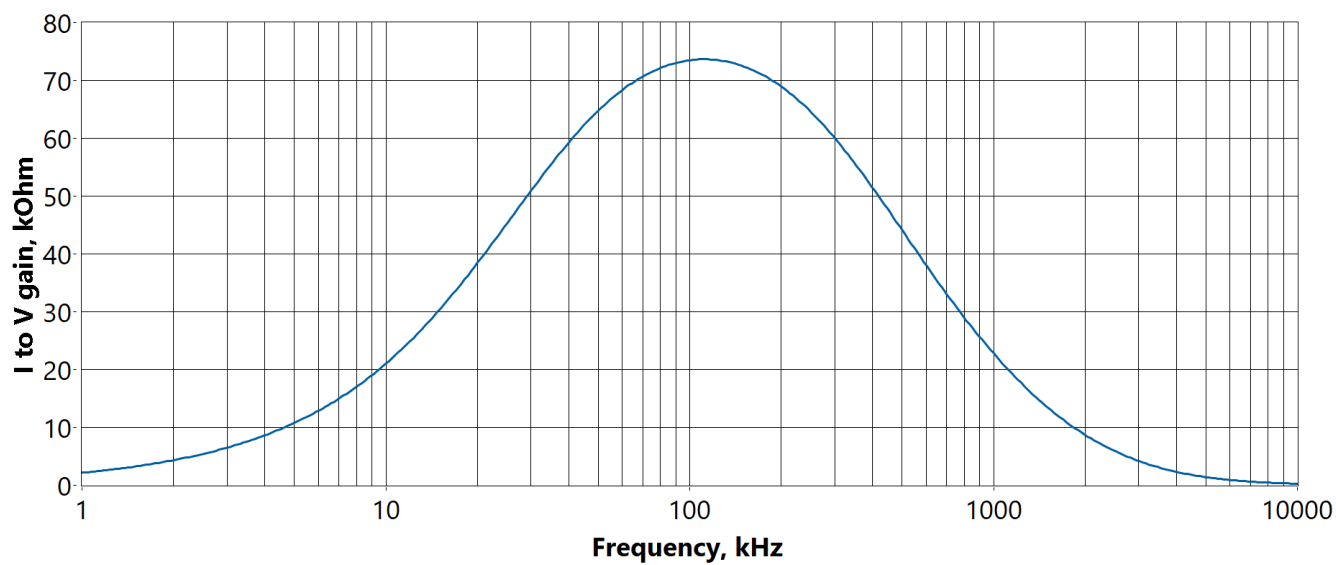

Рисунок 4.16 – Эквивалентное сопротивление предусилителя реперного канала.

АЧХ и ФЧХ предусилителя, разработанного для применения в реперном канале, представлены на рисунке 4.17.

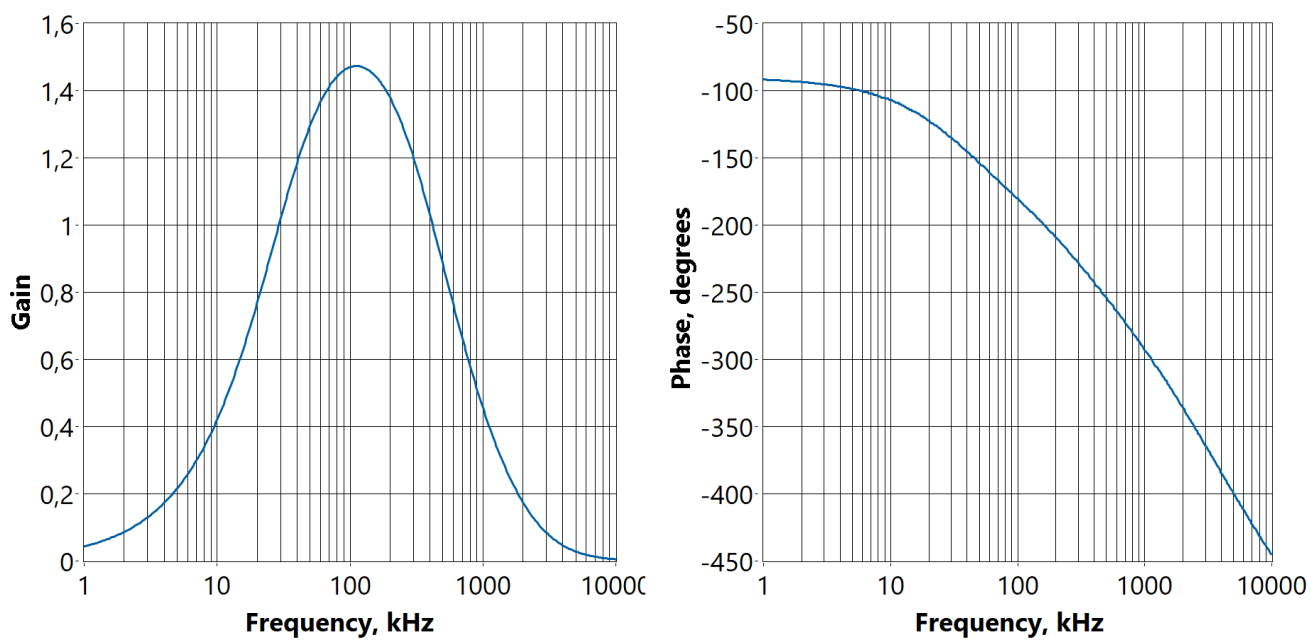

Рисунок 4.17 – АЧХ (слева) и ФЧХ (справа) реперного канала.

### 4.2.3. Управляющее программное обеспечение

Для работы с газоанализатором в среде разработки LabView было создано программное обеспечение для управления прибором, считывания данных, записанных на microSD-карту, и их обработки. Управление прибором реализовано посредством отправки команд длиной 408 байт, содержащих информацию о задаваемых токе накачки и температуре лазерного диода, параметрах ПИД-регулятора элемента Пельтье, встроенного в корпус лазера и служебную информацию. Панель управления настройками прибора, содержащая отправляемую команду и принимаемые от прибора пакеты в формате HEX для упрощения отладки, приведена на рисунке 4.18.

Рисунок 4.18 – Панель управления настройками прибора ГИМЛИ.

Принимаемые пакеты, декодируемые разработанным ПО, содержат 781 байт информации, содержащей показания фотоприемников основного и реперного каналов, значения внутреннего таймера, показания встроенного в корпус лазерного диода термистора и дополнительного термистора, определяющего температуру окружающего воздуха. Панель, отображающая сигналы обоих каналов прибора и температуры ЛД и окружающего воздуха приведена на рисунке 4.19.

Сигналы с фотодиодов основного и реперного каналов усиливаются в соответствующих полосах для дальнейшей работы с *1f*- и *2f*-сигналами основного канала прибора и *3f*-сигналом реперного канала. По *3f*-сигналу реперного канала реализован алгоритм стабилизации по пику выбранной линии поглощения метана, поскольку точка его симметрии совпадает по частоте с центром линии поглощения, как говорилось в предыдущей главе.

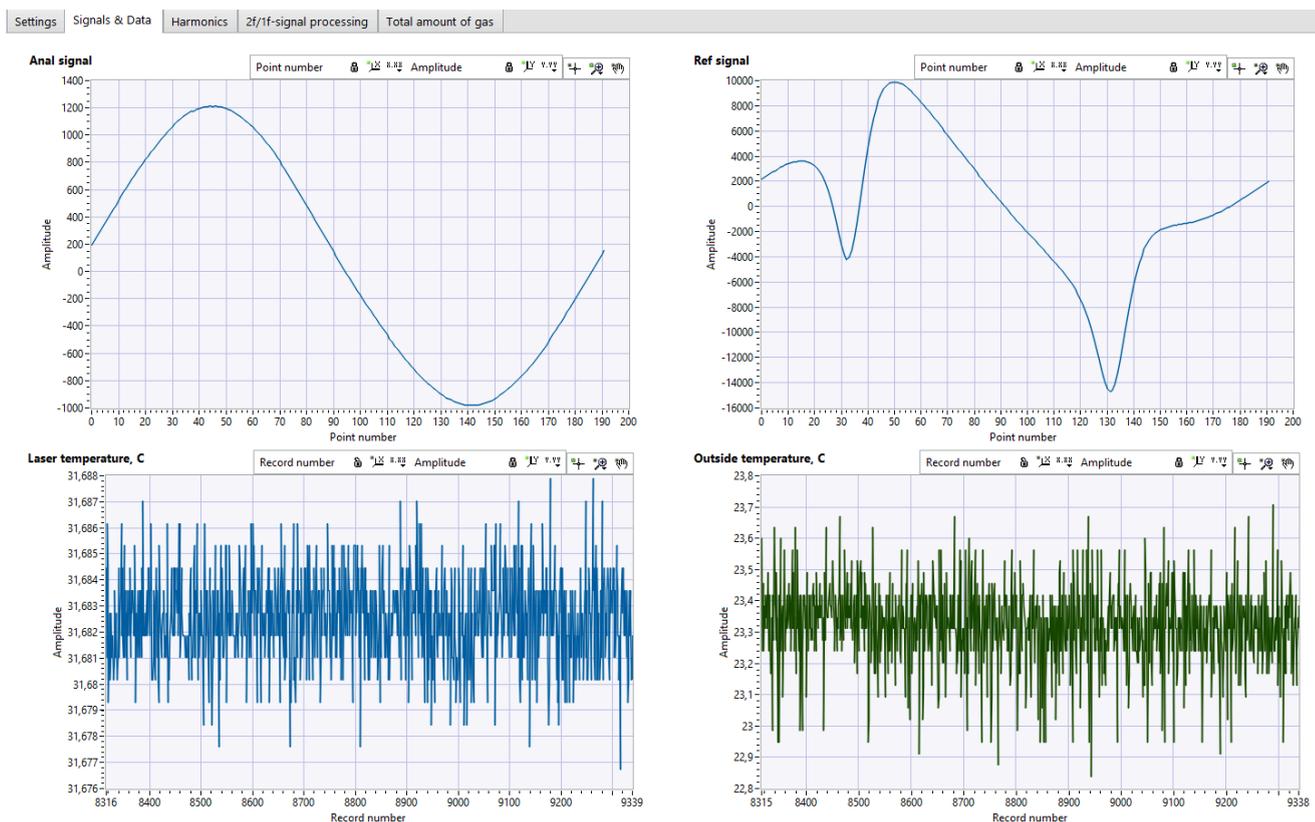

Рисунок 4.19 – Панель отображения сигналов аналитического (слева сверху) и реперного (справа сверху) каналов прибора и температур ЛД (снизу слева) и окружающего воздуха (снизу справа).

Однако данный алгоритм стабилизации является двухступенчатым. Для корректной работы используемого алгоритма стабилизации лазерного излучения по пику линии поглощения с использованием точки симметрии третьей гармоники принимаемого в реперном канале сигнала необходимо предварительно вывести частоту генерации лазера в сравнительно узкий диапазон шириной несколько сотых долей см$^{-1}$ около пика линии поглощения. Это требуется для предотвращения возможности ухода центральной частоты излучения лазера с экстремума *3f*-сигнала в направлении крыльев линии поглощения (см. рисунок 3.8с).

По этой причине после запуска прибора сперва запускается алгоритм стабилизации генерируемого лазером излучения по температуре кристалла. И требования к настройке ПИД-регулятора для этого этапа стабилизации оказываются высоки. После этого через заданное время включается алгоритм стабилизации по *3f*-сигналу.

Панель ПО с отображением значений первых трех гармонических составляющих принимаемого в аналитическом и реперном каналах сигналов показана на рисунке 4.20. Можно заметить, что *3f*-сигнал реперного канала с высокой точностью равен нулю.

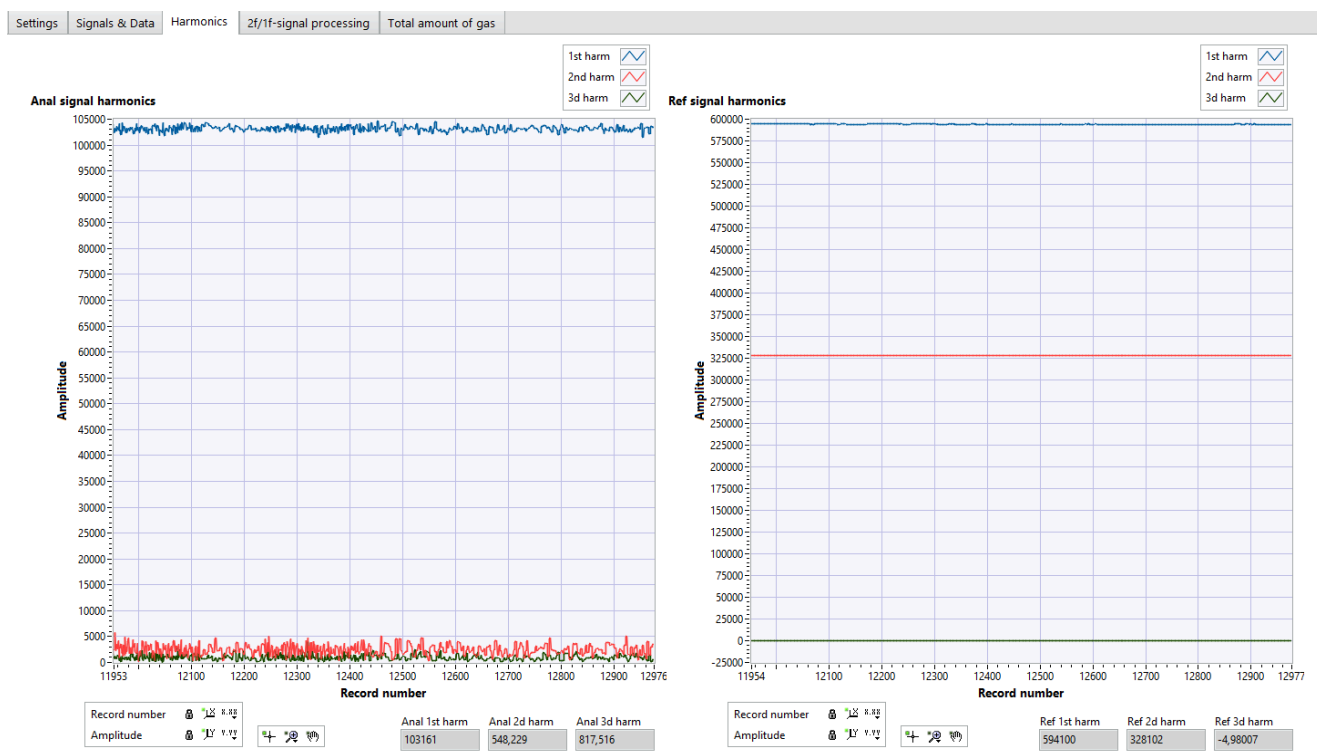

Рисунок 4.20 – Панель отображения значений первых трех гармонических составляющих принимаемого в аналитическом (слева) и реперном (справа) каналах сигналов.

Возможно использование альтернативного двухступенчатого алгоритма динамической стабилизации излучения по пику выбранной линии поглощения метана с таким же этапом предварительной стабилизации по температуре вблизи центра линии поглощения. Для второго этапа используется следующее соображение. При сканировании пика линии поглощения синусоидальным сигналом в отсутствии сдвига фазы совпадение центральной частоты лазерного излучения с положением центра линии поглощения и будет достижением требуемого условия стабилизации. Таким образом пик сканируемой линии поглощения будет соответствовать 0, $\pi$ и $2\pi$, модельный сигнал реперного канала, соответствующий такому случаю представлен на рисунке 4.21 синей кривой. Модельный сигнал при сдвиге фазы сигнала, обусловленном ФЧХ предусилителя реперного канала, что соответствует наблюдаемому сигналу в реперном канале прототипа газоанализатора, представлен на рисунке 4.21 красной кривой.

Можно заметить, что вне зависимости от сдвига фазы сигнала будет выполняться условие на интервал между соседними пиками сканируемой спектральной линии, соответствующий $\pi$ или половине периода синусоидального сигнала. Это условие можно так же, как и описанное ранее условие стабилизации по $3f$-сигналу, использовать для стабилизации частоты лазерного излучения по пику линии поглощения с точностью лучше $10^{-3}$ см$^{-1}$, чего вполне достаточно при ширине выбранной линии поглощения фонового метана в атмосферном воздухе на полувысоте ~0.145 см$^{-1}$.

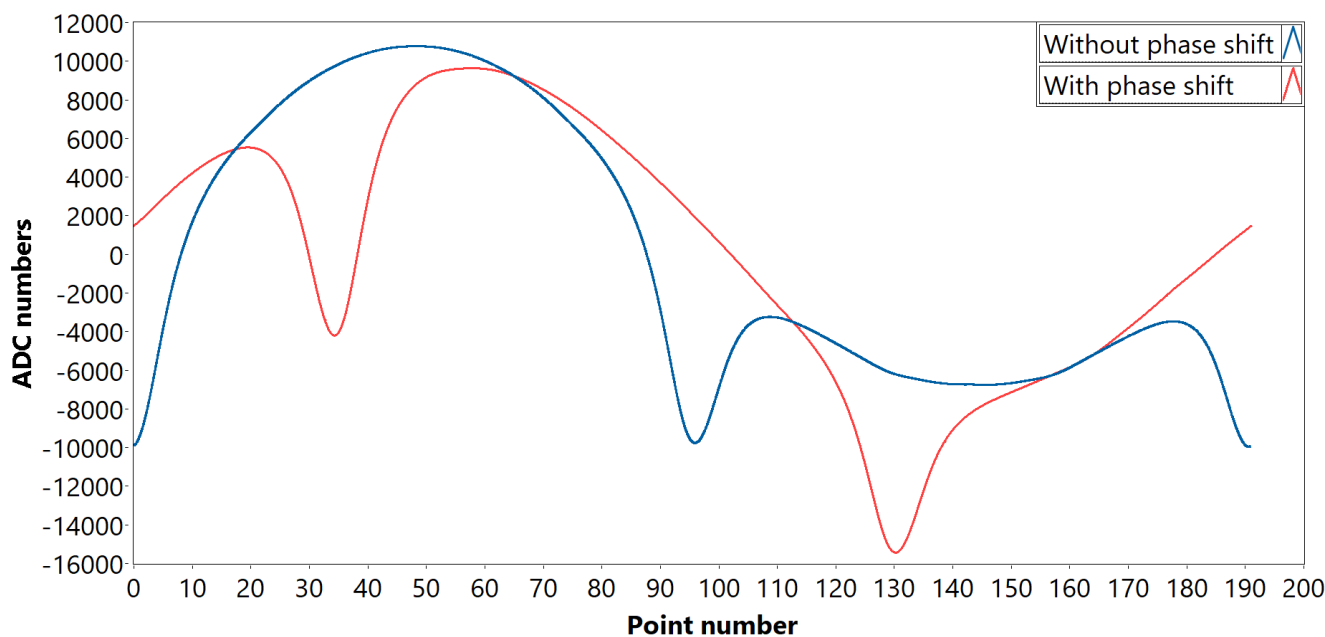

Рисунок 4.21 – Модельный сигнал реперного канала при совпадении центральной частоты лазерного излучения с положением центра линии поглощения в отсутствии сдвига фазы сигнала, обусловленного ФЧХ предусилителя реперного канала (синяя кривая) и при сдвиге фазы сигнала, соответствующего наблюдаемому сигналу в реперном канале (красная кривая).

Как было описано в предыдущей главе отношение *2f-* к *1f-*сигналу дает величину, пропорциональное значению интегрального содержания выбранного газа на удвоенной дистанции от прибора до рассеивающей излучение поверхности. Корректный пересчет значений указанного отношения в единицы ppm·м возможен при помощи калибровки прибора или создания модели цифрового двойника прибора, учитывающего все факторы окружающей среды, а также специфическое влияние электроники на получаемые результаты.

Попытки создания такой модели позволили лучше разобраться во многих аспектах обработки сигнала и воздействии внешних условий на регистрируемый сигнал, однако ввиду многократных изменений в электронных схемах прибора, влияющих на форму и интенсивность обрабатываемого сигнала, в процессе его отладки добиться полного соответствия модельных результатов с экспериментом не удалось.

Однако в будущих версиях газоанализатора при возможности отдельного изучения влияния выбранных схемотехнических решений на параметры анализируемого сигнала создание цифрового двойника может оказаться решаемой задачей – в таком случае прибор перейдет в класс оборудования, не требующего калибровки.

В описываемом прототипе газоанализатора в связи с указанными сложностями потребовалось проведение калибровки прибора по независимым измерениям другого устройства для измерения содержания метана в атмосферном воздухе. Методика проведенной калибровки будет описана далее, здесь же укажем, что для отображения результатов измерения в размерности ppm·м требуется линейное преобразование отношения *2f*- к *1f*-сигналу. На рисунке 4.22 показана панель отображения содержания метана на дистанции от прибора до рассеивающей поверхности в ppm·м.

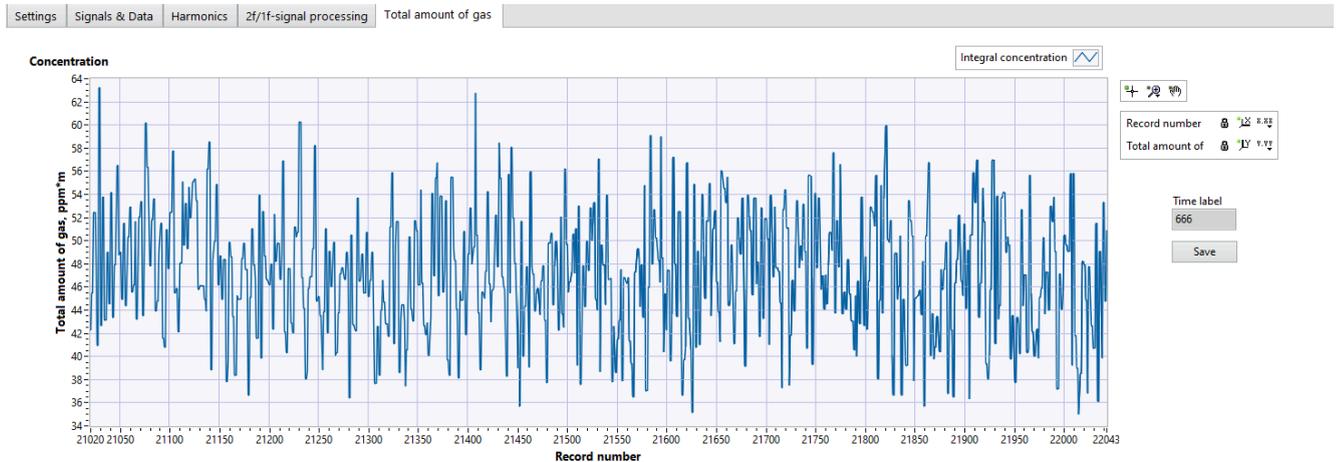

Рисунок 4.22 – Панель отображения содержания метана на дистанции от прибора до рассеивающей поверхности в ppm·м.

Описанное выше ПО позволяет контролировать весь ход измерений в случае установки газоанализатора на треногу и наведении блока 2 на вертикальную поверхность. В таком режиме работы предусмотрена возможность записи данных на ПК, минуя стадию записи на MicroSD-карту прибора. Пример организации такого эксперимента показан на рисунке 4.23.

Однако при работе с БПЛА описанное ПО требуется для вывода газоанализатора на рабочий режим и завершении записи данных после окончания полета. В отсутствии возможности в данном случае записи данных на ПК, происходит запись на установленную в прибор MicroSD-карту с частотой ~20 Гц. Записанные таким образом данные считываются и анализируются при помощи другого ПО, также разработанного в среде LabView, показавшей более высокую скорость считывания данных по протоколу RS-485 в сравнении с другими средами разработки ПО.

Разработанное для анализа считанных данных ПО позволяет помимо описанного выше применять к сигналу фильтрацию Савицкого-Голая, вычислять дистанцию до рассеивающей излучение поверхности в выбранный момент времени, а также проводить анализ дисперсии Аллана для выбранных фрагментов измерений.

Таким образом, разработанный комплекс ПО позволяет проводить полный цикл измерений, считывание и обработку данных как в реальном времени, так и постфактум, оставляя разработчикам возможность проведения отладки устройства. Также возможна реализация передачи данных с установленного на БПЛА прибора в реальном времени по протоколу беспроводной связи LoRaWAN, реализующему передачу данных на дистанцию до 15 км.

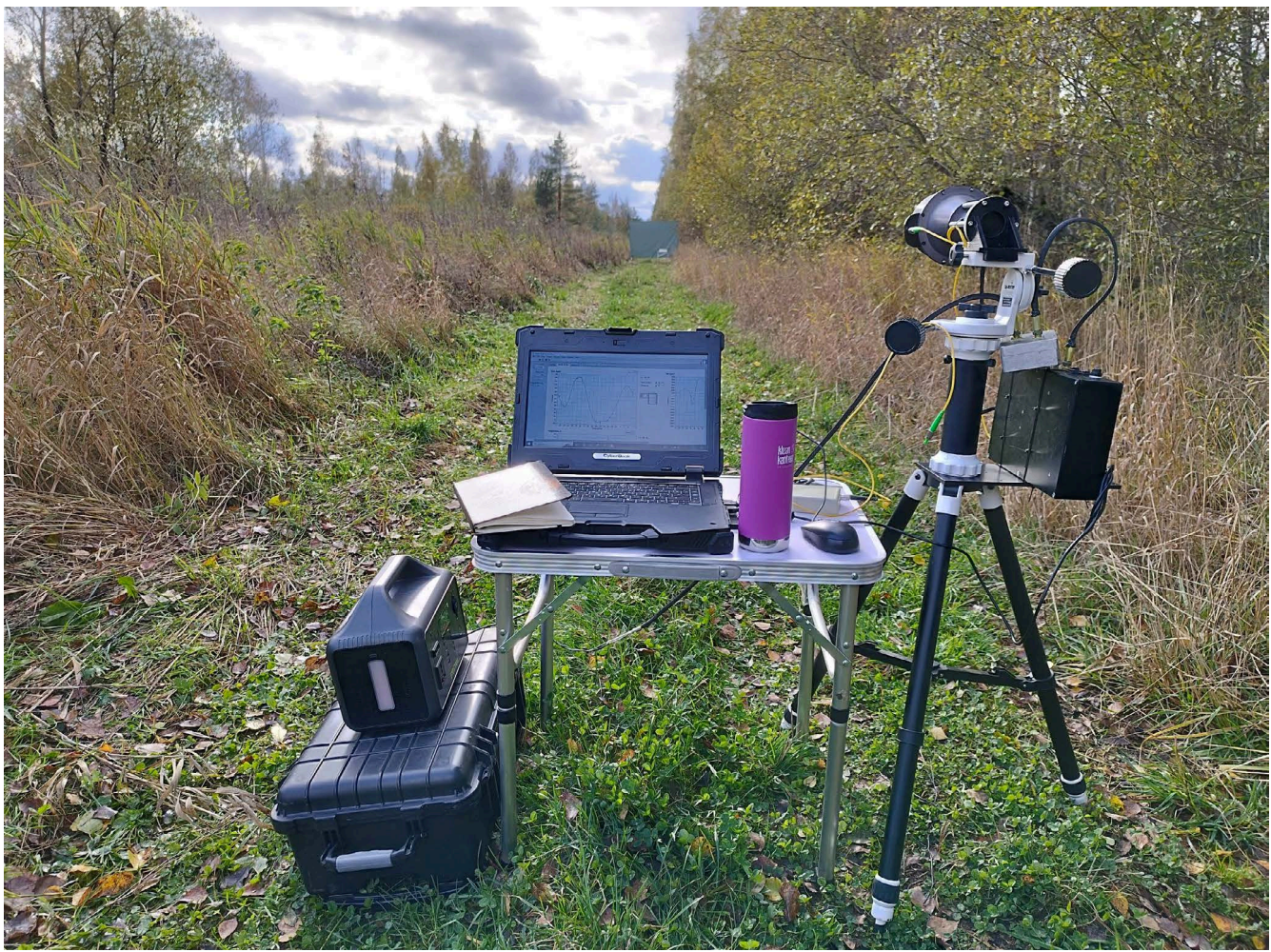

Рисунок 4.23 – Пример организации эксперимента с использованием прототипа газоанализатора ГИМЛИ, установленного на треногу, в горизонтальном режиме с наведением внешнего оптического блока прибора на вертикально установленный экран.

## 4.3. Результаты проведения полевых испытаний прототипа газоанализатора

За время проведения испытаний прототипа газоанализатора ГИМЛИ в полевых условиях была проанализирована его работа при разных схемах экспериментов – с установкой на борт БПЛА и в горизонтальном режиме с наведением внешнего оптического блока прибора на вертикально установленный экран при помощи треноги – во все сезоны, характерные для средних широт, в температурном интервале около $40℃$, при сухом воздухе и небольших осадках, с разными рассеивающими излучение поверхностями. Несколько иллюстраций хода испытаний приведены на рисунке 4.24.

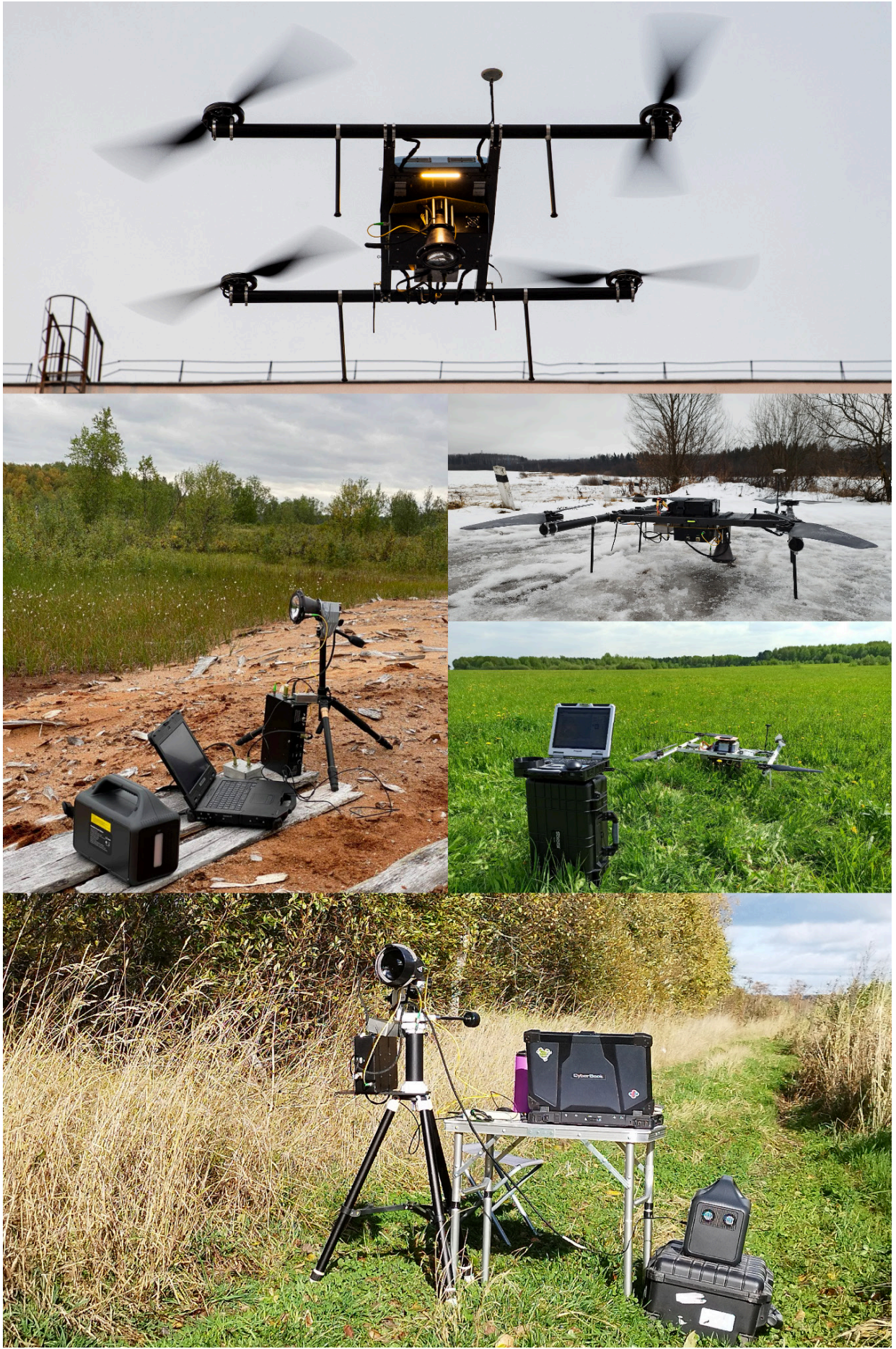

Рисунок 4.24 – Испытания прибора в разные сезоны при разных схемах экспериментов.

Была выявлена упомянутая зависимость сигнала ФД от температуры окружающей среды, требующая алгоритмического учета при обработке сигнала, выливающаяся в падение максимальной рабочей дистанции прибора со ~120 м при 26℃ до ~70 м при -11℃, которое изначально неверно объяснялось значительным поглощением снежного покрова в ближнем ИК-диапазоне.

Был подобран оптимальный алгоритм стабилизации частоты излучения лазера по выбранной спектральной линии поглощения метана. Также была обнаружена проблема засветок при попадании в принимающую оптическую систему прибора яркого солнечного света, предложенное решение которой будет описано далее.

На примере измерения фонового метана при температуре окружающего воздуха 26℃ рассмотрим поведение различных характеристик сигнала в зависимости от дистанции до рассеивающей поверхности.

Как уже упоминалось, в оптическом блоке прибора волоконный коллиматор лазерного излучения и приемный тракт рассеянного от подстилающей поверхности излучения смонтированы несоосно. С этим связано наличие минимальной рабочей дистанции между газоанализатором и рассеивающей поверхностью ~40 м, обусловленное проведенной юстировкой приемо-передающего оптического тракта для дистанции 50 м. Сигнал в аналитическом канале прибора при дистанции до рассеивающей поверхности 50 м, усредненный за 10 сек при частоте сэмплирования 20 Гц, приведен на рисунке 4.25.

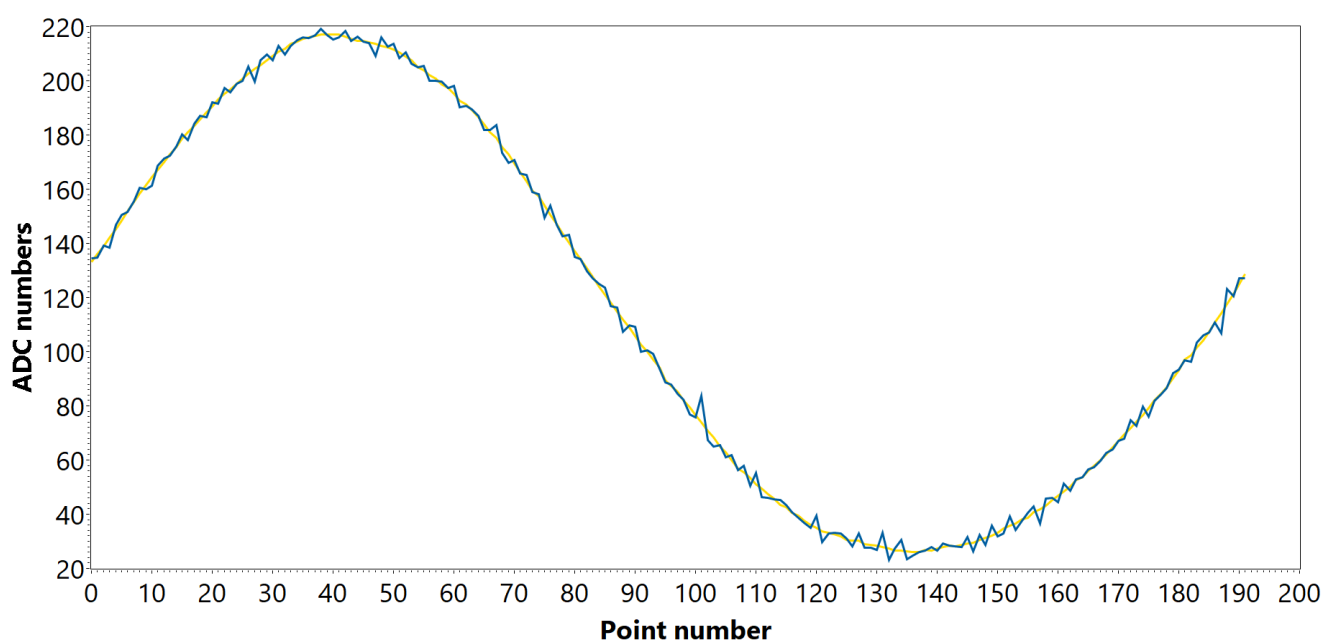

Рисунок 4.25 – Сигнал в аналитическом канале прибора при дистанции до рассеивающей поверхности 50 м, усредненный за 10 сек при частоте сэмплирования 20 Гц.

Прогнозируемый основной режим работы прибора заключается в измерении интегральной концентрации фонового метана в атмосферном воздухе с борта БПЛА, движущегося на высоте ~50 м со скоростью ~10-15 м/с. В таком случае возможные времена усреднения сигнала при пространственном разрешении не хуже десятков метров не будут превышать единицы секунд. Зависимость отклонения Аллана [300] для сигнала, полученного при дистанции до рассеивающей поверхности 50 м, от времени накопления и ее сравнение с той же зависимостью для белого шума показана на рисунке 4.26 в двойном логарифмическом масштабе.

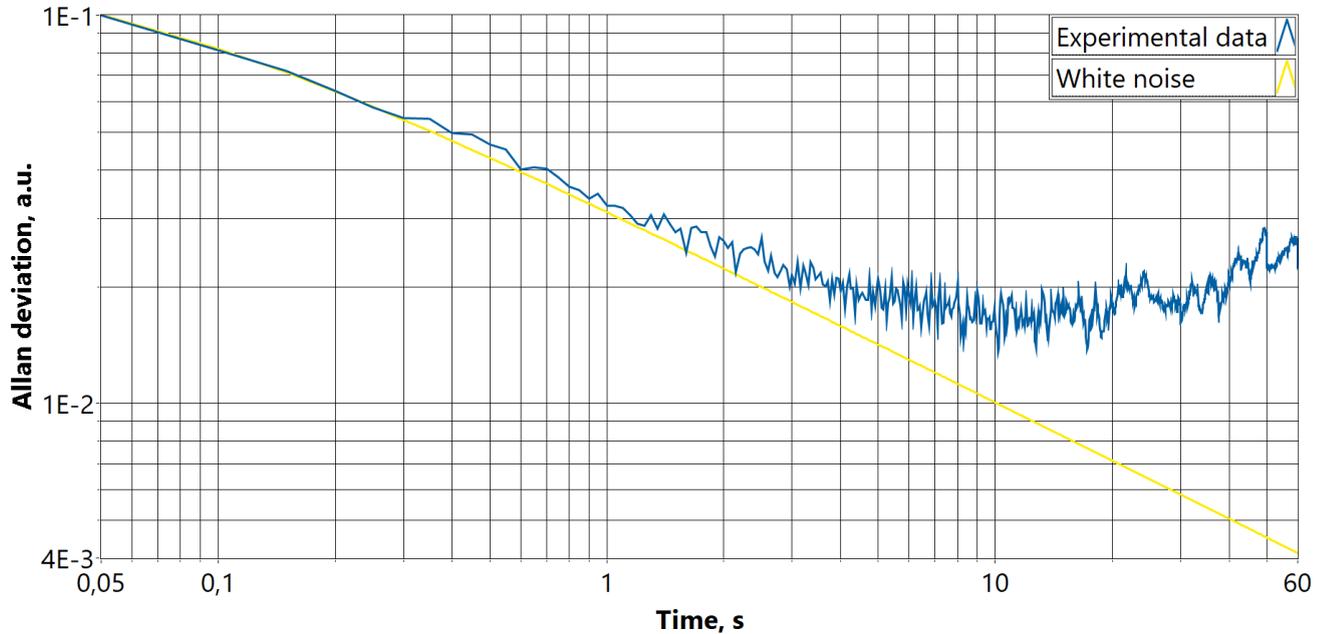

Рисунок 4.26 – Сравнение зависимости отклонения Аллана для сигнала, полученного при дистанции до рассеивающей поверхности 50 м, от времени накопления (синяя кривая) с той же зависимостью для белого шума (желтая кривая).

Как видно, значимые отличия от поведения отклонения Аллана белого шума наблюдаются для усреднений в течение 10 с и более, чего достаточно для описанных выше условий применения прибора, при этом отклонение Аллана для экспериментально полученных данных ведет себя классическим образом даже на временах накопления порядка двух минут.

Первичная характеристика регистрируемого сигнала, влияющая на результаты дальнейшего анализа – интенсивность или удвоенная амплитуда. Интенсивность регистрируемого рассеянного излучения будет уменьшаться обратно пропорционально квадрату дистанции до рассеивающей поверхности согласно формуле (4.2).

Экспериментальная зависимость детектируемой удвоенной амплитуды сигнала от дистанции до рассеивающей поверхности и ее приближение функцией вида $1/x^2$, полученное методом наименьших квадратов при помощи алгоритма на основе поворота Гивенса, приведены на рисунке 4.27.

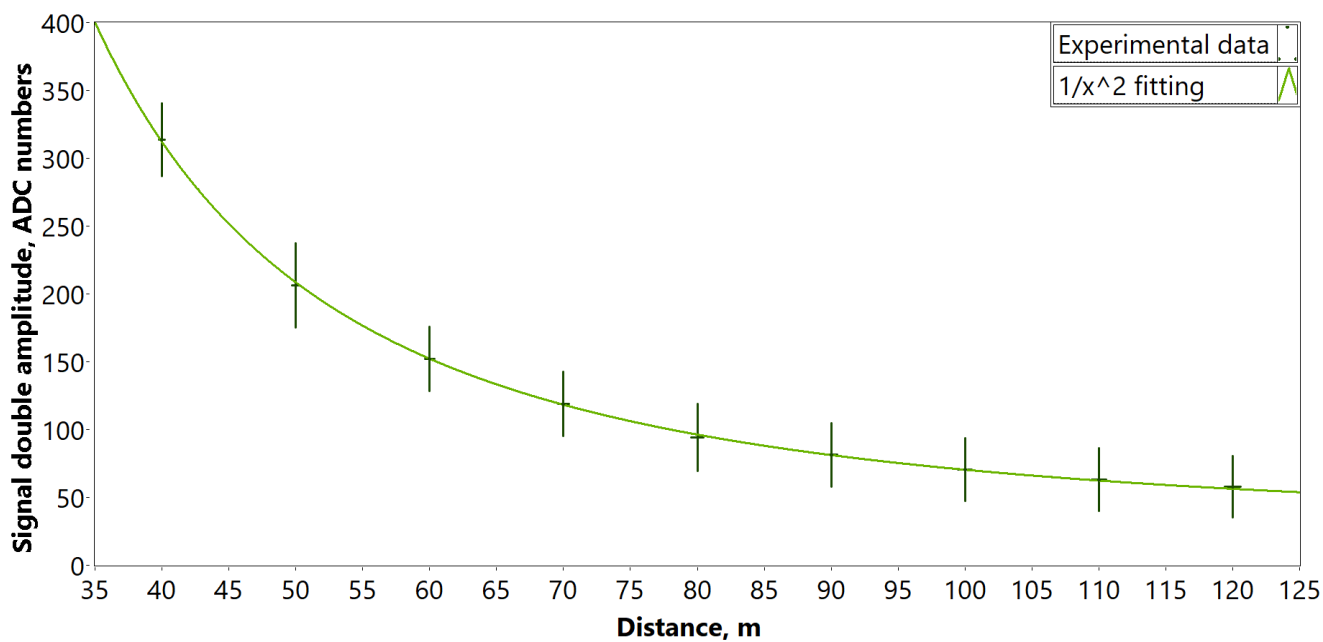

Рисунок 4.27 – Зависимость удвоенной амплитуды сигнала от дистанции до рассеивающей поверхности (зеленые точки) и ее приближение функцией вида 1/x² (салатовая кривая).

Сигнал, представленный на рисунке 4.25, имеет *n* гармонических составляющих на частотах от *f* до *nf*. Методика квадратурного детектирования позволяет выделять интересующие гармоники сигнала, необходимые для последующего анализа согласно формулам (3.22-3.24). *1f*-сигнал, имеющий физический смысл общей интенсивности принимаемого излучения, будет зависеть от дистанции до рассеивающей поверхности аналогично удвоенной амплитуде регистрируемого сигнала, как видно из рисунка 4.28.

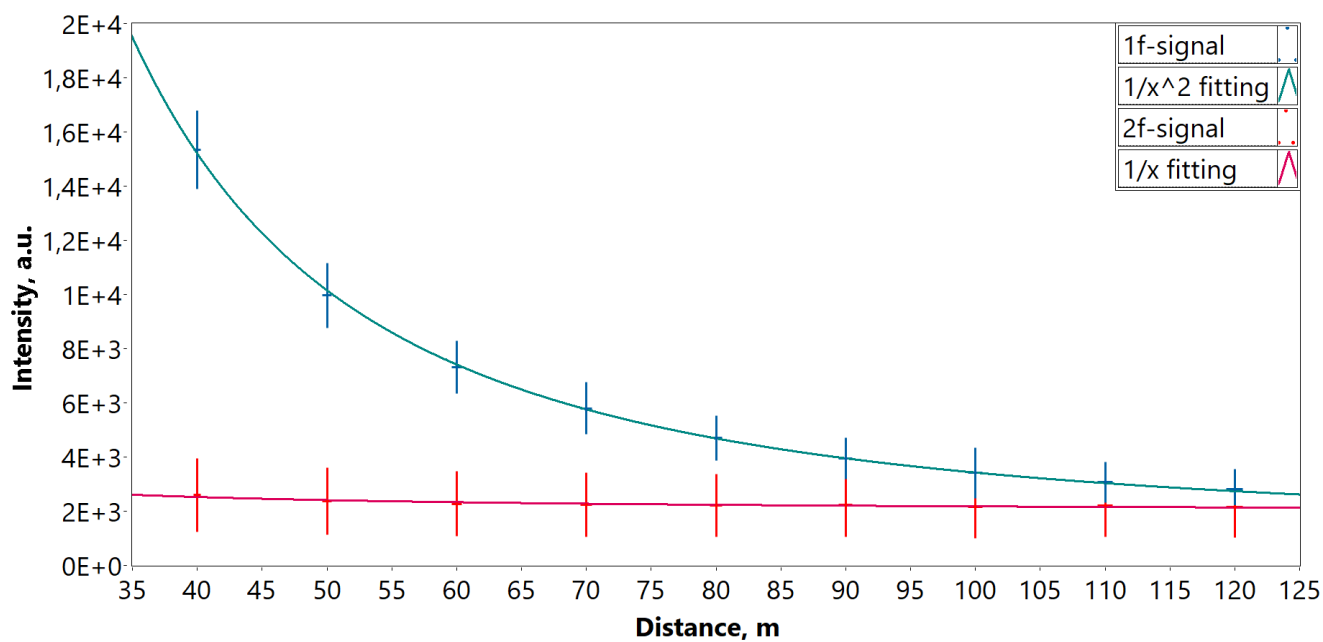

Рисунок 4.28 – Зависимости *1f*-сигнала (синие точки) и *2f*-сигнала (красные точки) от дистанции до рассеивающей поверхности и их приближения функциями вида 1/x² (бирюзовая кривая) и 1/x (розовая кривая) соответственно.

*2f*-сигнал, имеющий физический смысл интенсивности поглощения выбранной спектральной линии, согласно формуле (3.16), где длина оптического пути $L$ равна удвоенной дистанции $2h$ между прибором и рассеивающей поверхностью из формулы (4.2), описывающей зависимость от дистанции $h$ коэффициента потерь излучения на всей длине оптического пути от источника излучения до фотоприемника $K$, в результате будет зависеть от дистанции до рассеивающей поверхности как 1/x, что также проиллюстрировано на рисунке 4.28.

Таким образом, зависимость отношения *2f*-сигнала к *1f*-сигналу от расстояния до рассеивающей лазерное излучение поверхности, как следует из формулы (3.17), должна быть линейной. Иллюстрация полученной зависимости отношения *2f*-сигнала к *1f*-сигналу от расстояния до рассеивающей лазерное излучение поверхности для прибора ГИМЛИ и линейное приближение полученной экспериментальной зависимости приведена на рисунке 4.29.

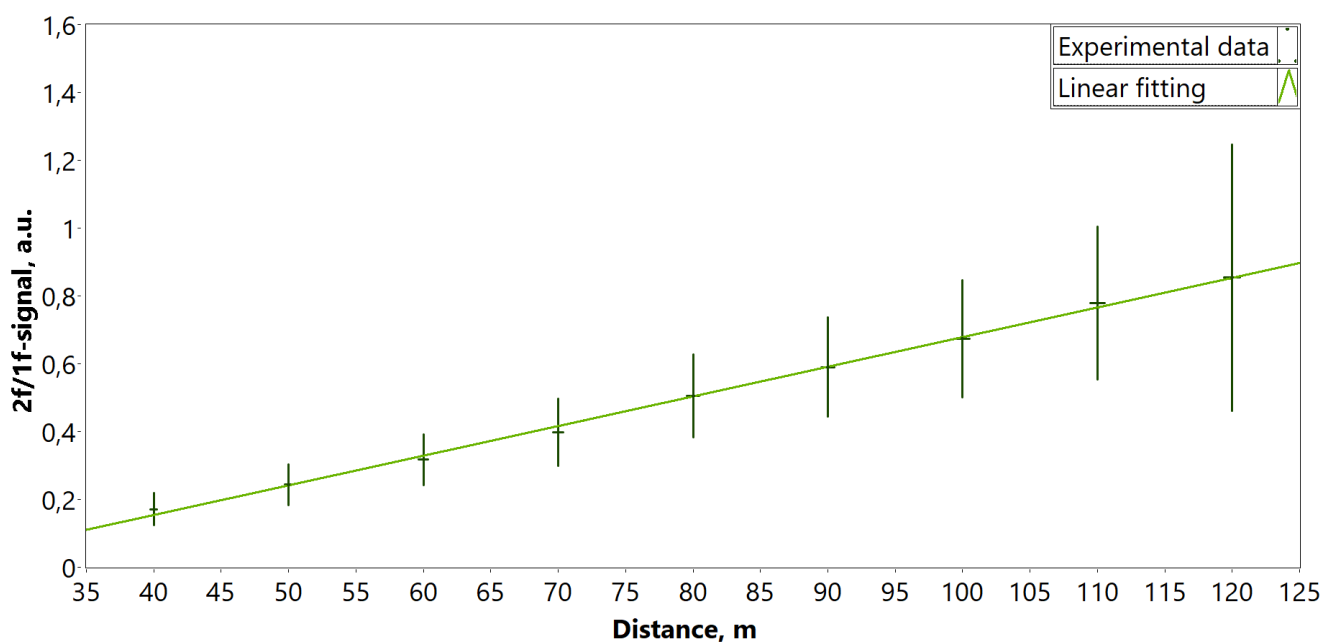

Рисунок 4.29 – Зависимость измеренных отношений *2f*-сигнала к *1f*-сигналу от дистанции между прибором и рассеивающей лазерное излучение поверхностью.

Как видно, линейное приближение хорошо описывает зависимость отношения *2f*-сигнала к *1f*-сигналу от расстояния до рассеивающей лазерное излучение поверхности.

### 4.3.1. Проведение калибровки прототипа прибора

С целью приведения результатов измерения содержания метана в атмосферном воздухе прототипа газоанализатора ГИМЛИ к размерности ppm·м была проведена калибровка прибора по независимым измерениям прецизионного газоанализатора LI-COR LI-7810, определяющего по заявлению производителя концентрацию CH₄ с точностью 2 ppb. Эксперимент проводился с установкой прототипа газоанализатора ГИМЛИ на треногу, позволяющую с хорошей точностью наводиться на экран из плотного хлопкового неокрашенного полотна, выставленного на определенную дистанцию от прибора. В ходе эксперимента дистанция менялась от 40 до 120 м с шагом 10 м. Температура окружающего воздуха почти не менялась и составляла ~26℃.

Устройство газоанализатора LI-7810 позволяет при помощи встроенного насоса прокачивать забираемый атмосферный воздух через аналитическую кювету для постоянного обновления анализируемой пробы. В ходе эксперимента оператор газоанализатора LI-7810 двигался вдоль оптического пути лазерного излучения прототипа газоанализатора ГИМЛИ, проходя дистанцию от треноги с прибором до экрана и обратно несколько раз для возможности последующего усреднения данных. Показания газоанализатора LI-7810 $N_{LI\text{-}COR}$, измеряемые в ppm, обновлялись с частотой $f_{LI\text{-}COR}$ = 1 Гц, исходя из чего по средней скорости движения оператора вдоль оптического пути $v_{op}$ можно получить величину содержания метана в воздухе в размерности ppm·м для дистанции $L$:

$$N_{ppm\cdot m} = \sum_{l=0}^{l=L} \left[ \frac{N_{LI\text{-}COR} \cdot v_{op}}{f_{LI\text{-}COR}} \right]. \tag{4.3}$$

Затем полученные значения усреднялись по числу проходов вдоль оптического пути. Найденная зависимость рассчитанных таким образом по показаниям газоанализатора LI-7810 интегральных концентраций метана от дистанции измерения $L$ при разных $L$ от 40 до 120 м с шагом 10 м показана на рисунке 4.30.

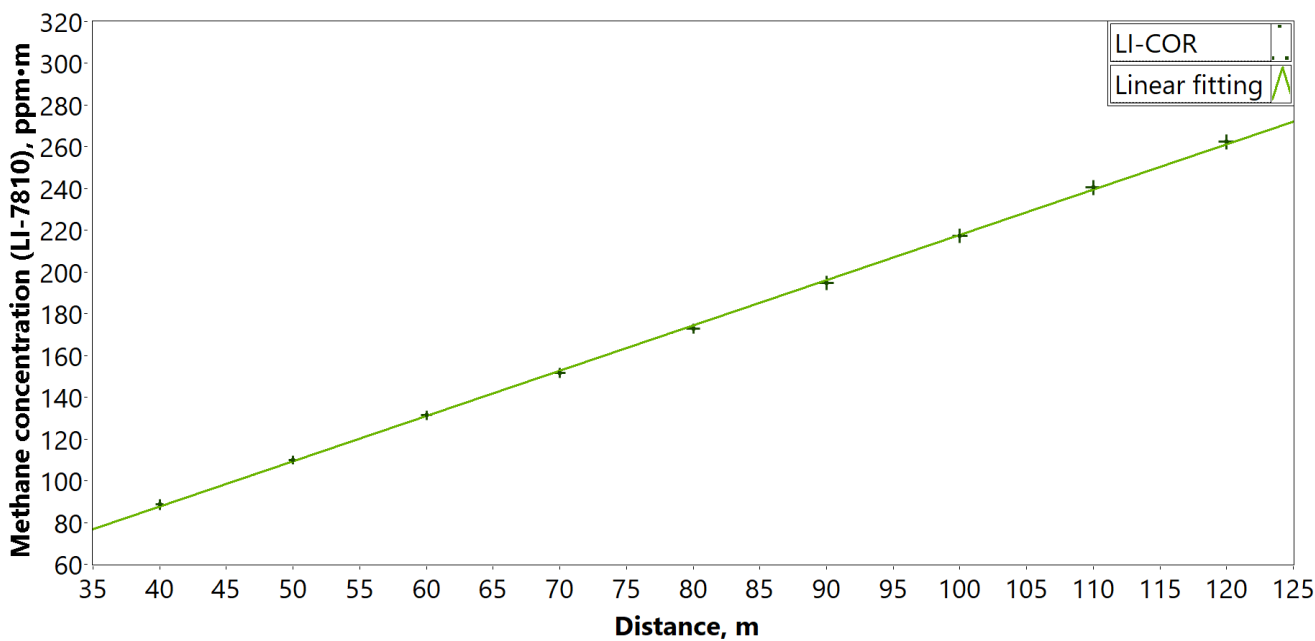

Рисунок 4.30 – Зависимость рассчитанных по показаниям газоанализатора LI-7810 интегральных концентраций метана от рабочей дистанции.

Ожидаемо ввиду хорошей перемешанности фонового метана в атмосферном воздухе зависимость оказалась линейной от расстояния. Основной вклад в погрешность определения величины $N_{ppm \cdot m}$ вносила точность измерения рабочей дистанции от оптической системы прототипа газоанализатора ГИМЛИ до экрана.

В случае каких-либо выбросов в показаниях газоанализатора LI-7810, связанных с изменениями направления движения воздуха, соответствующий набор данных отбраковывался при последующем анализе.

Для калибровки прототипа газоанализатора ГИМЛИ использовались данные, записанные за те же промежутки времени, что и неотбракованные данные газоанализатора LI-7810. По итогу обработки данных прототипа газоанализатора ГИМЛИ была получена зависимость отношения *2f*- к *1f*-сигналу от дистанции до рассеивающего экрана, представленная на рисунке 4.31.

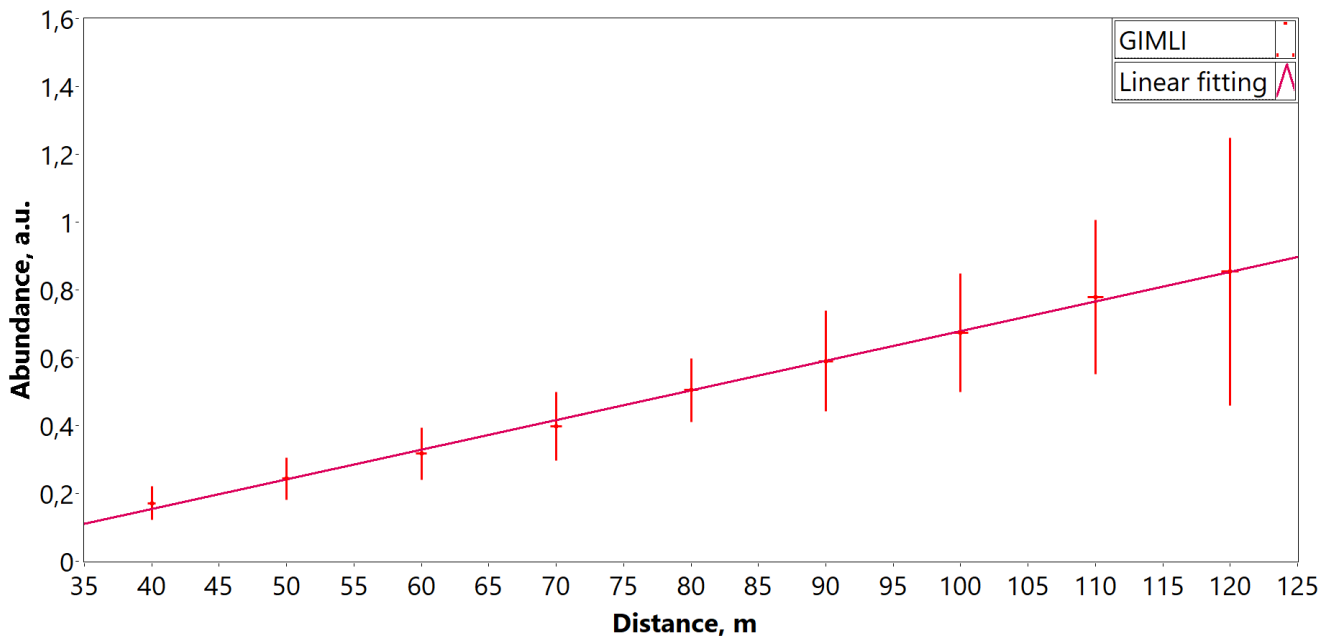

Рисунок 4.31 – Зависимость отношения *2f*- к *1f*-сигналу от дистанции до рассеивающего экрана. Экспериментально полученные значения с указанием погрешности их измерения и определения дистанции (красные точки) и линейное приближение экспериментальных данных (розовая линия).

Как видно, эта зависимость также хорошо описывается линейным приближением, что согласуется с равномерным распределением содержания метана в атмосферном воздухе вдоль трассы оптического пути. Тогда можно построить зависимость значений отношения *2f*- к *1f*-сигналу от величины $N_{ppm \cdot m}$ для соответствующих дистанций, показанную на рисунке 4.32.

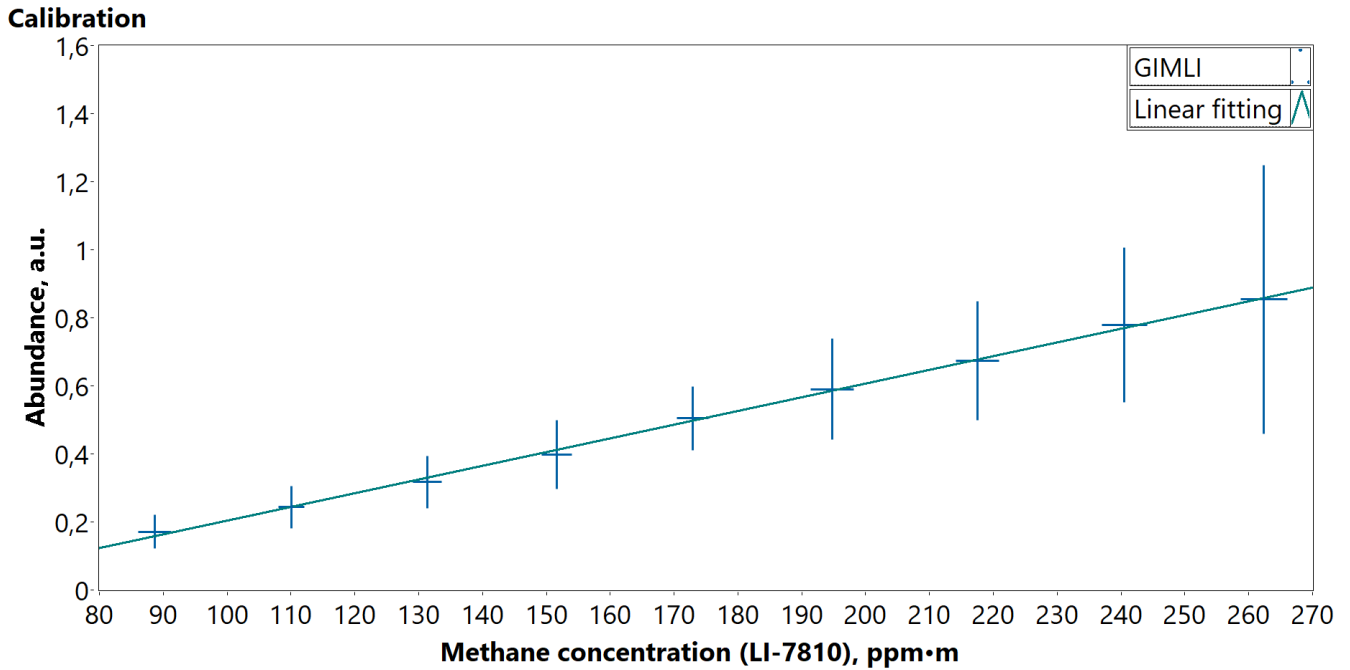

Рисунок 4.32 – Зависимость отношения *2f*- к *1f*-сигналу от рассчитанных по показаниям газоанализатора LI-7810 интегральных концентраций метана для разных дистанций. Экспериментально полученные значения с указанием погрешности измерения (синие точки) и линейное приближение полученной зависимости (бирюзовая линия).

Коэффициенты полученного линейного приближения зависимости значений отношения *2f*- к *1f*-сигналу от величины $N_{ppm\cdot m}$ для соответствующих дистанций были приняты в качестве калибровочных для последующих измерений содержания метана в атмосферном воздухе при помощи прототипа газоанализатора ГИМЛИ.

Также по результатам проведенной калибровки прибора была получена зависимость чувствительности прибора от дистанции до рассеивающей поверхности. При определении чувствительности прототипа газоанализатора по фоновой концентрации метана в атмосферном воздухе на разных дистанциях до рассеивающей излучение поверхности использовались данные, записанные в течение 10 с для каждой дистанции. По вычисленным значениям среднего $M$ и среднеквадратичного отклонения отношения *2f*- к *1f*-сигналу чувствительность для каждой дистанции определялась согласно формуле:

$$D_{GIMLI} = N_{ppm\cdot m} \cdot \frac{STD\left[\frac{2f-signal}{1f-signal}\right]}{M\left[\frac{2f-signal}{1f-signal}\right]}, \qquad (4.4)$$

где $N_{ppm\cdot m}$ – концентрация метана в единицах ppm·м по независимым измерениям прецизионного газоанализатора LI-COR LI-7810.

Интервал рабочих дистанций прибора де-факто определяется юстировкой оптической системы прибора на 50 м снизу и превышением интенсивностью полезного сигнала уровня шумов прибора, обусловленных преимущественно используемыми электронными схемами и наводками между двумя каналами прибора, – то есть отношением сигнала к шуму $SNR$ >1.5 – сверху. Отношение среднеквадратичного отклонения к среднему значению отношения $2f$- к $1f$-сигналу остается постоянным на всем интервале рабочих дистанций. Таким образом, чувствительность прибора ГИМЛИ должна линейно зависеть от дистанции до рассеивающей поверхности, поскольку в силу равномерности перемешивания фонового метана в атмосферном воздухе параметр $N_{ppm\cdot m}$ в формуле (4.4) будет зависеть от расстояния линейно.

Сравнение определенной для представленного газоанализатора зависимости чувствительности от дистанции между прибором и рассеивающей излучение поверхностью с заявляемыми чувствительностями коммерчески доступных газоанализаторов метана, пригодных для установки на борт БПЛА, показано на рисунке 4.33.

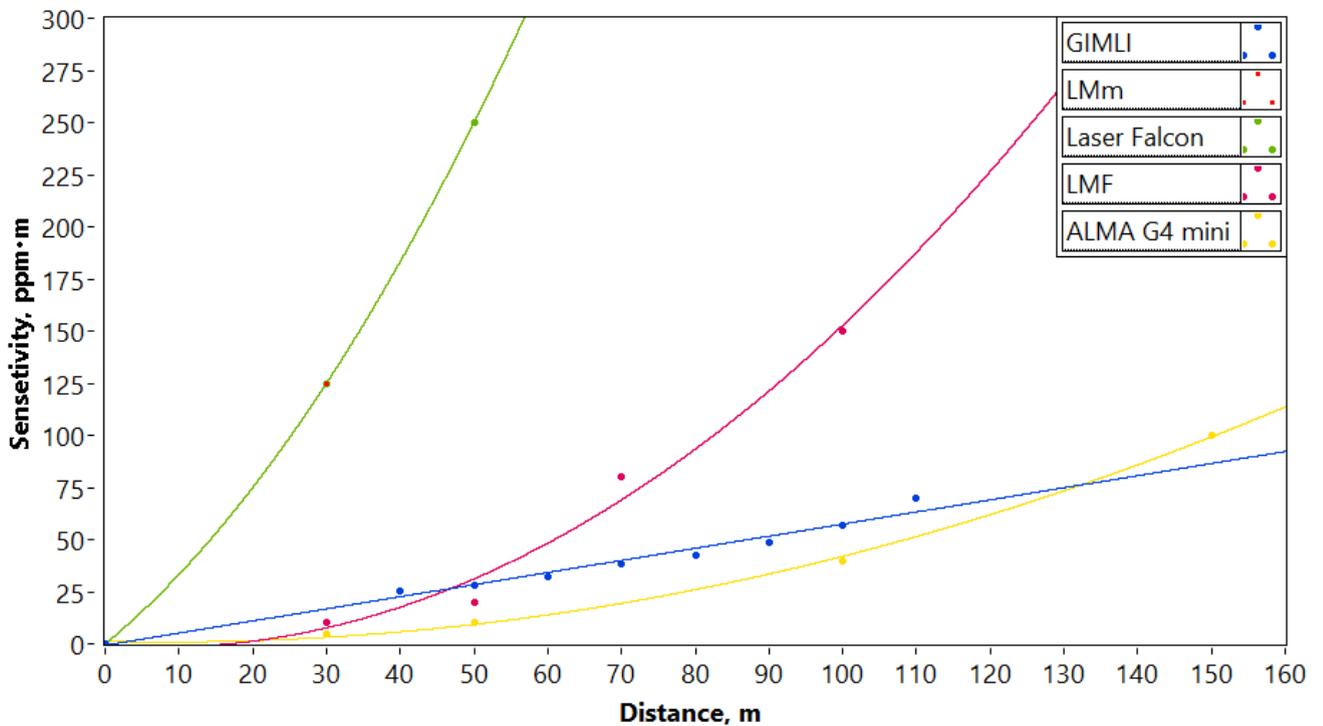

Рисунок 4.33 – Сравнение зависимости чувствительности от дистанции между прибором и рассеивающей излучение поверхностью прибора ГИМЛИ с заявляемыми чувствительностями коммерчески доступных газоанализаторов метана, пригодных для установки на борт БПЛА.

Как показала реализация предложенной методики модуляционной спектроскопии с квадратурным детектированием сигнала, формула (3.29) для вычисления расстояния от прибора до рассеивающей излучение лазера поверхности верна с точностью до коэффициента пропорциональности и свободного члена – сдвига фазы, обусловленного ФЧХ предусилителя аналитического канала. Таким образом, определение дистанции оказывается возможным при проведении калибровки по независимо измеренным расстояниям. Результаты такой калибровки представлены на рисунке 4.34.

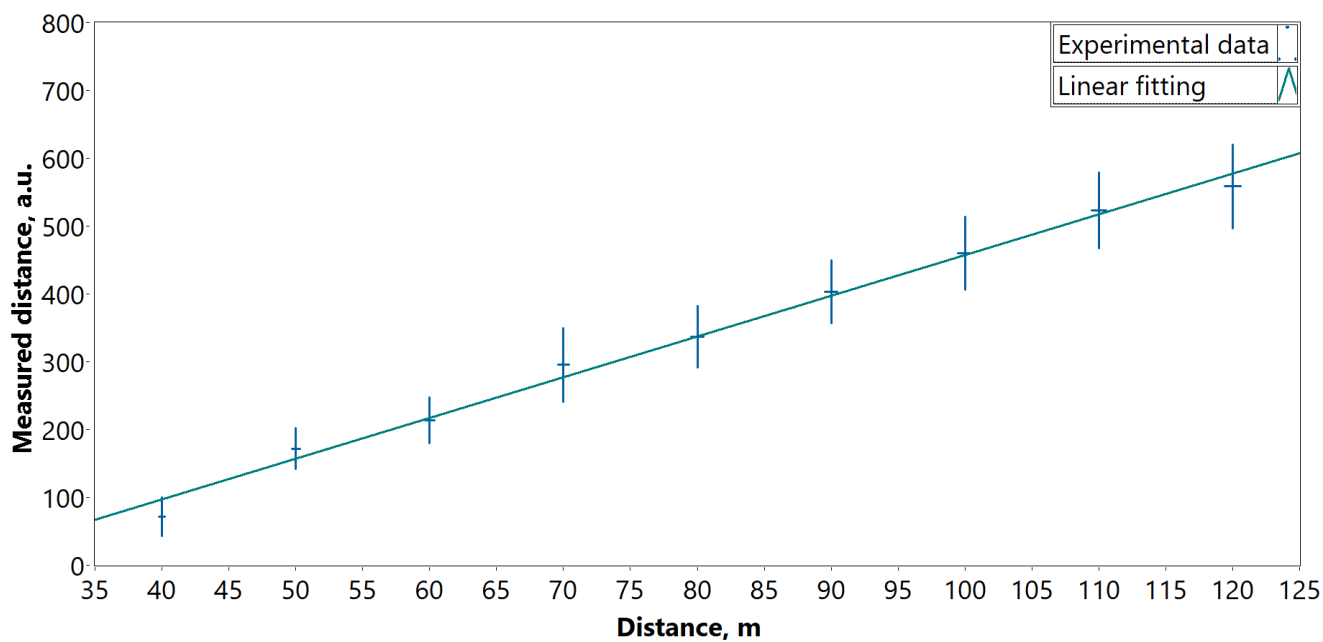

Рисунок 4.34 – Результаты калибровки методики измерения дистанций до рассеивающей поверхности по независимо измеренным расстояниям.

Таким образом, разработанный прототип газоанализатора ГИМЛИ может применяться для работ по мониторингу содержания метана в атмосферном воздухе, измеряя интегральную концентрацию метана в единицах ppm·м при независимом определении дистанции до рассеивающей поверхности с известной чувствительностью.

### 4.3.2. Особенности фотонных компонентов прототипа газоанализатора

Как упоминалось ранее, для регистрации рассеянного от подстилающей поверхности излучения был выбран InGaAs-фотодиод ThorLabs FGA21 с размером светочувствительной площадки 2 мм. В ходе работ с разработанным прототипом газоанализатора наблюдалось несоответствие получаемых значений отношения *2f*- к *1f*-сигналу при разной температуре окружающего воздуха. После проведения ряда исследований стало понятно, что такая зависимость от температуры связана с особенностью выбранного фотодиода аналитического канала. На рисунке 4.35 приведена зависимость чувствительности *R* фотодиода FGA21 от длины волны принимаемого излучения. Эта характеристика справедлива для комнатной температуры.

Указанный производителем параметр эквивалентной мощности шумов (Noise Equivalent Power, NEP) для длины волны 1550 нм можно пересчитать для рабочей длины волны 1651 нм как

$$NEP_{1651nm} = NEP_{1550nm} \frac{R_{1550nm}}{R_{1651nm}} \approx 6.4 \cdot 10^{-14}. \qquad (4.5)$$

При понижении температуры резкий спад чувствительности в области 1650-1700 нм сдвигается в сторону меньших длин волн. Таким образом, при работе на длине волны 1651 нм, отмеченной на рисунке 4.35 вертикальной линией, наблюдается явно выраженная зависимость чувствительности или уровня эквивалентной мощности шумов ФД от температуры.

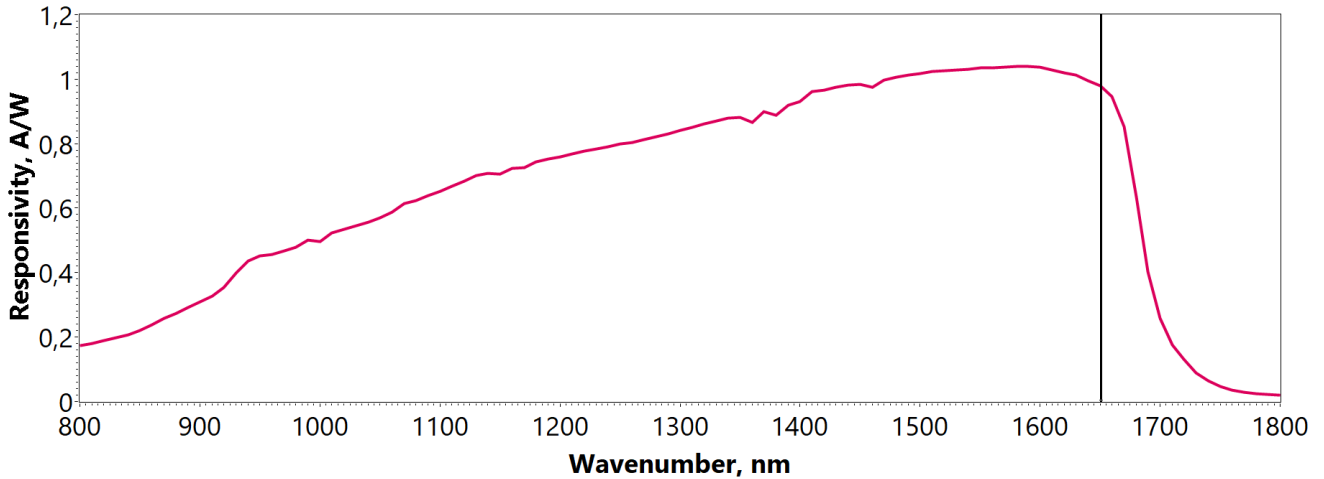

Рисунок 4.35 – Зависимость чувствительности фотодиода FGA21 от длины волны излучения.

Была проведена серия экспериментов с целью изучения отклика фотоприемного устройства при температурах окружающего воздуха от -11℃ до 26℃. При температуре -11℃ усредненный сигнал, регистрируемый в аналитическом канале при дистанции до рассеивающей поверхности 50 м, имеет размах в значениях АЦП ~54, а при 26℃ ~191 (рисунок 4.36).

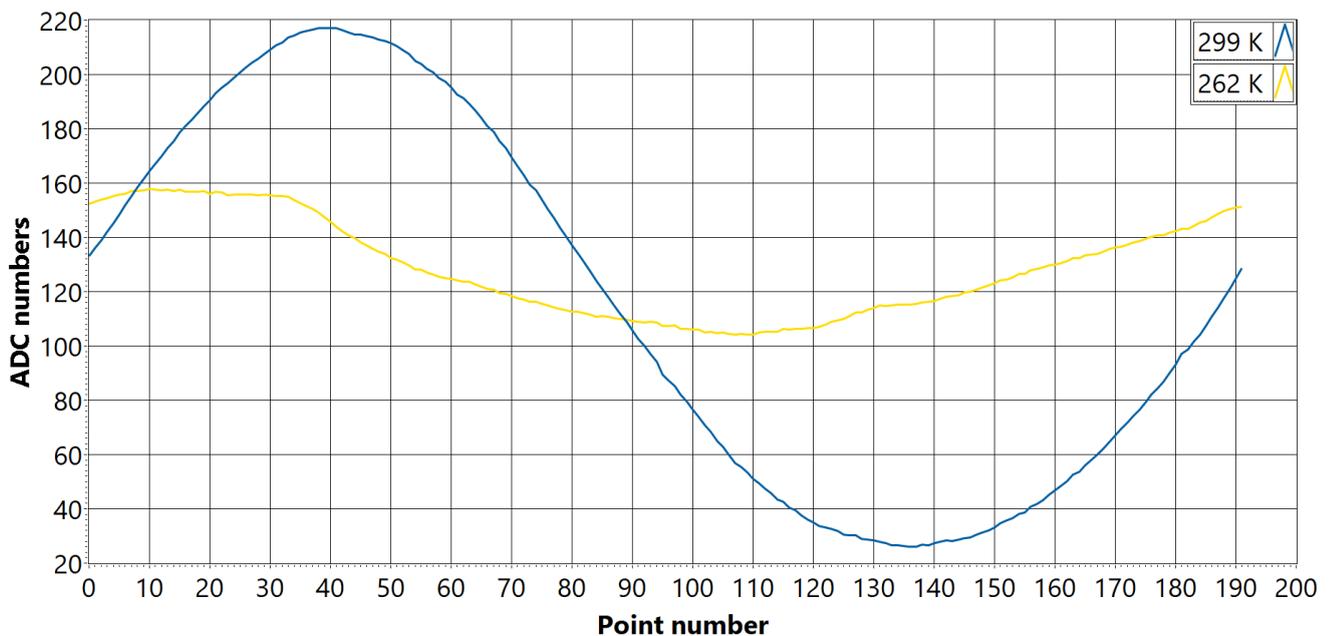

Рисунок 4.36 – Усредненный сигнал, регистрируемый в аналитическом канале при расстоянии до рассеивающей поверхности 50 м, снятый при температуре окружающего воздуха 26℃ (синяя кривая) и -11℃ (желтая кривая).

То есть при одинаковой фоновой концентрации метана и одинаковой дистанции до рассеивающей поверхности при изменении температуры окружающего воздуха на 37℃ отличие в амплитуде сигнала фотоприемника будет в 3.5 раза. Пренебречь таким фактором не представляется возможным.

Зависимость интенсивности регистрируемого в аналитическом канале сигнала при дистанции до рассеивающей поверхности 50 м от температуры окружающего воздуха в интервале от -11℃ до 26℃ имеет вид, представленный на рисунке 4.37.

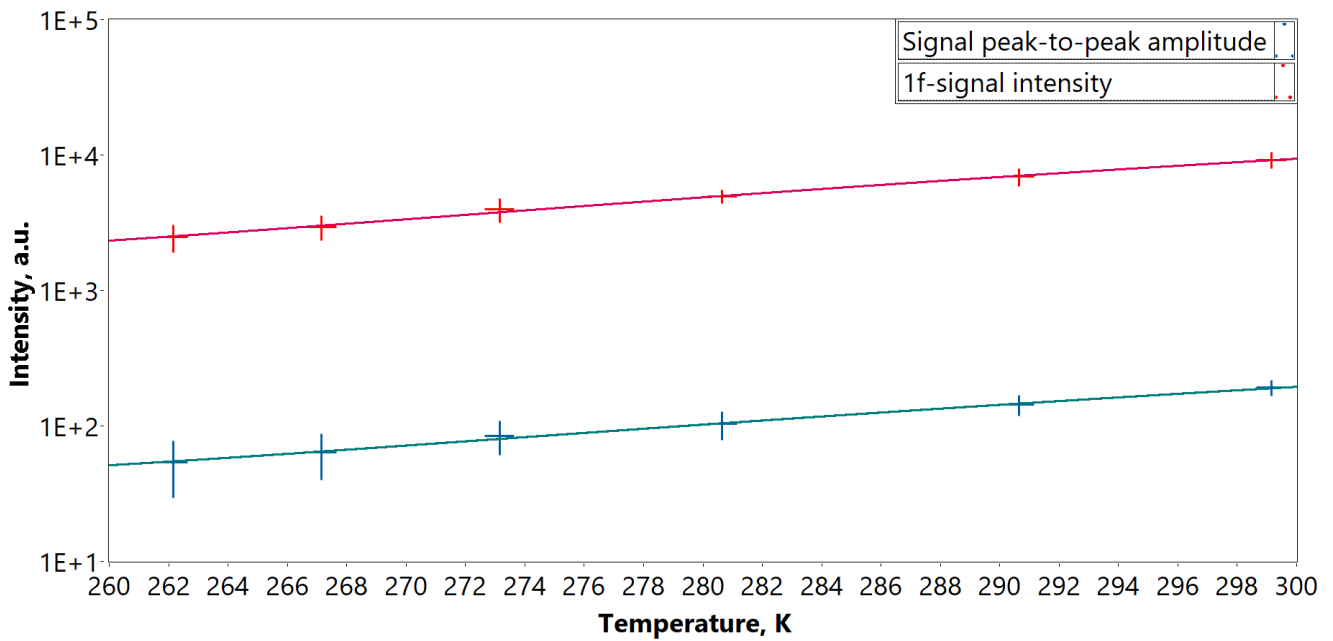

Рисунок 4.37 – Зависимость удвоенной амплитуды сигнала, регистрируемого в аналитическом канале (бирюзовая кривая), и интенсивности *1f*-сигнала (розовая кривая) от температуры окружающего воздуха в интервале от -11℃ до 26℃.

Как видно, интенсивность сигнала нелинейно растет с ростом температуры, и вблизи 25℃ ФД выходит на режим, описанный производителем. Важным фактором в случае применения алгоритмов обработки квадратурного детектирования синусоидально модулированного сигнала будет поведение вычисляемых гармонических составляющих сигнала при такой зависимости интенсивности регистрируемого в аналитическом канале сигнала от температуры окружающего воздуха.

*1f*-сигнал, имеющий физический смысл общей интенсивности принимаемого излучения, как видно из рисунка 4.37, будет меняться при изменении температуры по тому же закону, что и интенсивность сигнала. При этом изменение *2f*-сигнала, соответствующего интенсивности поглощения выбранной спектральной линии, будет находиться в пределах интервала ±0.5σ.

Таким образом, зависимость отношения *2f*-сигнала к *1f*-сигналу от температуры будет иметь обратный характер относительно интенсивности регистрируемого сигнала, что видно на рисунке 4.38.

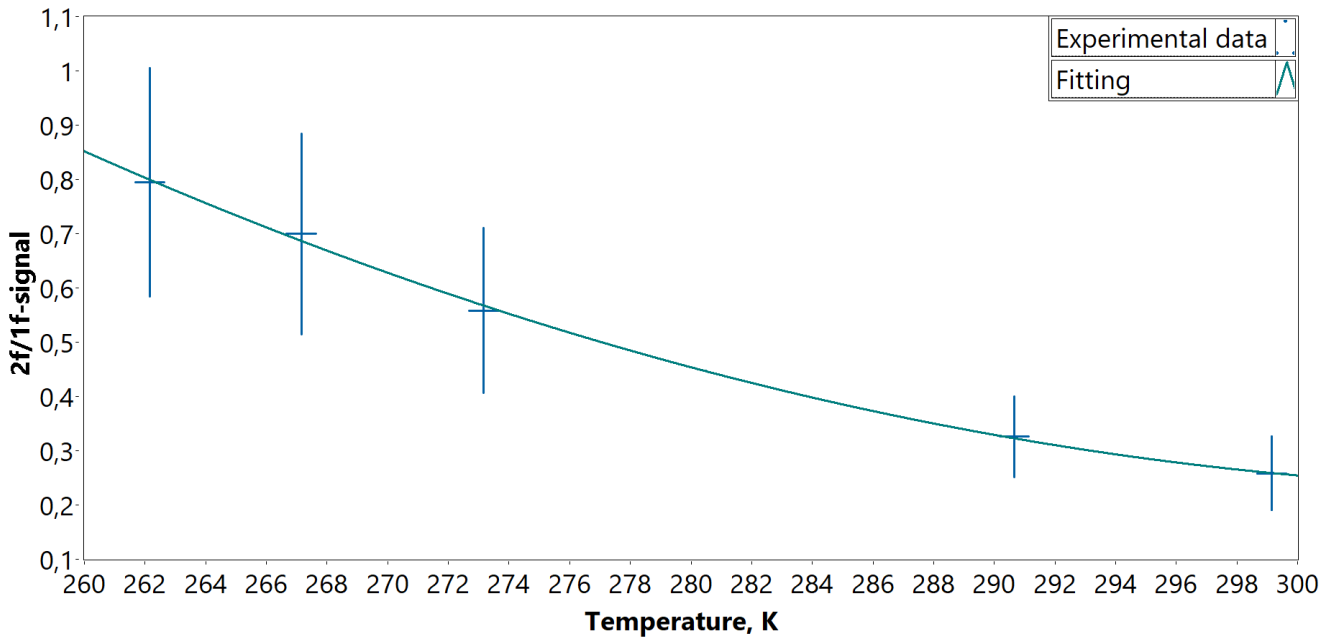

Рисунок 4.38 – Зависимость отношения *2f*-сигнала к *1f*-сигналу от температуры окружающего воздуха в интервале от -11℃ до 26℃.

На рисунке 4.39 представлено поведение зависимости отношения *2f*-сигнала к *1f*-сигналу, пропорционального измеряемой интегральной концентрации метана, от дистанции до рассеивающей поверхности при разных температурах окружающего воздуха.

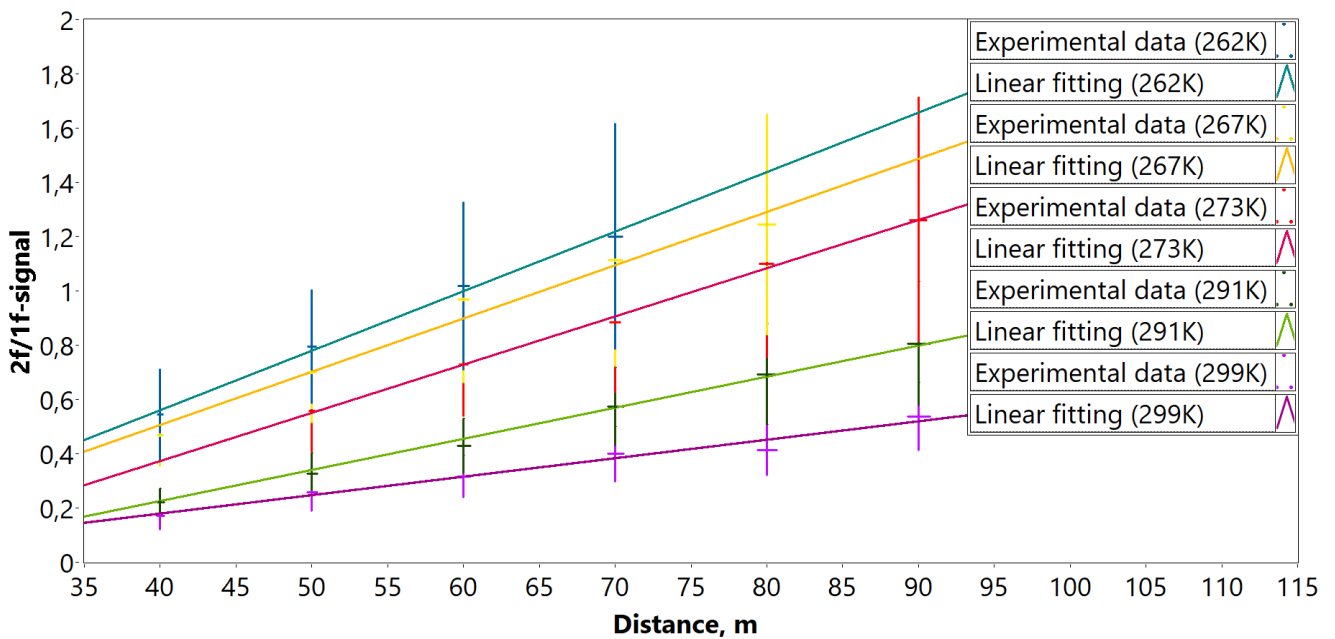

Рисунок 4.39 – Поведение зависимости отношения *2f*-сигнала к *1f*-сигналу от дистанции до рассеивающей поверхности при разных температурах окружающего воздуха.

Можно заметить, что изменение наклона приведенных зависимостей выглядит как линейная функция от температуры окружающего воздуха. На рисунке 4.40 отражено поведение углового коэффициента линейной зависимости отношения *2f*-сигнала к *1f*-сигналу от дистанции до рассеивающей поверхности при изменении температуры окружающего воздуха.

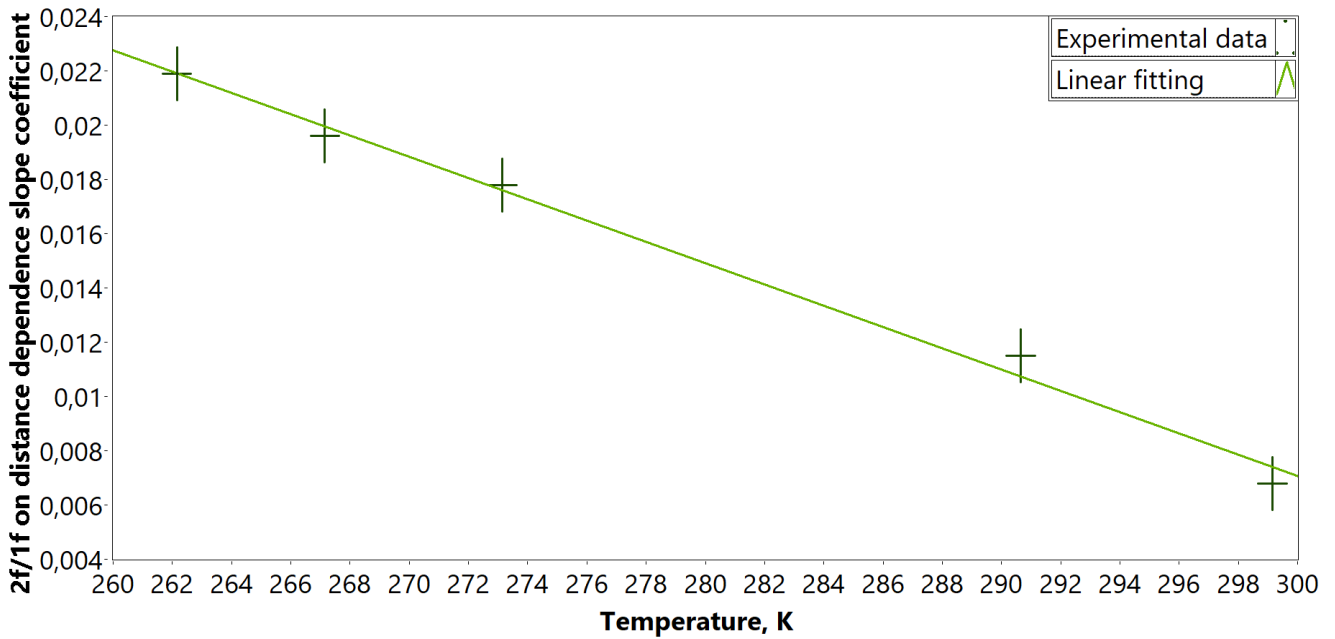

Рисунок 4.40 – Поведение углового коэффициента линейной зависимости отношения *2f*-сигнала к *1f*-сигналу от дистанции до рассеивающей поверхности при изменении температуры окружающего воздуха.

Как видно из рисунка 4.40, изменение наклона зависимости отношения *2f*-сигнала к *1f*-сигналу от дистанции до рассеивающей поверхности действительно хорошо описывается линейной функцией от температуры окружающего воздуха. Таким образом, последствия влияния зависимости чувствительности фотодиода ThorLabs FGA21 на длине волны 1651 нм от температуры окружающего воздуха на вычисляемое отношение *2f*-сигнала к *1f*-сигналу, пропорциональное измеряемой интегральной концентрации метана на дистанции до рассеивающей лазерное излучение поверхности, можно учитывать алгоритмически.

Замена текущего ФД на другой неохлаждаемый ФД, рассчитанный на используемый спектральный диапазон, не имеет смысла из-за меньшей эквивалентной мощности собственных шумов и меньшей ёмкости используемого ФД по сравнению с другими неохлаждаемыми ФД с тем же размером фоточувствительной площадки, являющейся оптимальной для предусмотренных задач. Замена на охлаждаемый InGaAs-ФД возможна в следующей версии газоанализатора.

При алгоритмическом учете описанных особенностей поведения используемого ФД максимальная рабочая дистанция до рассеивающей поверхности будет повторять зависимость интенсивности регистрируемого в аналитическом канале сигнала от температуры окружающего воздуха и меняться в изученном диапазоне температур от ~70 м при -11℃ до ~120 м при 26℃. При необходимости эту проблему можно компенсировать путем добавления схемы аналогового ПИД-регулятора нагревателя ФД. Такая схема планируется к реализации.

### 4.3.3. Методы борьбы с солнечными засветками

Результаты проведенных экспериментов среди прочего указывают на одну существенную проблему применения разработанного прототипа прибора. Попадание прямого или в некоторых случаях рассеянного яркого солнечного света в фотоприемный канал вводит его в режим насыщения, тем самым приводя к некорректным показаниям прибора. Кажущееся естественным в такой ситуации решение – использование оптического метода фильтрации принимаемого излучения – не удовлетворяет требованиям к компактному устройству малой массы.

В случае использования оптического фильтра нет возможности установить малогабаритную пластину перед фотодиодом, поскольку в таком случае фильтр окажется на пути сходящегося пучка излучения, фокусируемого на чувствительную площадку фотодиода. Установка такого фильтра возможна только в параллельном пучке.

В таком случае при установке фильтра перед коллимирующей на фотодиод излучение входной линзой его диаметр окажется равным диаметру линзы. При типичных толщинах оптических фильтров 3 мм и диаметре собирающей линзы в следующей версии прибора 30 см получим фильтр массой более 0.5 кг, что достаточно критичная добавка к массе прибора, которая не должна превышать 4-5 кг для возможности нормальной эксплуатации в качестве полезной нагрузки БПЛА.

Однако существует возможность использования фильтрации принимаемого излучения на этапе прохождения фототоком цепи усиления сигнала. Следует отметить, что компенсация внешней засветки в электрических цепях возможна лишь в тех случаях, когда полезный сигнал и внешняя засветка находятся в разных частотных диапазонах. В случае разработанного газоанализатора полезный сигнал находится в диапазоне 25-75 кГц, соответственно, возможно эффективное подавление внешней засветки на частотах ниже единиц кГц.

Неверно разработанная схема может решить проблему внешней засветки, но повлечет возникновение других проблем: существенное увеличение шумов, изменение параметров фотодиода в зависимости от уровня внешней засветки и др.

В качестве варианта, позволяющего добиться результата моделирования, наиболее близкого к желаемому, была подобрана схема с биполярным транзистором, включенным по схеме с общим эмиттером, где в обратную связь трансимпедансного каскада включен инвертирующий интегратор.

Полная схема, учитывающая использование однополярного питания +5 В, необходимость дополнительного усилителя, фильтров верхних и нижних частот, а также выходного дифференциального драйвера, приведена на рисунке 4.41. Усилитель рассчитан на фотодиод FGA21 с размером чувствительной площадки 2 мм и емкостью 150 пФ при обратном смещении 2 В.

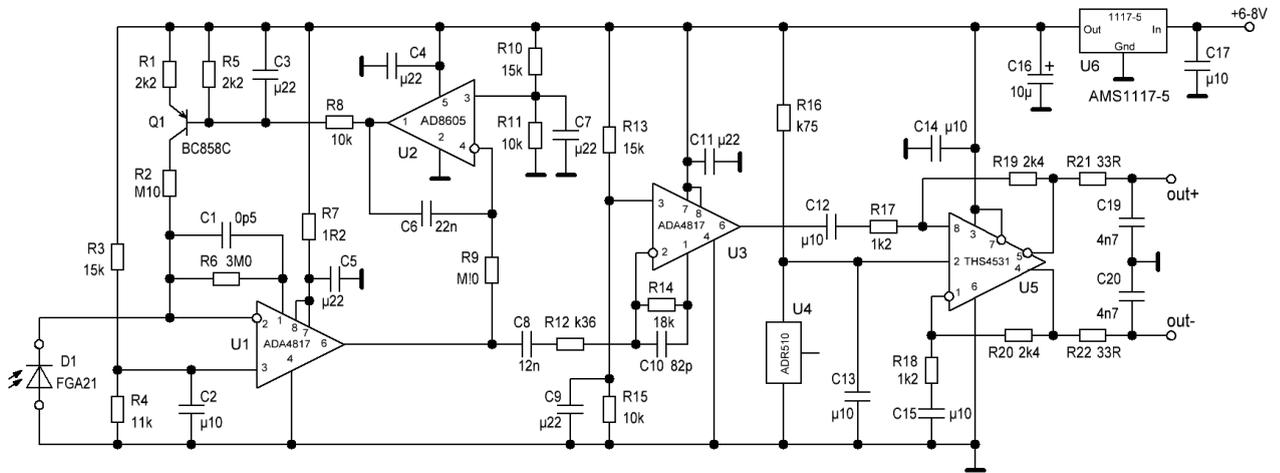

Рисунок 4.41 – Полная схема усилителя с компенсацией внешней засветки.

На рисунке 4.42 показана зависимость эквивалентного сопротивления усилителя от частоты. Коэффициент передачи для рабочего диапазона частот вблизи 50 кГц составляет около 200 МОм.

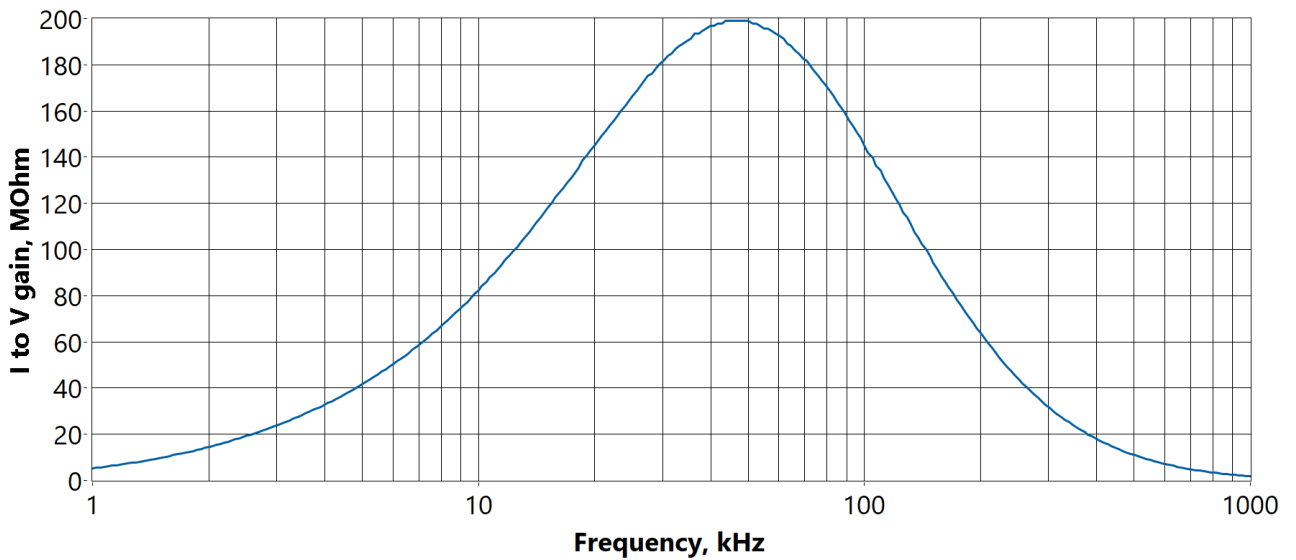

Рисунок 4.42 – Эквивалентное сопротивление предусилителя фототока для борьбы с солнечными засветками.

Подавление низкочастотных сигналов засветки в разработанной схеме, как видно, окажется весьма существенным, поскольку усиление составляющих принимаемого сигнала на частотах до ~400 Гц не превышает значений в сотни кОм, как видно из рисунка 4.43, демонстрирующего поведение эквивалентного сопротивления для низких частот в двойном логарифмическом масштабе.

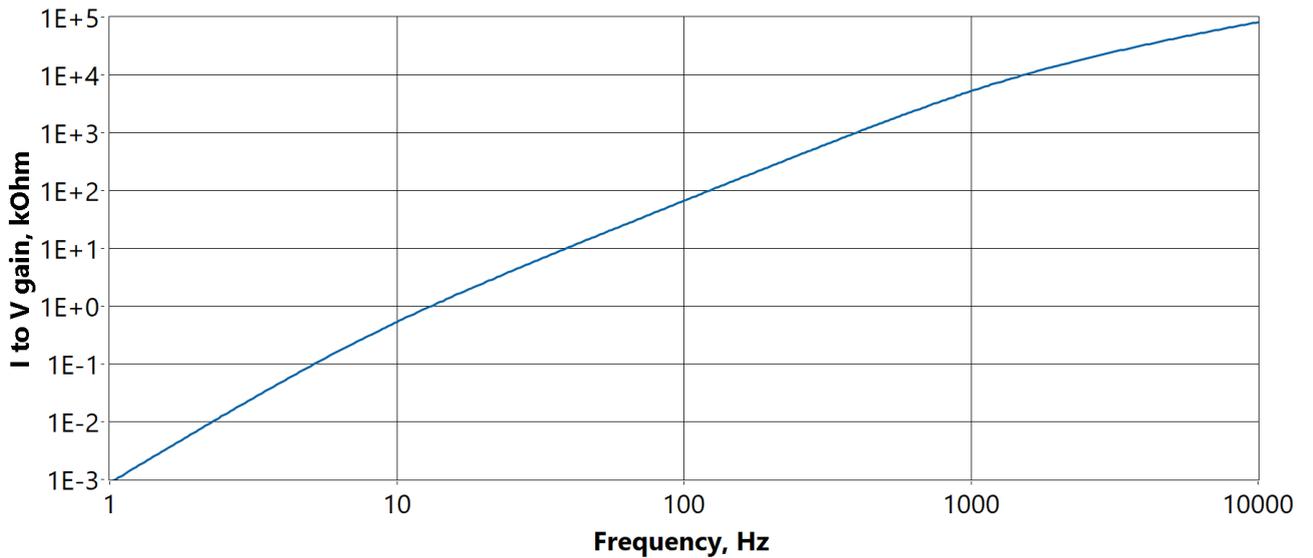

Рисунок 4.43 – Поведение эквивалентного сопротивления в области низких частот.

Амплитудно-частотные характеристики (АЧХ) и фазово-частотные характеристики (ФЧХ) разработанного предусилителя с компенсацией солнечных засветок представлены на рисунке 4.44.

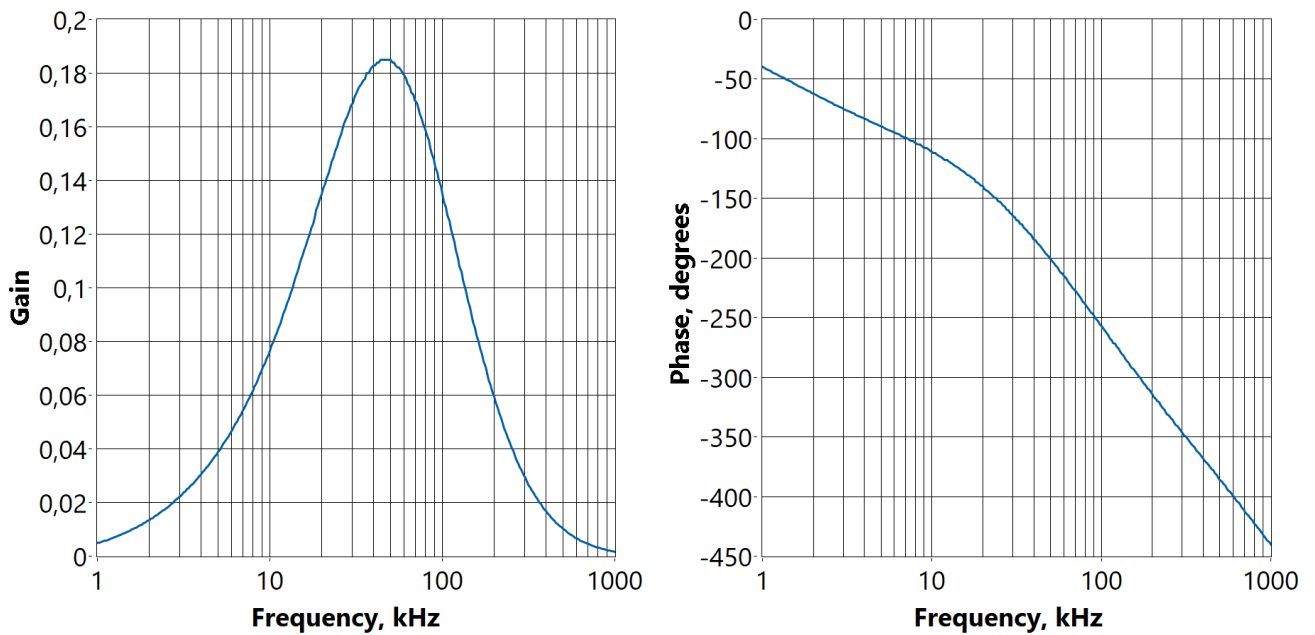

Рисунок 4.44 – АЧХ (слева) и ФЧХ (справа) разработанного предусилителя фототока для борьбы с солнечными засветками.

Как упоминалось в главе 3, ошибки измерения расстояния по фазе первой гармоники могут быть связаны с зависимостью ФЧХ от амплитуды сигнала. Проведенное моделирование показало, что на частоте 26 кГц при изменении выходного сигнала от 1 мВ до 2 В фаза этого сигнала меняется от -148.755° до -148.759°, то есть всего на 0.004°.

Описанный подход к фильтрации солнечных засветок, требующий разработки соответствующей схемы предусилителя фотодиода аналитического канала, был реализован при проектировании новой версии газоанализатора ГИМЛИ.

## 4.4. Разработка новой версии прибора

С учетом сделанных выводов по итогам отладки прототипа газоанализатора ГИМЛИ было начато проектирование новых версий прибора. Для возможности вариативности предельных параметров чувствительности прибора было разработано два варианта корпусирования прибора с использованием модульной структуры и частичном применении общих элементов для обоих вариантов. На рисунке 4.45 представлена модель варианта прибора с использованием в качестве собирающей на фотодиод аналитического канала оптики акриловой линзы Френеля диаметром 350 мм.

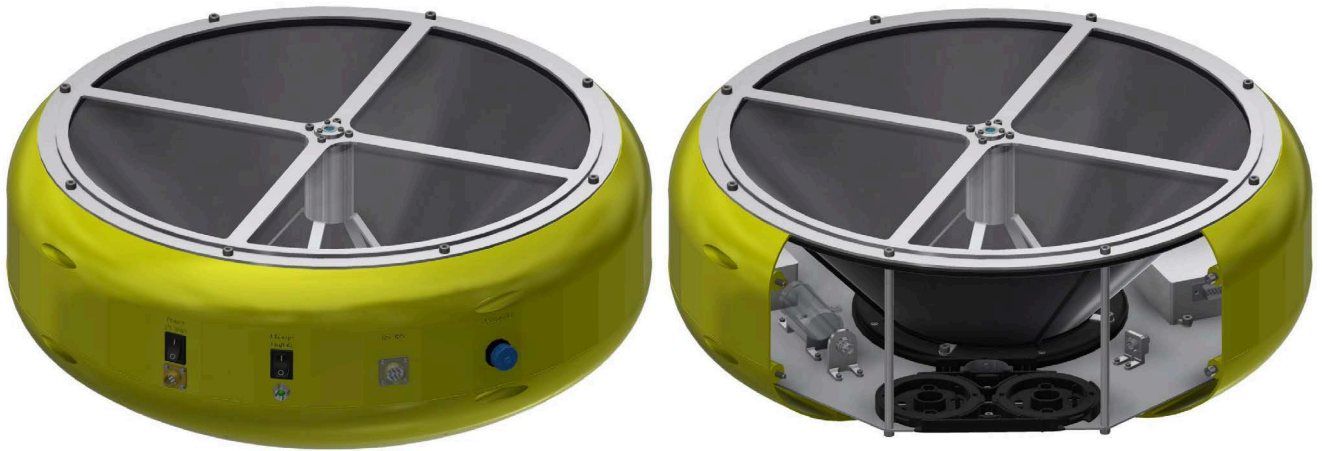

Рисунок 4.45 – Модель корпуса прибора с акриловой линзой Френеля диаметром 350 мм.

Масса такой линзы – 211 г, что на 30 г меньше массы линзы диаметром 100 мм, используемой в прототипе прибора, при этом поглощение акрила на используемой длине волны для толщины выбранной линзы 3 мм составляет ~20%, что не драматически сказывается на итоговой собираемой мощности излучения. Однако из-за большого диаметра линзы фокусное расстояние также не может быть слишком малым для корректной работы сборки приемника излучения, в данном случае F = 185 мм, что с учетом компоновки внутри корпуса электронных плат приводило к слишком большой высоте корпуса прибора. По этой причине была выбрана оптическая схема с использованием плоского зеркала с металлическим напылением для уменьшения высоты оптической схемы аналитического канала прибора почти вдвое. При достигнутых габаритах корпус подходит для установки на такой же БПЛА, как используемый для полетов с прототипом прибора, что показано на рисунке 4.46.

При этом по результатам моделирования аэродинамическое сопротивление набегающему потоку воздуха у новой версии прибора значительно меньше, чем у прототипа прибора, благодаря обтекаемой форме корпуса.

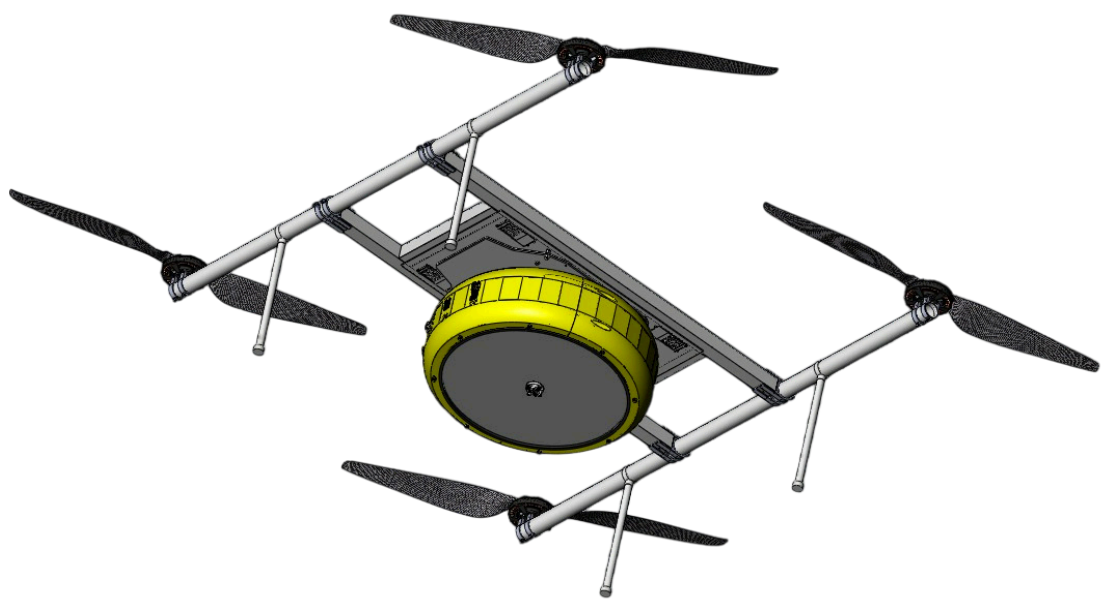

Рисунок 4.46 – Модель корпуса газоанализатора с принимающей линзой диаметром 350 мм, установленного на квадрокоптер «Ирбис-432».

Как было сказано выше, для вариативности предельных параметров чувствительности прибора был разработан корпус прибора на базе линзы Френеля той же толщины с диаметром 200 мм и фокусным расстоянием 120 мм, что позволило упростить оптическую схему прибора, разместив фотодиод аналитического канала в фокусе линзы без использования зеркал. В обоих вариантах корпусирования используется унифицированные элементы: ряд корпусных деталей, используемые лазерные диоды и фотодиоды, плата предусилителя фотодиода реперного канала и основная аналого-цифровая электронная плата управления (рисунок 4.47), позволившая избавиться от связанных с соединительными кабелями между цифровой платой управления и аналоговой платой драйвера лазера наводок, наблюдаемых в прототипе прибора.

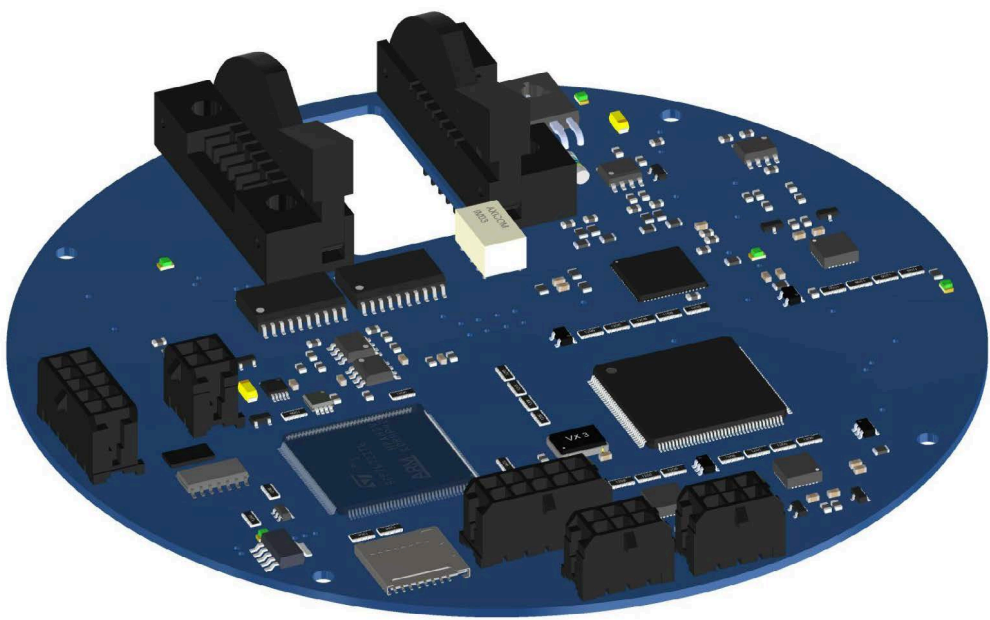

Рисунок 4.47 – Модель основной аналого-цифровой электронной платы управления.

Второй корпус имеет такую же обтекаемую форму, но заметно меньший диаметр и немного бо́льшую высоту. Благодаря бо́льшим диаметрам используемых собирающих рассеянное излучение линз нежели в прототипе прибора уровень полезного сигнала на той же дистанции от рассеивающей поверхности в новых версиях прибора будет в 4 и 12 раз выше соответственно для версии с 200 мм и 350 мм линзой Френеля. Вариант корпуса с линзой Френеля диаметром 200 мм представлен на рисунке 4.48.

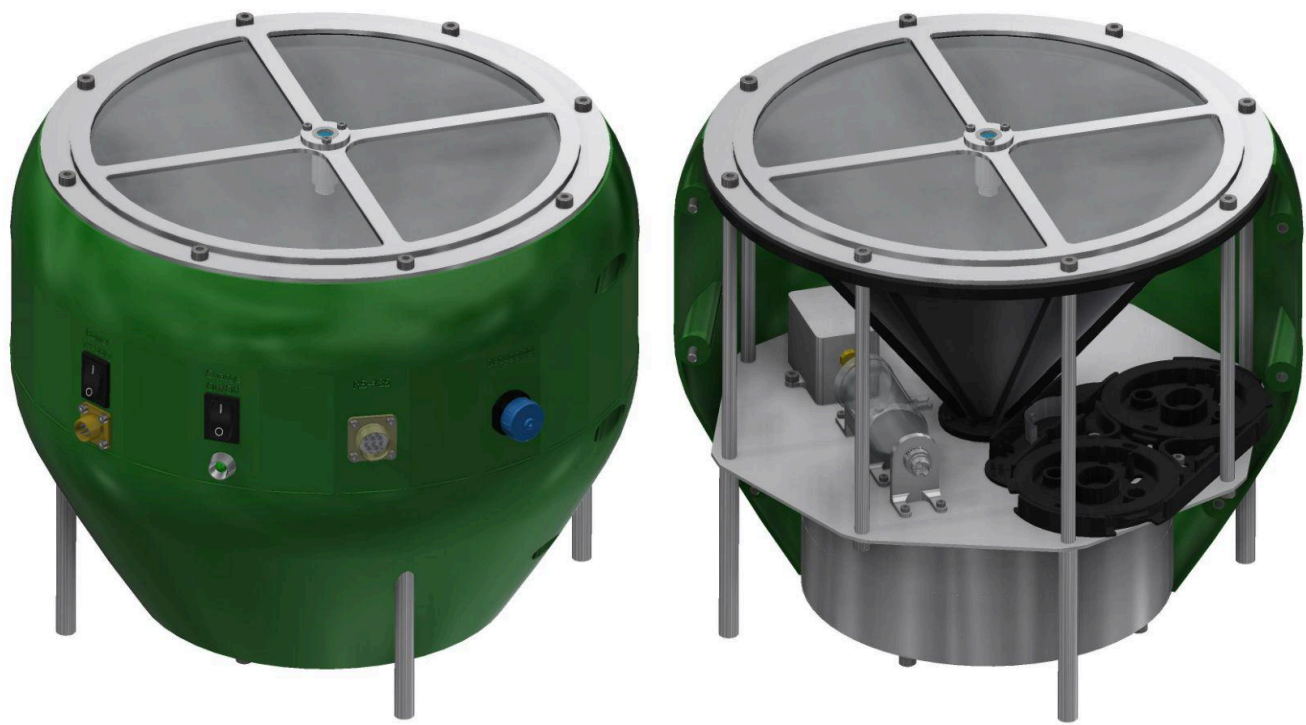

Рисунок 4.48 – Модель корпуса прибора с акриловой линзой Френеля диаметром 200 мм.

Также для новых версий газоанализатора ГИМЛИ предусмотрено применение диодных РОС-лазеров компании AeroDiode (Франция) 1650LD-3 или 1650LD-4 с выходной мощностью излучения 40 мВт или 100 мВт соответственно. Это в свою очередь увеличит уровень полезного сигнала еще в ~6-15 раз. Таким образом, увеличение чувствительности новых версий прибора может быть в ~25-185 раз относительно показателей для прототипа газоанализатора.

На рисунке 4.49 представлено сравнение экспериментально установленной чувствительности прототипа прибора в зависимости от дистанции до рассеивающей излучение поверхности и оценок чувствительности новых версий прибора – случая худшей чувствительности при использовании схемы с 200 мм линзой Френеля и выходной мощностью лазерного излучения 40 мВт и лучшей при использовании схемы с 350 мм линзой Френеля и выходной мощностью лазерного излучения 100 мВт. Для расстояния 50 м эти значения составят соответственно ~28 ppm·м, ~1.1 ppm·м и ~0.15 ppm·м.

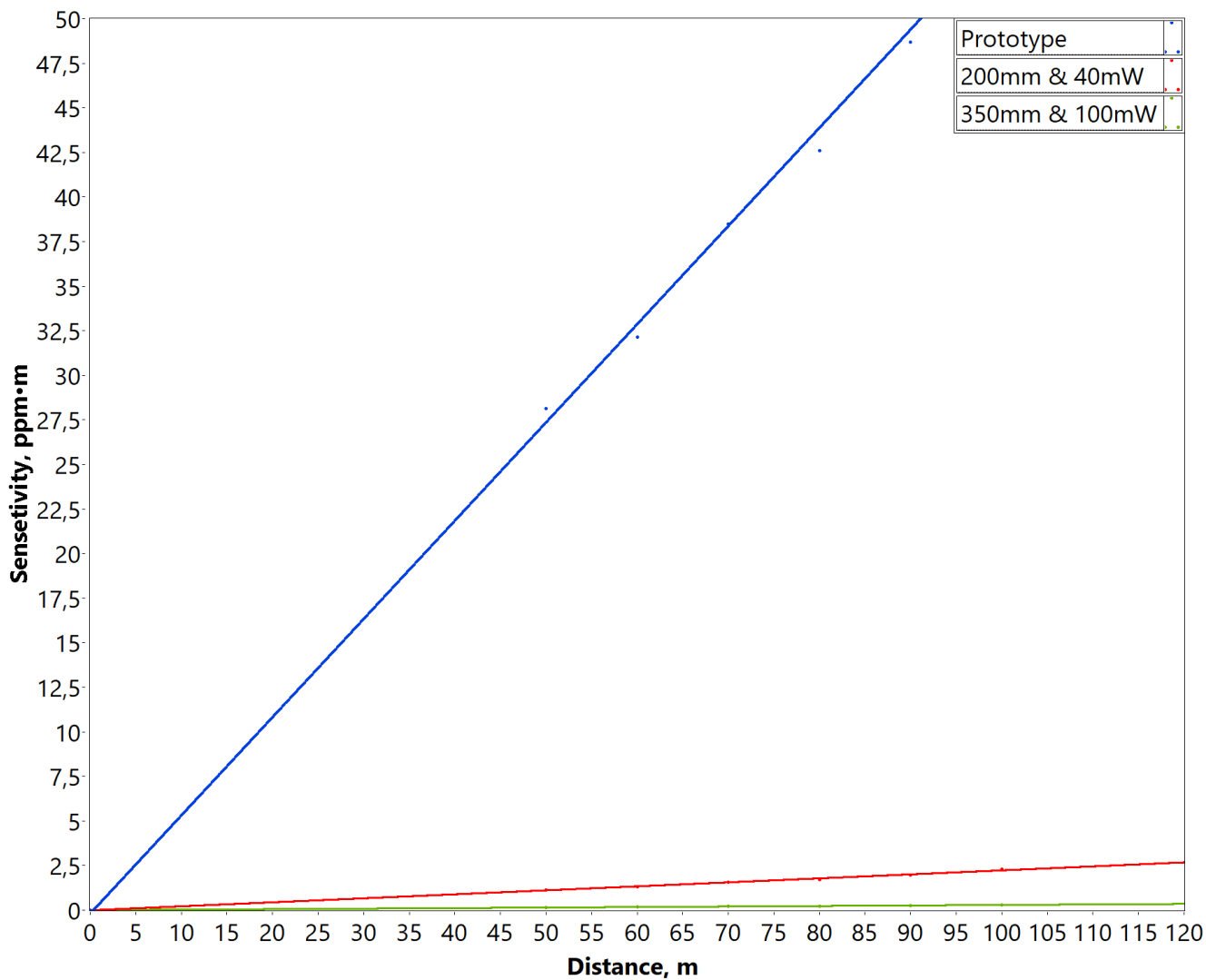

Рисунок 4.49 – Зависимость чувствительности прототипа газоанализатора ГИМЛИ от дистанции до рассеивающей лазерное излучение поверхности (синяя линия) и оценки таких зависимостей для новых версий прибора в случае худшей (красная линия) и лучшей возможной чувствительности (зеленая линия).

При этом, несмотря на значительное увеличение чувствительности, масса новых версий прибора мало отличается от массы прототипа прибора, поскольку в данном случае при разработке корпусов не использовались готовые решения, все корпусные детали были спроектированы под задачу минимизации массы при условии удовлетворения предельным значениям габаритов для возможности установки на БПЛА, а также увеличении обтекаемости корпуса.

С этой целью часть корпусных элементов, не являющихся каркасными или экранирующими, были изготовлены при помощи аддитивных технологий из пластика ПЭТГ и ПЛА. Внешний вид собранного корпуса прибора с 350 мм линзой Френеля представлен на рисунке 4.50.

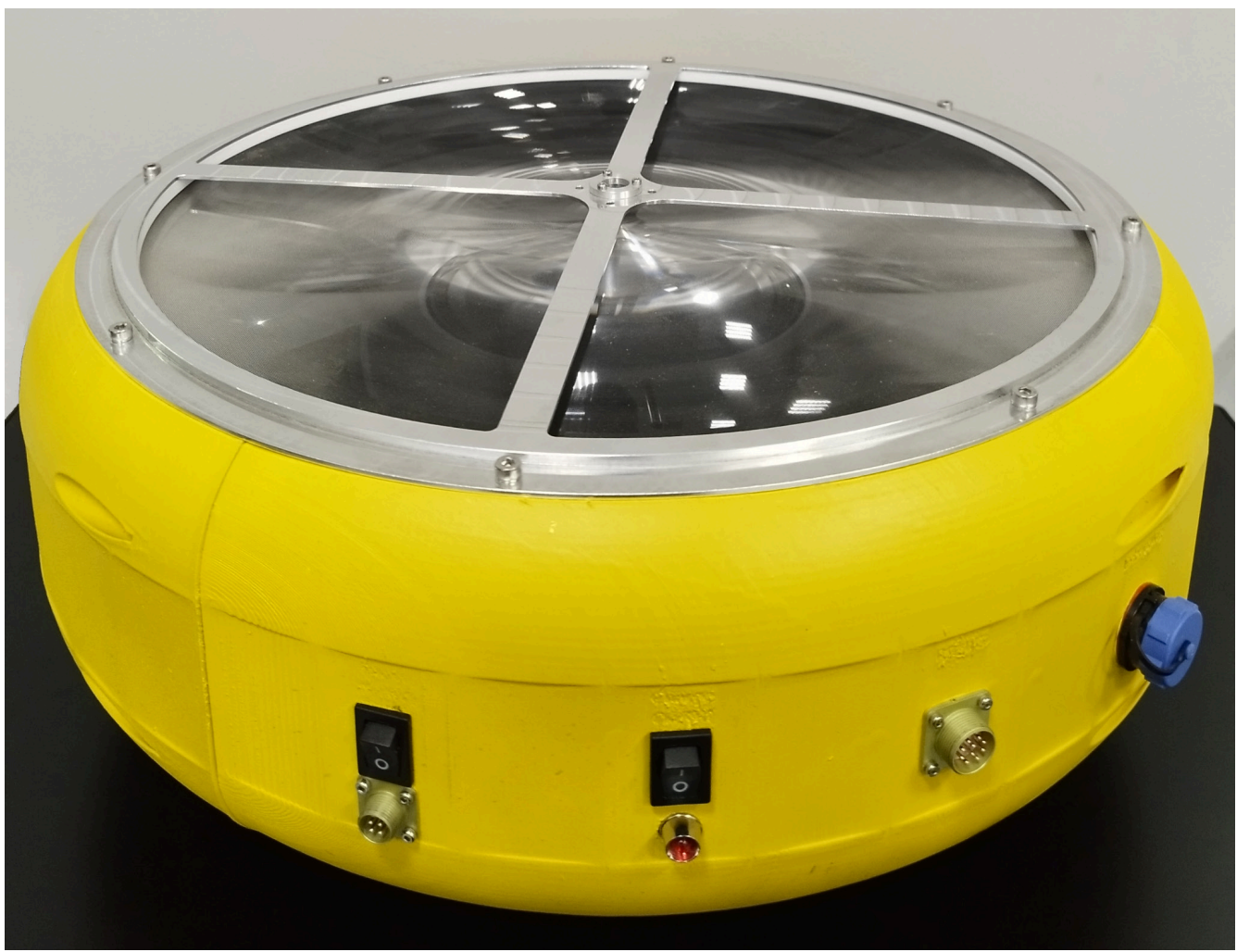

Рисунок 4.50 – Внешний вид версии прибора с акриловой линзой Френеля диаметром 350 мм.

Общие характеристики разработанных новых версий газоанализатора ГИМЛИ приведены в таблице 4.3.

Таблица 4.3 – Спецификация новых версий газоанализатора ГИМЛИ.

| Масса | ~3-4 кг |
|---|---|
| Габариты (ШВГ) | ⌀270/420 мм, высота 220/190 мм |
| Апертура принимающей оптики | 200/350 мм |
| Длина волны излучения | 1651 нм |
| Мощность лазерного излучения | 40/100 мВт |
| Частота сэмплирования | ~20 Гц |
| Потребление (стандартное/пиковое) | 12/35 Вт |
| Источник питания | встроенный аккумулятор/бортовой источник |
| Время работы от аккумулятора | ~3 ч |
| Рабочий диапазон температур | от -20℃ до +40℃ |

На момент написания заканчивается сборка новых версий прибора, проведена предварительная отладка алгоритма работы. Начало проведения полевых испытаний новых версий газоанализатора ГИМЛИ планируются на первую половину 2025 года.

## Выводы к главе 4

По результатам, изложенным в главе 4, можно сформулировать следующие основные выводы:

1. Показана возможность реализации устройства, основанного на методе абсорбционной диодно-лазерной спектроскопии с токовой гармонической модуляцией лазерного излучения при динамической стабилизации его центральной частоты по положению выбранной спектральной линии поглощения с квадратурным детектированием рассеянного от удаленной поверхности сигнала, для дистанционного мониторинга выбранного газа атмосферного воздуха в полевых условиях;

2. Продемонстрированы результаты работы и калибровки прототипа газоанализатора метана, основанного на применении методики модуляционной лазерной абсорбционной спектроскопии в комбинации с квадратурным приемом сигнала, описаны проблемы устройства, решенные на данном этапе алгоритмически, продемонстрированы методы решения, планируемые к реализации в новых версиях прибора;

3. Рассмотрено поведение различных характеристик сигнала в рамках предложенной методики при изменении дистанции до рассеивающей лазерное излучение поверхности, продемонстрирована линейная зависимость от дистанции результатов измерения интегральной концентрации метана и чувствительности разработанного прибора, связанная с равномерным распределением метана в атмосферном воздухе;

4. Габариты, масса и энергопотребление разработанного прибора, основанного на предложенных принципах, соответствуют характеристикам полезной нагрузки легких БПЛА, применение которых упростит и удешевит поиск утечек из магистральных газопроводов, мониторинг воздушной среды вблизи опасных производств, мусорных полигонов и иных антропогенных источников выбросов метана, а также в районах естественной эмиссии метана – в арктических регионах, районах с болотистой местностью и т.д;

5. Показано, что разработанный прототип газоанализатора метана не уступает по чувствительности существующим компактным лазерным газоанализаторам, применяемым для мониторинга метана в атмосферном воздухе с борта БПЛА, а на более безопасных для пилотирования высотах выше 50 м превосходит аналоги по этому показателю;

6. Продемонстрировано, что предложенная методика, лежащая в основе разработанного прототипа прибора, позволит осуществлять измерения интегральных значений концентрации выбранного газа с много большей чувствительностью, чем у существующих лазерных спектрометров лидарного типа, за счет алгоритма высокоточной стабилизации длины волны генерируемого лазерным источником излучения по пику линии поглощения, детектируемому в реперном канале прибора, а также применения более мощных лазерных источников и принимающей рассеянное излучение оптики бо́льших диаметров, что заложено в описанные новые версии газоанализатора;

7. Показано, что реализация данной методики позволяет независимо определять дистанцию до рассеивающей лазерное излучение поверхности, что необходимо для корректного вычисления усредненной концентрации анализируемой газовой составляющей атмосферного воздуха в единицах ppm;

8. Применение газоанализатора, основанного на предложенной методике, для дистанционного мониторинга метана в районах естественных и антропогенных источников его эмиссии в атмосферный воздух позволит улучшить методы измерения выбросов метана, что необходимо для решения важной проблемы регулирования выбросов парниковых газов в атмосферу – недостаточно высокого качества методов их учета.

# ЗАКЛЮЧЕНИЕ

В представленной работе описаны результаты исследования химического и изотопного состава газовых смесей на основе предложенных методов лазерной спектроскопии поглощения, полученные автором в рамках создания алгоритмов обработки спектральных данных Многоканального диодно-лазерного спектрометра ДЛС-Л в составе научной аппаратуры по изучению лунного реголита Газового аналитического комплекса станции «Луна-27», а также при создании газоанализатора для дистанционного мониторинга метана в атмосферном воздухе ГИМЛИ. Основные полученные результаты заключаются в следующем:

1. Показано, что использование метода диодно-лазерной абсорбционной спектроскопии дает возможность создавать газоаналитические сенсоры для изотопного анализа, удовлетворяющие требованиям космического приборостроения, на примере разработанного впервые газоаналитического сенсора для *in situ* исследования изотопных отношений содержащихся в лунном грунте химически связанных $H_2O$ и $CO_2$;

2. Продемонстрировано, что спектрометр ДЛС-Л, отвечающий требованиям космического приборостроения – обладающий малой массой, малыми габаритами и низким энергопотреблением, соответствующий критериям электромагнитной совместимости с оборудованием всего научного комплекса и радиационной стойкости, – работает с бо́льшим числом изотопологов молекул воды и углекислого газа, чем доступно для масс-спектрометров лабораторного класса;

3. Разработан программный комплекс для анализа спектральных данных, позволяющий успешно обрабатывать и анализировать экспериментальные данные прибора ДЛС-Л с учетом специфических особенностей эксперимента – дрейфа сигнала, оптической интерференции, шумов электроники и проч., проводить изотопный анализ для представляющих наибольший фундаментальный интерес в вопросах о процессах формирования Луны и источниках летучих веществ на ней изотопологов молекул воды и углекислого газа с точностью от 1‰ до 1%;

4. Продемонстрированы результаты работы лабораторного макета устройства, основанного на применении методики модуляционной лазерной абсорбционной спектроскопии с токовой гармонической модуляцией лазерного излучения при стабилизированной частоте генерации лазерного кристалла по положению выбранной спектральной линии поглощения в комбинации с квадратурным приемом сигнала, показывающие, что данный метод позволяет решить задачу базовой линии с точностью, достаточной для дистанционного мониторинга выбранного газа в атмосферном воздухе;

5. Показана возможность реализации устройства, основанного на методе абсорбционной диодно-лазерной спектроскопии с гармонической модуляцией лазерного излучения током накачки при динамической стабилизации его центральной частоты по положению выбранной спектральной линии поглощения с квадратурным детектированием рассеянного от удаленной поверхности сигнала, для дистанционного мониторинга заданных газовых составляющих атмосферного воздуха в полевых условиях с возможностью независимого определения дистанции до удаленной поверхности;

6. Продемонстрированы результаты работы и калибровки прототипа газоанализатора метана, основанного на применении методики модуляционной лазерной абсорбционной спектроскопии в комбинации с квадратурным приемом сигнала, описаны проблемы устройства, продемонстрированы методы их решения, рассмотрено поведение различных характеристик сигнала в рамках предложенной методики при изменении дистанции до рассеивающей лазерное излучение поверхности, продемонстрирована линейная зависимость от дистанции результатов измерения интегральной концентрации метана и чувствительности разработанного прибора, что соответствует гипотезе о равномерном распределении метана в приземном слое атмосферы;

7. Показано, что массогабаритные характеристики и энергопотребление разработанного прибора, основанного на предложенных принципах, соответствуют характеристикам полезной нагрузки легких БПЛА, что по чувствительности представленный прототип не уступает существующим компактным лазерным газоанализаторам, применяемым для мониторинга метана в атмосферном воздухе с борта БПЛА, а на более безопасных для пилотирования высотах выше 50 м превосходит аналоги по этому показателю, что при применении более мощных лазерных источников и принимающей рассеянное излучение оптики бóльших диаметров, что заложено в описанные новые версии газоанализатора, станет возможным осуществлять измерения интегральных значений концентрации выбранного газа с чувствительностью, превосходящей характеристики существующих лазерных спектрометров лидарного типа.

# СПИСОК СОКРАЩЕНИЙ И УСЛОВНЫХ ОБОЗНАЧЕНИЙ

**А**

АН – адсорбционный накопитель

АЦП – аналого-цифровой преобразователь

АЧХ – амплитудно-частотная характеристика

**Б**

БУНИ – блок управления научной информацией

БПЛА – беспилотный летательный аппарат

**В**

ВИП – вторичный источник питания

ВР – волоконный разветвитель

**Г**

ГАК – газовый аналитический комплекс

ГЗУ – грунтозаборное устройство

ГИМЛИ – газоанализатор для измерения метана лидарный инфракрасный

ГХ-Л – газовый хроматограф

**Д**

ДЛ – диодный лазер

ДЛС – диодно-лазерная спектроскопия

ДЛС-Л – диодно-лазерный спектрометр лунный

**З**

ЗИР – затухание излучения в резонаторе

ЗС – зависимый от скорости

**И**

ИК – инфракрасный

ИКИ РАН – Институт космических исследований российской академии наук

**К**

КА – космический аппарат

КБ – клапанный блок

КДО – конструкторско-доводочный образец

КК – капиллярные колонки

КО – коллимирующая оптика

**Л**

ЛМК – лунный манипуляторный комплекс

**М**

МГК – многоходовая газовая кювета

М-ДЛС – марсианский многоканальный диодно-лазерный спектрометр

МДУ – многоразовое десорбционное устройство

МК – микроконтроллер

ММС – многоходовая матричная система

МО – мерный объём

МФТИ – Московский физико-технический институт

**О**

ОА – оптико-акустический

ОВ – оптическое волокно

ОУ – операционный усилитель

**П**

ПГ – парниковый газ

ПК – персональный компьютер

ПЛИС – программируемая логическая интегральная схема

ПО – программное обеспечение

ПШПВ – полуширина на полувысоте

ПЯ – пиролитическая ячейка

**Р**

РОС – распределенная обратная связь

РК – реперная кювета

РУП – резонаторное увеличение поглощения

РЭ – рассеивающий экран

**С**

СИС – столкновения с изменением скорости

СТК – система торговли квотами

**Т**

ТА-Л – термический анализатор

ТИУ – трансимпедансный усилитель

**Ф**

ФД – фотодиод

ФП – фотоприемник

ФПУ – фотоприемное устройство

ФЧХ – фазо-частотная характеристика

**Ц**

ЦАП – цифро-аналоговый преобразователь

**Ч**

ЧФ – частотный фильтр

**Ш**

ШО – штатный образец

**С**

CEAS – cavity-enhanced absorption spectroscopy

CRDS – cavity ring-down spectroscopy

CW-CRDS – continuous-wave cavity ring-down spectroscopy

**E**

ESA – European Space Agency

**F**

FMS – frequency modulation spectroscopy

**I**

ICOS – integrated cavity output spectroscopy

**L**

LCROSS – Lunar CRater Observation and Sensing Satellite

LRO – Lunar Reconnaissance Orbiter

LEND – Lunar Exploration Neutron Detector

LOD – limit of detection

**N**

NASA – National Aeronautics and Space Administration

NEP – noise equivalent power

**P**

PS-CRDS – phase-shift cavity ring-down spectroscopy

ppm – particles per million

ppb – particles per billion

**T**

TCCON – Total Carbon Column Observing Network

**V**

VPDB – Vienna Pee Dee Belemnite

VSMOW – Vienna Standard Mean Ocean Water

**W**

WMS – wavelength modulation spectroscopy

# СПИСОК ПУБЛИКАЦИЙ ПО ТЕМЕ ДИССЕРТАЦИИ

# СПИСОК ЛИТЕРАТУРЫ

# Приложение А. Описание патента на изобретение способа и устройства для автономного дистанционного определения концентрации атмосферных газовых составляющих

РОССИЙСКАЯ ФЕДЕРАЦИЯ

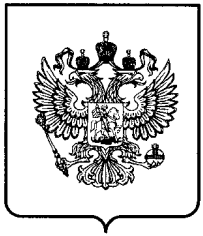

ФЕДЕРАЛЬНАЯ СЛУЖБА
ПО ИНТЕЛЛЕКТУАЛЬНОЙ СОБСТВЕННОСТИ

(19) **RU** (11) **2 736 178** (13) **C1**

(51) МПК
*G01N 21/61* (2006.01)

(52) СПК
*G01N 21/61* (2020.08)

(12) **ОПИСАНИЕ ИЗОБРЕТЕНИЯ К ПАТЕНТУ**

Статус: действует (последнее изменение статуса: 02.11.2022)
Пошлина: учтена за 4 год с 10.06.2023 по 09.06.2024. Установленный срок для уплаты пошлины за 5 год: с 10.06.2023 по 09.06.2024. При уплате пошлины за 5 год в дополнительный 6-месячный срок с 10.06.2024 по 09.12.2024 размер пошлины увеличивается на 50%.

(72) Автор(ы):
**Спиридонов Максим Владимирович (RU),
Мещеринов Вячеслав Вячеславович (RU),
Казаков Виктор Алексеевич (RU),
Газизов Искандер Шамилевич (RU)**

(73) Патентообладатель(и):
**федеральное государственное автономное образовательное учреждение высшего образования "Московский физико-технический институт (национальный исследовательский университет)" (RU)**

(54) **Способ и устройство для автономного дистанционного определения концентрации атмосферных газовых составляющих**

(57) Реферат:

Изобретение относится к области газоанализа и касается устройства для автономного дистанционного определения концентрации атмосферных газовых составляющих. Устройство включает в себя оптический блок, содержащий диодный лазер, обеспечивающий зондирующий луч и модулирующий амплитуду сигнала, приемную систему аналитического канала, реперную кювету и блок управления, приема и обработки данных. Блок управления, приема и обработки данных детектирует зондирующее излучение, прошедшее сквозь исследуемый газ, извлекает сигнал на первой высшей гармонической составляющей частоты модуляции f и на вторичной составляющей 2f и вычисляет степень поглощения зондирующего луча измеряемым газом по отношению первой гармонической составляющей ко второй. Стабилизация диодного лазера осуществляется по положению пика линии поглощения измеряемого газа по составляющей на частоте 3f с точностью до $10^{-4}$ см$^{-1}$. По сдвигу фаз вычисляется расстояние до рассеивающей излучение лазера поверхности. Технический результат заключается в увеличении чувствительности, уменьшении массогабаритных характеристик и обеспечении возможности установки газоанализатора на беспилотные летательные аппараты. 2 н. и 3 з.п. ф-лы, 6 ил.